\documentclass[aps,prb,twocolumn,amsmath,amssymb,superscriptaddress,floatfix,eqsecnum]{revtex4-1}
\usepackage{amsmath}% if you are using this package,
\usepackage{amsthm}
\usepackage{mathtools}
\usepackage{graphicx}
\usepackage{amssymb}
\usepackage{bm} % bold math
\usepackage{slashed} 
\usepackage{color} 
\usepackage{bbm}
\usepackage{amsfonts}
\usepackage[breaklinks=true,colorlinks,
citecolor=amar,linkcolor=frblue,urlcolor=brred]{hyperref}
\usepackage{cases}
\usepackage{float}
\usepackage{url}
\usepackage{framed}
\definecolor{blush}{rgb}{0.87, 0.36, 0.51}
\usepackage{cancel,soul}
\usepackage[normalem]{ulem} 
\usepackage{appendix}
\definecolor{bleu}{rgb}{0.0, 0.5, 0.69}

\definecolor{amar}{rgb}{0.9, 0.17, 0.31}

\definecolor{brred}{rgb}{0.8, 0.25, 0.33}
\definecolor{frblue}{rgb}{0.0, 0.45, 0.73}

\def\=={\raisebox{0.35pt}{$\mathrm{:}$}\!\!=}
\def\df{\,\raisebox{1.20pt}{.}\hspace{-3.1pt}\raisebox{3.35pt}{.}\!\!=\,}
\def\dfr{\,=\!\!\raisebox{0.95pt}{.}\hspace{-2.78pt}\raisebox{2.95pt}{.}\,}

\newcommand{\bea}{\begin{eqnarray}}
\newcommand{\eea}{\end{eqnarray}}

\newcommand{\ket}[1]{\lvert #1 \rangle}
\newcommand{\bra}[1]{\langle #1 \rvert}
\newcommand{\pket}[1]{\lvert #1 )}
\newcommand{\pbra}[1]{( #1 \rvert}
\newcommand{\braket}[2]{\langle #1 \lvert #2 \rangle}

\theoremstyle{definition}

\newtheorem{defn}{Definition}

\newtheorem{thm}{Theorem}

\theoremstyle{remark}

\begin{document}

\title{Lieb-Schultz-Mattis type theorems for Majorana models with discrete symmetries}
\author{\"{O}mer M. Aksoy}
\affiliation{Condensed Matter Theory Group, Paul Scherrer Institute, CH-5232 Villigen PSI, Switzerland}

\author{Apoorv Tiwari}
\affiliation{Condensed Matter Theory Group, Paul Scherrer Institute, CH-5232 Villigen PSI, Switzerland}
\affiliation{Department of Physics, University of Z\"urich, Winterthurerstrasse 190, 8057 Z\"urich, Switzerland}

\author{Christopher Mudry}
\affiliation{Condensed Matter Theory Group, Paul Scherrer Institute, CH-5232 Villigen PSI, Switzerland}
\affiliation{Institut de Physique, EPF Lausanne, Lausanne, CH-1015, Switzerland}

\begin{abstract}
We prove two Lieb-Schultz-Mattis type theorems that apply
to any  translationally invariant and local
fermionic $d$-dimensional lattice Hamiltonian
for which fermion-number conservation is broken down to
the conservation of fermion parity.
We show that when the internal symmetry group $G^{\,}_{f}$
is realized locally (in a repeat unit cell of the lattice)
by a nontrivial projective representation, then 
the ground state cannot be simultaneously nondegenerate, 
symmetric (with respect to lattice translations and $G^{\,}_{f}$), and gapped. 
We also show that when the repeat unit cell hosts
an odd number of Majorana degrees of freedom 
and the cardinality of the lattice is even,
then the ground state cannot be simultaneously nondegenerate,
gapped, and translation symmetric.
\end{abstract}

\date{\today}

\maketitle

\tableofcontents

\section{Introduction}

In 1961, Lieb, Schultz, and Mattis (LSM) proved
a theorem on the low-lying excited states of 
the nearest-neighbor antiferromagnetic quantum spin-1/2 chain\,
\cite{Lieb1961}. 
Accordingly, a quantum spin chain
with an odd number
of spin-1/2 degrees of freedom per repeat unit cell
that is simultaneously
translation and SO(3)-spin-rotation symmetric cannot
realize a gapped and symmetric ground state in the thermodynamic limit.

Since its original formulation,
the LSM theorem has been generalized in a number of
ways including extensions to higher dimensions,
other global continuous symmetry groups,
and different kinds of spatial symmetries~
\cite{
Affleck1986,Aizenman1994,Oshikawa1997,Yamanaka1997, 
Koma2000,Oshikawa2000,Hastings2004,Hastings2005,Roy2012,Parameswaran2013,Watanabe2015,
Watanabe2016,Qi2017,
Watanabe2018,Tasaki2018,Metlitski2018,Kobayashi2019,
He2020,Bachmann2020,Dubinkin2021}.
For instance, it has been understood that the SO(3)-spin-rotation symmetry
is not an essential requirement for an LSM constraint. In fact,
LSM{-type} theorems for U(1)-number-conserving Hamiltonians
{have been established} in arbitrary dimensions.
These theorems state that systems with noninteger filling
fraction $\nu$, defined as the average number of particles per unit
cell, cannot have a translationally invariant, nondegenerate,
and short-range entangled ground state\,
\cite{Affleck1986,Aizenman1994,Oshikawa1997,Yamanaka1997,Koma2000,Hastings2004,%
Hastings2005,Tasaki2018}.
Similar constraints have also been worked out for
number-conserving Hamiltonians that have non-symmorphic or magnetic space
group symmetries\,
\cite{Parameswaran2013,Roy2012,Watanabe2015,Watanabe2018}.

A number of LSM-type theorems pertaining to discrete
internal symmetries combined with crystallographic
symmetries have also been worked out~\cite{
Chen2011b,Parameswaran2013,Watanabe2015,Cheng2016,Po2017,Watanabe2018,Yang2018,
Ogata2019,Ogata2021}.
In the context of spin-chains with discrete symmetries, 
LSM-type theorems were proved by Ogata \emph{et al}.\ in Refs.%
~\onlinecite{Ogata2019,Ogata2020a,Ogata2021}.
They found that a translationally invariant spin chain
with half-integer spin at each site that possesses {either}
time-reversal
or $\mathbb{Z}^{\,}_{2}\times\mathbb{Z}^{\,}_{2}$-rotation symmetries
(rotations by $\pi$ around two axis, say $x$ and $z$), cannot have a
nondegenerate, gapped, and symmetric ground state. 
The proof of this
statement is based on the fact that a nondegenerate gapped ground state
of a local Hamiltonian satisfies the so called \textit{split property}\,\cite{Matsui2001,Matsui2013}.
LSM-type no-go theorems are then derived for states satisfying the split property by using operator algebra techniques and Gelfand–Naimark–Segal (GNS) construction.
Alternatively, similar no-go constraints are obtained 
in Refs.\ \onlinecite{Prakash2020,Tasaki2020} within the framework of matrix product states (MPS)\,\cite{Fannes1992,Klumper1992,Garcia2007}.
These derivations are based on the fact that one can approximate\,\cite{Schuch2008}
the nondegenerate gapped ground 
states of local Hamiltonians by injective MPS\,\cite{Verstraete2006,Hastings2007}.
There are two hypothesis common to many of these LSM-type theorems
with discrete symmetries. It is presumed that there exist
local (on-site) degrees of freedom that span a local Hilbert space
and realize
a nontrivial projective representation of the global symmetry.
{It is also presumed that}
the global Hilbert space is obtained by postulating that the
local degrees of freedom commute when separated in space.
The resulting LSM-type theorems are applicable to bosonic systems.
Generalizations to bosonic quantum systems in arbitrary dimension
with crystallographic symmetries and general discrete Abelian symmetries
have been proposed
using the notion of lattice homotopy\,
\cite{Po2017}.

There exists similarities between Hamiltonians obeying
LSM{-type} constraints
and the boundary modes of a symmetry-protected-topological (SPT)
insulator.  An SPT insulator has a nondegenerate, symmetric, and gapped
ground state when periodic boundary conditions are imposed.
When open boundary conditions are imposed, the effective low-energy
quantum Hamiltonian governing the dynamics of the boundary modes
of an SPT insulator supports a ground state that is either
(i) gapless,
(ii) symmetry-broken,
(iii) or topologically ordered
if the boundary is no less than two dimensional \cite{Chen2011a, Chen2011b, pollmann2012, Chen2013, Vishwanath2013, Gu2014}.
From a low-energy perspective,
both the effective boundary Hamiltonian of an SPT insulator
and the {bulk} Hamiltonian satisfying an LSM{-type}
constraint display quantum anomalies\cite{Cho2017, Metlitski2018, Else2020}.
For the former case, the quantum anomaly is typically
that for a global symmetry that acts locally on the boundary \cite{Ryu2012, kapustin2014symmetry, Witten2016, Hsieh2016, Han2017, Hsieh2016}.
In the latter case,
there is a mixed quantum anomaly between a global symmetry that
acts locally and a spatial symmetry such as translation.
These parallels led to the formulation of so-called
weak-SPT-LSM{-type}
theorems\,
\cite{Cho2017,Cheng2016,Kobayashi2019,Cheng2019, Jian2018}.
In particular, Ref.\ \onlinecite{Cheng2019}
has conjectured an LSM-type constraint for $d$-dimensional fermionic
lattice Hamiltonians with the help of a bulk-boundary correspondence.
The fermionic $d$-dimensional lattice Hamiltonian is interpreted
as the low-energy effective theory of a fermionic
$d+1$-dimensional lattice Hamiltonian that is gapped but supports midgap
boundary states such that
(i) they can be localized at each site of the $d$-dimensional lattice
(ii) where they span a local fermionic Fock space.
The parent fermionic $(d+1)$-dimensional lattice Hamiltonian
is an example of a weak-SPT fermionic insulator. LSM-like constraints
on the ground states of the $d$-dimensional lattice Hamiltonian
are inherited from the symmetries that protect the boundary states
of the $(d+1)$-dimensional lattice Hamiltonian.

As compared to LSM{-type} theorems for bosonic and U(1)-charge-conserving
fermionic Hamiltonians, LSM{-type} theorems for fermionic Hamiltonians
{without any U(1)-conserving symmetries} are much less explored~
\cite{Hsieh2016,Cheng2019}. {These LSM-type constraints
would be relevant for any long-range superconducting order
with fully broken SU(2)-spin-rotation that coexists with
some additional long-range order.}
Such fermionic Hamiltonians always admit a formulation in terms of
Majorana degrees of freedom.
To the best of our knowledge, there are no
proofs of LSM-type constraints relevant to
fermionic lattice Hamiltonians with translation symmetry and
some discrete internal symmetry (such as time-reversal symmetry, say)
for which fermion-number conservation is broken
down to the conservation of fermion parity.

In the present work, we state and prove two LSM-type theorems.
They apply to translationally invariant lattice Hamiltonians
acting on a fermionic Fock space spanned by Majorana degrees of freedom.
The lattice is embedded in $d$-dimensional Euclidean space.
For Theorem \ref{thm:LSM Theorem 1}, there also exists
a global symmetry associated to a symmetry group $G^{\,}_{f}$ 
that can be realized locally,
i.e., the number of Majorana degrees of freedom in a repeat unit cell
of the lattice is even. We prove within the framework of fermionic MPS (FMPS)
that whenever the Majorana degrees
of freedom within a single repeat unit cell realize
a nontrivial projective representation of $G^{\,}_{f}$,
then the lattice Hamiltonian cannot have a nondegenerate, gapped, and
symmetric ground state that can be described by an even- 
or odd-parity injective FMPS (for $d>1$ we must assume that
$G^{\,}_{f}$ is Abelian and all its elements
are represented by unitary operators),
{in agreement with the conjecture made by Cheng in Ref.\
\onlinecite{Cheng2019} when the fermion number is conserved}.
For Theorem \ref{thm:LSM Theorem 2},
it is only assumed that the repeat unit cell
supports an odd number of Majorana modes, the cardinality of the lattice is
even, and translation symmetry holds.
It then follows that the ground state cannot be simultaneously
nondegenerate, gapped, and translation symmetric. 

The rest of the paper is organized as follows. 
We introduce the main results of the present work in the form of 
Theorems \ref{thm:LSM Theorem 1} and \ref{thm:LSM Theorem 2}
in Sec.\ \ref{sec:thm}.
We present an overview of the internal symmetry group $G^{\,}_{f}$
and its projective representations
in Sec.\ \ref{sec:Symmetries}.
We introduce the framework for FMPS and present {a} FMPS-based proof 
of Theorem \ref{thm:LSM Theorem 1} in Sec.\ \ref{sec:1D_LSM}
when the dimension $d$ of space is one.
Theorem \ref{thm:LSM Theorem 2} when $d=1$ 
is then proved by making use of
Theorem \ref{thm:LSM Theorem 1}.
An independent proof of Theorem \ref{thm:LSM Theorem 2}
is given for any dimensions $d$ of space
in Sec.\ \ref{sec:higher_LSM}. A weaker version
of Theorem \ref{thm:LSM Theorem 1}
for any $d\geq1$ is also {provided in} Sec.\ \ref{sec:higher_LSM}. 
The latter proof is based on symmetry twisted
boundary conditions on twisted lattices. Finally, we collect several
examples in Sec.\ \ref{sec:examples} and conclude the main body of the
paper with a summary in Sec.\ \ref{sec:summary}. We present
details about group {cohomology}, FMPS construction and proof of
Theorem \ref{thm:LSM Theorem 1} in Appendices \ref{appsec:Group Cohomology},
\ref{appsec:Construction of fermionic matrix product states (FMPS)}
and \ref{appsec:Proof of LSM}, respectively.

\section{Main Results}
\label{sec:thm}

The notion of a global fermionic symmetry group $G^{\,}_{f}$ with a
local action plays a central role in this paper. What we have in mind
with this terminology is any lattice model obtained by discrete
translations of a repeat unit cell. The same even integer number
$n=2m$ of Majorana degrees of freedom (flavors), i.e., a local
fermionic Fock space of dimension $2^{m}$, is attached to any
repeat unit cell. The fermion parity can then be defined for any
repeat unit cell. Even though the fermion parity is generically not
conserved locally, it must be conserved globally. Hence, the symmetry
group $G^{\,}_{f}$ necessarily contains as a subgroup the cyclic group
of two elements generated by the fermion parity. The action of any
other element from the symmetry group $G^{\,}_{f}$ can be defined
locally, i.e., its action is represented by the same polynomial
on the algebra generated by the Majorana operators
within the repeat unit cell for any repeat unit cell of the lattice.
As with the fermion parity, this symmetry element need not be conserved
locally but must be conserved globally. A translation along the basis
that generates the lattice relates different repeat unit cells. Similarly, any
crystalline symmetry lies outside of the symmetry group $G^{\,}_{f}$.

The motivation for Theorem \ref{thm:LSM Theorem 1} is the following.
Given is a local lattice Hamiltonian $\widehat{H}$ that is 
symmetric under the group $G^{\,}_{\mathrm{trsl}}\times G^{\,}_{f}$,
where $G^{\,}_{\mathrm{trsl}}$ denotes the group of lattice translations.
Assume that the ground state $\ket{\Psi}$ is
symmetric under both $G^{\,}_{\mathrm{trsl}}$ and $G^{\,}_{f}$,
i.e., the ground state can change by no more than
a multiplicative phase factor (combined with
complex conjugation when the symmetry is represented by
an antiunitary operator).
Assume that a gap separates the ground state from all excited states.
Are there sufficient and necessary conditions for $\ket{\Psi}$
to be degenerate?

A sufficient condition
for all energy eigenvalues to be degenerate is that at least
one generator from $G^{\,}_{\mathrm{trsl}}$ and one element $g$ from $G^{\,}_{f}$
are represented globally by operators that commute projectively,
i.e., when passing the translation operator from the left to
the right of the operator representing globally the element $g$,
a multiplicative phase factor that cannot be gauged to unity arises.
Another sufficient condition for the case when $G^{\,}_{f}$
has an antiunitarily represented 
$\mathbb{Z}^{\,}_{2}$ subgroup (such as time-reversal symmetry)
is that the symmetry generator $g$
is represented globally by an operator that squares to minus the identity
(Kramers' theorem).
However, none of these conditions are necessary for the ground state
to be degenerate in the thermodynamic limit.
Even if all operators representing $G^{\,}_{\mathrm{trsl}}$
and all global operators representing $G^{\,}_{f}$ commute pairwise
and if all global operators representing $G^{\,}_{f}$ square to unity,
a gapped and symmetric ground state might still be degenerate
in the thermodynamic limit.

To develop an intuition for this last claim,
we consider a one-dimensional lattice with the topology of a ring
and assume that $G^{\,}_{f}$ is an Abelian group such that
any element $g\in G^{\,}_{f}$ is
represented globally by the unitary operator
$\widehat{U}^{\,}_{g}$.
Let $\widehat{H}^{\,}_{\mathrm{pbc}}$
be any local Hamiltonian that is invariant under the actions of 
the unitary operators $\widehat{T}^{\,}_{1}$ representing
a translation by one repeat unit cell and $\widehat{U}^{\,}_{g}$.
Assume that the spectrum of $\widehat{H}^{\,}_{\mathrm{pbc}}$
shows a gap between its ground and excited states.
We define the operator $\widehat{T}^{\,}_{1,h}$ 
which translates all lattice repeat unit cells by one to the right and
acts locally with the local unitary operator representing $h\in G^{\,}_{f}$
on the last repeat unit cell of the chain.
Suppose that there exists a Hamiltonian $\widehat{H}^{\,}_{\mathrm{twist}}$
that is constructed by deforming sub-extensively
many local terms from
$\widehat{H}^{\,}_{\mathrm{pbc}}$ 
and such that $\widehat{H}^{\,}_{\mathrm{twist}}$
is invariant under the actions of 
both $\widehat{T}^{\,}_{1,h}$ and $\widehat{U}^{\,}_{g}$.
As long as the spectral gaps at the end points $\lambda=0$ and $\lambda=1$ of
\begin{align}
\widehat{H}(\lambda) 
\df 
\lambda\,\widehat{H}^{\,}_{\mathrm{pbc}} 
+ 
(1-\lambda)\,\widehat{H}^{\,}_{\mathrm{twist}}
\label{eq:interpolation pbc twist}
\end{align}
does not close for $0\leq\lambda\leq1$, $\widehat{H}^{\,}_{\mathrm{pbc}}$ 
and $\widehat{H}^{\,}_{\mathrm{twist}}$
must share the same ground-state degeneracy.
Note that generically Hamiltonian
$\widehat{H}(\lambda)$ is not invariant under the actions of
$\widehat{T}^{\,}_{1}$ or $\widehat{T}^{\,}_{1,h}$ when $0<\lambda<1$.
This scenario is plausible, for $\widehat{H}^{\,}_{\mathrm{pbc}}$ 
and $\widehat{H}^{\,}_{\mathrm{twist}}$
differ by sub-extensively
many terms in the thermodynamic limit.
Now, we denote with
$\ket{\exp(\mathrm{i}K^{\,}_{h}),\exp(\mathrm{i}U^{\,}_{g})}$ 
a many-body ground state of $\widehat{H}^{\,}_{\mathrm{twist}}$ that is a
simultaneous eigenstate of $\widehat{T}^{\,}_{1,h}$ and $\widehat{U}^{\,}_{g}$
with the eigenvalues $\exp(\mathrm{i}K^{\,}_{h})$ and
$\exp(\mathrm{i}U^{\,}_{g})$, respectively. The many-body state 
$\widehat{T}^{\,}_{1,h}\,
\ket{\exp(\mathrm{i}K^{\,}_{h}),\exp(\mathrm{i}U^{\,}_{g})}$ 
is then also a ground state of $\widehat{H}^{\,}_{\mathrm{twist}}$.
A sufficient condition for
$\widehat{T}^{\,}_{1,h}\,
\ket{\exp(\mathrm{i}K^{\,}_{h}),\exp(\mathrm{i}U^{\,}_{g})}$
to be orthogonal to
$\ket{\exp(\mathrm{i}K^{\,}_{h}),\exp(\mathrm{i}U^{\,}_{g})}$
is that the product 
$\widehat{U}^{\,}_{g}\,\widehat{T}^{\,}_{1,h}$
differs from the product
$\widehat{T}^{\,}_{1,h}\,\widehat{U}^{\,}_{g}$
by a multiplicative phase
$\exp(\mathrm{i}\chi^{\,}_{g,h})\neq1$
that cannot be gauged away.
If so the ground state of $\widehat{H}^{\,}_{\mathrm{twist}}$
is necessarily degenerate.
If the gap never closes as a function of $\lambda$ in the interval
$\lambda \in [0,1]$ in Eq.\ 
\eqref{eq:interpolation pbc twist}, then 
$\widehat{H}^{\,}_{\mathrm{pbc}}$ and
$\widehat{H}^{\,}_{\mathrm{twis}}$
share the same ground-state degeneracy even though 
$\widehat{T}^{\,}_{1}$ and $\widehat{U}^{\,}_{g}$
commute for all $g\in G^{\,}_{f}$.

We are ready to state
Theorems \ref{thm:LSM Theorem 1} and \ref{thm:LSM Theorem 2}
for which $G^{\,}_{\mathrm{trsl}}$
stands for the group of lattice translations while
$G^{\,}_{f}$ stands for the global fermionic symmetry group with a 
local action (i.e., it is an internal symmetry group).

\begin{thm}
\label{thm:LSM Theorem 1}
Any one-dimensional lattice Hamiltonian that is local and
admits the symmetry group $G^{\,}_{\mathrm{trsl}}\times G^{\,}_{f}$
cannot have a nondegenerate, gapped, and
$G^{\,}_{\mathrm{trsl}}\times G^{\,}_{f}$-symmetric ground state
that can be described by an even- or odd-parity injective FMPS
if $G^{\,}_{f}$ is realized by a nontrivial 
projective representation on the local Fock space. 
\end{thm}

\begin{thm}
\label{thm:LSM Theorem 2}
A local Majorana Hamiltonian
with an odd number of Majorana degrees of freedom per repeat unit cell 
that is invariant under the symmetry group
$G^{\,}_{\mathrm{trsl}}\times G^{\,}_{f}$,
cannot have a nondegenerate, gapped and translationally invariant ground state.
\end{thm}

\noindent
\textbf{Comment 1.} 
The thermodynamic limit is implicit in both theorems.

\noindent
\textbf{Comment 2.}
Theorem \ref{thm:LSM Theorem 1} is only predictive when
$G^{\,}_{f}$ is realized by a nontrivial projective representation
on the local Fock space. When $G^{\,}_{f}$ is a Lie group its
projective representation on the local Fock space can be trivial.
If so, Theorem \ref{thm:LSM Theorem 1} is not predictive.
However, one can use complementary arguments such as the adiabatic
threading of a gauge flux to decide if the ground state is degenerate.
It is when $G^{\,}_{f}$ is a finite group that the full power of
Theorem \ref{thm:LSM Theorem 1} is unleashed.

\noindent
\textbf{Comment 3.} Theorem \ref{thm:LSM Theorem 1} is proved 
within the FMPS framework in Sec.\ \ref{sec:1D_LSM}. 
A weaker form of Theorem \ref{thm:LSM Theorem 1}
holds in any dimension if it is assumed that $G^{\,}_{f}$
is Abelian and can be realized locally using unitary operators.
The weaker version of Theorem \ref{thm:LSM Theorem 1}
that is valid in any dimension is proved using tilted and twisted
boundary conditions in Sec.\ \ref{sec:higher_LSM}.

\noindent
\textbf{Comment 4.} Theorem \ref{thm:LSM Theorem 2}
applies in any dimension of space without any restriction
on the internal fermionic symmetry group $G^{\,}_{f}$.

\noindent
\textbf{Comment 5.} 
The direct product structure of the symmetry group 
$G^{\,}_{\mathrm{trsl}}\times G^{\,}_{f}$ is crucial in 
Theorems \ref{thm:LSM Theorem 1} and \ref{thm:LSM Theorem 2},
and their generalizations to higher dimensions.
Indeed, it has been shown that when the total symmetry group 
does not have a direct product structure, such as is the case with 
magnetic translation symmetries, a symmetric, non-degenerate, gapped,
and short-range entangled ground state is not ruled out
when closed-boundary conditions are imposed\,\cite{Yang2018,Lu2017,Jiang2021}. 
However, such a short-range entangled ground state must then necessarily 
support gapless symmetry-protected boundary states when 
open boundary conditions are imposed.

\section{Projective representations of symmetries obeyed by Majoranas}
\label{sec:Symmetries}

Theorems \ref{thm:LSM Theorem 1} and \ref{thm:LSM Theorem 2}
relate the quantum dynamics obeyed by Majoranas to
the symmetries they obey. The smallest symmetry group 
associated to Majoranas originates from
the conservation of the parity (evenness or oddness)
of the total number of fermions. This symmetry
is associated to a cyclic group of order
two that we shall denote with $\mathbb{Z}^{F}_{2}$.
Other symmetries are possible, say time-reversal symmetry
or spin rotation symmetry.
All such additional symmetries define
a second group $G$. The first question to be answered
is how many different ways are there to marry into a group $G^{\,}_{f}$
the intrinsic symmetry group $\mathbb{Z}^{F}_{2}$ of Majoranas
with the model-dependent symmetry group $G$.
This problem in known in group theory as the
central extension of $G$ by $\mathbb{Z}^{F}_{2}$.
It delivers a family of distinct equivalence classes
with each equivalence class $[\gamma]$ in one-to-one correspondence with the
cohomology group $H^{2}\left(G,\mathbb{Z}^{F}_{2}\right)$. This result is
motivated in Sec.\
\ref{subsec:Marrying the fermion parity with the symmetry group G}.

Once a representative symmetry group $G^{\,}_{f}$
has been selected from
$[\gamma]\in H^{2}\left(G,\mathbb{Z}^{F}_{2}\right)$,
its representation on the Fock space spanned by
all the local quantum degrees of freedom,
a set that includes Majorana operators, must be constructed. Hereto,
there are many possibilities. Their enumeration amounts to
classifying the inequivalent projective representations of the
group $G^{\,}_{f}$.
All the inequivalent projective representations
of any one of the groups $G^{\,}_{f}$ obtained in Sec.\
\ref{subsec:Marrying the fermion parity with the symmetry group G}
are in one-to-one correspondence with the cohomology group
$H^{2}\big(G^{\,}_{f},\mathrm{U(1)}^{\,}_{\mathfrak{c}}\big)$.
This result is motivated in Sec.\ \ref{subsec:Projective representations of the group Gf: I}.
{The computation of $H^{2}\big(G^{\,}_{f},\mathrm{U(1)}^{\,}_{\mathfrak{c}}\big)$
is done in Sec.\ \ref{subsec:Projective representations of the group Gf: I}
and Sec.\ \ref{sec:examples}.}

\medskip \noindent Given two projective representations, a third one can be obtained
from a graded tensor product as is explained in Sec.\
\ref{subsec:Stacking rules}. We shall describe the stacking rules
used to construct nontrivial projective representations.

\subsection{Marrying the fermion parity with the symmetry group $G$}
\label{subsec:Marrying the fermion parity with the symmetry group G}

For quantum systems built out of an even number of local Majorana operators,
it is always possible to express all Majorana operators as the real
and imaginary parts of local fermionic creation or annihilation operators.
The parity {(evenness or oddness)}
of the total fermion number is always
a constant of the motion. If $\widehat{F}$ denotes the
operator whose eigenvalues counts the total number of
local fermions in the Fock space, then the parity operator
$(-1)^{\widehat{F}}$ necessarily commutes with the
Hamiltonian that dictates the quantum dynamics, even though
$\widehat{F}$ might not, as is the case in any mean-field
treatment of superconductivity.

We denote the group of two elements $e$ and $p$
\begin{equation}
\mathbb{Z}^{F}_{2}\df
\left\{e,p\,|\, e\, p=p\,e=p,\quad e=e\,e=p\,p\right\},
\end{equation}
whereby $e$ is the identity element and we shall interpret
the quantum representation of $p$ as the fermion parity operator.
It is because of this interpretation of the group element $p$ that
we attach the upper index $F$ to the cyclic group $\mathbb{Z}^{\,}_{2}$.
In addition to the symmetry group $\mathbb{Z}^{F}_{2}$,
we assume the existence of a second symmetry group
$G$ with the composition law $\cdot$ and the identity element
$\mathrm{id}$.
We would like to construct a new
symmetry group $G^{\,}_{f}$
out of the two groups $G$ and $\mathbb{Z}^{F}_{2}$. Here,
the symmetry group $G^{\,}_{f}$ inherits the ``fermionic'' label
$f$ from its center $\mathbb{Z}^{F}_{2}$. One possibility is to consider the Cartesian product
\begin{subequations}\label{eq:Cartesian product}
\begin{equation}
G\times\mathbb{Z}^{F}_{2}\df
\left\{(g,h)\ |\ g\in G,\quad h\in\mathbb{Z}^{F}_{2}\right\}  
\label{eq:Cartesian product a}
\end{equation}
with the composition rule
\begin{equation}
(g^{\,}_{1},h^{\,}_{1})\circ
(g^{\,}_{2},h^{\,}_{2})\df
(g^{\,}_{1}\cdot g^{\,}_{2},h^{\,}_{1}\ h^{\,}_{2}).
\label{eq:Cartesian product b}
\end{equation}
\end{subequations}
The resulting group $G^{\,}_{f}$ is the direct product of $G$
and $\mathbb{Z}^{F}_{2}$.
However, the composition rule (\ref{eq:Cartesian product b})
is not the only one compatible with the existence
of a neutral element, inverse, and associativity.
To see this, we assume first the existence of the map
\begin{subequations}
\label{eq:def ZF2}
\begin{equation}
\begin{split}
\gamma\colon
G\times G\ \to&\ \mathbb{Z}^{F}_{2},
\\
(g^{\,}_{1},g^{\,}_{2})\ \mapsto&\ \gamma(g^{\,}_{1},g^{\,}_{2}),
\end{split}
\label{eq:def ZF2 a}
\end{equation}
whereby we impose the conditions
\begin{equation}	
\gamma(\mathrm{id},g)=
\gamma(g,\mathrm{id})=
e,
\quad
\gamma(g^{-1},g)=
\gamma(g,g^{-1}),
\label{eq:def ZF2 b}
\end{equation}
for all $g\in G$ and
\begin{equation}
\gamma(g^{\,}_{1},g^{\,}_{2})\,
\gamma(g^{\,}_{1}\cdot g^{\,}_{2},g^{\,}_{3})=
\gamma(g^{\,}_{1}, g^{\,}_{2}\cdot g^{\,}_{3})\,
\gamma(g^{\,}_{2},g^{\,}_{3}),
\label{eq:def ZF2 c}
\end{equation}
for all $g^{\,}_{1},g^{\,}_{2},g^{\,}_{3}\in G$.
Second, we define $G^{\,}_{f}$ to be the set of all pairs
$(g,h)$ with $g\in G$ and $h\in\mathbb{Z}^{F}_{2}$
obeying the composition rule
\begin{equation}
\begin{split}
\underset{\gamma}{\circ}\colon&
\left(G\times\mathbb{Z}^{F}_{2}\right)
\times
\left(G\times\mathbb{Z}^{F}_{2}\right)
\ \to\
G\times\mathbb{Z}^{F}_{2},
\\
&
\Big(
(g^{\,}_{1},\,h^{\,}_{1}),
(g^{\,}_{2},\,h^{\,}_{2})
\Big)
\ \mapsto\,
(g^{\,}_{1},\,h^{\,}_{1})
\underset{\gamma}{\circ}
(g^{\,}_{2},\,h^{\,}_{2}),
\end{split}
\label{eq:def ZF2 d}
\end{equation}
where
\begin{equation}
(g^{\,}_{1},\,h^{\,}_{1})
\underset{\gamma}{\circ}
(g^{\,}_{2},\,h^{\,}_{2})
\df
\Big(
g^{\,}_{1}\cdot g^{\,}_{2},
h^{\,}_{1}
\,
h^{\,}_{2}
\,
\gamma(g^{\,}_{1},g^{\,}_{2})
\Big).
\label{eq:def ZF2 e}
\end{equation}
\end{subequations}
One verifies the following properties.
First, the order within the composition
$h^{\,}_{1}\, h^{\,}_{2}\,\gamma(g^{\,}_{1},g^{\,}_{2})$
is arbitrary since $\mathbb{Z}^{F}_{2}$ is Abelian.
Second, conditions (\ref{eq:def ZF2 b}) and (\ref{eq:def ZF2 c})
ensure that $G^{\,}_{f}$ is a group with the
neutral element
\begin{subequations}
\begin{equation}
(\mathrm{id},e),
\end{equation}
the inverse to $(g,h)$ is
\begin{equation}
(g^{-1},[\gamma(g,g^{-1})]^{-1}\,h^{-1}),
\end{equation}
and the center
(those elements of the group that commute with all group elements)
given by
\begin{equation}
(\mathrm{id},\mathbb{Z}^{F}_{2}),
\end{equation}
\end{subequations} 
i.e.,
the group $G^{\,}_{f}$ is a central extension of $G$ by $\mathbb{Z}^{F}_{2}$.
Third, the map $\gamma$ can be equivalent to a map $\gamma^{\prime}$
of the form (\ref{eq:def ZF2 a})
in that they generate two isomorphic groups.
This is true if there exists the one-to-one map
\begin{subequations}
\begin{equation}
\begin{split}
\widetilde{\kappa}\colon
G\times\mathbb{Z}^{F}_{2}\ \to&\ G\times\mathbb{Z}^{F}_{2},
\\
(g,h)\ \mapsto&\ (g,\kappa(g)\,h)
\end{split}
\label{eq:def map kappa a}
\end{equation}
induced by the map
\begin{equation}
\begin{split}
\kappa\colon
G\ \to&\ \mathbb{Z}^{F}_{2},
\\
g\ \mapsto&\ \kappa(g),
\end{split}
\label{eq:def map kappa b}
\end{equation}
\end{subequations}
such that the condition
\begin{equation}
\widetilde{\kappa}
\Big(
(g^{\,}_{1},\,h^{\,}_{1})
\,\underset{\gamma}{\circ}\,
(g^{\,}_{2},\,h^{\,}_{2})
\Big)=
\widetilde{\kappa}
\Big(
(g^{\,}_{1},\,h^{\,}_{1})
\Big)
\,\underset{\gamma'}{\circ}\,
\widetilde{\kappa}
\Big(
(g^{\,}_{2},\,h^{\,}_{2})
\Big)
\label{eq:condition on widetilde kappa for group isomorphism}
\end{equation}
holds for all $(g^{\,}_{1},\,h^{\,}_{1}),(g^{\,}_{2},\,h^{\,}_{2})
\in G\times\mathbb{Z}^{F}_{2}$.
In other words, 
$\gamma$ and $\gamma'$
generate two isomorphic groups
if the identity
\begin{equation}
\kappa(g^{\,}_{1}\cdot g^{\,}_{2})\cdot\gamma(g^{\,}_{1},g^{\,}_{2})=
\kappa(g^{\,}_{1})\cdot\kappa(g^{\,}_{2})\cdot\gamma^{\prime}(g^{\,}_{1},g^{\,}_{2})
\label{eq:condition on kappa for group isomorphism}
\end{equation}
holds for all $g^{\,}_{1},g^{\,}_{2}\in G$.
This group isomorphism defines an equivalence relation.
We say that the group $G^{\,}_{f}$
obtained by extending the group $G$ with the group $\mathbb{Z}^{F}_{2}$
through the map $\gamma$ splits when 
a map (\ref{eq:def map kappa b}) exists such that
\begin{equation}
\kappa(g^{\,}_{1}\cdot g^{\,}_{2})\cdot\gamma(g^{\,}_{1},g^{\,}_{2})=
\kappa(g^{\,}_{1})\cdot\kappa(g^{\,}_{2})
\end{equation}
holds for all $g^{\,}_{1},g^{\,}_{2}\in G$, i.e.,
$G^{\,}_{f}$ splits when it is isomorphic to the direct product
(\ref{eq:Cartesian product}).

The task of classifying all the non-equivalent central
extensions of $G$ by $\mathbb{Z}^{F}_{2}$ through $\gamma$
is achieved by enumerating all the elements of the second cohomology group 
$H^{2}\left(G,\mathbb{Z}^{F}_{2}\right)$, see Appendix
\ref{appsec:Group Cohomology}.
We define an index $[\gamma]\in H^{2}\left(G,\mathbb{Z}^{F}_{2}\right)$
to represent such an equivalence class, whereby the index $[\gamma]=0$ 
is assigned to the case when $G^{\,}_{f}$ splits.

\subsection{Projective representations of the group $G^{\,}_{f}$}
\label{subsec:Projective representations of the group Gf: I}

We denote with
$\Lambda$ a $d$-dimensional lattice with
$\bm{j}\in\mathbb{Z}^{d}$ labeling the repeat unit cells.
We are going to attach to $\Lambda$ a Fock space on
which projective representations of
the group $G^{\,}_{f}$ constructed in
Sec.\ \ref{subsec:Marrying the fermion parity with the symmetry group G}
are realized. This will be done using four assumptions.

\noindent \textbf{Assumption 1.}
We attach to each repeat unit cell $\bm{j}\in\Lambda$
the local Fock space $\mathcal{F}^{\,}_{\bm{j}}$.
This step requires that the number of Majorana degrees of freedom
in each repeat unit cell is even.
It is then possible to define the local fermion number operator
$\hat{f}^{\,}_{\bm{j}}$ and the local fermion-parity operator
\begin{equation}
\hat{p}^{\,}_{\bm{j}}\df(-1)^{\hat{f}^{\,}_{\bm{j}}}.
\label{eq: def parity operator}
\end{equation}
We assume that all local Fock spaces
$\mathcal{F}^{\,}_{\bm{j}}$
with $\bm{j}\in\Lambda$
are ``identical,'' in particular they 
share the same dimensionality $\mathcal{D}$.
This assumption is a prerequisite to imposing translation symmetry.

\noindent\textbf{Assumption 2.}
Each repeat unit cell $\bm{j}\in\Lambda$
is equipped with a representation
$\hat{u}^{\,}_{\bm{j}}(g)$ of $G^{\,}_{f}$
through the conjugation
\begin{subequations}\label{eq: def uj(g)}
\begin{equation}
\hat{o}^{\,}_{\bm{j}}\ \mapsto\
\hat{u}^{\,}_{\bm{j}}(g)\,\hat{o}^{\,}_{\bm{j}}\,\hat{u}^{\dag}_{\bm{j}}(g),
\qquad
[\hat{u}^{\,}_{\bm{j}}(g)]^{-1}=\hat{u}^{\dag}_{\bm{j}}(g),
\label{eq: def uj(g) a}
\end{equation}
of any operator $\hat{o}^{\,}_{\bm{j}}$ acting
on the local Fock space $\mathcal{F}^{\,}_{\bm{j}}$.
The representation (\ref{eq: def uj(g) a}) of $g\in G^{\,}_{f}$
can either be unitary or antiunitary. 
More precisely, let
\begin{equation}
\begin{split}
\mathfrak{c}: G^{\,}_{f} \ \to&\  \left\{1,-1\right\},
\\
g\ \mapsto&\ \mathfrak{c}(g),  
\end{split}
\label{eq: def uj(g) b}
\end{equation}
be a homomorphism. We then have the decomposition
\begin{align}
\hat{u}^{\,}_{\bm{j}}(g)\df
\begin{cases}
\hat{v}^{\,}_{\bm{j}}(g),
&
\text{ if $\mathfrak{c}(g)=+1$,}
\\&\\
\hat{v}^{\,}_{\bm{j}}(g)\,
\mathsf{K},
&
\text{ if $\mathfrak{c}(g)=-1$,}
\end{cases}
\label{eq: def uj(g) c}
\end{align}
where
\begin{equation}
\hat{v}^{-1}_{\bm{j}}(g)=
\hat{v}^{\dag}_{\bm{j}}(g),
\qquad
\hat{p}^{\,}_{\bm{j}}\,
\hat{v}^{\,}_{\bm{j}}(g)\,
\hat{p}^{\,}_{\bm{j}}=
(-1)^{\rho(g)}\,
\hat{v}^{\,}_{\bm{j}}(g),
\label{eq: def uj(g) d}
\end{equation}
is a unitary operator with the fermion parity
$\rho(g)\in\{0,1\}\equiv\mathbb{Z}^{\,}_{2}$
acting linearly on $\mathcal{F}^{\,}_{\bm{j}}$ and
$\mathsf{K}$ denotes complex conjugation on
the local Fock space $\mathcal{F}^{\,}_{\bm{j}}$.
Accordingly, the homomorphism $\mathfrak{c}(g)$
dictates if the representation of the element $g\in G^{\,}_{f}$ is
implemented through a unitary operator [$\mathfrak{c}(g)=1$]
or an antiunitary operator [$\mathfrak{c}(g)=-1$].
Finally, we always choose to represent locally the fermion parity
$p\in\mathbb{Z}^{F}_{2}$
by the Hermitian operator $\hat{p}^{\,}_{\bm{j}}$, 
\begin{equation}
\hat{u}^{\,}_{\bm{j}}(p)\df
\hat{p}^{\,}_{\bm{j}}\equiv (-1)^{\hat{f}^{\,}_{\bm{j}}}.
\label{eq: def uj(g) e}
\end{equation} 
\end{subequations}

\noindent\textbf{Assumption 3.}
For any two elements $g,h\in G^{\,}_{f}$
[to simplify notation, $g\,\underset{\gamma}{\circ}\,h\equiv g\,h$
for all $g,h\in G^{\,}_{f}$],
whereby $e=g\,g^{-1}=g^{-1}\,g$ denotes the neutral element and
$g^{-1}\in G^{\,}_{f}$
the inverse of $g\in G^{\,}_{f}$,
we postulate the projective representation 
\begin{subequations}\label{eq:def projective rep I}
\begin{align}
&
\hat{u}^{\,}_{\bm{j}}(e)=\hat{\openone}^{\,}_{\mathcal{D}},
\label{eq:def projective rep I a}
\\
&
\hat{u}^{\,}_{\bm{j}}(g)\,\hat{u}^{\,}_{\bm{j}}(h)=
e^{\mathrm{i}\phi(g,h)}\,
\hat{u}^{\,}_{\bm{j}}(g\,h),
\label{eq:def projective rep I b}
\\
&
\big[\hat{u}^{\,}_{\bm{j}}(g)\,\hat{u}^{\,}_{\bm{j}}(h)\big]\hat{u}^{\,}_{\bm{j}}(f)=
\hat{u}^{\,}_{\bm{j}}(g)\big[\hat{u}^{\,}_{\bm{j}}(h)\,\hat{u}^{\,}_{\bm{j}}(f)\big],
\label{eq:def projective rep I c}
\end{align}
\end{subequations}
whereby the identity operator acting on $\mathcal{F}^{\,}_{\bm{j}}$ is denoted
$\hat{\openone}^{\,}_{\mathcal{D}}$
and the function
\begin{subequations}\label{eq:def projective rep II}
\begin{equation}
\begin{split}
\phi: G^{\,}_{f}\times G^{\,}_{f}\ \to&\ [0,2\pi),
\\
(g,h)\ \mapsto&\ \phi(g,h),
\end{split}
\label{eq:def projective rep II a}
\end{equation}
must be compatible with the associativity in $G^{\,}_{f}$, i.e.,
\begin{align}
\phi(g,h)+\phi(g\,h,f)=
\phi(g,h\,f)+\mathfrak{c}(g)\,\phi(h,f),
\label{eq:def projective rep II b}
\end{align}
for all $g,h,f\in G^{\,}_{f}\,$
\footnote{
One recognizes that Eq.\ (\ref{eq:def projective rep II b})
is a generalization of Eq.\ (\ref{eq:def ZF2 c})
if one identifies the exponential of $\phi$ in
Eq.\ (\ref{eq:def projective rep II b})
with $\gamma$ in Eq.\ (\ref{eq:def ZF2 c})
[up to the homomorphism (\ref{eq: def uj(g) b})].}.
The map $\phi$ taking values in $[0,2\pi)$ and 
satisfying \eqref{eq:def projective rep II b}
is an example of a \textit{2-cocycle} with the
group action specified by the $\mathbb{Z}^{\,}_{2}$-valued
homomorphism $\mathfrak{c}$. In the vicinity of the value 0,
$\phi$ generates the Lie algebra $\mathfrak{u}(1)$.
The associated Lie group is denoted U(1).
Given the neutral element $e\in G^{\,}_{f}$,
a \textit{normalized 2-cocycle} obeys the additional constraint
\begin{align}
\phi(e,g)=
\phi(g,e)=0
\label{eq:def projective rep II c}
\end{align}
\end{subequations}
for all $g\in G^{\,}_{f}$.
Two 2-cocycles $\phi(g,h)$ and $\phi'(g,h)$ are said to be
equivalent if they can be consistently related through a map
\begin{equation}
\begin{split}
\xi\colon G^{\,}_{f}\ \to&\ [0,2\pi),
\\
g\ \mapsto&\ \xi(g),
\end{split}
\end{equation}
as follows. The equivalence relation
$\phi\sim \phi'$ holds if the transformation
\begin{subequations}\label{eq:U(1) gauge equivalence}
\begin{align}
\hat{u}(g) = e^{\mathrm{i}\xi(g)}\, \hat{u}'(g),
\label{eq:U(1) gauge equivalence a}
\end{align}
implies the relation
\begin{align}
\phi(g,h)-\phi'(g,h) = 
\xi(g)
+
\mathfrak{c}(g)\,
\xi(h)
-
\xi(g\,h),
\label{eq:U(1) gauge equivalence b}
\end{align}
\end{subequations}
between the 2-cocycle $\phi(g,h)$
associated to the projective representation $\hat{u}(g)$  
and the 2-cocycle $\phi'(g,h)$
associated to the projective representation $\hat{u}'(g)\,$
\footnote{%
One recognizes that Eq.\
(\ref{eq:U(1) gauge equivalence b})
is a generalization of Eq.\
(\ref{eq:condition on kappa for group isomorphism})
if one identifies the exponential of $\xi$ in
Eq.\ (\ref{eq:U(1) gauge equivalence b})
with $\kappa$ in Eq.\
(\ref{eq:condition on kappa for group isomorphism})
[up to the homomorphism (\ref{eq: def uj(g) b})].
}.
In particular, $\hat{u}$ is equivalent to an
ordinary representation (a trivial projective representation)
if $\phi'(g,h) = 0$ for all $g,h\in G^{\,}_{f}$. 
Any $\phi\sim 0$ is called a coboundary.
For any coboundary $\phi$ there must exist
a $\xi$ such that
\begin{align}
\phi(g,h)=
\xi(g)
+
\mathfrak{c}(g)\,\xi(h)
-
\xi(g\,h).
\label{eq:coboundary condition}
\end{align}
The space of equivalence classes of projective representations 
is obtained by taking the quotient of 2-cocycles
\eqref{eq:def projective rep II}
by coboundaries \eqref{eq:coboundary condition}.  
The resulting set is the
second cohomology group $H^{2}\big(G^{\,}_{f},\mathrm{U(1)}^{\,}_{\mathfrak{c}}\big)$,
which has an additive group structure. 
Appendix \ref{appsec:Group Cohomology} gives more details on
group cohomology.

\noindent\textbf{Assumption 4.}
We attach to $\Lambda$ the global Fock space
$\mathcal{F}^{\,}_{\Lambda}$ by taking the appropriate product over
$\bm{j}$ of the local Fock spaces $\mathcal{F}^{\,}_{\bm{j}}$.
This means that we impose some algebra on all local operators
differing by their repeat unit cell labels.

\noindent
\textbf{Example 1,} any two local fermion number operators
$\hat{f}^{\,}_{\bm{j}}$
and
$\hat{f}^{\,}_{\bm{j}'}$
must commute
\begin{equation}
\left[\hat{f}^{\,}_{\bm{j}},\hat{f}^{\,}_{\bm{j}'}\right]=0
\end{equation}
for any two distinct repeat unit cell $\bm{j}\neq\bm{j}'\in\Lambda$.
The total fermion and fermion-parity numbers are 
\begin{equation}
\widehat{F}^{\,}_{\Lambda}\df
\sum_{\bm{j}\in\Lambda}
\hat{f}^{\,}_{\bm{j}},
\qquad
\widehat{P}^{\,}_{\Lambda}\df
(-1)^{\widehat{F}_{\Lambda}},
\end{equation}
respectively.

\noindent
\textbf{Example 2,}
Any two Majorana operators labeled by
$\bm{j}\neq\bm{j}'\in\Lambda$ must anticommute.

\noindent
\textbf{Example 3,}
The algebra
\begin{equation}
\hat{u}^{\,}_{\bm{j}}(g)\,
\hat{u}^{\,}_{\bm{j}'}(g')=
(-1)^{\rho(g)\,\rho(g')}\,
\hat{u}^{\,}_{\bm{j}'}(g')\,
\hat{u}^{\,}_{\bm{j}}(g)
\end{equation}
holds for any distinct $\bm{j}\neq\bm{j}'\in\Lambda$ and any
$g,g'\in G^{\,}_{f}$
because of
Eq.\ (\ref{eq: def uj(g) d}). We then define the operator
\begin{equation}
\widehat{U}(g)\df
\begin{cases}
\prod\limits_{\bm{j}\in \Lambda}\hat{v}^{\,}_{\bm{j}}(g),
&
\text{ if $\mathfrak{c}(g)=+1$,}
\\&\\
\left[\prod\limits_{\bm{j}\in \Lambda}\hat{v}^{\,}_{\bm{j}}(g)\right]\,
K,
&
\text{ if $\mathfrak{c}(g)=-1$,}
\end{cases}
\label{eq: def U(g)}
\end{equation}
that implements globally on the Fock space $\mathcal{F}^{\,}_{\Lambda}$
the operation corresponding to the group element $g\in G^{\,}_{f}$.

\begin{figure}[t]
\begin{center}
\includegraphics[angle=0,width=0.4\textwidth]{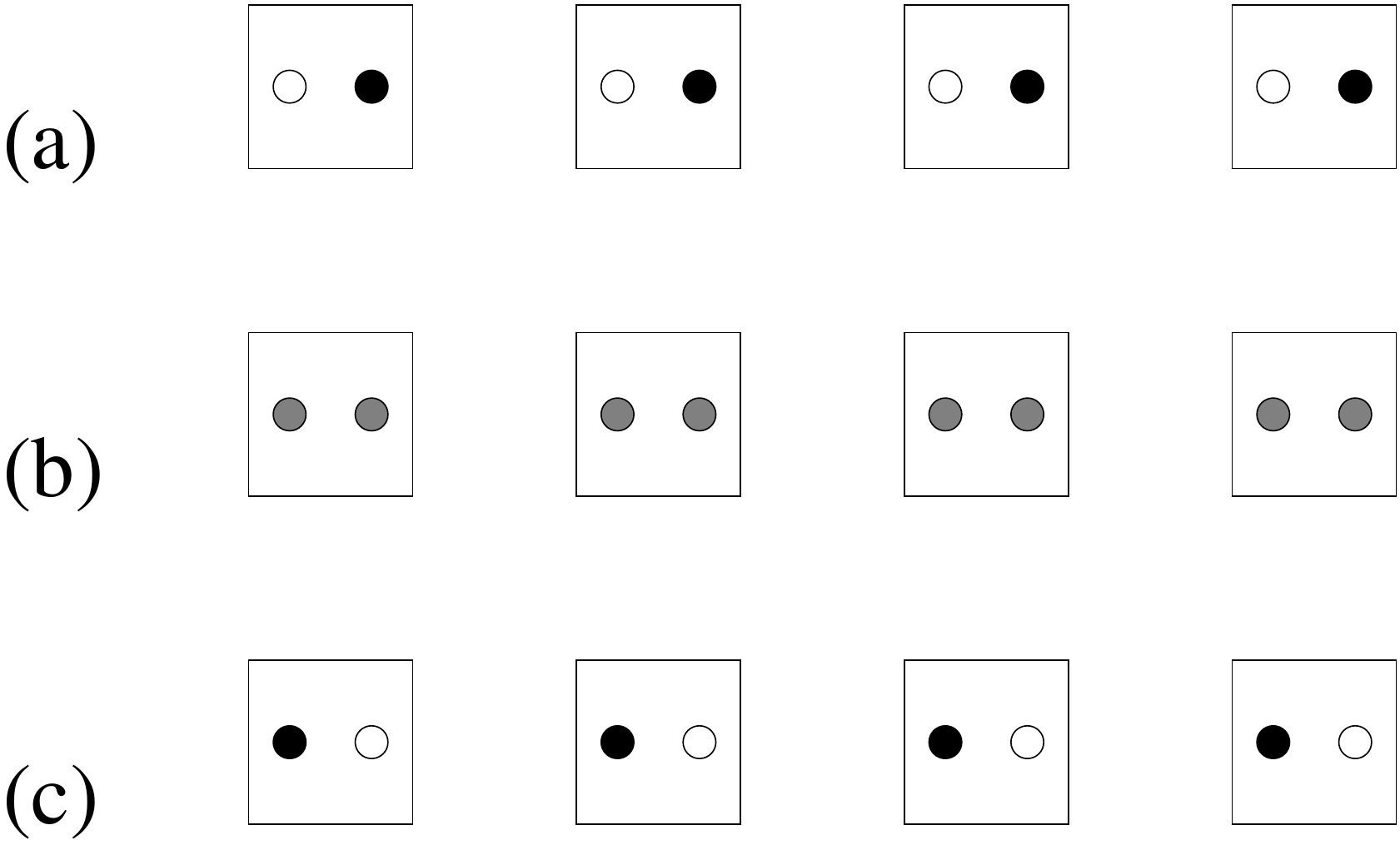}
\end{center}
\caption{
The repeat unit cells of a lattice $\Lambda$ are represented
pictorially by squares.  The lattice $\Lambda$ is chosen for
simplicity to be one dimensional.  (a) The repeat unit cell is
decorated with two circles, one empty, the other filled.  If periodic
boundary conditions are imposed, translations by one repeat unit cell
are symmetries. The permutation of the empty and filled circle within
all repeat unit cells is not a symmetry.  (b) If the filling pattern
is smoothly tuned ({through an on-site potential whose magnitude is color
coded, say})
so that both circles in an repeat unit cell have the
same filling, then the permutation of the left and right circles
within all repeat unit cells is a symmetry. One may then choose a
smaller repeat unit cell, a square centered about one circle only.
(c) Image of (a) under the permutation of the empty and filled circle
within all repeat unit cells.}
\label{Fig: even versus odd number of Majorana in repeat unit cell}
\end{figure}

%\subsection{Projective representations of the group $G^{\,}_{f}$: II}
%\label{subsec:Projective representations of the group Gf: II}

Theorem \ref{thm:LSM Theorem 1} refers to nontrivial projective representations
of the symmetry group $G^{\,}_{f}$ constructed in Sec.\
\ref{subsec:Marrying the fermion parity with the symmetry group G}.
We are going to characterize the projective representations
constructed in Sec.\
\ref{subsec:Projective representations of the group Gf: I}
by two indices $[(\nu,\rho)]$
and show that the second cohomology group $H^{2}(G^{\,}_{f},\mathrm{U(1)}^{\,}_{\mathfrak{c}})$
is equivalent to the equivalence classes $[(\nu,\rho)]$ of a tuple $(\nu,\rho)$.

Theorem \ref{thm:LSM Theorem 1} presumes the existence of
a local projective representation of the symmetry group $G^{\,}_{f}$.
This is only possible if the local Fock space
$\mathcal{F}^{\,}_{\bm{j}}$
defined in Sec.\ \ref{subsec:Projective representations of the group Gf: I}
is spanned by an even number of Majorana operators.
This hypothesis precludes a situation in which a fermion number
operator is well defined globally but not locally, for example
when the lattice $\Lambda$ is made of an even number of repeat unit cells,
but a repeat unit cell is assigned an odd number of Majorana operators.
(This can happen upon {changing the} parameters governing the quantum
dynamics as is illustrated in Fig.\
\ref{Fig: even versus odd number of Majorana in repeat unit cell}.)
We introduce the index $\mu=0,1$ to distinguish both possibilities.
The case $\mu=0$ applies
when the number of local Majorana operators at site $\bm{j}\in\Lambda$
is even, in which case the number of repeat unit cells in $\Lambda$
is any positive integer. The case $\mu=1$ applies
when the number of local Majorana operators at site $\bm{j}\in\Lambda$
is odd, in which case the number of repeat unit cell in $\Lambda$
must necessarily be an even positive integer.
The triplet
\begin{equation}
([(\nu,\rho)],\mu)\df
\begin{cases}
([(\nu,\rho)],0),&\hbox {if $\mu=0$, }
\\&\\
(0,0,1),&\hbox {if $\mu=1$, }
\end{cases}
\end{equation}
of indices allows to treat
Theorem \ref{thm:LSM Theorem 1}
and
\ref{thm:LSM Theorem 2}
together, as we are going to explain.

A similar set of three
indices also appears in the classification of
one-dimensional fermionic SPT phases. This is not an accident,
for fermionic SPT phases can also be classified in terms of
the projective representations
realized by the global symmetries after projection on the boundaries\,
\cite{Fidkowski2011,Bultinck2017,%
Williamson2016,Turzillo2019,Wang2020}.

\subsubsection{Indices $[(\nu,\rho)]$}
\label{subsec:Indices}

Theorem \ref{thm:LSM Theorem 1} is only predictive once it is established that
a projective representation of $G^{\,}_{f}$ is non trivial.
We recall that (i) the group $G^{\,}_{f}$ is a central extension
of the group $G$ by $\mathbb{Z}^{F}_{2}$ through the map $\gamma$ defined in
Eq.\ (\ref{eq:def ZF2}) and (ii) only the equivalence classes
$[\gamma]\in H^{2}\left(G,\mathbb{Z}^{F}_{2}\right)$ deliver non-isomorphic
groups. Choosing an element
$[\gamma]\in H^{2}\left(G,\mathbb{Z}^{F}_{2}\right)$
specifies $G^{\,}_{f}$. In turn,
a projective representation of $G^{\,}_{f}$ is specified by choosing
an element $\phi$ from the equivalence class
$[\phi]\in H^{2}\big(G^{\,}_{f},\mathrm{U(1)}^{\,}_{\mathfrak{c}}\big)$
where the 2-cocycle $\phi$ was defined in Eq.\
(\ref{eq:def projective rep II}). By convention,
a projective representation of $G^{\,}_{f}$ is trivial if
$\phi\sim0$, i.e., 
\begin{equation}
[\phi]\equiv \bm{0}
\end{equation}
is the trivial {(i.e., neutral)} element of
$H^{2}\big(G^{\,}_{f},\mathrm{U(1)}^{\,}_{\mathfrak{c}}\big)$
equipped with the addition as group composition.

Deciding if a projective representation of $G^{\,}_{f}$ is non trivial
amounts to calculating $H^{2}\big(G^{\,}_{f},\mathrm{U(1)}^{\,}_{\mathfrak{c}}\big)$.
This task is facilitated by the fact that
$H^{2}\big(G^{\,}_{f},\mathrm{U(1)}^{\,}_{\mathfrak{c}}\big)$
can be related to
$H^{2}\big(G,\mathbb{Z}^{F}_{2}\big)$,
$C^{2}\big(G,\mathrm{U(1)}\big)$,
and
$C^{1}\big(G,\mathbb{Z}^{\,}_{2}\big)$ 
together with a set of appropriately defined cocycle and coboundary conditions,
(see Appendix \ref{appsec:Group Cohomology} for definitions of these sets and conditions)
as was shown in the physics literature by
Turzillo and You in Ref.\ \onlinecite{Turzillo2019}. 
When the equivalence class $[\gamma]\in H^{2}(G,\mathbb{Z}^{F}_{2})$ is
trivial, $[\gamma]=0$, the second cohomology group $H^{2}\big(G^{\,}_{f},\mathrm{U(1)}^{\,}_{\mathfrak{c}}\big)$
reduces to the direct product of the cohomology groups
$H^{2}\big(G,\mathrm{U(1)}^{\,}_{\mathfrak{c}}\big)$
and $H^{1}\big(G,\mathbb{Z}^{\,}_{2}\big)$
as we shall explain.

To understand this claim, we consider first the simpler case when
the central extension $G^{\,}_{f}$ of the group $G$ by $\mathbb{Z}^{F}_{2}$
splits, i.e., when
\begin{equation}
[\gamma]=0\ \Longleftrightarrow\ G^{\,}_{f}=G\times\mathbb{Z}^{F}_{2}
\end{equation}
according to Sec.\
\ref{subsec:Marrying the fermion parity with the symmetry group G}.
If so, we can apply the Kuenneth formula for group cohomology,
\begin{align}
H^{2}\left(G^{\,}_{f},\mathrm{U(1)}^{\,}_{\mathfrak{c}}\right)=&\,
H^{2}\left(G\times\mathbb{Z}^{F}_{2},\mathrm{U(1)}^{\,}_{\mathfrak{c}}\right)
\nonumber\\
=&\,
H^{2}\left(G,\mathrm{U(1)}^{\,}_{\mathfrak{c}}\right)
\times
H^{1}\left(G,\mathbb{Z}^{\,}_{2}\right).
\label{eq:application Kuenneth formula}
\end{align}
Equation (\ref{eq:application Kuenneth formula})
states that the Abelian group
$H^{2}\big(G^{\,}_{f},\mathrm{U(1)}^{\,}_{\mathfrak{c}}\big)$
is the direct product between the Abelian group
$H^{2}\big(G,\mathrm{U(1)}^{\,}_{\mathfrak{c}}\big)$
and the Abelian group
$H^{1}\big(G,\mathbb{Z}^{\,}_{2}\big)$.

The cohomology group $H^{2}({G},\mathrm{U(1)}^{\,}_{\mathfrak{c}})$
is obtained by restricting the domain of definition of the
function $\phi$ in Eq.\ (\ref{eq:def projective rep II})
to the domain of definition $G\times G$.
We will reserve the letter $\nu$ to denote such a function.
An element of the Abelian group
$H^{2}\left(G,\mathrm{U(1)}^{\,}_{\mathfrak{c}}\right)$
with the addition as group composition
is the equivalence class $[\nu]$
with the neutral element
\begin{equation}
[\nu]=\mathsf{0}.
\end{equation}
The presence of the cohomology group $H^{1}\left(G,\mathbb{Z}^{\,}_{2}\right)$
on the right-hand side of
Eq.\ (\ref{eq:application Kuenneth formula})
can be understood as follows. Choose
$h$ in Eq.\ (\ref{eq:def projective rep I b})
to be the fermion parity $p$ (by the inclusion map).
We then have 
\begin{align}
\hat{u}(g)\,\hat{u}(p)=
e^{\mathrm{i}\phi(g,p)}\,
\hat{u}(g\,p),
\quad
\hat{u}(p)\,\hat{u}(g)=
e^{\mathrm{i}\phi(p,g)}\,
\hat{u}(p\,g).
\end{align}
Because $p$ belongs to the center of $G^{\,}_{f}$, $g\,p=p\,g$ implies that
\begin{align}
\hat{u}(g)\,\hat{u}(p)=
e^{\mathrm{i}[\phi(g,p)-\phi(p,g)]}\,
\hat{u}(p)\,\hat{u}(g).
\end{align}
Because the eigenvalues of the fermion parity operator are $0$ or $1$,
\begin{equation}
\phi(g,p)-\phi(p,g)=n\,\pi
\label{eq: phi(g,p)-phi(p,g)=npi}
\end{equation}
for some integer $n$, i.e., the projective representation of the
parity operator $\hat{u}(p)$ either commutes ($n$ even) or anticommutes
($n$ odd) with the projective representation $\hat{u}(g)$ of any
element of $G$. Hence, we may define the map
\begin{equation}
\begin{split}
\rho\colon G\ \to&\ \mathbb{Z}^{\,}_{2},
\\
g\ \mapsto&\, \rho(g)\df\frac{\phi(g,p)-\phi(p,g)}{\pi}\hbox{ mod 2},
\end{split}
\label{eq: def rho(g)}
\end{equation} 
whose equivalence classes $[\rho]$ under the gauge transformation
induced by Eq.\ (\ref{eq:U(1) gauge equivalence a}) define
$H^{1}\left(G,\mathbb{Z}^{\,}_{2}\right)$.
We recognize that the map \eqref{eq: def rho(g)} is the fermion parity
$\rho(g)\in\{0,1\}\equiv\mathbb{Z}^{\,}_{2}$ defined in
Eq.\ (\ref{eq: def uj(g) d}).
Even though the subgroup $G$ of $G\times\mathbb{Z}^{F}_{2}$
commutes with the subgroup $\mathbb{Z}^{F}_{2}$, this need
not be true under a projective representation. This possibility
is captured by the presence of
$H^{1}\left(G,\mathbb{Z}^{\,}_{2}\right)$
on the right-hand side of Eq.\ (\ref{eq:application Kuenneth formula}).
The neutral element of
$H^{1}\left(G,\mathbb{Z}^{\,}_{2}\right)$
equipped with the addition is
\begin{equation}
[\rho]=\mathtt{0}.
\end{equation}
All told, when the central extension $G^{\,}_{f}$ of the group $G$
by $\mathbb{Z}^{F}_{2}$ splits, the Kuenneth formula
(\ref{eq:application Kuenneth formula})
predicts that
\begin{subequations}\label{eq:indices if  Kuenneth formula holds}
\begin{equation}
[\phi]\equiv\left([\nu],[\rho]\right)
\label{eq:indices if  Kuenneth formula holds a}
\end{equation}
with the trivial projective representation defined by the condition
\begin{equation}
\bm{0}=[\phi]=\left([\nu],[\rho]\right)=(\mathsf{0},\mathtt{0}).
\label{eq:indices if  Kuenneth formula holds b}
\end{equation}
\end{subequations}
When the central extension $G^{\,}_{f}$ of the group $G$
by $\mathbb{Z}^{F}_{2}$
does not split, i.e., when
\begin{equation}
[\gamma]\neq 0
\label{eq: nonsplit condition on gamma}
\end{equation}
according to Sec.\
\ref{subsec:Marrying the fermion parity with the symmetry group G},
then the identification \eqref{eq:indices if  Kuenneth formula holds a} is no longer correct.
Turzillo and You in Ref.\ \onlinecite{Turzillo2019} have shown
in the context of SPT phases of matter that, the equivalence classes $[(\nu,\rho)]$ of 
the pair $(\nu,\rho)\in C^{2}(G,\mathrm{U(1)})\times C^{1}(G,\mathbb{Z}^{\,}_{2})$ that satisfy the conditions 
\begin{subequations}
\label{eq:no split condition on gamma nu and rho}
\begin{align}
\left(
\delta \nu -\pi\,\rho \smile \gamma,
\delta \rho
\right)
=
(0,0),
\label{eq:no split condition on gamma nu and rho a}
\end{align}
are in one to one correspondence with the second cohomology classes 
$[\phi]\in H^{2}(G^{\,}_{f},\mathrm{U(1)}^{\,}_{\mathfrak{c}})$.
Here, the operation $\delta$ is defined
in Eq.\ (\ref{appeq:definition delta n}),
while $\smile$ denotes the cup product defined in
Eq.\ (\ref{eq:def cup product}). 
Two pairs $(\nu,\rho)$ and $(\nu',\rho')$ are equivalent to each other if 
there exists another pair $(\alpha,\beta)$, whereby
\begin{align}
\begin{split}
\alpha\colon G \to&\ [0,2\pi),
\\
g \mapsto&\ \alpha(g),
\end{split}
\label{eq:no split condition on gamma nu and rho b}
\end{align}
while $\beta\in\mathbb{Z}^{\,}_{2}$ such that
\begin{align}
(\nu,\rho)	
=
(\nu',\rho')
+
\left(
\delta \alpha
+
\pi
\beta\smile\gamma,
\delta \beta
\right).
\label{eq:no split condition on gamma nu and rho c}
\end{align}
(See Appendix \ref{appsubsec: Classification of projective representations} for a detailed discussion.)
Hence, the second cohomology class is identified with
\begin{align}
[\phi]\equiv [(\nu,\rho)]
\label{eq:no split condition on gamma nu and rho d}
\end{align}
\end{subequations}
It can be seen from Eqs.\ \eqref{eq:no split condition on gamma nu and rho a} and \eqref{eq:no split condition on gamma nu and rho c}
that if $[\gamma]=0$ 
then $\delta \nu =0$, which is the defining condition for $\nu$ to be a 2-cocycle. 
We may then identify the gauge equivalent classes $[\nu]$
with the elements of the Abelian group
$H^{2}\left(G,\mathrm{U(1)}^{\,}_{\mathfrak{c}}\right)$ and
the identification \eqref{eq:no split condition on gamma nu and rho d} reduces to the one in 
Eq.\ \eqref{eq:indices if  Kuenneth formula holds a}.
However, when $[\gamma]\neq0$, the function
$\nu\colon G\times G\ \to\ \mathrm{U(1)}$
is called a 2-cochain and belongs to the set
$C^{2}\left(G,\mathrm{U(1)}\right)$.
In practice, Eq.\
(\ref{eq:no split condition on gamma nu and rho})
ties the 2-cochain $\nu$
to the 2- and 1-cochains $\gamma$ and $\rho$ that belong to the 
sets $C^{2}\left(G,\mathbb{Z}^{F}_{2}\right)$ and $C^{1}\left(G,\mathbb{Z}^{\,}_{2}\right)$, respectively.
From now on, we shall use the notation 
\eqref{eq:no split condition on gamma nu and rho d}
to trade in general the second cohomology class $[\phi]$
with the equivalence class $[(\nu,\rho)]$ and reserve the notation
$([\nu],[\rho])$ of Eq.\
\eqref{eq:indices if  Kuenneth formula holds a}
for the case when the underlying fermionic symmetry 
group $G^{\,}_{f}$ splits (i.e., $[\gamma]=0$).

\subsection{Stacking rules}
\label{subsec:Stacking rules}

In this section, we review the so-called stacking rules,
according to which the indices
$[(\nu,\rho)]$ and $\mu$ 
classifying an LSM constraint can be ``added.'' 

Given two Fock spaces 
$\mathcal{F}^{(1)}_{\bm{j}}$ and $\mathcal{F}^{(2)}_{\bm{j}}$,
let $\hat{u}^{(1)}_{\bm{j}}$ and $\hat{u}^{(2)}_{\bm{j}}$
be two local projective representations of $G^{\,}_{f}$ 
with indices 
$\big([(\nu^{\,}_{1},\,\rho^{\,}_{1})],0\big)$
and
$\big([(\nu^{\,}_{2},\,\rho^{\,}_{2})],0\big)$, respectively.
Stacking $\mathcal{F}^{(1)}_{\bm{j}}$
and
$\mathcal{F}^{(2)}_{\bm{j}}$
refers to taking the graded tensor product\,
\footnote{
To account for the fermionic statistics,
instead of the standard tensor product one must use a
$\mathbb{Z}^{\,}_{2}$ graded one. Fermionic Fock spaces then carry 
the structure of a $\mathbb{Z}^{\,}_{2}$ graded vector space,
also called a supervector space. See Appendix
\ref{appsec:Construction of fermionic matrix product states (FMPS)}
for more details on this construction.}
\begin{align}
\mathcal{F}^{(1\otimes^{\,}_{\mathfrak{g}}2)}_{\bm{j}}\df
\mathcal{F}^{(1)}_{\bm{j}}
\otimes^{\,}_{\mathfrak{g}}
\mathcal{F}^{(2)}_{\bm{j}}.
\end{align}
The local representation of $G^{\,}_{f}$
over $\mathcal{F}^{(1\otimes^{\,}_{\mathfrak{g}}2)}_{\bm{j}}$ is
\begin{subequations}
\begin{align}
\hat{u}^{(1\otimes^{\,}_{\mathfrak{g}}2)}_{\bm{j}}(g)\df
\begin{cases}
\hat{v}^{(1)}_{\bm{j}}(g)\,
\hat{v}^{(2)}_{\bm{j}}(g),
&
\hbox{if $\mathfrak{c}(g)=+1$},
\\&\\
\hat{v}^{(1)}_{\bm{j}}(g)\,
\hat{v}^{(2)}_{\bm{j}}(g)\,
\mathsf{K},
&
\hbox{if $\mathfrak{c}(g)=-1$}.
\end{cases}
\label{eq:def stacked operators}
\end{align}
This representation satisfies the composition rule
\begin{align}
\hat{u}^{(1\otimes^{\,}_{\mathfrak{g}}2)}_{\bm{j}}(g)\,
\hat{u}^{(1\otimes^{\,}_{\mathfrak{g}}2)}_{\bm{j}}(h)=
e^{\mathrm{i}\phi(g,h)}
\hat{u}^{(1\otimes^{\,}_{\mathfrak{g}}2)}_{\bm{j}}(g\,h),
\end{align}
where the phase $\phi(g,h){\in[0,2\pi)}$ is
\begin{align}
\label{eq:stacked rep phase rel}
\phi(g,h)=
\phi^{\,}_{1}(g,h)+\phi^{\,}_{2}(g,h)
+
\pi \rho^{\,}_{1}(h)\,\rho^{\,}_{2}(g).
\end{align}
\end{subequations}
The first two terms arise due to the composition rule of the 
two representations $\hat{u}^{(1)}_{\bm{j}}$ and $\hat{u}^{(2)}_{\bm{j}}$.
The last term encodes the fact that  representation $(1)$ of element
$h$ and representation $(2)$ of element $g$ anticommute if both of them 
have odd fermion parity. 
Equation \eqref{eq:stacked rep phase rel} can be decomposed into 
the stacking relations
\begin{equation}
\big([\left(\nu,\rho\right)],0\big)=
\big(
[
(
\nu^{\,}_{1}
+
\nu^{\,}_{2}
+
\pi\,
\rho^{\,}_{1}
\smile
\rho^{\,}_{2},\,
\rho^{\,}_{1}
+
\rho^{\,}_{2}
)
]
,0
\big).
\label{eq:stacking rel for indices}
\end{equation}
Here, $\rho^{\,}_{1}\smile\rho^{\,}_{2}$ is the cup product of 
the two 1-cochains $\rho^{\,}_{1}$ and $\rho^{\,}_{2}$
(see Appendix  \ref{appsec:Group Cohomology}
for the definition of the cup product).
Turzillo and You in Ref.\ \onlinecite{Turzillo2019}
have shown that a similar relationship holds in the context of
SPT phases of matter. The graded tensor product also allows to stack
two projective representations
$(\mathsf{0},\mathtt{0},\mu^{\,}_{1})$
and
$(\mathsf{0},\mathtt{0},\mu^{\,}_{2})$
with $\mu^{\,}_{1}=\hbox{1 mod 2}$ and $\mu^{\,}_{2}=\hbox{1 mod 2}$.
The result is a projective representation of the form
$\big([(\nu,\rho)],0\big)$
since it is then possible to define a local Fock space.

\section{Majorana LSM Theorem in one dimension}
\label{sec:1D_LSM}

In this section, we sketch the proofs for Theorems
\ref{thm:LSM Theorem 1} and \ref{thm:LSM Theorem 2}
in 1D space using the machinery of
fermionic matrix product states (FMPS).
We relegate some intermediate steps 
and technical details to Appendix 
\ref{appsec:Proof of LSM}. We begin with a definition of FMPS in Sec.\
\ref{sec:FMPS} (further background can be found in Appendix
\ref{appsec:Construction of fermionic matrix product states (FMPS)}.
and Refs.\ 
\onlinecite{Bultinck2017,Williamson2016,Kapustin2018,Turzillo2019}).

The strategy that we follow is to prove that
so-called injective FMPS are only compatible
with a trivial projective representation of the
symmetry group $G^{\,}_{f}$ discussed in Sec.\
\ref{sec:Symmetries}. 
The main steps of the proof for Theorems \ref{thm:LSM Theorem 1} and 
\ref{thm:LSM Theorem 2},
the first main results of this paper,
are provided in Sec.\ \ref{sec:thm LSM proof}
and \ref{sec:LSM mu 1}, respectively.
We close by discussing parallels with the
SPT phases in Sec.\
\ref{subsec:LSM constraints and classification of 1D fermionic SPTs}.

\subsection{Fermionic Matrix Product States}
\label{sec:FMPS}

Consider a one-dimensional lattice $\Lambda\cong\mathbb Z^{\,}_{N}$.
{At the repeat unit cell  $j=1,\cdots,N$,}
the local fermion number operator is denoted $\hat{f}^{\,}_{j}$  
and the local Fock space of dimension $\mathcal{D}^{\,}_{j}$
is denoted $\mathcal{F}^{\,}_{j}\cong \mathbb{C}^{\mathcal{D}^{\,}_{j}}$. 
We define with 
\begin{subequations}\label{eq:basis for local fermionic Fock space}
\begin{equation}
\ket{\psi^{\,}_{\sigma^{\,}_{j}}},
\qquad
\sigma^{\,}_{j}=1,\cdots,\mathcal{D}^{\,}_{j},
\label{eq:basis for local fermionic Fock space a}
\end{equation}
an orthonormal basis of $\mathcal{F}^{\,}_{j}$ such that
\begin{align}
(-1)^{\hat{f}^{\,}_{j}}\,\ket{\psi^{\,}_{\sigma^{\,}_{j}}}=
(-1)^{|\sigma^{\,}_{j}|}\ket{\psi^{\,}_{\sigma^{\,}_{j}}}.
\label{eq:basis for local fermionic Fock space b}
\end{align}
\end{subequations}
The fermion parity eigenvalue of the basis element 
$\ket{\psi^{\,}_{\sigma^{\,}_{j}}}$ is thus denoted $(-1)^{|\sigma^{\,}_{j}|}$
with $|\sigma^{\,}_{j}|\equiv0,1$.
The local Fock space $\mathcal{F}^{\,}_{j}$
admits the direct sum decomposition
\begin{subequations}\label{eq:direct sum decomposition local Fock space}
\begin{equation}
\mathcal{F}^{\,}_{j}=
\mathcal{F}^{(0)}_{j}
\oplus
\mathcal{F}^{(1)}_{j}
\label{eq:direct sum decomposition local Fock space a}
\end{equation}
where, given $p=0,1$,
\begin{equation}
\mathcal{F}^{(p)}_{j}\df
\mathrm{span}
\left\{
\ket{\psi^{\,}_{\sigma^{\,}_{j}}},\ \sigma^{\,}_{j}=1,\cdots,\mathcal{D}^{\,}_{j}
\ \Big|\
|\sigma^{\,}_{j}|=p
\right\}.
\label{eq:direct sum decomposition local Fock space b}
\end{equation}
\end{subequations}
One verifies that $\mathrm{dim}\,\mathcal{F}^{(0)}_{j}=
\mathrm{dim}\,\mathcal{F}^{(1)}_{j}=
\mathcal{D}^{\,}_{j}/{2}$.
%\begin{equation}
%\mathrm{dim}\,\mathcal{F}^{(0)}_{j}=
%\mathrm{dim}\,\mathcal{F}^{(1)}_{j}=
%\frac{\mathcal{D}^{\,}_{j}}{2}.
%\label{eq:direct sum decomposition local Fock space d}
%\end{equation}
To construct the Fock space $\mathcal{F}^{\,}_{\Lambda}$
for the lattice $\Lambda$, we demand that the direct sum
(\ref{eq:direct sum decomposition local Fock space})
also holds for $\mathcal{F}^{\,}_{\Lambda}$.
This is achieved with the help of the $\mathbb{Z}^{\,}_{2}$
tensor product $\otimes^{\,}_{\mathfrak{g}}$.
This tensor product preserves the $\mathbb{Z}^{\,}_{2}$-grading structure.
We define the reordering rule
\begin{equation}
\ket{\psi^{\,}_{\sigma^{\,}_{j}}}
\otimes^{\,}_{\mathfrak{g}}
\ket{\psi^{\,}_{\sigma^{\,}_{j'}}}\equiv
(-1)^{|\sigma^{\,}_{j}|\,|\sigma^{\,}_{j'}|}
\ket{\psi^{\,}_{\sigma^{\,}_{j'}}}
\otimes^{\,}_{\mathfrak{g}}
\ket{\psi^{\,}_{\sigma^{\,}_{j}}}
\label{eq:action Z2 graded tensor product}
\end{equation}
on any two basis elements
$\ket{\psi^{\,}_{\sigma^{\,}_{j}}}$
and
$\ket{\psi^{\,}_{\sigma^{\,}_{j'}}}$
of $\mathcal{F}^{\,}_{j}$ and $\mathcal{F}^{\,}_{j'}$
for any two distinct sites $j\in\Lambda$ and $j'\in\Lambda$,
respectively.
The rule (\ref{eq:action Z2 graded tensor product})
guarantees that states are antisymmetric under the 
exchange of an odd number of fermions on site $j$ with an odd number
of fermions on site $j'$ while symmetric otherwise.
We then define the fermionic Fock space $\mathcal{F}^{\,}_{\Lambda}$
for the lattice $\Lambda$ to be
\begin{equation}
\begin{split}
&
\mathcal{F}^{\,}_{\Lambda}\df
\mathrm{span}
\Bigg\{
\ket{\Psi^{\,}_{\bm{\sigma}}}
\ \Bigg|\
\ket{\Psi^{\,}_{\bm{\sigma}}}\equiv
\bigotimes\limits_{j=1}^{N}{}^{\,}_{\mathfrak{g}}\,
\ket{\psi^{\,}_{\sigma^{\,}_{j}}},
\\
&
\bm{\sigma}\equiv\left(\sigma^{\,}_{1},\cdots,\sigma^{\,}_{N}\right)\in
\{1,\cdots,\mathcal{D}^{\,}_{1}\}
\times\cdots\times
\{1,\cdots,\mathcal{D}^{\,}_{N}\}
\Bigg\}.
\end{split}
\label{eq:Fock space for Lambda}
\end{equation}
As the parity $|\sigma^{\,}_{j}|$
of the state $\ket{\psi^{\,}_{\sigma^{\,}_{j}}}$
can be generalized to the parity $|\bm{\sigma}|$
of the state $\ket{\Psi^{\,}_{\bm{\sigma}}}$
through the action of the global fermion number operator
\begin{equation}
\widehat{F}^{\,}_{\Lambda}\df
\sum_{j=1}^{N}\hat{f}^{\,}_{j},
\qquad
|\bm{\sigma}|\equiv
\sum_{j=1}^{N}|\sigma^{\,}_{j}|\hbox{ mod 2},
\end{equation}
the Fock space (\ref{eq:Fock space for Lambda})
inherits the direct sum decomposition
(\ref{eq:direct sum decomposition local Fock space a}),
\begin{equation}
\mathcal{F}^{\,}_{\Lambda}=
\mathcal{F}^{(0)}_{\Lambda}
\oplus
\mathcal{F}^{(1)}_{\Lambda}.
\label{eq:direct sum decomposition global Fock space}
\end{equation}
Any state $\ket{\Psi}\in\mathcal{F}^{\,}_{\Lambda}$
has the expansion
\begin{subequations}
\begin{equation}
\ket{\Psi}= 
\sum_{\bm{\sigma}}
c^{\,}_{\bm{\sigma}}\,
\ket{\Psi^{\,}_{\bm{\sigma}}}
\end{equation}
\end{subequations}
with the expansion coefficient $c^{\,}_{\bm{\sigma}}\in\mathbb{C}$.
Such a state is homogeneous if it belongs to either 
$\mathcal{F}^{(0)}_{\Lambda}$ or $\mathcal{F}^{(1)}_{\Lambda}$,
in which case it has a definite parity $|\Psi|\equiv0,1$.
From now on, we assume that all local Fock spaces are pairwise
isomorphic, i.e.,
\begin{equation}
\mathcal{D}^{\,}_{j}=\mathcal{D},
\qquad
\mathcal{F}^{\,}_{j}\cong \mathcal{F}^{\,}_{j'}  
\qquad
1\leq j<j'\leq N.
\end{equation}
This assumption is needed to impose translation symmetry below. We 
describe the construction of two families of states that
lie in $\mathcal{F}^{(0)}_{\Lambda}$ and $\mathcal{F}^{(1)}_{\Lambda}$,
respectively. To this end, we choose the positive integer $M$,
denote with $\openone^{\,}_{M}$ the unit $M\times M$ matrix
and define the following pair of $2M\times2M$ matrices
\begin{align}
P\df
\begin{pmatrix}
\openone^{\,}_{M} & 0
\\
0 & -\openone^{\,}_{M} 
\end{pmatrix},
\qquad
Y\df
\begin{pmatrix}
0 & \openone^{\,}_{M}
\\
-\openone^{\,}_{M} & 0
\end{pmatrix}.
\end{align}
The $2\times2$ grading that is displayed is needed
to represent the $\mathbb{Z}^{\,}_{2}$ grading in
Eq.\ (\ref{eq:direct sum decomposition global Fock space})
as will soon become apparent. The anticommuting matrices
$P$ and $Y$ belong to the set
$\mathrm{Mat}(2M,\mathbb{C})$
of all $2M\times2M$ matrices. This set is a
$4M^{2}$-dimensional vector space over the complex numbers.
\footnote{
The set $\mathrm{Mat}(2M,\mathbb{C})$
of all $2M\times2M$ matrices is a
$8M^{2}$-dimensional vector space over the real numbers.
         }
For any $\sigma^{\,}_{j}=1,\cdots,\mathcal{D}$ with $j\in\Lambda$,
we choose the matrices
\begin{subequations}\label{eq:def A0 and A1}
\begin{equation}
B^{\,}_{\sigma^{\,}_{j}},
C^{\,}_{\sigma^{\,}_{j}},
D^{\,}_{\sigma^{\,}_{j}},
E^{\,}_{\sigma^{\,}_{j}},
G^{\,}_{\sigma^{\,}_{j}}\in\mathrm{Mat}(M,\mathbb{C})
\label{eq:def A0 and A1 a}
\end{equation}
with the help of which we define the matrices
\begin{equation}
A^{(0)}_{\sigma^{\,}_{j}}\df
\begin{cases}
\begin{pmatrix}
B^{\,}_{\sigma^{\,}_{j}} & 0
\\
0 & C^{\,}_{\sigma^{\,}_{j}}
\end{pmatrix},
&
\hbox{ if } |\sigma^{\,}_{j}|=0,
\\&\\
\begin{pmatrix}
0 & D^{\,}_{\sigma^{\,}_{j}}
\\
E^{\,}_{\sigma^{\,}_{j}} & 0
\end{pmatrix},
&
\hbox{ if }
|\sigma^{\,}_{j}|=1,
\end{cases}
\label{eq:def A0 and A1 b}
\end{equation}
and
\begin{equation}
A^{(1)}_{\sigma^{\,}_{j}}\df
\begin{cases}
\begin{pmatrix}
G^{\,}_{\sigma^{\,}_{j}} & 0
\\
0 & G^{\,}_{\sigma^{\,}_{j}}
\end{pmatrix},
&
\hbox{ if } |\sigma^{\,}_{j}|=0,
\\&\\
\begin{pmatrix}
0 & G^{\,}_{\sigma^{\,}_{j}}
\\
-G^{\,}_{\sigma^{\,}_{j}} & 0
\end{pmatrix},
&
\hbox{ if }
|\sigma^{\,}_{j}|=1,
\end{cases}
\label{eq:def A0 and A1 c}
\end{equation}
\end{subequations}
from $\mathrm{Mat}(2M,\mathbb{C})$.
Observe that Eq.\ (\ref{eq:def A0 and A1 c})
is a special case of Eq.\ (\ref{eq:def A0 and A1 b}).
For any $\sigma^{\,}_{j}=1,\cdots,\mathcal{D}$ with $j\in\Lambda$,
the matrix $P$ commutes (anticommutes) with $A^{(p)}_{\sigma^{\,}_{j}}$ 
when $|\sigma^{\,}_{j}|=0$ ($|\sigma^{\,}_{j}|=1$),
\begin{equation}
P\,A^{(p)}_{\sigma^{\,}_{j}}=
(-1)^{|\sigma^{\,}_{j}|}\,
A^{(p)}_{\sigma^{\,}_{j}}\,P
\end{equation}
for both $p=0,1$. In contrast,
the matrix $Y$ commutes with $A^{(1)}_{\sigma^{\,}_{j}}$
\begin{equation}
Y\,A^{(1)}_{\sigma^{\,}_{j}}=
A^{(1)}_{\sigma^{\,}_{j}}\,Y
\end{equation}
for all $\sigma^{\,}_{j}=1,\cdots,\mathcal{D}$ with $j\in\Lambda$.

We are ready to define the FMPS.
We define states with either periodic boundary conditions (PBC) or antiperiodic
boundary conditions (APBC) denoted by the parameter $b=0$ or $1$, respectively.
They are
\begin{subequations}\label{eq:def FMPS}
\begin{equation}
\ket{\{A^{(0)}_{\sigma^{\,}_{j}}\};b}\df
\sum_{\bm{\sigma}}
\mathrm{tr}
\left[
P^{b+1}
A^{(0)}_{\sigma^{\,}_{1}}\,
\cdots
A^{(0)}_{\sigma^{\,}_{N}}\,
\right]
\ket{\Psi^{\,}_{\bm{\sigma}}}
\label{eq:def FMPS a}
\end{equation}
and
\begin{equation}
\ket{\{A^{(1)}_{\sigma^{\,}_{j}}\};b}\df
\sum_{\bm{\sigma}}
\mathrm{tr}
\left[
P^{b}\,Y\,
A^{(1)}_{\sigma^{\,}_{1}}\,
\cdots
A^{(1)}_{\sigma^{\,}_{N}}\,
\right]
\ket{\Psi^{\,}_{\bm{\sigma}}}
\label{eq:def FMPS b}
\end{equation}
\end{subequations}
for any choice of the matrices
(\ref{eq:def A0 and A1 b})
and
(\ref{eq:def A0 and A1 c}),
respectively, and with the basis
(\ref{eq:Fock space for Lambda})
of the Fock space $\mathcal{F}^{\,}_{\Lambda}$.
The following properties follow from the cyclicity of the trace
and from the fact that $Y$ is traceless.

\noindent \textbf{Property 1.}
The FMPS $\ket{\{A^{(p)}_{\sigma^{\,}_{j}}\};b}$ is homogeneous and
belongs to $\mathcal{F}^{(p)}_{\Lambda}$ for $p=0,1$.
This claim is a consequence of the identities
\begin{subequations}
\begin{align}
&
\sum_{j=1}^{N}|\sigma^{\,}_{j}|=1\hbox{ mod 2}
\implies
\mathrm{tr}
\left(
P^{b}\,P\,
A^{(0)}_{\sigma^{\,}_{1}}\,
\cdots
A^{(0)}_{\sigma^{\,}_{N}}\,
\right)
=0, 
\\
&
\sum_{j=1}^{N}|\sigma^{\,}_{j}|=0\hbox{ mod 2}
\implies
\mathrm{tr}
\left(
P^{b}\,Y\,
A^{(1)}_{\sigma^{\,}_{1}}\,
\cdots
A^{(1)}_{\sigma^{\,}_{N}}\,
\right)
=0.
\end{align}
\end{subequations}

\noindent \textbf{Property 2.}
The FMPS $\ket{\{A^{(p)}_{\sigma^{\,}_{j}}\};b}$ changes by a multiplicative
phase under a translation
by one repeat unit cell.
Indeed, one verifies that
\begin{equation}
\widehat{T}^{\,}_{b}\,
\ket{\{A^{(p)}_{\sigma^{\,}_{j}}\};b}=
e^{\mathrm{i}\pi (2k+b)/N}\,
\ket{\{A^{(p)}_{\sigma^{\,}_{j}}\};b},
\end{equation}
where  $\widehat{T}^{\,}_{b}$ is the generator of translation by one
repeat unit cell with boundary conditions $b=0,1$
and $k\in \mathbb Z$. 

\noindent \textbf{Property 3.}
The FMPS (\ref{eq:def FMPS a}) and (\ref{eq:def FMPS b})
are not uniquely specified by the choices
$\{A^{(p)}_{\sigma^{\,}_{j}}\}$ for $p=0,1$, respectively.
For example, the similarity transformation
\begin{equation}
A^{(0)}_{\sigma^{\,}_{j}}\mapsto
U\,A^{(0)}_{\sigma^{\,}_{j}}\, U^{-1},
\quad
\sigma^{\,}_{j}=1,\cdots,D,
\quad
j=1,\cdots,N,
\end{equation}
with $U$ any matrix that commutes with $P$ leaves the trace
unchanged. Another example occurs if there exists a
nonvanishing matrix $Q=Q^{2}\in\mathrm{Mat}(2M,\mathbb{C})$
such that
\begin{equation}
Q\,A^{(0)}_{\sigma^{\,}_{j}}=
Q\,A^{(0)}_{\sigma^{\,}_{j}}\,Q,
\qquad
\sigma^{\,}_{j}=1,\cdots,\mathcal{D}.
\label{eq:reducibility condition on a matrix}
\end{equation}
Indeed, one verifies that
Eq.\ (\ref{eq:reducibility condition on a matrix}) implies the identity
\begin{subequations}\label{eq:reducible versus decomposbale matrix}
\begin{equation}
\mathrm{tr}
\left[P^{b+1}\,
A^{(0)}_{\sigma^{\,}_{1}}\cdots A^{(0)}_{\sigma^{\,}_{N}}\right]=
\mathrm{tr}
\left[P^{b+1}\,
\tilde{A}^{(0)}_{\sigma^{\,}_{1}}\cdots 
\tilde{A}^{(0)}_{\sigma^{\,}_{N}}\right]
\label{eq:reducible versus decomposbale matrix a}
\end{equation}
with $\tilde{A}^{(0)}_{\sigma^{\,}_{j}}$ the matrix
\begin{equation}
\tilde{A}^{(0)}_{\sigma^{\,}_{j}}\df
Q\,A^{(0)}_{\sigma^{\,}_{j}}\,Q
+
(\openone^{\,}_{M}-Q)\,A^{(0)}_{\sigma^{\,}_{j}}\,(\openone^{\,}_{M}-Q).
\label{eq:reducible versus decomposbale matrix b}
\end{equation}
\end{subequations}
While conditions
(\ref{eq:reducibility condition on a matrix})
imply that all matrices
$A^{(0)}_{1},\cdots,A^{(0)}_{\mathcal{D}}$
are reducible, conditions
(\ref{eq:reducible versus decomposbale matrix b})
imply that all matrices
$\tilde{A}^{(0)}_{1},\cdots,\tilde{A}^{(0)}_{\mathcal{D}}$ 
are decomposable into the same block diagonal form. 
A necessary and sufficient condition on the $\mathcal{D}$ matrices
$A^{(0)}_{1},\cdots,A^{(0)}_{\mathcal{D}}$
to prevent that
Eq.\ (\ref{eq:reducibility condition on a matrix})
holds for some $Q\in\mathrm{Mat}(2M,\mathbb{C})$
is to demand
that there exists an integer $1\leq \ell^{\star}\leq N $ such that
the vector space spanned by the $\mathcal{D}^{\ell^{\star}}$ matrix products
\begin{subequations}
\begin{equation}
A^{(0)}_{\sigma^{\,}_{1}}\cdots A^{(0)}_{\sigma^{\,}_{\ell^{\star}}},
\qquad
\sigma^{\,}_{1},\cdots,\sigma^{\,}_{\ell^{\star}}=1,\cdots,\mathcal{D},
\label{eq:overcomplete basis of Mat(2M,C) if injective a}
\end{equation}
is $\mathrm{Mat}(2M,\mathbb{C})$\,
\footnote{
If $\mathcal{D}=2$, $M=1$, $A^{(0)}_{1}$ is $\mathrm{i}$ times
the second Pauli matrix,
and $A^{(0)}_{2}$ is the third Pauli matrix, then $\ell^{\star}=4$.
         }. More precisely, for any $A$ $\in$ $\mathrm{Mat}(2M,\mathbb{C})$,
it is possible to find $\mathcal{D}^{\ell^{\star}}$ coefficients 
$a^{(0)}_{\sigma^{\,}_{1},\cdots,\sigma^{\,}_{\ell^{\star}}}\in\mathbb{C}$
such that\,
\footnote{
The basis (\ref{eq:overcomplete basis of Mat(2M,C) if injective a})
is in general overcomplete owing to
the condition $\mathcal{D}^{\ell^{\star}}\geq4M^{2}$.
}
\begin{equation}
A=
\sum_{\sigma^{\,}_{1},\cdots,\sigma^{\,}_{\ell^{\star}}=1}^{\mathcal{D}}
a^{(0)}_{\sigma^{\,}_{1},\cdots,\sigma^{\,}_{\ell^{\star}}}\,
A^{(0)}_{\sigma^{\,}_{1}}\cdots A^{(0)}_{\sigma^{\,}_{\ell^{\star}}}.
\label{eq:overcomplete basis of Mat(2M,C) if injective b}
\end{equation}
\end{subequations}
In order to restrict the redundancy in the choice of
the matrices (\ref{eq:def A0 and A1})
that enter the FMPS (\ref{eq:def FMPS}),
we make the following definitions.

\begin{defn}
\label{def:injectivity even}
The even-parity FMPS (\ref{eq:def FMPS a})
is \textit{injective} if there exists an integer $\ell^{\star}\geq 1$
such that the $\mathcal{D}^{\ell^{\star}}$ products 
$A^{(0)}_{\sigma^{\,}_{1}}\cdots A^{(0)}_{\sigma^{\,}_{\ell^{\star}}}$
of $2M\times2M$ matrices span $\mathrm{Mat}(2M,\mathbb{C})$.
\end{defn}

\begin{defn}
\label{def:injectivity odd}
The odd-parity FMPS (\ref{eq:def FMPS b})
is \textit{injective} if there exists an integer $\ell^{\star}\geq 1$
such that the $\mathcal{D}^{\ell^{\star}}$ products 
$G^{\,}_{\sigma^{\,}_{1}}\cdots G^{\,}_{\sigma^{\,}_{\ell^{\star}}}$
of $M\times M$ matrices span $\mathrm{Mat}(M,\mathbb{C})$.
\end{defn}

The need to distinguish the definitions of injectivity
for even- and odd-parity FMPS stems from the fact that
for an odd-parity FMPS the
matrix $Y$ commutes with $A^{(1)}_{1},\cdots,A^{(1)}_{\mathcal{D}}$.
In other words, $Y$ is in the center of the algebra closed by products of 
$A^{(1)}_{1},\cdots,A^{(1)}_{\mathcal{D}}$. Injectivity requires
this center to be generated by $\openone^{\,}_{2M}$ and $Y$, i.e.,
the algebra closed by products of 
matrices $A^{(1)}_{1},\cdots,A^{(1)}_{\mathcal{D}}$
is a $\mathbb{Z}^{\,}_{2}$-graded simple algebra. 
For the center to be generated by no more than
$\openone^{\,}_{2M}$ and $Y$, the product of 
$G^{\,}_{1},\cdots,G^{\,}_{\mathcal{D}}$ must close a simple algebra
of $M\times M$ matrices, which is precisely 
the Definition \ref{def:injectivity odd}. The following properties of FMPS are essential to the proofs
of Theorems
\ref{thm:LSM Theorem 1}
and
\ref{thm:LSM Theorem 2}.

\noindent \textbf{Property 4.}
Let $\ell\geq\ell^{\star}$.
The $\mathcal{D}^{\ell}$ products 
$A^{(0)}_{\sigma^{\,}_{1}}\cdots A^{(0)}_{\sigma^{\,}_{\ell}}$
of $2M\times2M$ matrices span $\mathrm{Mat}(2M,\mathbb{C})$
for any injective even-parity FMPS.
The $\mathcal{D}^{\ell}$ products 
$G^{\,}_{\sigma^{\,}_{1}}\cdots G^{\,}_{\sigma^{\,}_{\ell}}$
of $M\times M$ matrices span $\mathrm{Mat}(M,\mathbb{C})$
for any $\ell\geq\ell^{\star}$ injective odd-parity FMPS.

\noindent \textbf{Property 5.}
If two sets of matrices 
$\{A^{(p)}_{\sigma^{\,}_{j}}\}$ and 
$\{\widetilde{A}^{(p)}_{\sigma^{\,}_{j}}\}$
generate the same injective FMPS,
there then exists an invertible matrix $U$
and a phase $\varphi^{\,}_{U}\in[0,2\pi)$
such that\,
\cite{Bultinck2017}  
\begin{subequations}\label{eq:FMPs gauge transformation}
\begin{align}
\widetilde{A}^{(p)}_{\sigma^{\,}_{j}}=
e^{\mathrm{i}\varphi^{\,}_{U}}\,
U\,
A^{(p)}_{\sigma^{\,}_{j}}\,
U^{-1},
\label{eq:FMPs gauge transformation a}
\end{align}
for any $\sigma^{\,}_{j}=1,\cdots,\mathcal{D}$,
and
\begin{equation}
P=\pm U\,P\,U^{-1},
\label{eq:FMPs gauge transformation b}
\end{equation}
for $p=0$, while
\begin{equation}
P=U\,P\,U^{-1},
\qquad
Y=\pm U\,Y\,U^{-1},
\label{eq:FMPs gauge transformation c}
\end{equation}
for $p=1$.
\end{subequations}
Here, the phase $\varphi^{\,}_{U}$ is needed to compensate for
the possibility that the matrix $U$ anticommutes with $P$ or $Y$.
We also observe that the index $\sigma^{\,}_{j}$ that labels the local
fermion number is preserved under the conjugation by $U$.
The transformation
(\ref{eq:FMPs gauge transformation})
that leaves an injective FMPS invariant
is called a gauge transformation.

\noindent \textbf{Property 6.}
Definitions \ref{def:injectivity even} and \ref{def:injectivity odd}
ensure that
the two-point correlation function of any pair of local operators
taken in an injective FMPS decays exponentially fast with their separation.
This provides an additional motivation to study them 
as they can be used to describe nondegenerate and gapped ground states\,
\cite{Bultinck2017}. 

With the formalism introduced in Secs.\ \ref{sec:Symmetries} and \ref{sec:FMPS},
we restate Theorem \ref{thm:LSM Theorem 1} as follows:
If an even- or odd-parity injective FMPS is translation-invariant 
and symmetric under a projective representation of
the symmetry group $G^{\,}_{f}$ defined in Sec.\ \ref{sec:Symmetries},
then the projective representation of  $G^{\,}_{f}$
must have a trivial second group cohomology class $[\phi]=0$.
We recover Theorem \ref{thm:LSM Theorem 1} by negating this statement.

\subsection{Proof of  Theorem \ref{thm:LSM Theorem 1}}
\label{sec:thm LSM proof}
Our strategy is inspired by the study of injective bosonic MPS
assumed to be $G^{\,}_{\mathrm{trsl}}\times G$-invariant
made by Tasaki in Ref.\ \onlinecite{Tasaki2020}.
For the fermionic case, we shall distinguish the cases of
even- and odd-parity FMPS, as each case demands
distinct conditions for injectivity. 
For the case of even-parity injective FMPS,
we shall establish the following identity between
\textit{any} matrix $A\in\mathrm{Mat}(2M,\mathbb{C})$
and a \textit{given} norm preserving
$W\in\mathrm{Mat}(2M,\mathbb{C})$
that is induced by a projective representation  
of the symmetry group $G^{\,}_{f}$.
There exists a phase $\delta\in[0,2\pi)$ and
a nonvanishing positive integer $\ell^{\star}$ such that
\begin{subequations}
\label{eq:desired relation lemma 1}
\begin{equation}
A=e^{\mathrm{i}\ell\,\delta}\, W^{-1}\,A\,W,
\end{equation}
holds for all $\ell$$=$$\ell^{\star},\ell^{\star}+1,\ell^{\star}+2,\cdots$
and all $A$$\in$$\mathrm{Mat}(2M,\mathbb{C})$.
This is only possible if
\begin{equation}
\delta=0,
\end{equation}
\end{subequations}
which obviously holds when $A$ is the identity matrix
$\openone^{\,}_{2M}$.
For the case of odd FMPS, we shall establish the same identity
as \eqref{eq:desired relation lemma 1} for any matrix
$A \in \mathrm{Mat}(2M,\mathbb{C})$ that commutes with matrix $Y$, i.e.,
$Y$ is in the center of the algebra spanned by
such matrices $A$. Theorem \ref{thm:LSM Theorem 1}
will follow from the interpretation of the condition
$\delta=0$ as the projective representation of 
$G^{\,}_{f}$ defined in Sec.\ \ref{sec:Symmetries}
to have trivial second group cohomology class. 

\subsubsection{Case of even-parity injective FMPS}
\label{subsubsec:Case of even-parity injective FMPS}

We start from the even-parity injective FMPS
\begin{equation}
\ket{\{A^{(0)}_{\sigma^{\,}_{j}}\};b}\df
\sum_{\bm{\sigma}}
\mathrm{tr}
\left[
P^{b+1}\,
A^{(0)}_{\sigma^{\,}_{1}}\,
\cdots
A^{(0)}_{\sigma^{\,}_{N}}\,
\right]
\ket{\Psi^{\,}_{\bm{\sigma}}}.
\label{eq:def injective FMPS aeven parity}
\end{equation}
Let $g$ be an element from $G^{\,}_{f}$
be represented by the operator
$\widehat{U}(g)$
as defined in Sec.\ 
\ref{subsec:Projective representations of the group Gf: I}.

On the one hand, we have the identity
\begin{subequations}
\begin{align}
\widehat{U}(g)\,\ket{\{A^{(0)}_{\sigma^{\,}_{j}}\};b}=&\,
\sum_{\bm{\sigma}}
\mathrm{tr}
\left[
P^{b+1}
A^{(0)}_{\sigma^{\,}_{1}}
\cdots
A^{(0)}_{\sigma^{\,}_{N}}
\right]
\widehat{U}(g)\,
\ket{\Psi^{\,}_{\bm{\sigma}}}
\nonumber\\
\equiv&\,
\sum_{\bm{\sigma}}
\mathrm{tr}
\left[
P^{b+1}
A^{(0)}_{\sigma^{\,}_{1}}(g)
\cdots
A^{(0)}_{\sigma^{\,}_{N}}(g)
\right]
\ket{\Psi^{\,}_{\bm{\sigma}}},
\end{align}
where
\begin{align}
&
A^{(0)}_{\sigma^{\,}_{j}}(g)\df
\sum_{\sigma^{\prime}_{j}=1}^{\mathcal{D}}
\left[\mathcal{U}(g)\right]^{\,}_{\sigma^{\,}_{j}\sigma^{\prime}_{j}}
\mathsf{K}^{\,}_{g}\left[A^{(0)}_{\sigma^{\prime}_{j}}\right],
\\
&
\left[\mathcal{U}(g)\right]^{\,}_{\sigma^{\,}_{j}\sigma^{\prime}_{j}}\df
\bra{\psi^{\,}_{\sigma^{\,}_{j}}}
\hat{v}^{\,}_{j}
\ket{\psi^{\,}_{\sigma^{\prime}_{j}}},
\\
&
\mathsf{K}^{\,}_{g}\left[A^{(0)}_{\sigma^{\,}_{j}}\right]
\df
\begin{cases}
A^{(0)}_{\sigma^{\,}_{j}},&\hbox{ if $\mathfrak{c}(g)=0$, }
\\&\\
\mathsf{K}\,
A^{(0)}_{\sigma^{\,}_{j}}\,
\mathsf{K},&\hbox{ if $\mathfrak{c}(g)=1$. }
\end{cases}
\end{align}
\end{subequations}
(Complex conjugation is denoted with $\mathsf{K}$.) On the other hand, we have the identity
\begin{align}
\widehat{U}(g)\,\ket{\{A^{(0)}_{\sigma^{\,}_{j}}\};b}=&\,
e^{\mathrm{i}\eta(g;b)}\,
\ket{\{A^{(0)}_{\sigma^{\,}_{j}}\};b}
\nonumber\\
=&\,
\ket{\{e^{\mathrm{i}\eta(g;b)/N}\,A^{(0)}_{\sigma^{\,}_{j}}\};b}
\end{align}
for some phase $\eta(g;b)$$\in$$[0,2\pi)$
if we assume that
$\widehat{U}(g)$$\ket{\{A^{(0)}_{\sigma^{\,}_{j}}\};b}$
is an eigenstate of the norm-preserving operator
$\widehat{U}(g)$, as it should be if $G^{\,}_{f}$ is a symmetry. By the assumption of injectivity, the matrices
$A^{(0)}_{\sigma^{\,}_{j}}(g)$ and
$e^{\mathrm{i}\eta(g;b)/N}\,A^{(0)}_{\sigma^{\,}_{j}}$
are related by a similarity transformation
(\ref{eq:FMPs gauge transformation}),
i.e., there exists an invertible matrix $U(g)$ and
a phase $\varphi^{(b)}_{U(g)}\in[0,2\pi)$ such that
\begin{equation}
e^{\mathrm{i}\eta(g;b)/N}\,A^{(0)}_{\sigma^{\,}_{j}}=
e^{\mathrm{i}\varphi^{(b)}_{U(g)}}\,
U(g)\,
A^{(0)}_{\sigma^{\,}_{j}}(g)\,
U^{-1}(g)
\label{eq:pre key step lemma}
\end{equation}
for any $\sigma^{\,}_{j}$. We massage Eq.\ (\ref{eq:pre key step lemma}) into
\begin{subequations}\label{eq:key step lemma}
\begin{equation}
e^{\mathrm{i}\theta(g;b)}\,
U^{\dag}(g)\,
A^{(0)}_{\sigma^{\,}_{j}}\,U(g)=
\sum_{\sigma^{\prime}_{j}=1}^{\mathcal{D}}
\left[\mathcal{U}(g)\right]^{\,}_{\sigma^{\,}_{j}\sigma^{\prime}_{j}}
\mathsf{K}^{\,}_{g}\left[A^{(0)}_{\sigma^{\prime}_{j}}\right],
\label{eq:key step lemma b}
\end{equation}
where we have introduced the phase
\begin{equation}
\theta(g;b)\df
\frac{\eta(g;b)}{N}
-
\varphi^{(b)}_{U(g)}.
\end{equation}
\end{subequations}
Consider a second element $h\in G^{\,}_{f}$ asides from
$g\in G^{\,}_{f}$. We can use the relation (\ref{eq:key step lemma b})
with $g$ replaced by the composition $g\,h$. We can also
iterate the relation (\ref{eq:key step lemma b})
by evaluating the composition
$\widehat{U}(g)\,\widehat{U}(h)\,\ket{\{A^{(0)}_{\sigma^{\,}_{j}}\};b}$.
After some algebra (Appendix \ref{appsubsec:Even-parity FMPS}),
one finds that (i) the phase
\begin{subequations}\label{eq:key step lemma iterated} 
\begin{equation}
\delta(g,h;b)\df
\mathfrak{c}(g)\,
\theta(h;b)
+
\theta(g;b)
-
\phi(g,h)
-
\theta(g\,h;b)
\label{eq:key step lemma iterated a}  
\end{equation}
that relates the normalized 2-cocycle defined in
Eqs.\
(\ref{eq:def projective rep I})
and
(\ref{eq:def projective rep II})
to the phase (\ref{eq:key step lemma b}),
(ii) the map represented by
\begin{equation}
V(g)\df
\begin{cases}
U(g),&\hbox{ if $\mathfrak{c}(g)=0$, }
\\&\\
U(g)\,\mathsf{K},&\hbox{ if $\mathfrak{c}(g)=1$, }
\end{cases}
\label{eq:key step lemma iterated b}  
\end{equation}
and the $\mathcal{D}$ matrices
$A^{(0)}_{\sigma^{\,}_{j}}$,
are related by
\begin{equation}
e^{\mathrm{i}\delta(g,h;b)}\,A^{(0)}_{\sigma^{\,}_{j}}\,W(g,h)=
W(g,h)\,A^{(0)}_{\sigma^{\,}_{j}}
\label{eq:key step lemma iterated c}  
\end{equation}
for any $\sigma^{\,}_{j}=1,\cdots,\mathcal{D}$,
where
\begin{equation}
W(g,h)\df
V(g)\,V(h)\,V^{-1}(g\,h).
\label{eq:key step lemma iterated d}  
\end{equation}
\end{subequations}
We are going to make use of the injectivity of the FMPS
a second time after massaging Eq.\ (\ref{eq:key step lemma iterated c})  
into
\begin{equation}
A^{(0)}_{\sigma^{\,}_{j}}=
e^{\mathrm{i}\delta(g,h;b)}\,
W^{-1}(g,h)\,A^{(0)}_{\sigma^{\,}_{j}}\,W(g,h)
\label{eq:final step proof lemma even FMPS: I}
\end{equation}
for any $\sigma^{\,}_{j}=1,\cdots,\mathcal{D}$.
For any integer $\ell=1,2,\cdots$,
iteration of Eq.\ (\ref{eq:final step proof lemma even FMPS: I})
gives
\begin{equation}
\prod_{j=1}^{\ell}A^{(0)}_{\sigma^{\,}_{j}}=
e^{\mathrm{i}\ell\,\delta(g,h;b)}\,
W^{-1}(g,h)\,\left[\prod_{j=1}^{\ell}A^{(0)}_{\sigma^{\,}_{j}}\right]\,W(g,h).
\label{eq:final step proof lemma even FMPS: II}
\end{equation}
When $\ell\geq\ell^{*}$, injectivity  of the FMPS implies that 
any matrix $A\in\mathrm{Mat}(2M,\mathbb{C})$
can be written as a linear superposition of all
the possible monomials $\prod_{j=1}^{\ell}A^{(0)}_{\sigma^{\,}_{j}}$
of order $\ell$, each of which obeys
Eq.\ (\ref{eq:final step proof lemma even FMPS: II})
[recall Eq.\ (\ref{eq:overcomplete basis of Mat(2M,C) if injective b})].
Hence, we arrive at the identity
\begin{equation}
A=
e^{\mathrm{i}\ell\,\delta(g,h;b)}\,
W^{-1}(g,h)\,A\,W(g,h),
\qquad\forall \ell\geq\ell^{\star},
\end{equation}
for any $A\in\mathrm{Mat}(2M,\mathbb{C})$,
which implies, in turn, that  $W(g,h)$
belongs to the center of the algebra spanned by monomials
$\prod_{j=1}^{\ell}A^{(0)}_{\sigma^{\,}_{j}}$.
For even-parity FMPS, this center is one-dimensional as it is
generated by the unit matrix $\openone^{\,}_{2M}$.
In particular, we can choose $A=\openone^{\,}_{2M}$ for which
\begin{subequations}
\begin{equation}
\openone^{\,}_{2M}=
e^{\mathrm{i}\ell\,\delta(g,h;b)}\,
\openone^{\,}_{2M},
\end{equation}
which implies that
\begin{equation}
\delta(g,h;b)=0,
\label{eq:coboundary condition for even-parity case}
\end{equation}
\end{subequations}
and, therefore, $[\phi]=0$
[recall Eq.\ (\ref{eq:coboundary condition})].

\subsubsection{Case of odd-parity injective FMPS}
\label{subsubsec:Case of odd-parity injective FMPS}

The odd-parity FMPS differs from the even-parity FMPS in that
the $\mathcal{D}^{\ell}$ products
$A^{(1)}_{\sigma^{\,}_{1}}\cdots A^{(1)}_{\sigma^{\,}_{\ell}}$ 
for any $\ell\geq\ell^{\star}$
span a subalgebra of $\mathrm{Mat}(2M,\mathbb{C})$ with 
the center spanned by $\openone^{\,}_{2M}$ and $Y$. 
This difference is of no consequence until
reaching the odd-parity counterpart to Eq.\
(\ref{eq:pre key step lemma}).
However, for the odd-parity counterpart to Eq.\
(\ref{eq:pre key step lemma})
multiplication of $U(g)$ from the left by any element from the center
generated by $\openone^{\,}_{2M}$ and $Y$,
\begin{equation}
\big[
a(g)\,\openone^{\,}_{2M}
+
b(g)\,Y
\big]
U(g),
\qquad
|a(g)|^{2} + |b(g)|^{2}=1,
\end{equation}
leaves Eq.\ (\ref{eq:pre key step lemma}) unchanged.
To fix this subtlety, we replace $U(g)$ in
Eq.\ (\ref{eq:pre key step lemma})
by $U^{(0)}(g)$ which is defined by
\begin{subequations}
\begin{align}
&
U(g)\df
\big[
a(g)\,\openone^{\,}_{2M}
+
b(g)\,Y
\big]\,
U^{(0)}(g),
\\
&
P\,
U^{(0)}(g)\,
P
=
U^{(0)}(g).
\end{align}
\end{subequations}
With this change in mind, all the steps leading to
Eq.\ \eqref{eq:key step lemma iterated}
for the even-parity case can be repeated for the odd-parity case.
The analog to the even-parity coboundary condition
(\ref{eq:coboundary condition for even-parity case})
then follows, thereby completing the proof of Theorem
\ref{thm:LSM Theorem 1}.

\subsection{Proof of Theorem \protect{\ref{thm:LSM Theorem 2}}}
\label{sec:LSM mu 1}

Theorem \ref{thm:LSM Theorem 1}
presumes the existence of a local fermionic Fock space,
i.e., of an even number of Majorana degrees of freedom per 
repeat unit cell. This hypothesis precludes
translation invariant lattice Hamiltonians
with odd number of Majorana operators per repeat unit cell
such as
\begin{subequations}\label{eq:def Kitaev chain at criticality}
\begin{align}
\widehat{H}^{\,}_{\mathrm{K}}\df
\sum_{j=1}^{2M}
\mathrm{i}
\hat{\gamma}^{\,}_{j}\,
\hat{\gamma}^{\,}_{j+1}.
\end{align}
Here, the Hermitian operators
$\{\hat{\gamma}^{\,}_{j}=\hat{\gamma}^{\dag}_{j}\}$
obey the Majorana algebra
\begin{equation}
\left\{
\hat{\gamma}^{\,}_{j},
\hat{\gamma}^{\,}_{j'}
\right\}=
2\delta^{\,}_{jj'},
\qquad
j,j'=1,\cdots,2M,
\end{equation}
\end{subequations}
and the total number $2M$ of repeat unit cell is an even integer.
Hamiltonian $\widehat{H}^{\,}_{\mathrm{K}}$ 
realizes the critical point between the two 
topologically distinct phases of the Kitaev chain. In the continuum limit, it
describes a helical pair of Majorana fields and has a gapless spectrum.

Motivated by this example, we now prove a separate LSM constraint
on Majorana lattice models with an odd Majorana flavors per repeat unit site.
We use Theorem \ref{thm:LSM Theorem 1} for the proof.

Let $n\geq 0$ be an integer and 
\begin{align}
\hat{\chi}^{\,}_{j}\df
\left(
\hat{\chi}^{\,}_{j,1},
\hat{\chi}^{\,}_{j,2},
\cdots,
\hat{\chi}^{\,}_{j,2m+1}
\right)^{\mathsf{T}}
\end{align} 
be the spinor made of $2m+1$ Majorana operators.
Let the Hamiltonian $\widehat{H}$ be 
local and translationally invariant.
We write
\begin{align}
\widehat{H}\equiv
\sum_{j=1}^{2M}
\hat{h}
\big(
\hat{\chi}^{\,}_{j-q},
\dots,
\hat{\chi}^{\,}_{j},
\dots,
\hat{\chi}^{\,}_{j+q}
\big),
\label{eq:Hamiltonian local mu 1}
\end{align}
where $\hat{h}$ is a Hermitian polynomial
of $2q$ Majorana spinors
$\left\{\hat{\chi}^{\,}_{j-q},\dots,
\hat{\chi}^{\,}_{j+q}\right\}$ with $q$ a positive integer.
The finiteness of $q$ renders $\widehat{H}$ local.
Hamiltonian \eqref{eq:Hamiltonian local mu 1} is defined over 
$2M$ sites, since an even number of Majorana operators are needed 
to have a well-defined Fock space. We
assume that $\widehat{H}$
has a nondegenerate gapped ground state
$\ket{\Psi^{\,}_{0}}$. We are going to deliver a contradiction
by making use of Theorem \ref{thm:LSM Theorem 1},
thereby proving Theorem \ref{thm:LSM Theorem 2}.

Define the Hamiltonian,
\begin{subequations}\label{eq:doubled Hamiltonian mu 1}
\begin{align}
\widehat{H}^{\prime}
\df
\sum_{j=1}^{2M}
\sum_{\alpha=1}^{2}
\hat{h}
\left(
\hat{\chi}^{(\alpha)}_{j-q},
\cdots,
\hat{\chi}^{(\alpha)}_{j+q}
\right),
\label{eq:doubled Hamiltonian mu 1 a}
\end{align}
which is the sum of two copies of Hamiltonian
\eqref{eq:Hamiltonian local mu 1}.
The repeat unit cell labeled by $j=1,\cdots,2M$ now contains
two Majorana spinors labeled by $\alpha=1,2$.
Hamiltonian (\ref{eq:doubled Hamiltonian mu 1}) thus
acts on a Fock space which is locally spanned by an
even number of Majorana flavors. At each site $j=1,\cdots,2M$
one can define a local fermionic Fock space. 
Since there is no coupling between the two copies $\alpha=1,2$ of
Majorana spinors,
$\widehat{H}^{\prime}$
inherits from $\widehat{H}$
the nondegenerate gapped ground state
\begin{equation}
\ket{\Psi^{\prime}_{0}}\df
\ket{\Psi_{0}}\otimes^{\,}_{\mathfrak{g}}\ket{\Psi_{0}}.
\label{eq:doubled Hamiltonian mu 1 b}
\end{equation}
\end{subequations}
Since at each site $j$, there is no term coupling the two copies
$\hat{\chi}^{(1)}_{j}$ and $\hat{\chi}^{(2)}_{j}$,
$\widehat{H}^{\prime}$
is invariant under any local permutation
\begin{subequations}
\begin{align}
\begin{pmatrix}
\hat{\chi}^{(1)}_{j}\\
\hat{\chi}^{(2)}_{j}
\end{pmatrix}
\mapsto
\begin{pmatrix}
\hat{\chi}^{(2)}_{j}\\
\hat{\chi}^{(1)}_{j}
\end{pmatrix}.
\label{eq:two copies symmetry mu 1}
\end{align}
The local representation of the fermion parity operator is
\begin{align}
\widehat{P}^{\,}_{j}\df
\prod_{l=1}^{2n+1}
\left[
\mathrm{i}\,
\hat{\chi}^{(1)}_{j,l}\,
\hat{\chi}^{(2)}_{j,l}
\right].
\end{align}
\end{subequations}
Under the transformation \eqref{eq:two copies symmetry mu 1},
the local fermion parity operator $\widehat{P}^{\,}_{j}$
acquires the phase $(-1)^{2n+1}=-1$. Therefore, the
symmetry transformation
\eqref{eq:two copies symmetry mu 1} anticommutes with
$\widehat{P}^{\,}_{j}$. This anticommutation 
relation implies a nontrivial
second group cohomology class $[\phi]\neq 0$
of $G^{\,}_{f}$,
independent of the group of onsite symmetries of Hamiltonian
\eqref{eq:Hamiltonian local mu 1}\,
\footnote{
This anticommutation relation implies nontrivial index 
$[\rho]$ which is defined in Sec.\
\ref{subsec:Indices}.}. 
Therefore by Theorem \ref{thm:LSM Theorem 1}
Hamiltonian $\widehat{H}^{\prime}$ 
cannot have a nondegenerate gapped ground state.
This is in contradiction with the initial assumption that Hamiltonian 
\eqref{eq:Hamiltonian local mu 1} has the nondegenerate 
gapped ground state $\ket{\Psi^{\,}_{0}}$.

We observe that dimensionality $d$ of space played no role
in the proof of Theorem \ref{thm:LSM Theorem 2} until
Theorem \ref{thm:LSM Theorem 1} was used. Hence,
Theorem \ref{thm:LSM Theorem 2} holds for any $d$
if Theorem \ref{thm:LSM Theorem 1} holds for any $d$.

One can interpret Theorem \ref{thm:LSM Theorem 2} as the 
inability to write down an injective FMPS for the ground state of 
translationally invariant Hamiltonians with an odd number of
Majorana flavors per repeat unit cell.
This is because one cannot define the matrices $A^{\,}_{\sigma^{\,}_{j}}$
as there is no well-defined Fock space at site $j$ to begin with.

\subsection{LSM constraints and classification of 1D fermionic SPTs}
\label{subsec:LSM constraints and classification of 1D fermionic SPTs}

In order to prove Theorem \ref{thm:LSM Theorem 1},
we have shown that $\delta(g,h;b)$ defined in Eq.\
(\ref{eq:key step lemma iterated a}) vanishes.
Consequently, $W(g,h;b)$ defined in Eq.\
(\ref{eq:key step lemma iterated d})
must be proportional to the unit matrix $\openone^{\,}_{2M}$.
If so, the similarity transformations
$V(g;b)$, $V(h;b)$, and $V^{-1}(g\,h;b)$
that enter $W(g,h;b)$ must also realize a projective representation
of $G^{\,}_{f}$. This observation allows us to draw a bridge to the
classification of 1D fermionic SPT phases.

It is known that group cohomology classes corresponding to 
representations of $G^{\,}_{f}$ induced by similarity transformations $V(g)$
classify bosonic\,
\cite{Chen2011a,Schuch2011,Chen2011b}
and fermionic\,
\cite{Fidkowski2011,Bultinck2017,Williamson2016,Turzillo2019,Wang2020,Bourne2021}
SPT phases.  Similarly, for a given
symmetry group $G^{\,}_{f}$ and in 1D,
fermionic SPT phases are classified by a
triplet of indices $\big([(\nu,\rho)],\mu\big)$.
Indices $[(\nu,\rho)]$ are related to
$[\phi]$$\in$$H^{2}\big(G^{\,}_{f},\mathrm{U(1)}^{\,}_{\mathfrak{c}}\big)$.
The component $\nu$ of the indices $[(\nu,\rho)]$ encodes the information
about the projective representations of
the symmetry group $G$.
The component $\rho$ of the indices $[(\nu,\rho)]$ encodes 
the algebra between the representations of a group
element $g\in G$ and the fermion parity $p\in \mathbb{Z}^{F}_{2}$.
Finally, the index $\mu\in\{0,1\}$
characterizes the total fermion parity of the SPT ground state, or
equivalently the parity of the total
number of boundary Majorana modes.

Although the same cohomology group
$H^{2}\big(G^{\,}_{f},\mathrm{U(1)}^{\,}_{\mathfrak{c}}\big)$
appears in the classification of 1D LSM-type constraints and
1D SPT phases, they have a different origin.
For the 1D LSM-type constraints,
$H^{2}\big(G^{\,}_{f},\mathrm{U(1)}^{\,}_{\mathfrak{c}}\big)$
arises when classifying the projective representations of
$G^{\,}_{f}$ on a local Fock space.
For the 1D SPT phases,
$H^{2}\big(G^{\,}_{f},\mathrm{U(1)}^{\,}_{\mathfrak{c}}\big)$
arises when classifying the boundary projective representations
of global symmetries.

\section{Majorana LSM theorems in higher dimensions}
\label{sec:higher_LSM}

In this section, we extend Theorem \ref{thm:LSM Theorem 1}
to any dimension $d$ of space when the symmetry group $G^{\,}_{f}$ is Abelian
and all elements $g\in G^{\,}_{f}$ are represented by unitary operators.
Our method is inspired by the one used recently in
Ref.\ \onlinecite{Yao2021}
for quantum spin Hamiltonians.

Consider a $d$-dimensional lattice $\Lambda$
with periodic boundary conditions in each linearly independent
direction $\hat{\mu}=\hat{1},\cdots,\hat{d}$
such that $\Lambda$ realizes a $d$-torus.
Let each repeat unit cell be labeled as 
$\bm{j}$ and host a local fermionic Fock space 
$\mathcal{F}^{\,}_{\bm{j}}$
that is generated by a Majorana spinor
$\hat{\chi}^{\,}_{\bm{j}}$ with $2n$ components $\hat{\chi}^{\,}_{\bm{j},l}$,
$l=1,\cdots,2n$.
The fermionic Fock space attached to the lattice $\Lambda$
is denoted by $\mathcal{F}^{\,}_{\Lambda}$.
We impose the global symmetry corresponding to the
central extension $G^{\,}_{f}$ of $G$
by $\mathbb{Z}^{F}_{2}$ as defined in Sec.\
\ref{subsec:Marrying the fermion parity with the symmetry group G},
whereby $G^{\,}_{f}$ is assumed to be Abelian.
We also impose translation symmetry.
If the $d$-dimensional lattice $\Lambda$
has $N^{\,}_{\hat{\mu}}$ repeat unit cell in the
$\hat{\mu}$-direction
and thus the cardinality
\begin{equation}
|\Lambda|\equiv \prod_{\hat{\mu}=\hat{1}}^{\hat{d}}N^{\,}_{\hat{\mu}},
\end{equation}
the translation group is
\begin{equation}
G^{\,}_{\mathrm{trsl}}\equiv
\mathbb{Z}^{\,}_{N^{\,}_{\hat{1}}}\times\mathbb{Z}^{\,}_{N^{\,}_{\hat{2}}}
\times\cdots\times\mathbb{Z}^{\,}_{N^{\,}_{\hat{d}}}.
\label{eq: def translation symmetry group}
\end{equation}
By assumption, the combined symmetry group is the Cartesian product group
\begin{equation}
G^{\,}_{\mathrm{total}}\equiv
G^{\,}_{\mathrm{trsl}}\times G^{\,}_{f}.
\label{eq: def Gtotal}
\end{equation}
The representation of the translation group
(\ref{eq: def translation symmetry group})
is generated by the unitary operator 
$\widehat{T}^{\,}_{\hat{\mu}}$
whose action on the Majorana spinors is
\begin{subequations}
\label{eq: def rep Gtrsl direct product}
\begin{align}
\widehat{T}^{\vphantom{-1}}_{\hat{\mu}}\,
\hat{\chi}^{\,}_{\bm{j}}\,
\widehat{T}^{-1}_{\hat{\mu}}=
\hat{\chi}^{\,}_{\bm{j}+\bm{e}^{\,}_{\hat{\mu}}},
\qquad
\widehat{T}^{-1}_{\hat{\mu}}=\widehat{T}^{\dag}_{\hat{\mu}},
\label{eq: def rep Gtrsl direct product a}
\end{align}
along the $\hat{\mu}$-direction
($\bm{e}^{\,}_{\hat{\mu}}$ is a basis-vector along the 
$\hat{\mu}$-direction).
Imposing periodic boundary conditions implies
\begin{equation}
\left(\widehat{T}^{\,}_{\hat{\mu}}\right)^{N^{\,}_{\hat{\mu}}+1}=
\widehat{T}^{\,}_{\hat{\mu}}.
\label{eq: def rep Gtrsl direct product b}
\end{equation}
\end{subequations}
The representation 
$\widehat{U}(g)$ of $g\in G^{\,}_{f}$ is defined 
in Sec.\ \ref{subsec:Projective representations of the group Gf: I}.
Any translationally and $G^{\,}_{f}$-invariant
local Hamiltonian acting on 
$\mathcal{F}^{\,}_{\Lambda}$ can be written in the form
\begin{subequations}\label{eq:gen_Ham}
\begin{align}
\widehat{H}^{\,}_{\mathrm{pbc}}\df
\sum_{\hat{\mu}=\hat{1}}^{\hat{d}}
\sum_{n^{\,}_{\hat{\mu}}=1}^{N^{\,}_{\hat{\mu}}}
\left(\widehat{T}^{\vphantom{\dag}}_{\hat{\mu}}\right)^{n^{\,}_{\hat{\mu}}}\, 
\hat{h}^{\,}_{\bm{j}}\,
\left(\widehat{T}^{\dag}_{\hat{\mu}}\right)^{n^{\,}_{\hat{\mu}}},
\label{eq:gen_Ham a}
\end{align}
where $\hat{h}^{\,}_{\bm{j}}$
is a local Hermitian operator centered at an arbitrary repeat unit cell
$\bm{j}$. More precisely, it is a finite-order polynomial in 
the Majorana operators centered at $\bm{j}$ that is also 
invariant under all the non-spatial symmetries, i.e.,
\begin{align}
\hat{h}^{\vphantom{\dag}}_{\bm{j}}=
\widehat{U}(g)\,
\hat{h}^{\vphantom{\dag}}_{\bm{j}}\,
\widehat{U}^{-1}(g)=
\left(\hat{h}^{\,}_{\bm{j}}\right)^{\dag}
\label{eq:gen_Ham b}
\end{align}
\end{subequations}
for any $g\in G^{\,}_{f}$.
Instead of extracting spectral properties
of Hamiltonian $\widehat{H}^{\,}_{\mathrm{pbc}}$ directly, we shall
do so with the family of Hamiltonians indexed by
$g\in G^{\,}_{f}$ and given by 
\begin{align}
\widehat{H}^{\mathrm{tilt}}_{\mathrm{twis}}(g)
\df
\sum_{a=1}^{|\Lambda|}
\left(\widehat{T}^{\,}_{\hat{1}}(g)\right)^{a}\, 
\hat{h}^{\mathrm{tilt}}_{1}\,
\left(\widehat{T}^{-1}_{\hat{1}}(g)\right)^{a},
\label{eq: def tilted+twisted H}
\end{align}
where $\hat{h}^{\mathrm{tilt}}_{1}$ is a $G^{\,}_{f}$-symmetric
and local Hermitian operator and $\widehat{T}^{\,}_{\hat{1}}(g)$
is the  ``$g$-twisted translation operator'' to be defined shortly.
We shall derive LSM-like constraints for
$\widehat{H}^{\mathrm{tilt}}_{\mathrm{twis}}(g)$
and then explain why those LSM-like constraints also
apply to $\widehat{H}^{\,}_{\mathrm{pbc}}$.
To this end, we will explain what is
meant by the upper index ``tilt'' for tilted
and the lower index ``twis'' for twisted
and how $\widehat{H}^{\mathrm{tilt}}_{\mathrm{twis}}(g)$
and $\widehat{H}^{\,}_{\mathrm{pbc}}$ differ.

\subsection{Case of a $d=1$-dimensional lattice}

As a warm up, we first consider the one-dimensional case, i.e.,
$\Lambda\cong\mathbb{Z}^{\,}_{N}$. We impose two assumptions in addition to
those previously assumed. These are that every element in $G^{\,}_{f}$
is unitarily represented (\textbf{Assumption 5}) and that $G^{\,}_{f}$
is an Abelian group (\textbf{Assumption 6}).
These two assumptions were superfluous when proving Theorem
\ref{thm:LSM Theorem 1} using injective FMPS in Sec.\ \ref{sec:1D_LSM}.
This drawback is compensated by the possibility to extend the proof that
follows to any dimension $d$ of space.

Twisted boundary conditions
are implemented by defining the symmetry
twisted translation operator 
\begin{subequations}\label{eq: def twisted BCS d=1}
\begin{align}
\widehat{T}^{\,}_{\hat{1}}(g)
\df
\hat{v}^{\,}_{1}(g)\,
\widehat{T}^{\,}_{\hat{1}}
\end{align}
through its action
\begin{align}
\widehat{T}^{\,}_{\hat{1}}(g)\,
\hat{\chi}^{\,}_{j}\,
\widehat{T}^{-1}_{\hat{1}}(g)= 
\begin{cases}
(-1)^{\rho(g)}\,
\hat{\chi}^{\,}_{j+1},
&
\hbox{if $j\neq N$,} 
\\&\\
\hat{v}^{\,}_{1}(g)\,
\hat{\chi}^{\,}_{1}\,
\hat{v}^{-1}_{1}(g),
&
\hbox{if $j=N$,} 
\end{cases}
\end{align}
\end{subequations}
for $j=1,\cdots, N$, where $\rho(g)\in\{0,1\}\equiv\mathbb{Z}^{\,}_{2}$
is defined in Eq.\ (\ref{eq: def uj(g) d})
[see also Eq.\ (\ref{eq: def rho(g)})].
We then consider any Hamiltonian of the form
(\ref{eq: def tilted+twisted H}) where  
the operator $\hat{h}^{\mathrm{tilt}}_{1}$
{in Eq.\ (\ref{eq: def tilted+twisted H})}
is nothing but the operator $\hat{h}^{\,}_{\bm{j}}$
in Eq.\ (\ref{eq:gen_Ham a})
with $\Lambda$ restricted to a one-dimensional lattice.
Such a twisted boundary condition is equivalent to coupling
the Majorana operators to a background Abelian
gauge field with a holonomy $g\in G^{\,}_{f}$
around the spatial cycle.
The effect of turning on such a background field is that it
delivers the operator algebra
(see Appendix
\ref{appsec:Proof Theorem 1 with twisted boundary conditions})
\begin{subequations}\label{eq:nontrivial_symm_alg} 
\begin{equation}
\left[\widehat{T}^{\,}_{\hat{1}}(g)\right]^{N}=
\widehat{U}(g),
\qquad
g\in G^{\,}_{f}
\end{equation}
and 
\begin{align}
\widehat{U}(h)^{-1}\,
\widehat{T}^{\,}_{\hat{1}}(g)\,
\widehat{U}(h)
=
e^{\mathrm{i}\chi(g,h)}\,
\widehat{T}^{\,}_{\hat{1}}(g),
\quad
h\in G^{\,}_{f},
\label{eq:nontrivial_symm_alg a}
\end{align}
where
\begin{align}
\chi(g,h)\df
\phi(h,g)
-
\phi(g,h)
+
(N-1)\pi\,\rho(h)[\rho(g)+1].
\label{eq:nontrivial_symm_alg b}
\end{align}
\end{subequations}
The same algebra with $\rho(g)\equiv0$ for all $g\in G^{\,}_{f}$
was obtained by Yao and Oshikawa in Refs.\
\onlinecite{Yao2020a,Yao2021}.
The phase $\chi(g,h)$ is vanishing if and only 
if the second cohomology class $[\phi]$ is trivial
[see Appendix \ref{appsec:Proof Theorem 1 with twisted boundary conditions}].
As explained in
Sec.\ \ref{subsec:Projective representations of the group Gf: I},
we can trade the index $[\phi]$
with the indices $[(\nu,\rho)]$.

If $\chi(g,h)\hbox{ mod $2\pi$}$ is nonvanishing, 
one-dimensional representations of
\eqref{eq:nontrivial_symm_alg}
are not allowed. The ground state
of any Hamiltonian of the form
(\ref{eq: def tilted+twisted H})
is either degenerate or spontaneously breaks the symmetry
in the thermodynamic limit. We have
rederived the Theorem \ref{thm:LSM Theorem 1}
for the Abelian group $G^{\,}_{f}$ that is represented 
unitarily when twisted boundary conditions apply.

If we assume that the choice of
boundary conditions cannot change the ground-state degeneracy
when all excited states are separated from the ground states by
an energy gap, then the Theorem \ref{thm:LSM Theorem 1}
applies to all boundary conditions compatible with translation symmetry
that are imposed on the one-dimensional chain $\Lambda$ and,
in particular, to Hamiltonians of the form
(\ref{eq:gen_Ham}) with $\Lambda$ restricted to a one-dimensional lattice
that obey periodic boundary conditions.
A necessary condition for this assumption to hold
is that all correlation functions between local operators
decay sufficiently fast, a condition known to be an attribute of
any Hamiltonian with gapped ground states\, \cite{Hastings2006}.

We emphasize that, 
in rederiving Theorem \ref{thm:LSM Theorem 1}, we have taken 
(i) the group $G^{\,}_{f}$ to be Abelian and (ii) representation
$\hat{u}^{\,}_{j}(g)$ to be unitary for all $g\in G^{\,}_{f}$.
There exist several challenges in relaxing both of these assumptions. 
When the group is taken to be non-Abelian, one cannot consistently 
define a twisted Hamiltonian \eqref{eq: def tilted+twisted H}
that is invariant under both global symmetry transformations
$\widehat{U}(h)$
and symmetry twisted translation operators
$\widehat{T}^{\,}_{\hat{1}}(g)$
without imposing stricter constraints on local operators
$\hat{h}^{\mathrm{tilt}}_{1}$
than Eq.\ \eqref{eq: def tilted+twisted H}.
The challenges with imposing antiunitary
twisted boundary conditions with the group element $g\in G^{\,}_{f}$
are the following.
First, complex conjugation is
applied on all the states in the Fock space $\mathcal{F}^{\,}_{\Lambda}$.
This means that Hamiltonian (\ref{eq:gen_Ham})
can differ from Hamiltonian (\ref{eq: def tilted+twisted H})
through an extensive number of terms when $\mathfrak{c}(g)=-1$,
in which case it is not obvious to us how to safely tie
some spectral properties of
Hamiltonians (\ref{eq: def tilted+twisted H})
and
(\ref{eq:gen_Ham}).
Second, not all representations of the group $G^{\,}_{f}$
are either even or odd under complex conjugation, in which case
conjugation of $\widehat{T}^{\,}_{\hat{1}}(g)$
by $\widehat{U}(h)^{-1}$
need not result anymore in a mere phase factor multiplying
$\widehat{T}^{\,}_{\hat{1}}(g)$
when $\mathfrak{c}(g)=-1$. In view of this difficulty with interpreting antiunitary
twisted boundary conditions, we observe that the FMPS construction
of LSM-type constraints is more general than the one
using twisted boundary conditions.

\begin{figure}[t]
\begin{center}
\includegraphics[angle=0,width=0.4\textwidth]{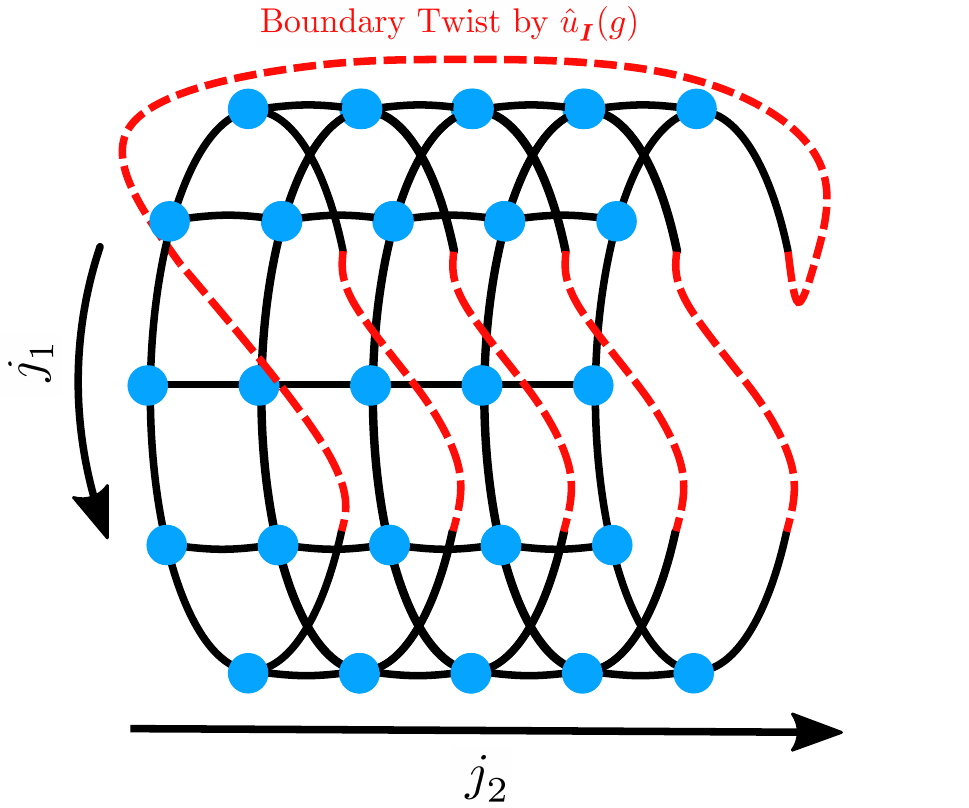}
\end{center}
\caption{
Example of a path that visits all the sites of
a two-dimensional lattice
that decorates the surface of a torus.
}
\label{Fig: tilted torus}
\end{figure}

\subsection{Case of a $d>1$-dimensional lattice}

We now assume that $\Lambda$ is a $d>1$-dimensional lattice.
We would like to generalize the twisted boundary conditions
(\ref{eq: def twisted BCS d=1})
obeyed by the Majorana operators to arbitrary spatial dimensions. 
There is no unique way for doing so. In what follows, we construct a 
group of translations $G^{\mathrm{tilt}}_{\mathrm{trsl}}$ that is cyclic.
This is achieved by imposing \textit{tilted} or 
\textit{sheared} boundary conditions.
After constructing 
$G^{\mathrm{tilt}}_{\mathrm{trsl}}$, we twist the boundary conditions in a particular 
way using the local representations of elements of
the on-site (internal) symmetry group $G^{\,}_{f}$. 
The operators representing translations on the lattice with
tilted and twisted boundary conditions
may not commute with the operators representing elements of $G^{\,}_{f}$,
even though all elements of $G^{\mathrm{tilt}}_{\mathrm{trsl}}$
commute with all elements of $G^{\,}_{f}$
by assumption (\ref{eq: def Gtotal}).
When this is so, the representation of
$G^{\mathrm{tilt}}_{\mathrm{total}}=G^{\mathrm{tilt}}_{\mathrm{trsl}}\times G^{\,}_{f}$
is necessarily larger than one dimensional, in which case
the ground states are either degenerate or the symmetry
group $G^{\mathrm{tilt}}_{\mathrm{total}}=G^{\mathrm{tilt}}_{\mathrm{trsl}}\times G^{\,}_{f}$
is spontaneously broken.

Our strategy is to construct the counterpart
of Eqs.\ (\ref{eq: def twisted BCS d=1})
and (\ref{eq:nontrivial_symm_alg}). To this end,
we are going to trade the translation symmetry group
(\ref{eq: def translation symmetry group}),
which is a polycyclic group when $d>1$,
for the cyclic group
\begin{equation}
G^{\mathrm{tilt}}_{\mathrm{trsl}}\equiv
\mathbb{Z}^{\,}_{N^{\,}_{\hat{1}}\cdots N^{\,}_{\hat{d}}}
\label{eq: def Gtrsl tilted}
\end{equation}
and define the combined symmetry group
\begin{equation}
G^{\mathrm{tilt}}_{\mathrm{total}}\equiv
G^{\mathrm{tilt}}_{\mathrm{trsl}}\times G^{\,}_{f}.
\label{eq: def Gtotal tilted}
\end{equation}
The intuition underlying the construction of the
tilted translation symmetry group
$G^{\mathrm{tilt}}_{\mathrm{trsl}}$
is provided by Fig.\ \ref{Fig: tilted torus}.
As a set, the elements of $G^{\mathrm{tilt}}_{\mathrm{trsl}}$
can be labeled by the elements of $G^{\,}_{\mathrm{trsl}}$,
namely
\begin{equation}
\begin{split}
G^{\mathrm{tilt}}_{\mathrm{trsl}}\df
\Big\{&
\Big(
(t^{\,}_{\hat{1}})^{n^{\,}_{\hat{1}}},
\cdots,
(t^{\,}_{\hat{d}})^{n^{\,}_{\hat{d}}}
\Big)\ \Big|\
\\
&\qquad
n^{\,}_{\hat{\mu}}=1,\cdots,N^{\,}_{\hat{\mu}},
\quad
\hat{\mu}=\hat{1},\cdots,\hat{d}
\Big\}.
\end{split}
\label{eq: def elements Gtilt,trsl as Cartesian product}
\end{equation}
However, as a group we would like to label the elements of
$G^{\mathrm{tilt}}_{\mathrm{trsl}}$
as those of the cyclic group with
$|\Lambda|$ elements, i.e.,
\begin{equation}
\label{eq:def titlted translation group}
\begin{split}
G^{\mathrm{tilt}}_{\mathrm{trsl}}\df
\Big\{
t^{n}
\ \Big|\
n=1,\cdots,|\Lambda|
\Big\}.
\end{split}
\end{equation}
This is achieved by carefully choosing the group composition
for the elements
(\ref{eq: def elements Gtilt,trsl as Cartesian product}), i.e.,
by iterating $d-1$ central extensions.

\noindent
\textbf{Step 1.} We consider
$\mathbb{Z}^{\,}_{N^{\,}_{\hat{1}}}$ generated by $t^{\,}_{\hat{1}}$
and extend it by 
$\mathbb{Z}^{\,}_{N^{\,}_{\hat{2}}}$ generated by $t^{\,}_{\hat{2}}$ through
the map
\begin{subequations}\label{eq: step 1 general d}
\begin{equation}
\begin{split}
&
\Theta^{\,}_{\hat{1}}:
\mathbb{Z}^{\,}_{N^{\,}_{\hat{1}}}\times
\mathbb{Z}^{\,}_{N^{\,}_{\hat{1}}}\to
\mathbb{Z}^{\,}_{N^{\,}_{\hat{2}}},
\\
&
\Theta^{\,}_{\hat{1}}
\Big((t^{\,}_{\hat{1}})^{a},(t^{\,}_{\hat{1}})^{b}\Big)\df
(t^{\,}_{\hat{2}})
^{\frac{1}{N^{\,}_{\hat{1}}}\left(a+b-[a+b]^{\,}_{N^{\,}_{\hat{1}}}\right)},
\end{split}
\end{equation}
for any $a,b=1,\cdots,N^{\,}_{\hat{1}}$, to obtain
$\mathbb{Z}^{\,}_{N^{\,}_{\hat{1}}\,N^{\,}_{\hat{2}}}$,
the group of translations on the tilted lattice restricted to
$\mathbb{R}^{2}$. Here, the notation $[a+b]^{\,}_{n}$
is used to denote addition modulo $n$. 
This group extension can be summarized by the short exact sequence   
\begin{align}
1
\rightarrow 
\mathbb{Z}^{\,}_{N^{\,}_{\hat{2}}}
\rightarrow 
\mathbb{Z}^{\,}_{N^{\,}_{\hat{1}}\,N^{\,}_{\hat{2}}} 
\rightarrow 
\mathbb{Z}^{\,}_{N^{\,}_{\hat{1}}} 
\rightarrow 
1
\end{align}
and is labeled by the extension classes
\begin{equation}
[\Theta^{\,}_{\hat{1}}]\in
H^{2}
\big(
\mathbb{Z}^{\,}_{N^{\,}_{\hat{1}}},\mathbb{Z}^{\,}_{N^{\,}_{\hat{2}}}
\big).
\end{equation}
\end{subequations}
Using this extension class and the standard expression for group
composition in an extended group, we may identify $t^{\,}_{\hat{1}}$
as the generator of $\mathbb{Z}^{\,}_{N^{\,}_{\hat{1}}\,N^{\,}_{\hat{2}}}$.

\noindent
\textbf{Step 2.} We consider
$\mathbb{Z}^{\,}_{N^{\,}_{\hat{1}}\,N^{\,}_{\hat{2}}}$ generated by
$t^{\,}_{\hat{1}}$
and extend it by 
$\mathbb{Z}^{\,}_{N^{\,}_{\hat{3}}}$ generated by $t^{\,}_{\hat{3}}$
through the map
\begin{subequations}\label{eq: step 2 general d}
\begin{equation}
\begin{split}
&
\Theta^{\,}_{\hat{2}}:
\mathbb{Z}^{\,}_{N^{\,}_{\hat{1}}\,N^{\,}_{\hat{2}}}\times
\mathbb{Z}^{\,}_{N^{\,}_{\hat{1}}\,N^{\,}_{\hat{2}}}\to
\mathbb{Z}^{\,}_{N^{\,}_{\hat{3}}},
\\
&
\Theta^{\,}_{\hat{2}}
\Big((t^{\,}_{\hat{1}})^{a},(t^{\,}_{\hat{1}})^{b}\Big)\df
(t^{\,}_{\hat{3}})
^{\frac{1}{N^{\,}_{\hat{1}}\,N^{\,}_{\hat{2}}}
\left(a+b-[a+b]^{\,}_{N^{\,}_{\hat{1}}\,N^{\,}_{\hat{2}}}\right)},
\end{split}
\end{equation}
for any $a,b=1,\cdots,N^{\,}_{\hat{1}}\,N^{\,}_{\hat{2}}$,
to obtain
$\mathbb{Z}^{\,}_{N^{\,}_{\hat{1}}\,N^{\,}_{\hat{2}}\,N^{\,}_{\hat{3}}}$
the group of translations on the tilted lattice restricted to
$\mathbb{R}^{3}$.
This group extension can be summarized by the short exact sequence   
\begin{align}
1
\rightarrow 
\mathbb{Z}^{\,}_{N^{\,}_{\hat{3}}}
\rightarrow 
\mathbb{Z}^{\,}_{N^{\,}_{\hat{1}}\,N^{\,}_{\hat{2}}\,N^{\,}_{\hat{3}}}  
\rightarrow 
\mathbb{Z}^{\,}_{N^{\,}_{\hat{1}}\,N^{\,}_{\hat{2}}} 
\rightarrow 
1
\end{align}
and is labeled by the extension classes
\begin{equation}
[\Theta^{\,}_{\hat{2}}]\in
H^{2}
\big(
\mathbb{Z}^{\,}_{N^{\,}_{\hat{1}}\,N^{\,}_{\hat{2}}},\mathbb{Z}^{\,}_{N^{\,}_{\hat{3}}}
\big).
\end{equation}
\end{subequations}
Using this extension class and the standard expression for group
composition in an extended group, we may identify $t^{\,}_{\hat{1}}$ as the
generator of
$\mathbb{Z}^{\,}_{N^{\,}_{\hat{1}}\,N^{\,}_{\hat{2}}\,N^{\,}_{\hat{3}}}$.

\noindent
\textbf{Step $d-1$.} We consider
$\mathbb{Z}^{\,}_{N^{\,}_{\hat{1}}\cdots N^{\,}_{\hat{d}-\hat{1}}}$
generated by $t^{\,}_{\hat{1}}$
and extend it by 
$\mathbb{Z}^{\,}_{N^{\,}_{\hat{d}}}$ generated by $t^{\,}_{\hat{d}}$ through
the map
\begin{subequations}\label{eq: step d-1 general d}
\begin{equation}
\begin{split}
&
\Theta^{\,}_{\hat{d}-\hat{1}}:
\mathbb{Z}^{\,}_{N^{\,}_{\hat{1}}\cdots N^{\,}_{\hat{d}-\hat{1}}}\times
\mathbb{Z}^{\,}_{N^{\,}_{\hat{1}}\cdots N^{\,}_{\hat{d}-\hat{1}}}\to
\mathbb{Z}^{\,}_{N^{\,}_{\hat{d}}},
\\
&
\Theta^{\,}_{\hat{d}-\hat{1}}
\Big((t^{\,}_{\hat{1}})^{a},(t^{\,}_{\hat{1}})^{b}\Big)\df
(t^{\,}_{\hat{d}})^{
\frac{1}{N^{\,}_{\hat{1}}\cdots N^{\,}_{\hat{d}-\hat{1}}}
\left(a+b-[a+b]^{\,}_{N^{\,}_{\hat{1}}\cdots N^{\,}_{\hat{d}-\hat{1}}}\right)
   },
\end{split}
\end{equation}
for any $a,b=1,\cdots,N^{\,}_{\hat{1}}\cdots N^{\,}_{\hat{d}-\hat{1}}$,
to obtain $\mathbb{Z}^{\,}_{N^{\,}_{\hat{1}}\cdots\,N^{\,}_{\hat{d}}}$
the group of translations on the tilted lattice $\Lambda$.
This group extension can be summarized by the short exact sequence   
\begin{align}
1
\rightarrow 
\mathbb{Z}^{\,}_{N^{\,}_{\hat{d}}
\rightarrow }
\mathbb{Z}^{\,}_{N^{\,}_{\hat{1}}\cdots N^{\,}_{\hat{d}}}  
\rightarrow 
\mathbb{Z}^{\,}_{N^{\,}_{\hat{1}}\cdots N^{\,}_{\hat{d}-\hat{1}}} 
\rightarrow 
1
\end{align}
and is labeled by the extension classes
\begin{equation}
[\Theta^{\,}_{\hat{d}-\hat{1}}]\in
H^{2}
\big(
\mathbb{Z}^{\,}_{N^{\,}_{\hat{1}}\cdots N^{\,}_{\hat{d}-\hat{1}}},
\mathbb{Z}^{\,}_{N^{\,}_{\hat{d}}}
\big).
\end{equation}
\end{subequations}
Using this extension class and the standard expression for group
composition in an extended group, we may identify $t^{\,}_{\hat{1}}$ as the
generator of $\mathbb{Z}^{\,}_{N^{\,}_{\hat{1}}\,\cdots N^{\,}_{\hat{d}}}$.

At the quantum level, we represent the cyclic group
(\ref{eq: def Gtrsl tilted})
by replacing Eq.\ (\ref{eq: def rep Gtrsl direct product}) with
\begin{subequations}\label{eq: def rep Gtrsl tilted}
\begin{align}
\widehat{T}^{\,}_{\hat{\mu}}\,
\hat{\chi}^{\,}_{\bm{j}}\,
\widehat{T}^{-1}_{\hat{\mu}}=
\hat{\chi}^{\,}_{t^{\,}_{\hat{\mu}}(\bm{j})},
\label{eq: def rep Gtrsl tilted a}
\end{align}
where $t^{\,}_{\hat{\mu}}(\bm{j})$ is the action of the cyclic group
$G^{\mathrm{tilt}}_{\mathrm{trsl}}$ on the repeat unit cell $\bm{j}\in\Lambda$.
The cyclicity of
$G^{\mathrm{tilt}}_{\mathrm{trsl}}\equiv
\mathbb{Z}^{\,}_{N^{\,}_{\hat{1}}\,\cdots N^{\,}_{\hat{d}}}\equiv
\mathbb{Z}^{\,}_{|\Lambda|}$
is enforced by
\begin{equation}
\left(\widehat{T}^{\,}_{\hat{\mu}}\right)^{N^{\,}_{\hat{\mu}}}=
\begin{cases}
\widehat{T}^{\,}_{\hat{\mu}+\hat{1}},&
\hbox{if $\hat{\mu}=\hat{1},\cdots,\hat{d}-\hat{1}$},
\\&\\
\widehat{\openone}^{\,}_{\Lambda},&
\hbox{if $\hat{\mu}=\hat{d}$}.
\end{cases}
\label{eq: def rep Gtrsl tilted b}
\end{equation}
\end{subequations}
With this convention,
$(\widehat{T}^{\,}_{\hat{1}})^{a}$
with $a=1,\cdots,|\Lambda|$
represents all the elements
$(t^{\,}_{\hat{1}})^{a}$ with $a=1,\cdots,|\Lambda|$
of $G^{\mathrm{tilt}}_{\mathrm{trsl}}$.
Equation (\ref{eq:gen_Ham})
is replaced by
\begin{subequations}\label{eq:gen_Ham bis}
\begin{align}
\widehat{H}^{\mathrm{tilt}}
\df
\sum_{a=1}^{|\Lambda|}
\left(\widehat{T}^{\,}_{\hat{1}}\right)^{a}\, 
\hat{h}^{\mathrm{tilt}}_{1}\,
\left(\widehat{T}^{-1}_{\hat{1}}\right)^{a},
\label{eq:gen_Ham bis a}
\end{align}
whereby
\begin{align}
\hat{h}^{\mathrm{tilt}}_{1}=
\widehat{U}(g)\,
\hat{h}^{\mathrm{tilt}}_{1}\,
\widehat{U}^{-1}(g)=
\left(\hat{h}^{\mathrm{tilt}}_{1}\right)^{\dag}
\label{eq:gen_Ham bis b}
\end{align}
\end{subequations}
holds for any $g\in G^{\,}_{f}$.
The locality of the polynomial
$\hat{h}^{\mathrm{tilt}}_{1}$
is no longer manifest when comparing the integers
that now label the local Majorana operators in
$\hat{h}^{\mathrm{tilt}}_{1}$.
The locality of $\hat{h}^{\mathrm{tilt}}_{1}$
is inherited from the fact
that $\hat{h}^{\,}_{\bm{j}}$
is local while
$\hat{h}^{\mathrm{tilt}}_{1}$
is nothing but a mere rewriting of
$\hat{h}^{\,}_{\bm{j}}$
in the cyclic representation of $\bm{j}\in\Lambda$.
Hamiltonian (\ref{eq:gen_Ham bis})
differs from Hamiltonian
(\ref{eq:gen_Ham})
by a sub-extensive number of terms
of order $|\Lambda|/N^{\,}_{\hat{1}}$.
The same number of terms would distinguish Hamiltonian
(\ref{eq:gen_Ham}) from the Hamiltonian
$\widehat{H}^{\,}_{\mathrm{twis}}(g)$ obtained by
replacing the periodic boundary conditions
(\ref{eq: def rep Gtrsl direct product b})
by twisted one, i.e., 
by multiplying the right-hand side of Eq.\
(\ref{eq: def rep Gtrsl direct product b})
with $\widehat{U}(g)$ for $g\neq e$ and $\hat{\mu}=\hat{1}$.

Because of the cyclicity of
$G^{\mathrm{tilt}}_{\mathrm{trsl}}\equiv
\mathbb{Z}^{\,}_{N^{\,}_{\hat{1}}\,\cdots N^{\,}_{\hat{d}}}\equiv
\mathbb{Z}^{\,}_{|\Lambda|}$
and of its quantum representation,
we can adapt the definition
(\ref{eq: def twisted BCS d=1})
for the twisted translation operator when $d=1$ to
that when $d>1$. We define for any $g\in G^{\,}_{f}$
with $\mathfrak{c}(g)=+1$
the generator of twisted translation
\begin{subequations}\label{eq: def twisted BCS d>1}
\begin{align}
\widehat{T}^{\,}_{\hat{1}}(g)\df
\hat{u}^{\,}_{\bm{I}}(g)\,
\widehat{T}^{\,}_{\hat{1}},
\qquad
\hat{u}^{-1}_{\bm{I}}(g)=
\hat{u}^{\dag}_{\bm{I}}(g),
\end{align}
through its action
\begin{align}
\widehat{T}^{\,}_{\hat{1}}(g)\,
\hat{\chi}^{\,}_{\bm{j}}\,
\widehat{T}^{-1}_{\hat{1}}(g)= 
\begin{cases}
(-1)^{\rho(g)}\,
\hat{\chi}^{\,}_{t^{\,}_{\hat{1}}(\bm{j})},
&
\hbox{if $\bm{j}\neq\bm{N}$,} 
\\&\\
\hat{u}^{\,}_{\bm{I}}(g)\,
\hat{\chi}^{\,}_{\bm{I}}\,
\hat{u}^{-1}_{\bm{I}}(g),
&
\hbox{if $\bm{j}=\bm{N}$,} 
\end{cases}
\end{align}
\end{subequations}
on any Majorana operator labeled by $\bm{j}\in\Lambda$.
Here, $\bm{I}\equiv (1,\cdots,1)\in\Lambda$,
$\bm{N}=(N^{\,}_{\hat{1}},\cdots,N^{\,}_{\hat{d}})\in\Lambda$,
and $\bm{j}=(n^{\,}_{\hat{1}},\cdots,n^{\,}_{\hat{d}})$
with $n^{\,}_{\hat{\mu}}=1,\cdots,N^{\,}_{\hat{\mu}}$.
One verifies that
these twisted translation operators satisfy the
twisted operator algebra
\begin{subequations}\label{eq:nontrivial_symm_alg d>1}
\begin{align}
\widehat{U}(h)^{-1}\,
\widehat{T}^{\,}_{\hat{1}}(g)\,
\widehat{U}(h)
=
e^{\mathrm{i}\chi(g,h)}\,
\widehat{T}^{\,}_{\hat{1}}(g),
\label{eq:nontrivial_symm_alg d>1 a}
\end{align}
where
\begin{align}
\chi(g,h)\df
\phi(g,h)
-
\phi(h,g)
+
(|\Lambda|-1)\pi\,\rho(h)[\rho(g)+1],
\label{eq:nontrivial_symm_alg d>1 b}
\end{align}
\end{subequations}
which is nothing but the algebra
(\ref{eq:nontrivial_symm_alg})
with the identification
$N\to|\Lambda|\equiv N^{\,}_{\hat{1}}\cdots N^{\,}_{\hat{d}}$.
Finally, we define the family of Hamiltonians
(\ref{eq: def tilted+twisted H})
that obey twisted boundary conditions.
The proof of Theorem \ref{thm:LSM Theorem 1} when $d>1$
for Hamiltonians of the form (\ref{eq: def tilted+twisted H})
is the same as that when $d=1$.
Because the family of Hamiltonians
(\ref{eq: def tilted+twisted H})
only differ from the family of Hamiltonians
(\ref{eq:gen_Ham})
obeying periodic boundary conditions by
a sub-extensive number of terms,
the LSM-like conditions
characterizing the existence of nondegenerate gapped ground states
valid for Hamiltonians of the form
(\ref{eq: def tilted+twisted H})
are conjectured to be also
valid for Hamiltonians of the form
(\ref{eq:gen_Ham}).

\subsection{Theorem \ref{thm:LSM Theorem 2} in $d>1$ dimensions}

We have extended Theorem \ref{thm:LSM Theorem 1} to any spatial
dimension $d$.  As discussed at the end of Sec.\ \ref{sec:LSM mu 1},
if Theorem \ref{thm:LSM Theorem 1} holds for any $d$, then so does
Theorem \ref{thm:LSM Theorem 2}.  It is nevertheless instructive to
provide an alternative
proof of Theorem \ref{thm:LSM Theorem 2} for any spatial dimension $d$
without relying on Theorem \ref{thm:LSM Theorem 1}.

We consider a $d$-dimensional lattice $\Lambda$ such that at each
repeat unit cell labeled by $\bm{j}\in \Lambda$, there exists a
Majorana spinor $\hat{\chi}^{\,}_{\bm{j}}$ with $2n+1$ components
$\hat{\chi}^{\,}_{\bm{j},l}$, $l=1,\cdots, 2n+1$.  To have a
well-defined total Fock space on lattice $\Lambda$, we set the total
number of sites $|\Lambda|$ in the lattice to be even.  On lattice
$\Lambda$, we impose the tilted translation symmetry group
$G^{\mathrm{tilt}}_{\mathrm{trsl}}$ defined in Eq.\
\eqref{eq:def titlted translation group}.
Let $\widehat{T}^{\,}_{\hat{1}}$ be the
representation of the generator of the cyclic group
$G^{\mathrm{tilt}}_{\mathrm{trsl}}$ with the action
\eqref{eq: def rep Gtrsl tilted}
on the Majorana spinors $\hat{\chi}^{\,}_{\bm{j}}$.

In terms of the Majorana spinors $\hat{\chi}^{\,}_{\bm{j}}$, 
the total fermion parity operator $\widehat{P}$ has
the representation
\begin{align}
\widehat{P}
\df
\mathrm{i}^{|\Lambda|/2}
\prod_{\bm{j}\in \Lambda}
\prod_{l =1}^{2n+1}
\hat{\chi}^{\,}_{\bm{j},l}.
\end{align}
Conjugation of the fermion parity operator $\widehat{P}$ by
the tilted translation operator $\widehat{T}^{\,}_{\hat{1}}$
delivers
\begin{align}
\label{eq:anticommutation parity transl}
\widehat{T}^{\vphantom{-1}}_{\hat{1}}\,
\widehat{P}\,
\widehat{T}^{-1}_{\hat{1}}
=
(-1)^{|\Lambda|-1}
\widehat{P}
=
-\widehat{P},
\end{align}
where we arrived at the last equality by noting that $|\Lambda|$ is an
even integer.  The factor $(-1)^{|\Lambda|-1}$ arises since each
spinor $\hat{\chi}^{\,}_{\bm{j}}$ consists of an odd number of Majorana
operators.  The nontrivial algebra
\eqref{eq:anticommutation parity transl}
implies that the ground state of any Hamiltonian that commutes with
$\widehat{P}$, the generators of the tilted translation group, and the
generators of $G^{\,}_{f}$
is either degenerate or spontaneously breaks translation
or $G^{\,}_{f}$ symmetry.  
If one assumes that the degeneracy of the ground states when gapped
is independent of the choice made for the boundary conditions,
we reproduce Theorem \ref{thm:LSM Theorem 2}.  We note that the algebra
\eqref{eq:anticommutation parity transl} was shown in
Ref.\ \onlinecite{Hsieh2016} for a one dimensional Majorana chain and
interpreted as the existence of Witten's quantum-mechanical
supersymmetry\, \cite{Witten1982}. 

\section{Examples}
\label{sec:examples}

All our results apply to any central extension $G^{\,}_{f}$
of the group $G$ by the group $\mathbb{Z}^{F}_{2}$
associated to the fermion parity.
Establishing LSM-type conditions requires
(i) constructing a projective representation of
$G^{\,}_{f}$
and 
(ii) verifying which one of the group cohomology classes
$[\phi]\in H^{2}\big(G^{\,}_{f},\mathrm{U(1)}^{\,}_{\mathfrak{c}}\big)$
is realized by this projective representation.
We have shown how the group cohomology classes
$[\phi]\in H^{2}\big(G^{\,}_{f},\mathrm{U(1)}^{\,}_{\mathfrak{c}}\big)$
are associated to the indices
$([(\nu,\rho)],\mu)$
with
$\nu\in C^{2}\big(G,\mathrm{U(1)}\big)$,
$\rho\in C^{1}\big(G,\mathbb{Z}^{\,}_{2}\big)$,
and
$\mu=0,1$ the evenness or oddness of the local number
of Majorana degrees of freedom (flavors).
It is impossible to proceed any further without
choosing the group $G$.

We shall choose the central extension $G^{\,}_{f}$
of the group $G$ by the group $\mathbb{Z}^{F}_{2}$
to be the split Abelian group
$G^{\,}_{f}=G\times\mathbb{Z}^{F}_{2}$ with $G=\mathbb{Z}^{T}_{2}$,
the split Abelian group
$G^{\,}_{f}=G\times\mathbb{Z}^{F}_{2}$
with $G=\mathbb{Z}^{\,}_{2}\times\mathbb{Z}^{\,}_{2}$,
and the nonsplit Abelian group
$G^{\,}_{f}=\mathbb{Z}^{FT}_{4}$
with $G=\mathbb{Z}^{T}_{2}$.
Their group cohomology is reviewed in
Appendix \ref{appsec:Group Cohomology}.
For the first two cases (split groups), we denote the corresponding indices
by $([\nu],[\rho],\mu)$ as explained in Sec.\ \ref{subsec:Indices}.

Given any one of these groups, we shall
define a global fermionic Fock space
$\mathcal{F}^{\,}_{\Lambda}=\mathcal{F}^{\,}_{0}\oplus\mathcal{F}^{\,}_{1}$
and construct a projective representation
that realizes the indices
$([(\nu,\rho)],\mu)$
labeling $H^{2}\big(G^{\,}_{f},\mathrm{U(1)}^{\,}_{\mathfrak{c}}\big)$.
The global fermionic Fock space
$\mathcal{F}^{\,}_{\Lambda}=\mathcal{F}^{\,}_{0}\oplus\mathcal{F}^{\,}_{1}$
is here always constructed from $n\,|\Lambda|$
Hermitian operators
\begin{subequations}
\begin{equation}
\hat{\chi}^{\,}_{\bm{j},a}=
\hat{\chi}^{\dag}_{\bm{j},a},
\qquad
\bm{j}\in\Lambda,
\qquad
a=1,\cdots,n,
\end{equation}
obeying the Majorana (Clifford) algebra
\begin{equation}
\left\{
\hat{\chi}^{\,}_{\bm{j},a},
\hat{\chi}^{\,}_{\bm{j}',a'}
\right\}=
2\delta^{\,}_{\bm{j},\bm{j}'}\,\delta^{\,}_{a,a'},
\
\bm{j},\bm{j}'\in\Lambda,
\
a,a'=1,\cdots,n.
\end{equation}
\end{subequations}
The index $\mu$ takes the value 0 when $n=2m$ is an even integer,
in which case the cardinality $|\Lambda|$ of the lattice $\Lambda$
is any positive integer and we may always define the fermionic creation and
annihilation operators
\begin{subequations}\label{eq:def fermions out Majoranas mu=0}
\begin{equation}
\hat{c}^{\dag}_{\bm{j},2b-1}\df
\frac{
\hat{\chi}^{\,}_{\bm{j},2b-1}
-
\mathrm{i}
\hat{\chi}^{\,}_{\bm{j},2b}
}{2},
\quad
\hat{c}^{\,}_{\bm{j},2b-1}\df
\frac{
\hat{\chi}^{\,}_{\bm{j},2b-1}
+
\mathrm{i}
\hat{\chi}^{\,}_{\bm{j},2b}
}{2},
\label{eq:def fermions out Majoranas mu=0 a}
\end{equation}
with $\bm{j}\in\Lambda$ and $b=1,\cdots,m$.
The local and global
fermionic Fock space
$\mathcal{F}^{\,}_{\bm{j}}$
and
$\mathcal{F}^{\,}_{\Lambda}$
are then
\begin{equation}
\begin{split}
\mathcal{F}^{\,}_{\bm{j}}\df
\mathrm{span}
\Bigg\{&
\prod_{b=1}^{m}
\left(\hat{c}^{\dag}_{\bm{j},2b-1}\right)^{n^{\,}_{\bm{j},2b-1}}\,\ket{0}
\ \Bigg|\
\\
&\quad
n^{\,}_{\bm{j},2b-1}=0,1,\quad
\hat{c}^{\,}_{\bm{j},2b-1}\,\ket{0}=0
\Bigg\}
\end{split}
\label{eq:def fermions out Majoranas mu=0 b}
\end{equation}
and
\begin{equation}
\begin{split}
\mathcal{F}^{\,}_{\Lambda}\df
\mathrm{span}
\Bigg\{&
\prod_{\bm{j}\in\Lambda}
\prod_{b=1}^{m}
\left(\hat{c}^{\dag}_{\bm{j},2b-1}\right)^{n^{\,}_{\bm{j},2b-1}}\,\ket{0}
\ \Bigg|\
\\
&\quad
n^{\,}_{\bm{j},2b-1}=0,1,\quad
\hat{c}^{\,}_{\bm{j},2b-1}\,\ket{0}=0
\Bigg\},
\end{split}
\label{eq:def fermions out Majoranas mu=0 c}
\end{equation}
\end{subequations}
respectively. In order to define the
operation of complex conjugation $\mathsf{K}$
on both the local and global Fock space, we define
\begin{subequations}
\begin{equation}
\mathsf{K}
\left(
z\,\hat{c}^{\dag}_{\bm{j},2b-1}
+
w\,\hat{c}^{\,}_{\bm{j}',2b'-1}
\right)
\mathsf{K}\df
z^{*}\,\hat{c}^{\dag}_{\bm{j},2b-1}
+
w^{*}\,\hat{c}^{\,}_{\bm{j}',2b'-1}
\end{equation}
for any pair of complex number $z,w\in\mathbb{C}$
(for any $\bm{j},\bm{j}'\in\Lambda$ and $b,b'=1,\cdots,m$)
and
\begin{equation}
\mathsf{K}\,\ket{0}\equiv\ket{0}.
\end{equation}
This implies the transformation law
\begin{equation}
\mathsf{K}\,\hat{\chi}^{\,}_{\bm{j},2b-1}\,\mathsf{K}=
+\hat{\chi}^{\,}_{\bm{j},2b-1},
\qquad
\mathsf{K}\,\hat{\chi}^{\,}_{\bm{j},2b}\,\mathsf{K}=
-\hat{\chi}^{\,}_{\bm{j},2b},
\end{equation}
\end{subequations}
for any $\bm{j}\in\Lambda$ and $b=1,\cdots,m$.

The index $\mu$ takes the value 1 when $n=2m+1$ is an odd integer,
in which case the cardinality $|\Lambda|$ of the lattice $\Lambda$
must be an even positive integer. For notational simplicity, we shall assume
that $\Lambda$ is bipartite, that is the disjoint union of two
interpenetrating sublattices
$\Lambda^{\,}_{A}$ and $\Lambda^{\,}_{B}$ such that all nearest neighbors
of the sites in $\Lambda^{\,}_{A}$ belong to $\Lambda^{\,}_{B}$ and vice versa.
We impose on $\Lambda$ the topology of a torus.
We define the fermionic creation and
annihilation operators

\begin{subequations}\label{eq:def fermions out Majoranas mu=1}
\begin{equation}
\hat{c}^{\dag}_{\bm{j},a}\df
\frac{
\hat{\chi}^{\,}_{\bm{j},a}
-
\mathrm{i}
\hat{\chi}^{\,}_{(\bm{j}+\bm{\mu}),a}
}{2},
\quad
\hat{c}^{\,}_{\bm{j},a}\df
\frac{
\hat{\chi}^{\,}_{\bm{j},a}
+
\mathrm{i}
\hat{\chi}^{\,}_{(\bm{j}+\bm{\mu}),a}
}{2},
\label{eq:def fermions out Majoranas mu=1 a}
\end{equation}
where $\bm{j}\in\Lambda^{\,}_{A}$, $a=1,\cdots,2m+1$,
and $\bm{\mu}$ is any fixed basis vectors spanning $\Lambda$.
The global fermionic Fock space
$\mathcal{F}^{\,}_{\Lambda}$
is then
\begin{equation}
\begin{split}
\mathcal{F}^{\,}_{\Lambda}\df
\mathrm{span}
\Bigg\{&
\prod_{\bm{j}\in\Lambda^{\,}_{A}}
\prod_{a=1}^{2m+1}
\left(\hat{c}^{\dag}_{\bm{j},a}\right)^{n^{\,}_{\bm{j},a}}\,\ket{0}
\ \Bigg|\
\\
&\quad
n^{\,}_{\bm{j},a}=0,1,\quad
\hat{c}^{\,}_{\bm{j},a}\,\ket{0}=0
\Bigg\}.
\end{split}
\label{eq:def fermions out Majoranas mu=1 b}
\end{equation}
\end{subequations}
In order to define the
operation of complex conjugation $\mathsf{K}$
on the global Fock space, we define
\begin{subequations}
\begin{equation}
\mathsf{K}
\left(
z\,\hat{c}^{\dag}_{\bm{j},a}
+
w\,\hat{c}^{\,}_{\bm{j}',a'}
\right)
\mathsf{K}\df
z^{*}\,\hat{c}^{\dag}_{\bm{j},a}
+
w^{*}\,\hat{c}^{\,}_{\bm{j}',a'}
\end{equation}
for any pair of complex number $z,w\in\mathbb{C}$
(for any $\bm{j},\bm{j}'\in\Lambda^{\,}_{A}$ and $a,a'=1,\cdots,2m+1$)
and
\begin{equation}
\mathsf{K}\,\ket{0}\equiv\ket{0}.
\end{equation}
This implies the transformation law
\begin{equation}
\mathsf{K}\,\hat{\chi}^{\,}_{\bm{j},a}\,\mathsf{K}=
+\hat{\chi}^{\,}_{\bm{j},a},
\qquad
\mathsf{K}\,\hat{\chi}^{\,}_{(\bm{j}+\bm{\mu}),a}\,\mathsf{K}=
-\hat{\chi}^{\,}_{(\bm{j}+\bm{\mu}),a},
\end{equation}
\end{subequations}
for any $\bm{j}\in\Lambda^{\,}_{A}$ and $a=1,\cdots,2m+1$.

Both for $\mu=0,1$, we shall assume a quantum dynamics governed by
Hamiltonians of the form
(\ref{eq:gen_Ham})
where $\hat{h}^{\,}_{\bm{j}}$
in Eq.\ (\ref{eq:gen_Ham b})
is a finite-order polynomial in the Majorana operators.
The order of each monomial entering this polynomial is necessarily even for 
$\mathbb{Z}^{F}_{2}$ to be a symmetry group.
The Hamiltonian is noninteracting if the order of $\hat{h}^{\,}_{\bm{j}}$
is two, interacting otherwise. The finiteness of the order guarantees
locality. We also introduce the notion of range of
$\hat{h}^{\,}_{\bm{j}}$ which is the largest separation
between the lattice indices of the Majorana operators
present in $\hat{h}^{\,}_{\bm{j}}$. If the range vanishes,
then Hamiltonian (\ref{eq:gen_Ham})
is the sum over $|\Lambda|$ commuting Hermitian operators,
in which case the spectrum of Hamiltonian (\ref{eq:gen_Ham})
is obtained by diagonalizing
$\hat{h}^{\,}_{\bm{j}}$.

For the split Abelian group
$G^{\,}_{f}=G\times\mathbb{Z}^{F}_{2}$ with $G=\mathbb{Z}^{T}_{2}$,
we find that only the projective representations of the group algebra
that belong to the trivial group cohomology class
can be realized by noninteracting fermions.
Any projective representation of the group algebra
that belong to a nontrivial group cohomology class
prohibits bilinear terms in the fermions in any
$G^{\,}_{f}$-symmetric Hamiltonian of the form (\ref{eq:gen_Ham}).
Such intrinsically interacting Hamiltonians are quantum perturbations of
classical Ising-type Hamiltonians.

For the split Abelian group
$G^{\,}_{f}=G\times\mathbb{Z}^{F}_{2}$
with $G=\mathbb{Z}^{\,}_{2}\times\mathbb{Z}^{\,}_{2}$
and the nonsplit group
$G^{\,}_{f}=\mathbb{Z}^{FT}_{4}$
with $G=\mathbb{Z}^{T}_{2}$,
we find that any projective representation of the group algebra
that belongs to a nontrivial group cohomology class
implies that any $G^{\,}_{f}$-symmetric Hamiltonian 
of the form (\ref{eq:gen_Ham})
is necessarily gapless when quadratic in the fermions.
For any one-dimensional lattice Hamiltonian,
Theorem \ref{thm:LSM Theorem 1}
then predicts that any
$G^{\,}_{f}$-symmetric interaction of the form (\ref{eq:gen_Ham})
that opens a spectral gap in the noninteracting spectrum
must break spontaneously at least one of the
symmetries responsible
for the noninteracting spectrum being gapless.

\subsection{One-dimensional space with the symmetry group
$\mathbb{Z}^{T}_{2}\times\mathbb{Z}^{F}_{2}$ for an even number of local
Majorana flavors}

The lattice is $\Lambda=\{1,\cdots,N\}$ with $N=2M$ an even integer
and the global fermionic Fock space $\mathcal{F}^{\,}_{\Lambda}$
is of dimension $2^{n\,M}$ with $n$ the number of local Majorana flavors.
By choosing the cardinality $|\Lambda|=2M$ to be even,
we make sure that the lattice is bipartite. This
allows to treat the two values of the index
$\mu\df\hbox{$n$ mod 2}$ in parallel.
The symmetry group
$G^{\,}_{f}\df\mathbb{Z}^{T}_{2}\times\mathbb{Z}^{F}_{2}$
is a split group. The group
$G\df\mathbb{Z}^{T}_{2}=\left\{e,\,t\right\}$ 
corresponds to reversal of time.

The local antiunitary representation $\hat{u}^{\,}_{j}(t)$
of reversal of time generates a projective representation of the group
$\mathbb{Z}^{T}_{2}$.
The local unitary representation $\hat{u}^{\,}_{j}(p)$
of the fermion parity $p$ generates a projective representation
of the group $\mathbb{Z}^{F}_{2}$.
According to Appendix \ref{appsubsec: ZT2 times ZF2}, 
all cohomologically distinct projective representations
of $\mathbb{Z}^{T}_{2}\times\mathbb{Z}^{F}_{2}$
are determined by the independent indices $[\nu]=0,1$ and $[\rho]=0,1$
through the relations\,
\footnote{
We have chosen the convention of always representing
the generator $p$ of $\mathbb{Z}^{F}_{2}$ by a Hermitian operator
according to Eq.\ (\ref{eq: def uj(g) e})}
\begin{subequations}\label{eq:indices for ZT2}
\begin{align}
&
\hat{u}^{\,}_{j}(t)\,
\hat{u}^{\,}_{j}(t)=
(-1)^{[\nu]}\,
\hat{u}^{\,}_{j}(e),
\label{eq:indices for ZT2 a}
\\
&
\hat{u}^{\,}_{j}(t)\,
\hat{u}^{\,}_{j}(p)=
(-1)^{[\rho]}\,
\hat{u}^{\,}_{j}(p)\,
\hat{u}^{\,}_{j}(t),
\label{eq:indices for ZT2 b}
\end{align}
This gives the four distinct group cohomology classes
\begin{equation}
([\nu],[\rho],0)\in
\left\{
(0,0,0),
(1,0,0),
(0,1,0),
(1,1,0)
\right\}.
\label{eq:indices for ZT2 c}
\end{equation}
\end{subequations}
All but the group cohomology class $([\nu],[\rho],0)=(1,0,0)$
can be realized using $n=2$ local Majorana flavors.
The group cohomology class $([\nu],[\rho],0)=(1,0,0)$ requires at least
$n=4$ local Majorana flavors for it to be realized.
We will start with the nontrivial projective representation
in the group cohomology class $([\nu],[\rho],0)=(1,1,0)$
that we shall represent using two local Majorana flavors.
We will then construct successively the 
projective representations
in the group cohomology classes
$([\nu],[\rho],0)=(1,0,0)$,
$([\nu],[\rho],0)=(0,1,0)$,
and
$([\nu],[\rho],0)=(0,0,0)$
by using the graded tensor product, i.e.,
by considering 4, 6, and 8 flavors of local Majoranas,
respectively. This will allow us to verify explicitly
the stacking rules of Sec.\ \ref{subsec:Stacking rules}
according to which Eq.\ (\ref{eq:stacking rel for indices})
simplifies to the rule
\begin{equation}
([\nu],[\rho],0)=
\left(
[\nu^{\,}_{1}]+[\nu^{\,}_{2}]+[\rho^{\,}_{1}][\rho^{\,}_{2}],
[\rho^{\,}_{1}]+[\rho^{\,}_{2}],0
\right)
\label{eq:Stacking rules Z2TxZ2F}
\end{equation}
when $G^{\,}_{f}=\mathbb{Z}^{T}_{2}\times\mathbb{Z}^{F}_{2}$.
The indices (\ref{eq:indices for ZT2 c})
will thus be shown to form the cyclic group
$\mathbb{Z}^{\,}_{4}$
with respect to the stacking rule (\ref{eq:Stacking rules Z2TxZ2F}).

\subsubsection{Group cohomology class $([\nu],[\rho],\mu)=(1,1,0)$}
\label{subsubsec:Cohomology class (1,1,0)}

The local fermionic Fock space $\mathcal{F}^{\,}_{j}$
of dimension $\mathcal{D}=2$
is generated by the doublet of Majorana operators
\begin{equation}
\hat{\chi}^{\,}_{j}\equiv
\begin{pmatrix}
\hat{\chi}^{\,}_{j,1}
\\
\hat{\chi}^{\,}_{j,2}
\end{pmatrix},
\qquad
j=1,\cdots,2M.
\label{eq:fundamental Majorana doublet for (1,1,0)}
\end{equation}
One verifies that
\begin{subequations}\label{eq:Z2T 110 rep}
\begin{align}
&    
\hat{u}^{\,}_{j}(t)\df
-
\mathrm{i}
\hat{\chi}^{\,}_{j,2}\,
\mathsf{K},
\label{eq:Z2T 110 rep a}
\\
&
\hat{u}^{\,}_{j}(p)\df
\mathrm{i}
\hat{\chi}^{\,}_{j,1}\,
\hat{\chi}^{\,}_{j,2},
\label{eq:Z2T 110 rep b}
\end{align}
realizes the projective algebra
(\ref{eq:indices for ZT2})
with
\begin{equation}
[\nu]=1,
\qquad
[\rho]=1.
\label{eq:Z2T 110 rep c}
\end{equation}
\end{subequations}
One verifies that the Majorana doublet
(\ref{eq:fundamental Majorana doublet for (1,1,0)})
is odd under conjugation by both $\hat{u}^{\,}_{j}(t)$ and $\hat{u}^{\,}_{j}(p)$.
Time-reversal symmetry forbids any Hermitian quadratic form
for the doublet (\ref{eq:fundamental Majorana doublet for (1,1,0)}).

The only Hamiltonian of the form (\ref{eq:gen_Ham})
that is of quartic order and of range $r=1$ is 
\begin{align}
\label{eq:Z2T 110 Ham}
\widehat{H}^{\,}_{\mathrm{pbc}}=
\lambda
\sum_{j=1}^{2M}
\hat{\chi}^{\,}_{j,1}\,
\hat{\chi}^{\,}_{j,2}\,
\hat{\chi}^{\,}_{j+1,1}\,
\hat{\chi}^{\,}_{j+1,2},
\qquad
\lambda\in\mathbb{R}.
\end{align}
This Hamiltonian is nothing but the sum 
\begin{subequations}\label{eq:Z2T 110 Ham bis}
\begin{align}
\widehat{H}^{\,}_{\mathrm{pbc}}=
\sum_{j=1}^{2M}
\hat{h}^{\,}_{j}
\label{eq:Z2T 110 Ham bis a}
\end{align}
over commuting operators
\begin{equation}
\hat{h}^{\,}_{j}\df
-
4\lambda
\left(
\hat{n}^{\,}_{j}-\frac{1}{2}
\right)
\left(
\hat{n}^{\,}_{j+1}-\frac{1}{2}
\right)
\label{eq:Z2T 110 Ham bis b}
\end{equation}
when expressed in terms of the fermion-number operator
\begin{equation}
\begin{split}
&
\hat{n}^{\,}_{j}=\hat{c}^{\dag}_{j}\,\hat{c}^{\,}_{j},
\qquad
\hat{n}^{\,}_{j}\,\hat{c}^{\dag}_{j'}\,\ket{0}=
\delta^{\,}_{j,j'}\,n^{\,}_{j'}\,\hat{c}^{\dag}_{j'}\,\ket{0},
\\
&
n^{\,}_{j}=0,1,
\qquad
j,j'\in\Lambda.
\end{split}
\label{eq:Z2T 110 Ham bis c}
\end{equation}
\end{subequations}
It is thus diagonalized in the fermion-number basis
\begin{subequations}
\begin{equation}
\mathcal{F}^{\,}_{\Lambda}=
\mathrm{span}
\{\ket{n^{\,}_{1},\cdots,n^{\,}_{2M}}\}
\end{equation}
in which it is represented by the classical Ising Hamiltonian
\begin{equation}
\widehat{H}^{(\mathrm{I})}_{\mathrm{pbc}}\df
-\lambda
\sum_{j=1}^{2M}
\sigma^{\,}_{j}\,
\sigma^{\,}_{j+1},
\qquad
\sigma^{\,}_{j}\df
2\,n^{\,}_{j}-1.
\label{eq:classical Ising Hamiltonian}
\end{equation}
\end{subequations}
The subspace $\mathcal{F}^{\,}_{\mathrm{gs}}$
in the global Fock space
$\mathcal{F}^{\,}_{\Lambda}=\mathcal{F}^{\,}_{0}\oplus\mathcal{F}^{\,}_{1}$
that is spanned by the linearly independent ground states
is twofold degenerate. It is spanned by either the ferromagnetic states
\begin{subequations}
\begin{equation}
\mathcal{F}^{\,}_{\mathrm{gs}}=
\mathrm{span}
\{
\ket{0,0,\cdots,0,0},
\ket{1,1,\cdots,1,1}
\}\subset\mathcal{F}^{\,}_{0}
\end{equation}
when $\lambda>0$ or the antiferromagnetic states
\begin{equation}
\mathcal{F}^{\,}_{\mathrm{gs}}=
\mathrm{span}
\{
\ket{1,0,\cdots,1,0},
\ket{0,1,\cdots,0,1}
\}\subset\mathcal{F}^{\,}_{M\,\mathrm{mod}\,2}
\end{equation}
\end{subequations}
when $\lambda<0$. Because we made sure that the lattice is bipartite
($|\Lambda|=N=2M$), $\mathcal{F}^{\,}_{\mathrm{gs}}$ is homogeneous\,%
\footnote{
This is not so when $|\Lambda|=N=2M+1$ is odd. 
}.
All ground states are separated from all excited states by the gap
$|2\lambda|$. The action of ``time reversal''
in the fermion representation is that of a ``particle-hole'' transformation
under which $\hat{n}^{\,}_{j}\mapsto(1-\hat{n}^{\,}_{j})$ and
$\sigma^{\,}_{j}\mapsto-\sigma^{\,}_{j}$. The action of parity
is trivial (the identity) in the fermion representation.
Reversal of time is broken spontaneously at zero temperature and 
in the thermodynamic limit in the sense that applying
either a uniform or staggered magnetic field that couples
to the Ising spins through a Zeeman coupling,
taking the thermodynamic limit, and switching off the
Zeeman coupling selects one of the two degenerate ground states
when $\lambda>0$ and $\lambda<0$, respectively.

The complexity of the Hamiltonian of the form (\ref{eq:gen_Ham})
that is of quartic order and of range $r=2$ increases dramatically.
For any cluster made of the three repeat unit cells $j-1$, $j$, and $j+1$,
one can construct three groups of five monomials. The first group
is made of
\begin{subequations}
\begin{align}
&
\widehat{X}^{\,}_{j;12|12|00}\df
\hat{\chi}^{\,}_{j-1,1}\,
\hat{\chi}^{\,}_{j-1,2}\,
\hat{\chi}^{\,}_{j,1}\,
\hat{\chi}^{\,}_{j,2},
\\
&
\widehat{X}^{\,}_{j;12|10|10}\df
\hat{\chi}^{\,}_{j-1,1}\,
\hat{\chi}^{\,}_{j-1,2}\,
\hat{\chi}^{\,}_{j,1}\,
\hat{\chi}^{\,}_{j+1,1},
\\
&
\widehat{X}^{\,}_{j;12|10|02}\df
\hat{\chi}^{\,}_{j-1,1}\,
\hat{\chi}^{\,}_{j-1,2}\,
\hat{\chi}^{\,}_{j,1}\,
\hat{\chi}^{\,}_{j+1,2},
\\
&
\widehat{X}^{\,}_{j;12|02|10}\df
\hat{\chi}^{\,}_{j-1,1}\,
\hat{\chi}^{\,}_{j-1,2}\,
\hat{\chi}^{\,}_{j,2}\,
\hat{\chi}^{\,}_{j+1,1},
\\
&
\widehat{X}^{\,}_{j;12|02|02}\df
\hat{\chi}^{\,}_{j-1,1}\,
\hat{\chi}^{\,}_{j-1,2}\,
\hat{\chi}^{\,}_{j,2}\,
\hat{\chi}^{\,}_{j+1,2}.
\end{align}
\end{subequations}
The second group is made of
$
\widehat{X}^{\,}_{j;12|00|12}
$,
$
\widehat{X}^{\,}_{j;10|10|12}
$,
$
\widehat{X}^{\,}_{j;10|02|12}
$,
$
\widehat{X}^{\,}_{j;02|10|12}
$,
and
$
\widehat{X}^{\,}_{j;02|02|12}
$.
The third group is made of
$
\widehat{X}^{\,}_{j;00|12|12}
$,
$
\widehat{X}^{\,}_{j;10|12|10}
$,
$
\widehat{X}^{\,}_{j;10|12|02}
$,
$
\widehat{X}^{\,}_{j;02|12|10}
$,
and
$
\widehat{X}^{\,}_{j;02|12|02}
$.
Because of translation invariance,
only one of the two monomials
$\widehat{X}^{\,}_{j;00|12|12}$
and
$\widehat{X}^{\,}_{j;12|12|00}$
needs to be accounted for.
We are left with the generic cluster Hamiltonian
\begin{align}
\begin{split}
\hat{h}^{\,}_{j}=&\,
\lambda^{\,}_{1}\,
\hat{\chi}^{\,}_{j-1,1}
\hat{\chi}^{\,}_{j-1,2}
\hat{\chi}^{\,}_{j,1}
\hat{\chi}^{\,}_{j,2}
\\
&\,
+
\lambda^{\,}_{2}\,
\hat{\chi}^{\,}_{j-1,1}
\hat{\chi}^{\,}_{j-1,2}
\hat{\chi}^{\,}_{j+1,1}
\hat{\chi}^{\,}_{j+1,2}
\\
&\,\qquad
+
\lambda^{\,}_{3}\,
\hat{\chi}^{\,}_{j-1,1}
\hat{\chi}^{\,}_{j-1,2}
\hat{\chi}^{\,}_{j,1}
\hat{\chi}^{\,}_{j+1,1}
\\
&\,\qquad\qquad\qquad\qquad\ddots
\\
&\,\qquad\qquad
+
\lambda^{\,}_{14}\,
\hat{\chi}^{\,}_{j-1,2}\,
\hat{\chi}^{\,}_{j,1}\,
\hat{\chi}^{\,}_{j,2}\,
\hat{\chi}^{\,}_{j+1,2}
\end{split}
\end{align}
with fourteen real-valued couplings.
If we set the twelve couplings 
$\lambda^{\,}_{3}=\cdots\lambda^{\,}_{14}=0$
to zero, we obtain the classical Ising model
\begin{equation}
\widehat{H}^{\mathrm{I}'}_{\mathrm{pbc}}\df
-
\sum_{j=1}^{2M}
\left(
\lambda^{\,}_{1}\,
\sigma^{\,}_{j}\,
\sigma^{\,}_{j+1}
+
\lambda^{\,}_{2}\,
\sigma^{\,}_{j}\,
\sigma^{\,}_{j+2}
\right)
\label{eq:classical Ising Hamiltonian bis}
\end{equation}
in the same fermion-number basis of the Fock space
that delivered the classical Ising Hamiltonian
(\ref{eq:classical Ising Hamiltonian}).
Any perturbation with one of the couplings
$\lambda^{\,}_{3}$, $\cdots$, $\lambda^{\,}_{14}$
breaks the fermion-number conservation down to the conservation
of the fermion-parity number. Any gapped phase in the
14-dimensional coupling space is predicted by
Theorem \ref{thm:LSM Theorem 1} 
to be either degenerate or break spontaneously time-reversal
or translation symmetry.
This prediction can be verified explicitly for
the classical Ising model (\ref{eq:classical Ising Hamiltonian bis})
with nearest- and next-nearest-neighbor interactions.

\subsubsection{Group cohomology class $([\nu],[\rho],\mu)=(1,0,0)$}
\label{subsubsec:Cohomology class (1,0,0)}

The local fermionic Fock space $\mathcal{F}^{\,}_{j}$
of dimension $\mathcal{D}=4$
is generated by the quartet of Majorana operators\,
\footnote{
It is not possible to represent the
group cohomology class $([\nu],[\rho],\mu)=(1,0,0)$
with a doublet of Majorana operators.
}
\begin{equation}
\hat{\chi}^{\,}_{j}\equiv
\begin{pmatrix}
\hat{\chi}^{\,}_{j,1}&
\cdots&
\hat{\chi}^{\,}_{j,4}
\end{pmatrix}^{\mathsf{T}},
\qquad
j=1,\cdots,2M.
\label{eq:Majorana quartet for (1,0,0)}
\end{equation}
One verifies that
\begin{subequations}\label{eq:Z2T 100 rep}
\begin{align}
&    
\hat{u}^{\,}_{j}(t)\df
-
\mathrm{i}
\hat{\chi}^{\,}_{j,2}\,
\hat{\chi}^{\,}_{j,4}\,
\mathsf{K},
\label{eq:Z2T 100 rep a}
\\
&
\hat{u}^{\,}_{j}(p)\df
\hat{\chi}^{\,}_{j,1}\,
\hat{\chi}^{\,}_{j,2}\,
\hat{\chi}^{\,}_{j,3}\,
\hat{\chi}^{\,}_{j,4},
\label{eq:Z2T 100 rep b}
\end{align}
realizes the projective algebra
(\ref{eq:indices for ZT2})
with
\begin{equation}
[\nu]=1,
\qquad
[\rho]=0.
\label{eq:Z2T 100 rep c}
\end{equation}
\end{subequations}
One verifies that the Majorana quartet
(\ref{eq:Majorana quartet for (1,0,0)})
is even under conjugation by $\hat{u}^{\,}_{j}(t)$
and odd under conjugation by $\hat{u}^{\,}_{j}(p)$.
Time-reversal symmetry forbids any Hermitian quadratic form
for the quartet (\ref{eq:Majorana quartet for (1,0,0)}).

The only Hamiltonian of the form (\ref{eq:gen_Ham})
that is of quartic order and of range $r=0$ is 
\begin{align}
\label{eq:Z2T 100 Ham}
\widehat{H}^{\,}_{\mathrm{pbc}}=
\lambda
\sum_{j=1}^{2M}
\hat{\chi}^{\,}_{j,1}\,
\hat{\chi}^{\,}_{j,2}\,
\hat{\chi}^{\,}_{j,3}\,
\hat{\chi}^{\,}_{j,4},
\qquad
\lambda\in\mathbb{R}.
\end{align}
This Hamiltonian is nothing but the sum 
\begin{subequations}\label{eq:Z2T 100 Ham bis}
\begin{align}
\widehat{H}^{\,}_{\mathrm{pbc}}=
\sum_{j=1}^{2M}
\hat{h}^{\,}_{j}
\label{eq:Z2T 100 Ham bis a}
\end{align}
over commuting operators
\begin{equation}
\hat{h}^{\,}_{j}\df
-
4\lambda
\left(
\hat{n}^{\,}_{j,1}-\frac{1}{2}
\right)
\left(
\hat{n}^{\,}_{j,3}-\frac{1}{2}
\right)
\label{eq:Z2T 100 Ham bis b}
\end{equation}
when expressed in terms of the fermion-number operator
\begin{equation}
\begin{split}
&
\hat{n}^{\,}_{j,2b-1}\df\hat{c}^{\dag}_{j,2b-1}\,\hat{c}^{\,}_{j,2b-1},
\\
&
\hat{n}^{\,}_{j,2b-1}\,\hat{c}^{\dag}_{j',2b'-1}\ket{0}=
\delta^{\,}_{j,j'}\,\delta^{\,}_{b,b'}\,
n^{\,}_{j',2b'-1}\,\hat{c}^{\dag}_{j',2b'-1}\ket{0},
\\
&
n^{\,}_{j,2b-1}=0,1,
\qquad
j,j'\in\Lambda,
\qquad
b,b'=1,2.
\end{split}
\label{eq:Z2T 100 Ham bis c}
\end{equation}
\end{subequations}
It is thus diagonalized in the fermion-number basis
\begin{subequations}
\begin{equation}
\mathcal{F}^{\,}_{\Lambda}=
\mathrm{span}
\left\{
\left|
\begin{tiny}
\begin{pmatrix}n^{\,}_{1,1}\\n^{\,}_{1,3}\end{pmatrix},
\cdots,
\begin{pmatrix}n^{\,}_{2M,1}\\n^{\,}_{2M,3}\end{pmatrix}
\end{tiny}
\right\rangle
\right\}    
\end{equation}
in which it is represented by two Ising chains
(labeled 1 and 3) coupled through
their rungs only (but not along the chains),
\begin{equation}
\begin{split}
&
\widehat{H}^{(\mathrm{Irung})}_{\mathrm{pbc}}\df
-\lambda
\sum_{j=1}^{2M}
\sigma^{\,}_{j,1}\,
\sigma^{\,}_{j,3},
\\
&
\sigma^{\,}_{j,2b-1}\df
2\,n^{\,}_{j,2b-1}-1,
\qquad
b=1,2.
\end{split}
\label{eq:classical Ising Hamiltonian 100}
\end{equation}
\end{subequations}
The subspace $\mathcal{F}^{\,}_{\mathrm{gs}}$
in the global Fock space
$\mathcal{F}^{\,}_{\Lambda}=\mathcal{F}^{\,}_{0}\oplus\mathcal{F}^{\,}_{1}$
that is spanned by the linearly independent ground states
is $2^{2M}$-fold degenerate. It is spanned by either
\begin{subequations}
\begin{equation}
\mathcal{F}^{\,}_{\mathrm{gs}}=
\mathrm{span}
\{
\ket{
\begin{tiny}
\begin{pmatrix}0\\0\end{pmatrix},\cdots,\begin{pmatrix}0\\0\end{pmatrix}
\end{tiny}},\cdots,
\ket{
\begin{tiny}
\begin{pmatrix}1\\1\end{pmatrix},\cdots,\begin{pmatrix}1\\1\end{pmatrix}
\end{tiny}}
\}\subset\mathcal{F}^{\,}_{0}
\end{equation}
when $\lambda>0$ or the antiferromagnetic states
\begin{equation}
\mathcal{F}^{\,}_{\mathrm{gs}}=
\mathrm{span}
\{
\ket{
\begin{tiny}
\begin{pmatrix}0\\1\end{pmatrix},\cdots,\begin{pmatrix}0\\1\end{pmatrix}
\end{tiny}},\cdots,
\ket{
\begin{tiny}
\begin{pmatrix}1\\0\end{pmatrix},\cdots,\begin{pmatrix}1\\0\end{pmatrix}
\end{tiny}}
\}\subset\mathcal{F}^{\,}_{0}
\end{equation}
\end{subequations}
when $\lambda<0$.
All ground states are separated from all excited states by the gap
$|2\lambda|$. The action of ``time reversal''
in the fermion representation is that of a ``particle-hole'' transformation
under which $\hat{n}^{\,}_{j,2b-1}\mapsto(1-\hat{n}^{\,}_{j,2b-1})$
and
$\sigma^{\,}_{j,2b-1}\mapsto-\sigma^{\,}_{j,2b-1}$. The action of parity
is trivial (the identity) in the fermion representation.

The complexity of the Hamiltonian of the form (\ref{eq:gen_Ham})
that is of quartic order and of range $r=1$ increases dramatically.
For any cluster made of the two repeat unit cells $j$ and $j+1$,
the generic cluster Hamiltonian $\hat{h}^{\,}_{j}$
that is summed over in Hamiltonian (\ref{eq:gen_Ham})
is the sum over 70 (choose 4 out of 8) monomials of the form
\begin{equation}
\hat{\chi}^{\,}_{j,a}\,
\hat{\chi}^{\,}_{j,b}\,
\hat{\chi}^{\,}_{j+1,c}\,
\hat{\chi}^{\,}_{j+1,d},
\quad
1\leq a<b\leq4,
\quad
1\leq c<d\leq4,
\end{equation}
each one weighted by a real-valued coupling. Of these 70 couplings,
69 are independent by translation symmetry. Three of the 69 monomials
are compatible with fermion-number conservation. All other monomials
break the fermion-number conservation down to the conservation of the
fermion-parity number. If these 66 couplings are set to zero
and the remaining three couplings are set to $\lambda$,
we obtain the classical Ising ladder
\begin{equation}
\begin{split}
\widehat{H}^{(\mathrm{Iladder})}_{\mathrm{pbc}}\df&\,
-
\lambda
\sum_{j=1}^{2M}
\left(
\sigma^{\,}_{j,1}\,
\sigma^{\,}_{j,3}
+
\sum_{b=1,2}
\sigma^{\,}_{j,2b-1}\,
\sigma^{\,}_{j,2b-1}
\right)
\end{split}
\label{eq:classical Ising Hamiltonian bis bis}
\end{equation}
in the same fermion-number basis of the Fock space
that delivered Hamiltonian
(\ref{eq:classical Ising Hamiltonian 100}).
Any gapped phase in the
66-dimensional coupling space is predicted by
Theorem \ref{thm:LSM Theorem 1} 
to be either degenerate or break spontaneously 
time-reversal or translation symmetry.
This prediction can be verified explicitly for
the classical Ising ladder (\ref{eq:classical Ising Hamiltonian bis bis})
for which the subspace $\mathcal{F}^{\,}_{\mathrm{gs}}$
in the global Fock space
$\mathcal{F}^{\,}_{\Lambda}=\mathcal{F}^{\,}_{0}\oplus\mathcal{F}^{\,}_{1}$
that is spanned by the linearly independent ground states
is twofold degenerate. It is spanned by either
\begin{subequations}
\begin{equation}
\mathcal{F}^{\,}_{\mathrm{gs}}=
\mathrm{span}
\{
\ket{
\begin{tiny}
\begin{pmatrix}0\\0\end{pmatrix},\cdots,\begin{pmatrix}0\\0\end{pmatrix}
\end{tiny}},
\ket{
\begin{tiny}
\begin{pmatrix}1\\1\end{pmatrix},\cdots,\begin{pmatrix}1\\1\end{pmatrix}
\end{tiny}}
\}\subset\mathcal{F}^{\,}_{0}
\end{equation}
when $\lambda>0$ or the antiferromagnetic states
\begin{equation}
\mathcal{F}^{\,}_{\mathrm{gs}}=
\mathrm{span}
\{
\ket{
\begin{tiny}
\begin{pmatrix}0\\1\end{pmatrix},\cdots,\begin{pmatrix}0\\1\end{pmatrix}
\end{tiny}},
\ket{
\begin{tiny}
\begin{pmatrix}1\\0\end{pmatrix},\cdots,\begin{pmatrix}1\\0\end{pmatrix}
\end{tiny}}
\}\subset\mathcal{F}^{\,}_{0}
\end{equation}
\end{subequations}
when $\lambda<0$.
Reversal of time is broken spontaneously at zero temperature and
in the thermodynamic limit in the sense that applying
either a uniform or staggered (within the repeat unit cell)
magnetic field that couples
to the Ising spins through a Zeeman coupling,
taking the thermodynamic limit, and switching off the
Zeeman coupling selects one of the two degenerate ground states
when $\lambda>0$ and $\lambda<0$, respectively.

\subsubsection{Group cohomology class $([\nu],[\rho],\mu)=(0,1,0)$}
\label{subsubsec:Cohomology class (0,1,0)}

The local fermionic Fock space $\mathcal{F}^{\,}_{j}$
of dimension $\mathcal{D}=8$
is generated by the sextet of Majorana operators
\begin{equation}
\hat{\chi}^{\,}_{j}\equiv
\begin{pmatrix}
\hat{\chi}^{\,}_{j,1}&
\cdots&
\hat{\chi}^{\,}_{j,6}
\end{pmatrix}^{\mathsf{T}},
\qquad
j=1,\cdots,2M.
\label{eq:Majorana quartet for (0,1,0)}
\end{equation}
One verifies that
\begin{subequations}\label{eq:Z2T 010 rep}
\begin{align}
&    
\hat{u}^{\,}_{j}(t)\df
-
\mathrm{i}
\hat{\chi}^{\,}_{j,2}\,
\hat{\chi}^{\,}_{j,4}\,
\hat{\chi}^{\,}_{j,6}\,
\mathsf{K},
\label{eq:Z2T 010 rep a}
\\
&
\hat{u}^{\,}_{j}(p)\df
\mathrm{i}
\hat{\chi}^{\,}_{j,1}\,
\hat{\chi}^{\,}_{j,2}\,
\hat{\chi}^{\,}_{j,3}\,
\hat{\chi}^{\,}_{j,4}\,
\hat{\chi}^{\,}_{j,5}\,
\hat{\chi}^{\,}_{j,6},
\label{eq:Z2T 010 rep b}
\end{align}
realizes the projective algebra
(\ref{eq:indices for ZT2})
with
\begin{equation}
[\nu]=0,
\qquad
[\rho]=1.
\label{eq:Z2T 010 rep c}
\end{equation}
\end{subequations}
One verifies that the Majorana sextet
(\ref{eq:Majorana quartet for (0,1,0)})
is odd under conjugation by both $\hat{u}^{\,}_{j}(t)$ and $\hat{u}^{\,}_{j}(p)$.
Time-reversal symmetry forbids any Hermitian quadratic form
for the quartet (\ref{eq:Majorana quartet for (0,1,0)}).

The generic Hamiltonian of the form (\ref{eq:gen_Ham})
that is of quartic order and of range $r=0$ is 
\begin{equation}
\widehat{H}^{\,}_{\mathrm{pbc}}\df
\sum_{j=1}^{2M}
\sum_{1\leq a^{\,}_{1}<a^{\,}_{2}<a^{\,}_{3}<a^{\,}_{4}\leq6}
\lambda^{\,}_{a^{\,}_{1}\,a^{\,}_{2}\,a^{\,}_{3}\,a^{\,}_{4}}\,
\prod_{i=1}^{4}
\hat{\chi}^{\,}_{j,a^{\,}_{i}}
\end{equation}
with $\lambda^{\,}_{a^{\,}_{1}\,a^{\,}_{2}\,a^{\,}_{3}\,a^{\,}_{4}}$
a real-valued coupling. Each monomial
\begin{equation}
\widehat{X}^{\,}_{j|a^{\,}_{1}\,a^{\,}_{2}\,a^{\,}_{3}\,a^{\,}_{4}}\df
\prod_{i=1}^{4}
\hat{\chi}^{\,}_{j,a^{\,}_{i}},
\quad
1\leq a^{\,}_{1}<a^{\,}_{2}<a^{\,}_{3}<a^{\,}_{4}\leq6,
\label{eq:Z2T 010 Ham}
\end{equation}
is a Hermitian operator with the eigenvalues $\pm1$
as it squares to the identity. Any two such monomials
on the right-hand side of Eq.\ (\ref{eq:Z2T 010 Ham})
either commute or anticommute.
In the fermion-number basis
(\ref{eq:def fermions out Majoranas mu=0}),
the 3 monomials
\begin{subequations}\label{eq:Ising monomials 010}
\begin{align}
&
\widehat{X}^{\,}_{j|1\,2\,3\,4}\df
-4
\left(
\hat{n}^{\,}_{j,1}-\frac{1}{2}
\right)
\left(
\hat{n}^{\,}_{j,3}-\frac{1}{2}
\right),
\\
&
\widehat{X}^{\,}_{j|1\,2\,5\,6}\df
-4
\left(
\hat{n}^{\,}_{j,1}-\frac{1}{2}
\right)
\left(
\hat{n}^{\,}_{j,5}-\frac{1}{2}
\right),
\\
&
\widehat{X}^{\,}_{j|3\,4\,5\,6}\df
-4
\left(
\hat{n}^{\,}_{j,3}-\frac{1}{2}
\right)
\left(
\hat{n}^{\,}_{j,5}-\frac{1}{2}
\right),
\end{align}
\end{subequations}
are the only ones that are compatible with
conservation of the fermion-number.
All remaining 12 monomials break the
conservation of the fermion-number in the basis
(\ref{eq:def fermions out Majoranas mu=0})
down to that of the fermion-parity number.
Any gapped phase in the
15-dimensional coupling space is predicted by
Theorem \ref{thm:LSM Theorem 1} 
to be either degenerate or break spontaneously time-reversal
or translation symmetry.
This prediction can be verified explicitly
by summing with the same weight the three monomials defined in
Eq.\ (\ref{eq:Ising monomials 010}).
One obtains three Ising chains coupled through their rungs only.
The ground state is then $2^{2M}$-fold degenerate. This macroscopic
degeneracy becomes twofold by increasing the range from $r=0$ to $r=1$
and considering the three Ising chains (labeled 1, 3, 5)
with the Hamiltonian [in the fermion-number basis
(\ref{eq:def fermions out Majoranas mu=0})]
\begin{equation}
\begin{split}
\widehat{H}^{(\mathrm{I3chains})}_{\mathrm{pbc}}\df&\,
-
\lambda
\sum_{b=1}^{3}\
\sum_{j=1}^{2M}
\sigma^{\,}_{j,2b-1}\,
\sigma^{\,}_{j+1,2b-1}
\\
&\,
-
\lambda
\sum_{j=1}^{2M}
\left(
\sigma^{\,}_{j,1}\,
\sigma^{\,}_{j,3}
+
\sigma^{\,}_{j,3}\,
\sigma^{\,}_{j,5}
\right)
\end{split}
\label{eq:classical Ising Hamiltonian bis bis bis}
\end{equation}
with $\lambda\in\mathbb{R}$, say.
The subspace $\mathcal{F}^{\,}_{\mathrm{gs}}$
in the global Fock space
$\mathcal{F}^{\,}_{\Lambda}=\mathcal{F}^{\,}_{0}\oplus\mathcal{F}^{\,}_{1}$
that is spanned by the linearly independent ground states
is either
\begin{subequations}
\begin{equation}
\mathcal{F}^{\,}_{\mathrm{gs}}=
\mathrm{span}
\left\{
\left|
\begin{tiny}
\begin{pmatrix}0\\0\\0\end{pmatrix},\cdots,\begin{pmatrix}0\\0\\0\end{pmatrix}
\end{tiny}
\right\rangle,
\left|
\begin{tiny}
\begin{pmatrix}1\\1\\1\end{pmatrix},\cdots,\begin{pmatrix}1\\1\\1\end{pmatrix}
\end{tiny}
\right\rangle
\right\}\subset\mathcal{F}^{\,}_{0}
\end{equation}
when $\lambda>0$ or the antiferromagnetic states
\begin{equation}
\mathcal{F}^{\,}_{\mathrm{gs}}=
\mathrm{span}
\left\{
\left|
\begin{tiny}
\begin{pmatrix}0\\1\\0\end{pmatrix},\cdots,\begin{pmatrix}0\\1\\0\end{pmatrix}
\end{tiny}
\right\rangle,
\left|
\begin{tiny}
\begin{pmatrix}1\\0\\1\end{pmatrix},\cdots,\begin{pmatrix}1\\0\\1\end{pmatrix}
\end{tiny}
\right\rangle
\right\}\subset\mathcal{F}^{\,}_{0}
\end{equation}
\end{subequations}
when $\lambda<0$.
Reversal of time is broken spontaneously at zero temperature and
in the thermodynamic limit in the sense that applying
either a uniform or staggered (within the repeat unit cell)
magnetic field that couples
to the Ising spins through a Zeeman coupling,
taking the thermodynamic limit, and switching off the
Zeeman coupling selects one of the two degenerate ground states
when $\lambda>0$ and $\lambda<0$, respectively.

\subsubsection{Group cohomology class $([\nu],[\rho],\mu)=(0,0,0)$}
\label{subsubsec:Cohomology class (0,0,0)}

The local fermionic Fock space $\mathcal{F}^{\,}_{j}$
of dimension $\mathcal{D}=16$
is generated by the octuplet of Majorana operators
\begin{equation}
\hat{\chi}^{\,}_{j}\equiv
\begin{pmatrix}
\hat{\chi}^{\,}_{j,1}&
\cdots&
\hat{\chi}^{\,}_{j,8}
\end{pmatrix}^{\mathsf{T}},
\qquad
j=1,\cdots,2M.
\label{eq:Majorana quartet for (0,0,0)}
\end{equation}
One verifies that
\begin{subequations}\label{eq:Z2T 000 rep}
\begin{align}
&    
\hat{u}^{\,}_{j}(t)\df
-
\mathrm{i}
\hat{\chi}^{\,}_{j,2}\,
\hat{\chi}^{\,}_{j,4}\,
\hat{\chi}^{\,}_{j,6}\,
\hat{\chi}^{\,}_{j,8}\,
\mathsf{K},
\label{eq:Z2T 000 rep a}
\\
&
\hat{u}^{\,}_{j}(p)\df
\hat{\chi}^{\,}_{j,1}\,
\hat{\chi}^{\,}_{j,2}\,
\hat{\chi}^{\,}_{j,3}\,
\hat{\chi}^{\,}_{j,4}\,
\hat{\chi}^{\,}_{j,5}\,
\hat{\chi}^{\,}_{j,6}\,
\hat{\chi}^{\,}_{j,7}\,
\hat{\chi}^{\,}_{j,8},
\label{eq:Z2T 000 rep b}
\end{align}
realizes the projective algebra
(\ref{eq:indices for ZT2})
with
\begin{equation}
[\nu]=0,
\qquad
[\rho]=0.
\label{eq:Z2T 000 rep c}
\end{equation}
\end{subequations}
One verifies that the Majorana octuplet
(\ref{eq:Majorana quartet for (0,0,0)})
is even under conjugation by $\hat{u}^{\,}_{j}(t)$ and
odd under conjugation by $\hat{u}^{\,}_{j}(p)$.
Time-reversal symmetry forbids any Hermitian quadratic form
for the quartet (\ref{eq:Majorana quartet for (0,0,0)}).

Theorem \ref{thm:LSM Theorem 1} is inoperative.
It is possible to find examples of both nondegenerate and degenerate
gapped Hamiltonians that are translation invariant and $G^{\,}_{f}$ invariant.

To prove this claim, it suffices to consider a
generic Hamiltonian of the form (\ref{eq:gen_Ham})
that is of quartic order and of range $r=0$. It is given by
\begin{equation}
\widehat{H}^{\,}_{\mathrm{pbc}}\df
\sum_{j=1}^{2M}
\sum_{1\leq a^{\,}_{1}<a^{\,}_{2}<a^{\,}_{3}<a^{\,}_{4}\leq8}
\lambda^{\,}_{a^{\,}_{1}\,a^{\,}_{2}\,a^{\,}_{3}\,a^{\,}_{4}}\,
\prod_{i=1}^{4}
\hat{\chi}^{\,}_{j,a^{\,}_{i}}
\end{equation}
with $\lambda^{\,}_{a^{\,}_{1}\,a^{\,}_{2}\,a^{\,}_{3}\,a^{\,}_{4}}$
a real-valued coupling. Each monomial
\begin{equation}
\widehat{X}^{\,}_{j|a^{\,}_{1}\,a^{\,}_{2}\,a^{\,}_{3}\,a^{\,}_{4}}\df
\prod_{i=1}^{4}
\hat{\chi}^{\,}_{j,a^{\,}_{i}},
\quad
1\leq a^{\,}_{1}<a^{\,}_{2}<a^{\,}_{3}<a^{\,}_{4}\leq8,
\label{eq:Z2T 000 Ham}
\end{equation}
is a Hermitian operator with the eigenvalues $\pm1$
as it squares to the identity. Its two degenerate eigenspaces are therefore
8-dimensional. Any two monomials
of the form (\ref{eq:Z2T 000 Ham})
either commute or anticommute. There are 70
(choose 4 out of 8) such monomials. The six
monomials
\begin{subequations}\label{eq:Ising monomials 000}
\begin{align}
&
\widehat{X}^{\,}_{j|1\,2\,3\,4}\df
-4
\left(
\hat{n}^{\,}_{j,1}-\frac{1}{2}
\right)
\left(
\hat{n}^{\,}_{j,3}-\frac{1}{2}
\right),
\\
&
\widehat{X}^{\,}_{j|1\,2\,5\,6}\df
-4
\left(
\hat{n}^{\,}_{j,1}-\frac{1}{2}
\right)
\left(
\hat{n}^{\,}_{j,5}-\frac{1}{2}
\right),
\\
&
\widehat{X}^{\,}_{j|1\,2\,7\,8}\df
-4
\left(
\hat{n}^{\,}_{j,1}-\frac{1}{2}
\right)
\left(
\hat{n}^{\,}_{j,7}-\frac{1}{2}
\right),
\\
&
\widehat{X}^{\,}_{j|3\,4\,5\,6}\df
-4
\left(
\hat{n}^{\,}_{j,3}-\frac{1}{2}
\right)
\left(
\hat{n}^{\,}_{j,5}-\frac{1}{2}
\right),
\\
&
\widehat{X}^{\,}_{j|3\,4\,7\,8}\df
-4
\left(
\hat{n}^{\,}_{j,3}-\frac{1}{2}
\right)
\left(
\hat{n}^{\,}_{j,7}-\frac{1}{2}
\right),
\\
&
\widehat{X}^{\,}_{j|5\,6\,7\,8}\df
-4
\left(
\hat{n}^{\,}_{j,5}-\frac{1}{2}
\right)
\left(
\hat{n}^{\,}_{j,7}-\frac{1}{2}
\right),
\end{align}
\end{subequations}
are the only ones that are compatible with
conservation of the fermion-number in the fermion-number basis
(\ref{eq:def fermions out Majoranas mu=0}). All remaining
64 monomials break the conservation of this fermion-number
down to conservation of the fermion-parity number.
Among these, the 16 monomials generated by expanding
\begin{equation}
\left(
\hat{\chi}^{\,}_{j,1}
+
\hat{\chi}^{\,}_{j,2}
\right)
\left(
\hat{\chi}^{\,}_{j,3}
+
\hat{\chi}^{\,}_{j,4}
\right)
\left(
\hat{\chi}^{\,}_{j,5}
+
\hat{\chi}^{\,}_{j,6}
\right)
\left(
\hat{\chi}^{\,}_{j,7}
+
\hat{\chi}^{\,}_{j,8}
\right)
\end{equation}
are special because they form the basis to represent all
16 terms of the form
\begin{equation}
\begin{split}
&
\widehat{A}^{\,}_{j|0000}\df
\hat{c}^{\dag}_{j,1}\,
\hat{c}^{\dag}_{j,3}\,  
\hat{c}^{\dag}_{j,5}\,
\hat{c}^{\dag}_{j,7},
\\
&
\widehat{A}^{\,}_{j|0001}\df
\hat{c}^{\dag}_{j,1}\,
\hat{c}^{\dag}_{j,3}\,  
\hat{c}^{\dag}_{j,5}\,
\hat{c}^{\,}_{j,7},
\cdots,
\widehat{A}^{\,}_{j|1000}\df
\hat{c}^{\,}_{j,1}\,
\hat{c}^{\dag}_{j,3}\,  
\hat{c}^{\dag}_{j,5}\,
\hat{c}^{\dag}_{j,7},
\\
&
\qquad\qquad\qquad\qquad\vdots
\\
&
\widehat{A}^{\,}_{j|1110}\df
\hat{c}^{\,}_{j,1}\,
\hat{c}^{\,}_{j,3}\,  
\hat{c}^{\,}_{j,5}\,
\hat{c}^{\dag}_{j,7},
\cdots,
\widehat{A}^{\,}_{j|0111}\df
\hat{c}^{\dag}_{j,1}\,
\hat{c}^{\,}_{j,3}\,  
\hat{c}^{\,}_{j,5}\,
\hat{c}^{\,}_{j,7},
\\
&
\widehat{A}^{\,}_{j|1111}\df
\hat{c}^{\,}_{j,1}\,
\hat{c}^{\,}_{j,3}\,  
\hat{c}^{\,}_{j,5}\,
\hat{c}^{\,}_{j,7},
\end{split}
\label{eq:16 operators taking any Ising configuration to time-reverses one}
\end{equation}
in the fermion-number basis
(\ref{eq:def fermions out Majoranas mu=0}).
In the 70-dimensional coupling space of Hamiltonian
(\ref{eq:Z2T 000 Ham}), there is room
to find gapped Hamiltonians with either a degenerate or a nondegenerate
ground state.

On the one hand, the local Hamiltonian
\begin{align}
\hat{h}^{(\mathrm{I4rungs})}_{j}\df&\,
\lambda
\left(
\widehat{X}^{\,}_{j|1\,2\,3\,4}
+
\widehat{X}^{\,}_{j|3\,4\,5\,6}
+
\widehat{X}^{\,}_{j|5\,6\,7\,8}
\right)
\nonumber\\
=&\,
-\lambda
\left(
\sigma^{\,}_{j,1}\,
\sigma^{\,}_{j,3}
+
\sigma^{\,}_{j,3}\,
\sigma^{\,}_{j,5}
+
\sigma^{\,}_{j,5}\,
\sigma^{\,}_{j,7}
\right)
\end{align}
in the fermion-number basis
(\ref{eq:def fermions out Majoranas mu=0})
is none but the classical nearest-neighbor Ising Hamiltonian 
for a rung of four Ising spins ordered from bottom to top as 1,3,5,7.
As such, the subspace $\mathcal{F}^{\,}_{j\,\mathrm{gs}}$
in the local Fock space
$\mathcal{F}^{\,}_{j}=\mathcal{F}^{\,}_{j\,0}\oplus\mathcal{F}^{\,}_{j\,1}$
that is spanned by the linearly independent ground states
of $\hat{h}^{(\mathrm{I4rungs})}_{j}$ is either
\begin{subequations}\label{eq:ground states Ising rung case 000}
\begin{equation}
\mathcal{F}^{\,}_{j\,\mathrm{gs}}=
\mathrm{span}
\left\{
\left|
\begin{tiny}
\begin{pmatrix}0\\0\\0\\0\end{pmatrix}
\end{tiny}
\right\rangle,
\left|
\begin{tiny}
\begin{pmatrix}1\\1\\1\\1\end{pmatrix}
\end{tiny}
\right\rangle
\right\}\subset\mathcal{F}^{\,}_{j\,0}
\label{eq:ground states Ising rung case 000 a}
\end{equation}
when $\lambda>0$ or
\begin{equation}
\mathcal{F}^{\,}_{j\,\mathrm{gs}}=
\mathrm{span}
\left\{
\left|
\begin{tiny}
\begin{pmatrix}0\\1\\0\\1\end{pmatrix}
\end{tiny}
\right\rangle,
\left|
\begin{tiny}
\begin{pmatrix}1\\0\\1\\0\end{pmatrix}
\end{tiny}
\right\rangle
\right\}\subset\mathcal{F}^{\,}_{j\,0}
\label{eq:ground states Ising rung case 000 b}
\end{equation}
\end{subequations}
when $\lambda<0$
in the fermion-number basis
(\ref{eq:def fermions out Majoranas mu=0}).

On the other hand, the local Hamiltonian
\begin{equation}
\hat{h}^{\,}_{j}\df
\lambda
\left(
\widehat{X}^{\,}_{j|1\,2\,3\,4}
+
\widehat{X}^{\,}_{j|3\,4\,5\,6}
+
\widehat{X}^{\,}_{j|5\,6\,7\,8}
+
\widehat{X}^{\,}_{j|1\,3\,5\,7}
\right)
\end{equation}
has a nondegenerate ground state.
This is so because the monomial $\widehat{X}^{\,}_{j|1\,3\,5\,7}$
is the sum over all 16 operators
(\ref{eq:16 operators taking any Ising configuration to time-reverses one})
with equal weight. Hence, its action on either the basis
(\ref{eq:ground states Ising rung case 000 a})
or the basis
(\ref{eq:ground states Ising rung case 000 b})
is to exchange the two basis states, thereby lifting their degeneracies.

The counterpart to this mechanism to lift the twofold degeneracy of a 2-rung
Ising Hamiltonian is not available in Sec.\
(\ref{subsubsec:Cohomology class (0,1,0)})
because of fermion-parity conservation
[to exchange the ferromagnetic ground states, one would need the odd-parity
perturbation
$\hat{c}^{\dag}_{j,1}\,\hat{c}^{\dag}_{j,3}\,\hat{c}^{\dag}_{j,5}+\mathrm{H.c.}$].
The same is true in Sec.\
(\ref{subsubsec:Cohomology class (1,0,0)}).
Lifting the twofold degeneracy of a 1-rung Ising Hamiltonian
with the help of the perturbation
$\hat{c}^{\dag}_{j,1}\,\hat{c}^{\dag}_{j,3}+\mathrm{H.c.}$
is not possible because time-reversal symmetry prohibits any
local quadratic term.

By identifying the set
\begin{subequations}
\begin{equation}
([\nu],[\rho],0)\in
\left\{
(1,1,0),\,
(1,0,0),\,
(0,1,0),\,
(0,0,0)
\right\}
\end{equation}
with the four distinct group cohomology classes
(\ref{eq:indices for ZT2})
and by defining a group operation using the stacking rules
(\ref{eq:Stacking rules Z2TxZ2F}),
we have justified the identification
\begin{equation}
\begin{split}
&
g\to(1,1,0),
\qquad
g^{2}\to(1,0,0),
\\
&
\qquad\qquad
g^{3}\to(0,1,0),
\qquad
g^{4}\to (0,0,0),
\end{split}
\end{equation}
where $g$ is the generator of the cyclic group
\begin{equation}
\mathbb{Z}^{\,}_{4}\df\left\{g,g^{2},g^{3},g^{4}\equiv e\right\}.
\end{equation}
\end{subequations}
The fact that the stacking rules
(\ref{eq:Stacking rules Z2TxZ2F})
obeyed by the indices $([\rho],[\nu],\mu=0)$
realize the cyclic group
$\mathbb{Z}^{\,}_{4}$
is reminiscent of the fact that the topological index
(the integer number of Majorana boundary zero modes)
of noninteracting fermions belonging to the symmetry class
BDI in one-dimensional space is to be replaced by one
belonging to the cyclic group $\mathbb{Z}^{\,}_{8}$
if interactions compatible with the symmetry class BDI
are allowed\, \cite{Fidkowski2010,Fidkowski2011}.
The difference between the cyclic group $\mathbb{Z}^{\,}_{4}$
for LSM-type constraints and the cyclic group $\mathbb{Z}^{\,}_{8}$
in Ref.\ \onlinecite{Fidkowski2011}
arises because we must set $\mu=0$ when defining locally
the Fock space, whereas there is no such constraint
at the boundary of an SPT phase.

\subsection{One-dimensional space with the symmetry group $\mathbb{Z}^{\,}_{2}
\times\mathbb{Z}^{\,}_{2}\times \mathbb{Z}^{F}_{2}$ for an even number of local
Majorana flavors}

The lattice is $\Lambda=\{1,\cdots,N\}$ with $N=2M$ an even integer
and the global fermionic Fock space $\mathcal{F}^{\,}_{\Lambda}$
is of dimension $2^{n\,M}$ with $n$ the number of local Majorana flavors.
By choosing the cardinality $|\Lambda|=2M$ to be even,
we make sure that the lattice is bipartite. This
allows to treat the two values of the index
$\mu\df\hbox{$n$ mod 2}$ in parallel.
The symmetry group
$G^{\,}_{f}\df G\times\mathbb{Z}^{F}_{2}$
is a split group. As usual, $\mathbb{Z}^{F}_{2}$ is generated by $p$.
We choose $G\df\mathbb{Z}^{\,}_{2}\times\mathbb{Z}^{\,}_{2}$.
The Abelian group $G$ has hence two generators $g^{\,}_{1}$
and $g^{\,}_{2}$ that commute pairwise,
while each of them squares to the identity.
We shall only consider the case when the local number of Majorana
flavors $n=2m$ is an even positive integer. The index $\mu$ then
takes the value $\mu=0$.

Any local projective representation of $G^{\,}_{f}$ can be
labeled by the pair of  indices
$[\nu]\in H^{2}\big(G,\mathrm{U(1)}^{\,}_{\mathfrak{c}}\big)$
and
$[\rho]=([\rho]^{\,}_{1},[\rho]^{\,}_{2})$
with
$[\rho]^{\,}_{1},[\rho]^{\,}_{2}\in H^{1}\big(G,\mathbb{Z}^{\,}_{2}\big)$
through the relations\,
\footnote{
We have chosen the convention of always representing
the generator $p$ of $\mathbb{Z}^{F}_{2}$ by a Hermitian operator
according to Eq.\ (\ref{eq: def uj(g) e})}
\begin{subequations}\label{eq:projective rep G=Z2xZ2}
\begin{align}
&
\hat{u}(g^{\,}_{1})\,\hat{u}(g^{\,}_{2})=
(-1)^{[\nu]}\,
\hat{u}(g^{\,}_{2})\,\hat{u}(g^{\,}_{1}),
\quad
[\nu]=0,1,
\label{eq:projective rep G=Z2xZ2 a}
\\
&
\hat{u}(g^{\,}_{i})\,
\hat{u}(p)
\,\,\,=
(-1)^{[\rho]^{\,}_{i}}\,
\hat{u}(p)\,
\hat{u}(g^{\,}_{i}),
\quad
[\rho]^{\,}_{i}=0,1.
\label{eq:projective rep G=Z2xZ2 b}
\end{align}
\end{subequations}
This gives the eight distinct group cohomology classes
\begin{equation}
\begin{split}
([\nu],[\rho],0)=
\Big\{
&
\big(0,(0,0),0\big),
\big(1,(0,0),0\big),
\\
&
\big(0,(0,1),0\big),
\big(1,(0,1),0\big),
\\
&
\big(0,(1,0),0\big),
\big(1,(1,0),0\big),
\\
&
\big(0,(1,1),0\big),
\big(1,(1,1),0\big)
\Big\}.
\end{split}
\end{equation}
Here, the group cohomology class
$\big(0,(0,0),0\big)$ is interpreted as the trivial representation.
Theorem \ref{thm:LSM Theorem 1} is predictive for any of the remaining
seven group cohomology classes.
It is shown in Appendix \ref{appsubsec: Z2 Z2 ZF2}
that these eight distinct group cohomology classes
form the (stacking) group 
$\mathbb{Z}^{\,}_{2}\times\mathbb{Z}^{\,}_{2}\times\mathbb{Z}^{\,}_{2}$,
whereby the group composition is defined by the stacking rule
\eqref{eq:Z2Z2ZF2 stacking rule}.
This (stacking) group is generated by the three group cohomology classes
$\big(1,(1,0),0\big)$, $\big(1,(0,1),0\big)$, and $\big(1,(0,0),0\big)$, 
as we are going to verify explicitly.

\subsubsection{Group cohomology class $([\nu],[\rho],0)=\big(1,(1,0),0\big)$}

The local fermionic Fock space $\mathcal{F}^{\,}_{j}$
of dimension $\mathcal{D}=4$
is generated by the quartet of Majorana operators\,
\footnote{
It is not possible to represent the
cohomology class $([\nu],[\rho],\mu)=\big(1,(1,0),0\big)$
with a doublet of Majorana operators.}        
\begin{equation}
\hat{\chi}^{\,}_{j}\equiv
\begin{pmatrix}
\hat{\chi}^{\,}_{j,1}&
\cdots&
\hat{\chi}^{\,}_{j,4}
\end{pmatrix}^{\mathsf{T}},
\qquad
j=1,\cdots,2M.
\label{eq:Majorana quartet for (1,(1,0),0)}
\end{equation}
One verifies that
\begin{subequations}\label{eq:Z2xZ2 1100 rep}
\begin{align}
&
\hat{u}^{\,}_{j}(g^{\,}_{1})\df
\hat{\chi}^{\,}_{j,1},
\label{eq:Z2xZ2 1100 rep a}
\\
&
\hat{u}^{\,}_{j}(g^{\,}_{2})\df
\hat{\chi}^{\,}_{j,1}\,
\hat{\chi}^{\,}_{j,3},
\label{eq:Z2xZ2 1100 rep b}
\\
&
\hat{u}^{\,}_{j}(p)
\,\,
\df
\hat{\chi}^{\,}_{j,1}\,
\hat{\chi}^{\,}_{j,2}\,
\hat{\chi}^{\,}_{j,3}\,
\hat{\chi}^{\,}_{j,4},
\label{eq:Z2xZ2 1100 rep c}
\end{align}
\end{subequations}
realizes the projective representation
(\ref{eq:projective rep G=Z2xZ2})
with
$([\nu],[\rho],\mu)=\big(1,(1,0),0\big)$.

An example of a translation- and $G^{\,}_{f}$-invariant
Hamiltonian of the form (\ref{eq:gen_Ham}),
order 2, and range $r=1$ is
\begin{equation}
\begin{split}
\widehat{H}^{\,}_{\mathrm{pbc}}\df&\,
\sum_{j=1}^{2M}
\sum_{a=1}^{4}
\lambda^{\,}_{a}\,
\mathrm{i}
\hat{\chi}^{\,}_{j,a}\,
\hat{\chi}^{\,}_{j+1,a}
\\
&\,
+
\sum_{j=1}^{2M}
\lambda^{\,}_{2,4}\,
\mathrm{i}
\hat{\chi}^{\,}_{j,2}\,
\hat{\chi}^{\,}_{j,4},
\qquad
\lambda^{\,}_{a},\lambda^{\,}_{2,4}\in\mathbb{R}.
\end{split}
\label{eq:Z2xZ2 1100 Ham}
\end{equation}
This Hamiltonian does not conserve the fermion-number in
the fermion-number basis
(\ref{eq:def fermions out Majoranas mu=0}).
It can be thought of as four Kitaev chains
each of which has an effective
index $\mu=1$. When $\lambda^{\,}_{24}=0$,
all Kitaev chains decouple and are fine-tuned to
their quantum critical point
(\ref{eq:def Kitaev chain at criticality})
between their symmetry-protected and
topologically inequivalent gapped phases.
Kitaev chains 2 and 4 are coupled by the on-site term
$\mathrm{i}\hat{\chi}^{\,}_{j,2}\,\hat{\chi}^{\,}_{j,4}$.
The on-site term
$\mathrm{i}\hat{\chi}^{\,}_{j,2}\,\hat{\chi}^{\,}_{j,4}$
gaps chains 2 and 4. The low-energy sector of the theory
is that of two decoupled quantum critical Kitaev chains
labeled 1 and 3. The quadratic term
$\mathrm{i}\hat{\chi}^{\,}_{j,1}\,\hat{\chi}^{\,}_{j,3}$
that would gap chains 1 and 3, thereby delivering a
nondegenerate gapped ground state, is odd under conjugation
by $\hat{u}^{\,}_{j}(g^{\,}_{1})$ and thus forbidden by symmetry.
The stability of this gapless phase to on-site quadratic perturbations
can thus be thought of as a consequence of Theorem
\ref{thm:LSM Theorem 1}. Theorem \ref{thm:LSM Theorem 1}
also predicts that any
$G^{\,}_{f}$-symmetric interaction of the form (\ref{eq:gen_Ham})
that opens a spectral gap in the noninteracting spectrum
must break spontaneously at least one of the
symmetries responsible
for the noninteracting spectrum being gapless.

When two copies of this $\big(1,(1,0),0\big)$ representation 
are stacked, the local fermionic Fock space $\mathcal{F}^{\,}_{j}$
of dimension $\mathcal{D}=16$
is generated by the octuplet of Majorana operators
\begin{equation}
\hat{\chi}^{\,}_{j}\equiv
\begin{pmatrix}
\hat{\chi}^{\,}_{j,1}&
\cdots&
\hat{\chi}^{\,}_{j,8}
\end{pmatrix}^{\mathsf{T}},
\qquad
j=1,\cdots,2M.
\label{eq:Majorana octuplet for (1,(1,0),0)x(1,(1,0),0)}
\end{equation}
One verifies that 
\begin{subequations}\label{eq:Z2xZ2 1100 stacked rep}
\begin{align}
&
\hat{u}^{\,}_{j}(g^{\,}_{1})\df
\hat{\chi}^{\,}_{j,1}\,
\hat{\chi}^{\,}_{j,5},
\\
&
\hat{u}^{\,}_{j}(g^{\,}_{2})\df
\hat{\chi}^{\,}_{j,1}\,
\hat{\chi}^{\,}_{j,3}\,
\hat{\chi}^{\,}_{j,5}\,
\hat{\chi}^{\,}_{j,7},
\label{eq:Z2xZ2 1100 stacked rep a}
\\
&
\hat{u}^{\,}_{j}(p)\df
\hat{\chi}^{\,}_{j,1}\,
\hat{\chi}^{\,}_{j,2}\,
\hat{\chi}^{\,}_{j,3}\,
\hat{\chi}^{\,}_{j,4}\,
\hat{\chi}^{\,}_{j,5}\,
\hat{\chi}^{\,}_{j,6}\,
\hat{\chi}^{\,}_{j,7}\,
\hat{\chi}^{\,}_{j,8},
\label{eq:Z2xZ2 1100 stacked rep b}
\end{align}
\end{subequations}
realizes the projective representation
(\ref{eq:projective rep G=Z2xZ2})
with
$([\nu],[\rho],\mu)=\big(0,(0,0),0\big)$,
i.e., the trivial projective representation. 
In this trivial representation,
any flavor $a=1,\cdots,8$ has an image $a'=(a+4)\hbox{ mod }8$
such that $\hat{\chi}_{j,a}$ and $\hat{\chi}_{j,a'}$
transform identically under $G^{\,}_{f}$.
All on-site terms
$\mathrm{i}\hat{\chi}_{j,a}\,\hat{\chi}_{j,a+4}$
with $a=1,\cdots,4$ are then $G^{\,}_{f}$ symmetric.
The ground-state degeneracy of any
translation- and $G^{\,}_{f}$-invariant Hamiltonian
of the form (\ref{eq:gen_Ham})
can be lifted by including the 4 on-site terms
$\mathrm{i}\hat{\chi}_{j,a}\,\hat{\chi}_{j,a+4}$
with $a=1,\cdots,4$ , i.e.,
Theorem \ref{thm:LSM Theorem 1} is not predictive.

\subsubsection{Group cohomology class $([\nu],[\rho],\mu)=\big(1,(0,1),0\big)$}

The local fermionic Fock space $\mathcal{F}^{\,}_{j}$
of dimension $\mathcal{D}=4$
is generated by the quartet of Majorana operators
(\ref{eq:Majorana quartet for (1,(1,0),0)})\,
\footnote{
It is not possible to represent the
cohomology class $([\nu],[\rho],\mu)=\big(1,(0,1),0\big)$
with a doublet of Majorana operators.
}.        
One verifies that 
\begin{subequations}\label{eq:Z2xZ2 1010 rep}
\begin{align}
&
\hat{u}^{\,}_{j}(g^{\,}_{1})\df
\hat{\chi}^{\,}_{j,1}\,
\hat{\chi}^{\,}_{j,3},
\\
&
\hat{u}^{\,}_{j}(g^{\,}_{2})\df
\hat{\chi}^{\,}_{j,1},
\label{eq:Z2xZ2 1010 stacked rep a}
\\
&
\hat{u}^{\,}_{j}(p)\df
\hat{\chi}^{\,}_{j,1}\,
\hat{\chi}^{\,}_{j,2}\,
\hat{\chi}^{\,}_{j,3}\,
\hat{\chi}^{\,}_{j,4},
\label{eq:Z2xZ2 1010 stacked rep b} 
\end{align}
\end{subequations}
realizes the projective representation
(\ref{eq:projective rep G=Z2xZ2})
with
$([\nu],[\rho],\mu)=\big(1,(0,1),0\big)$.

Equation (\ref{eq:Z2xZ2 1010 rep})
differs from Eq.\ (\ref{eq:Z2xZ2 1100 rep})
by interchanging $g^{\,}_{1}$ and $g^{\,}_{2}$.
This difference does not affect the reasoning
leading to to the conclusion that the gapless
Hamiltonian (\ref{eq:Z2xZ2 1100 Ham})
is the most general translation-invariant,
$G^{\,}_{f}$-invariant, order 2, and range $r=1$
Hamiltonian of the form (\ref{eq:gen_Ham})
whose hopping is diagonal with respect to the flavor index.
This difference also implies that stacking 
two copies of the $\big(1,(0,1),0\big)$ representation
(\ref{eq:Z2xZ2 1010 rep})
delivers the trivial projective representation
$\big(0,(0,0),0\big)$
encoded by Eqs.\
(\ref{eq:Majorana octuplet for (1,(1,0),0)x(1,(1,0),0)})
and
(\ref{eq:Z2xZ2 1100 stacked rep}),
for which Theorem
\ref{thm:LSM Theorem 1}
is not predictive anymore.

\subsubsection{Group cohomology class $([\nu],[\rho],\mu)=(1,(0,0),0)$}

The local fermionic Fock space $\mathcal{F}^{\,}_{j}$
of dimension $\mathcal{D}=4$
is generated by the quartet of Majorana operators
(\ref{eq:Majorana quartet for (1,(1,0),0)})\,
\footnote{
It is not possible to represent the
cohomology class $([\nu],[\rho],\mu)=\big(1,(0,0),0\big)$
with a doublet of Majorana operators.
}.        
One verifies that
\begin{subequations}\label{eq:Z2xZ2 1000 rep}
\begin{align}
&
\hat{u}^{\,}_{j}(g^{\,}_{1})\df
\hat{\chi}^{\,}_{j,2}\,
\hat{\chi}^{\,}_{j,3},
\\
&
\hat{u}^{\,}_{j}(g^{\,}_{2})\df
\hat{\chi}^{\,}_{j,1}\,
\hat{\chi}^{\,}_{j,3},
\\
&
\hat{u}^{\,}_{j}(p)\df
\hat{\chi}^{\,}_{j,1}\,
\hat{\chi}^{\,}_{j,2}\,
\hat{\chi}^{\,}_{j,3}\,
\hat{\chi}^{\,}_{j,4},
\end{align}
\end{subequations}
realizes the projective representation
(\ref{eq:projective rep G=Z2xZ2})
with
$([\nu],[\rho],\mu)=\big(1,(0,0),0\big)$.

The most general translation- and $G^{\,}_{f}$-invariant
Hamiltonian of the form (\ref{eq:gen_Ham})
of order 2, range $r=1$, and
whose hopping is diagonal with respect to the flavor index
is then 
\begin{align}
\widehat{H}^{\,}_{\mathrm{pbc}}=
\sum_{j=1}^{N}
\sum_{a=1}^{4}
\lambda^{\,}_{a}\,
\mathrm{i}
\hat{\chi}^{\,}_{j,a}\,
\hat{\chi}^{\,}_{j+1,a},
\qquad
\lambda^{\,}_{a}\in\mathbb{R},
\end{align}
i.e., four decoupled Kitaev chains that are fine-tuned to
their quantum critical point
(\ref{eq:def Kitaev chain at criticality})
between their symmetry-protected and topologically inequivalent gapped phases.
No on-site quadratic term is allowed by the symmetries.
The stability of this gapless phase to on-site quadratic perturbations
can thus be thought of as a consequence of Theorem
\ref{thm:LSM Theorem 1}. Theorem \ref{thm:LSM Theorem 1}
also predicts that any
$G^{\,}_{f}$-symmetric interaction of the form (\ref{eq:gen_Ham})
that opens a spectral gap in the noninteracting spectrum
must break spontaneously at least one of the
symmetries responsible
for the noninteracting spectrum being gapless.

When two copies of this $\big(1,(0,0),0\big)$ representation 
are stacked, the local fermionic Fock space $\mathcal{F}^{\,}_{j}$
of dimension $\mathcal{D}=16$
is generated by the octuplet of Majorana operators
(\ref{eq:Majorana octuplet for (1,(1,0),0)x(1,(1,0),0)})
with the projective representation
\begin{subequations}\label{eq:Z2xZ2 1000 stacked rep}
\begin{align}
&
\hat{u}^{\,}_{j}(g^{\,}_{1})\df
\hat{\chi}^{\,}_{j,2}\,
\hat{\chi}^{\,}_{j,3}\,
\hat{\chi}^{\,}_{j,6}\,
\hat{\chi}^{\,}_{j,7},
\\
&
\hat{u}^{\,}_{j}(g^{\,}_{2})\df
\hat{\chi}^{\,}_{j,1}\,
\hat{\chi}^{\,}_{j,3}\,
\hat{\chi}^{\,}_{j,5}\,
\hat{\chi}^{\,}_{j,7},
\label{eq:Z2xZ2 1000 stacked rep a}
\\
&
\hat{u}^{\,}_{j}(p)\df
\hat{\chi}^{\,}_{j,1}\,
\hat{\chi}^{\,}_{j,2}\,
\hat{\chi}^{\,}_{j,3}\,
\hat{\chi}^{\,}_{j,4}\,
\hat{\chi}^{\,}_{j,5}\,
\hat{\chi}^{\,}_{j,6}\,
\hat{\chi}^{\,}_{j,7}\,
\hat{\chi}^{\,}_{j,8},
\label{eq:Z2xZ2 1000 stacked rep b}
\end{align}
\end{subequations}
that realizes the group cohomology class
$([\nu],[\rho],\mu)=\big(0,(0,0),0\big)$,
i.e., the trivial group cohomology class.
In this trivial representation,
any flavor $a=1,\cdots,8$ has an image $a'=(a+4)\hbox{ mod }8$
such that $\hat{\chi}_{j,a}$ and $\hat{\chi}_{j,a'}$
transform identically under $G^{\,}_{f}$.
All on-site terms
$\mathrm{i}\hat{\chi}_{j,a}\,\hat{\chi}_{j,a+4}$
with $a=1,\cdots,4$ are then $G^{\,}_{f}$ symmetric.
The ground-state degeneracy of any
translation- and $G^{\,}_{f}$-invariant Hamiltonian
of the form (\ref{eq:gen_Ham})
can be lifted by including the 4
on-site terms
$\mathrm{i}\hat{\chi}_{j,a}\,\hat{\chi}_{j,a+4}$
with $a=1,\cdots,4$ , i.e.,
Theorem \ref{thm:LSM Theorem 1} is not predictive.

\subsubsection{Group cohomology class $([\nu],[\rho],\mu)=\big(1,(1,1),0\big)$}

When representations (\ref{eq:Z2xZ2 1100 rep})
and (\ref{eq:Z2xZ2 1010 rep}) are stacked,
the local fermionic Fock space $\mathcal{F}^{\,}_{j}$
of dimension $\mathcal{D}=16$
is generated by the octuplet of Majorana operators
(\ref{eq:Majorana octuplet for (1,(1,0),0)x(1,(1,0),0)}).
One verifies that
\begin{subequations}\label{eq:Z2xZ2 1110 rep}
\begin{align}
&
\hat{u}^{\,}_{j}(g^{\,}_{1})\df
\hat{\chi}^{\,}_{j,1}\,
\hat{\chi}^{\,}_{j,5}\,
\hat{\chi}^{\,}_{j,7},
\\
&
\hat{u}^{\,}_{j}(g^{\,}_{2})\df
\hat{\chi}^{\,}_{j,1}\,
\hat{\chi}^{\,}_{j,3}\,
\hat{\chi}^{\,}_{j,5},
\\
&
\hat{u}^{\,}_{j}(p)\df
\hat{\chi}^{\,}_{j,1}\,
\hat{\chi}^{\,}_{j,2}\,
\hat{\chi}^{\,}_{j,3}\,
\hat{\chi}^{\,}_{j,4}\,
\hat{\chi}^{\,}_{j,5}\,
\hat{\chi}^{\,}_{j,6}\,
\hat{\chi}^{\,}_{j,7}\,
\hat{\chi}^{\,}_{j,8},
\end{align}
\end{subequations}
realizes the projective representation
(\ref{eq:projective rep G=Z2xZ2})
with
$([\nu],[\rho],\mu)=\big(1,(1,1),0\big)$.

An example of a translation- and $G^{\,}_{f}$-invariant
Hamiltonian of the form (\ref{eq:gen_Ham}),
order 2, and range $r=1$ is
\begin{equation}
\begin{split}
\widehat{H}^{\,}_{\mathrm{pbc}}\df&\,
\sum_{j=1}^{N}
\Bigg[
\sum_{a=1}^{8}
\lambda^{\,}_{a}\,
\mathrm{i}
\hat{\chi}^{\,}_{j,a}\,
\hat{\chi}^{\,}_{j+1,a}
+
\lambda^{\,}_{1,5}\,
\mathrm{i}
\hat{\chi}^{\,}_{j,1}\,
\hat{\chi}^{\,}_{j,5}
\\
&\,
+
\lambda^{\,}_{2,4}\,
\mathrm{i}
\hat{\chi}^{\,}_{j,2}\,
\hat{\chi}^{\,}_{j,4}
+
\lambda^{\,}_{6,8}\,
\mathrm{i}
\hat{\chi}^{\,}_{j,6}\,
\hat{\chi}^{\,}_{j,8}
\Bigg]
\end{split}
\end{equation}
with $\lambda^{\,}_{a},\lambda^{\,}_{a,b}\in\mathbb{R}$ for $a,b=1,\cdots,8$.
By construction, this Hamiltonian is gapless since
the quantum critical Kitaev chains 3 and 7 are decoupled from all other
Kitaev chains. This gapless phase is stable to any on-site quadratic 
perturbation since the only on-site quadratic terms
$
\mathrm{i}
\hat{\chi}^{\,}_{j,3}\,
\hat{\chi}^{\,}_{j,a}
$
and
$
\mathrm{i}
\hat{\chi}^{\,}_{j,7}\,
\hat{\chi}^{\,}_{j,a}
$
that could gap the quantum critical Kitaev chains 3 and 7
are odd under $G$ for any $a\neq3,7$.
The stability of this gapless phase to on-site quadratic perturbations
can thus be thought of as a consequence of Theorem
\ref{thm:LSM Theorem 1}. Theorem \ref{thm:LSM Theorem 1}
also predicts that any
$G^{\,}_{f}$-symmetric interaction of the form (\ref{eq:gen_Ham})
that opens a spectral gap in the noninteracting spectrum
must break spontaneously at least one of the
symmetries responsible
for the noninteracting spectrum being gapless.

When two copies of this $\big(1,(1,1),0\big)$ representation 
are stacked, the local fermionic Fock space $\mathcal{F}^{\,}_{j}$
of dimension $\mathcal{D}=256$ is generated by $16$ Majorana operators.
One verifies that
\begin{subequations}\label{eq:Z2xZ2 1110 stacked rep}
\begin{align}
&
\hat{u}^{\,}_{j}(g^{\,}_{1})\df
\hat{\chi}^{\,}_{j,1}\,
\hat{\chi}^{\,}_{j,5}\,
\hat{\chi}^{\,}_{j,7}\,
\hat{\chi}^{\,}_{j,9}\,
\hat{\chi}^{\,}_{j,13}\,
\hat{\chi}^{\,}_{j,15},
\\
&
\hat{u}^{\,}_{j}(g^{\,}_{2})\df
\hat{\chi}^{\,}_{j,1}\,
\hat{\chi}^{\,}_{j,3}\,
\hat{\chi}^{\,}_{j,5}\,
\hat{\chi}^{\,}_{j,9}\,
\hat{\chi}^{\,}_{j,11}\,
\hat{\chi}^{\,}_{j,13},
\\
&
\hat{u}^{\,}_{j}(p)\df
\prod_{a=1}^{16}
\hat{\chi}^{\,}_{j,a},
\end{align}
\end{subequations}
realizes the group cohomology class
$([\nu],[\rho],\mu)=\big(0,(0,0),0\big)$,
i.e., the trivial group cohomology class.
There exists a bijective map $a\mapsto a'\df(a+8)\hbox{ mod }16$ such that
all on-site terms
$\mathrm{i}\hat{\chi}_{j,a}\,\hat{\chi}_{j,a'}$
with $a=1,\cdots,8$ can be shown to be $G^{\,}_{f}$ symmetric.
The ground-state degeneracy of any translation- and $G^{\,}_{f}$-invariant
Hamiltonian of the form (\ref{eq:gen_Ham})
can be lifted by including the 8 on-site terms
$\mathrm{i}\hat{\chi}_{j,a}\,\hat{\chi}_{j,a+8}$ with $a=1,\cdots,8$, i.e.,
Theorem \ref{thm:LSM Theorem 1} is not predictive.

\subsubsection{Group cohomology class $([\nu],[\rho],\mu)=\big(0,(1,0),0\big)$}

When representations (\ref{eq:Z2xZ2 1000 rep})
and (\ref{eq:Z2xZ2 1100 rep}) are stacked,
the local fermionic Fock space $\mathcal{F}^{\,}_{j}$
of dimension $\mathcal{D}=16$
is generated by the octuplet of Majorana operators
(\ref{eq:Majorana octuplet for (1,(1,0),0)x(1,(1,0),0)}).
One verifies that
\begin{subequations}\label{eq:Z2xZ2 0100 rep}
\begin{align}
&
\hat{u}^{\,}_{j}(g^{\,}_{1})\df
\hat{\chi}^{\,}_{j,2}\,
\hat{\chi}^{\,}_{j,3}\,
\hat{\chi}^{\,}_{j,5},
\\
&
\hat{u}^{\,}_{j}(g^{\,}_{2})\df
\hat{\chi}^{\,}_{j,1}\,
\hat{\chi}^{\,}_{j,3}\,
\hat{\chi}^{\,}_{j,5}\,
\hat{\chi}^{\,}_{j,7},
\\
&
\hat{u}^{\,}_{j}(p)\df
\hat{\chi}^{\,}_{j,1}\,
\hat{\chi}^{\,}_{j,2}\,
\hat{\chi}^{\,}_{j,3}\,
\hat{\chi}^{\,}_{j,4}\,
\hat{\chi}^{\,}_{j,5}\,
\hat{\chi}^{\,}_{j,6}\,
\hat{\chi}^{\,}_{j,7}\,
\hat{\chi}^{\,}_{j,8},
\end{align}
\end{subequations}
realizes the projective representation
(\ref{eq:projective rep G=Z2xZ2})
with
$([\nu],[\rho],\mu)=\big(0,(1,0),0\big)$.

An example of a translation- and $G^{\,}_{f}$-invariant
Hamiltonian of the form (\ref{eq:gen_Ham}),
order 2, and range $r=1$ is
\begin{equation}
\begin{split}
\widehat{H}^{\,}_{\mathrm{pbc}}\df&\,
\sum_{j=1}^{N}
\Bigg[
\sum_{a=1}^{8}
\lambda^{\,}_{a}\,
\mathrm{i}
\hat{\chi}^{\,}_{j,a}\,
\hat{\chi}^{\,}_{j+1,a}
+
\lambda^{\,}_{1,7}\,
\hat{\chi}^{\,}_{j,1}\,
\hat{\chi}^{\,}_{j,7}
\\
&\,
+
\lambda^{\,}_{4,8}\,
\hat{\chi}^{\,}_{j,4}\,
\hat{\chi}^{\,}_{j,8}
+
\lambda^{\,}_{3,5}\,
\hat{\chi}^{\,}_{j,3}\,
\hat{\chi}^{\,}_{j,5}
\Bigg]
\end{split}
\label{eq:Z2xZ2 0100 rep example gapless}
\end{equation}
with $\lambda^{\,}_{a},\lambda^{\,}_{a,b}\in\mathbb{R}$ for $a,b=1,\cdots,8$.
By construction, this Hamiltonian is gapless since
the quantum critical Kitaev chains 2 and 6 are decoupled from all other
Kitaev chains. This gapless phase is stable to any on-site and quadratic
perturbation since the only on-site quadratic terms
$
\mathrm{i}
\hat{\chi}^{\,}_{j,2}\,
\hat{\chi}^{\,}_{j,a}
$
and
$
\mathrm{i}
\hat{\chi}^{\,}_{j,6}\,
\hat{\chi}^{\,}_{j,a}
$
that could gap the quantum critical Kitaev chains 2 and 6
are odd under $G$ for any $a\neq2,6$.
The stability of this gapless phase to on-site quadratic perturbations
can thus be thought of as a consequence of Theorem
\ref{thm:LSM Theorem 1}. Theorem \ref{thm:LSM Theorem 1}
also predicts that any
$G^{\,}_{f}$-symmetric interaction of the form (\ref{eq:gen_Ham})
that opens a spectral gap in the noninteracting spectrum
must break spontaneously at least one of the
symmetries responsible
for the noninteracting spectrum being gapless.

When two copies of this $\big(0,(1,0),0\big)$ representation 
are stacked, the local fermionic Fock space $\mathcal{F}^{\,}_{j}$
of dimension $\mathcal{D}=256$ is generated by $16$ Majorana operators.
One verifies that
\begin{subequations}\label{eq:Z2xZ2 0100 stacked rep}
\begin{align}
&
\hat{u}^{\,}_{j}(g^{\,}_{1})\df
\hat{\chi}^{\,}_{j,2}\,
\hat{\chi}^{\,}_{j,3}\,
\hat{\chi}^{\,}_{j,5}\,
\hat{\chi}^{\,}_{j,10}\,
\hat{\chi}^{\,}_{j,11}\,
\hat{\chi}^{\,}_{j,13},
\\
&
\hat{u}^{\,}_{j}(g^{\,}_{2})\df
\hat{\chi}^{\,}_{j,1}\,
\hat{\chi}^{\,}_{j,3}\,
\hat{\chi}^{\,}_{j,5}\,
\hat{\chi}^{\,}_{j,7}\,
\hat{\chi}^{\,}_{j,9}\,
\hat{\chi}^{\,}_{j,11}\,
\hat{\chi}^{\,}_{j,13}\,
\hat{\chi}^{\,}_{j,15},
\\
&
\hat{u}^{\,}_{j}(p)\df
\prod_{a=1}^{16}
\hat{\chi}^{\,}_{j,a},
\end{align}
\end{subequations}
realizes the group cohomology class
$([\nu],[\rho],\mu)=\big(0,(0,0),0\big)$,
i.e., the trivial group cohomology class.
There exists a bijective map $a\mapsto a'\df(a+16)\hbox{ mod }16$ such that
all on-site terms
$\mathrm{i}\hat{\chi}_{j,a}\,\hat{\chi}_{j,a'}$
with $a=1,\cdots,8$ can be shown to be $G^{\,}_{f}$ symmetric.
The ground-state degeneracy of any translation- and $G^{\,}_{f}$-invariant
Hamiltonian of the form (\ref{eq:gen_Ham})
can be lifted by including the 8 on-site terms
$\mathrm{i}\hat{\chi}_{j,a}\,\hat{\chi}_{j,a+8}$ with $a=1,\cdots,8$,
i.e., Theorem \ref{thm:LSM Theorem 1} is not predictive.

\subsubsection{Group cohomology class $([\nu],[\rho],\mu)=\big(0,(0,1),0\big)$}

When representations (\ref{eq:Z2xZ2 1000 rep})
and (\ref{eq:Z2xZ2 1010 rep}) are stacked,
the local fermionic Fock space $\mathcal{F}^{\,}_{j}$
of dimension $\mathcal{D}=16$
is generated by the octuplet of Majorana operators
(\ref{eq:Majorana octuplet for (1,(1,0),0)x(1,(1,0),0)}).
One verifies that
\begin{subequations}\label{eq:Z2xZ2 0010 rep}
\begin{align}
&
\hat{u}^{\,}_{j}(g^{\,}_{1})\df
\hat{\chi}^{\,}_{j,1}\,
\hat{\chi}^{\,}_{j,3}\,
\hat{\chi}^{\,}_{j,5}\,
\hat{\chi}^{\,}_{j,7},
\\
&
\hat{u}^{\,}_{j}(g^{\,}_{2})\df
\hat{\chi}^{\,}_{j,2}\,
\hat{\chi}^{\,}_{j,3}\,
\hat{\chi}^{\,}_{j,5},
\\
&
\hat{u}^{\,}_{j}(p)\df
\hat{\chi}^{\,}_{j,1}\,
\hat{\chi}^{\,}_{j,2}\,
\hat{\chi}^{\,}_{j,3}\,
\hat{\chi}^{\,}_{j,4}\,
\hat{\chi}^{\,}_{j,5}\,
\hat{\chi}^{\,}_{j,6}\,
\hat{\chi}^{\,}_{j,7}\,
\hat{\chi}^{\,}_{j,8},
\end{align}
\end{subequations}
realizes the projective representation
(\ref{eq:projective rep G=Z2xZ2})
with
$([\nu],[\rho],\mu)=\big(0,(0,1),0\big)$.

Equation (\ref{eq:Z2xZ2 0010 rep})
differs from Eq.\ (\ref{eq:Z2xZ2 0100 rep})
by interchanging $g^{\,}_{1}$ and $g^{\,}_{2}$.
This difference does not affect the reasoning
leading to to the conclusion that the gapless
Hamiltonian (\ref{eq:Z2xZ2 0100 rep example gapless})
is the most general translation-invariant,
$G^{\,}_{f}$-invariant, order 2, and range $r=1$
Hamiltonian of the form (\ref{eq:gen_Ham})
whose hopping is diagonal with respect to the flavor index.
This difference also implies that stacking 
two copies of the $\big(1,(0,1),0\big)$ representation
(\ref{eq:Z2xZ2 0010 rep})
delivers the trivial projective representation
$\big(0,(0,0),0\big)$
encoded by Eqs.\
(\ref{eq:Z2xZ2 0100 stacked rep}),
for which Theorem
\ref{thm:LSM Theorem 1}
is not predictive anymore.

\subsubsection{Group cohomology class $([\nu],[\rho],\mu)=\big(0,(1,1),0\big)$}

When representations (\ref{eq:Z2xZ2 1000 rep})
and (\ref{eq:Z2xZ2 1110 rep}) are stacked,
the local fermionic Fock space $\mathcal{F}^{\,}_{j}$
of dimension $\mathcal{D}=64$
is generated by 12 Majorana operators.
One verifies that
\begin{subequations}\label{eq:Z2xZ2 0110 rep}
\begin{align}
&
\hat{u}^{\,}_{j}(g^{\,}_{1})\df
\mathrm{i}\,
\hat{\chi}^{\,}_{j,2}\,
\hat{\chi}^{\,}_{j,3}\,
\hat{\chi}^{\,}_{j,5}\,
\hat{\chi}^{\,}_{j,9}\,
\hat{\chi}^{\,}_{j,11},
\\
&
\hat{u}^{\,}_{j}(g^{\,}_{2})\df
\mathrm{i}\,
\hat{\chi}^{\,}_{j,1}\,
\hat{\chi}^{\,}_{j,3}\,
\hat{\chi}^{\,}_{j,5}\,
\hat{\chi}^{\,}_{j,7}\,
\hat{\chi}^{\,}_{j,9},
\\
&
\hat{u}^{\,}_{j}(p)\df
\prod_{a=1}^{12}
\hat{\chi}^{\,}_{j,a},
\end{align}
\end{subequations}
realizes the projective representation
(\ref{eq:projective rep G=Z2xZ2})
with
$([\nu],[\rho],\mu)=\big(0,(1,1),0\big)$.

An example of a translation- and $G^{\,}_{f}$-invariant
Hamiltonian of the form (\ref{eq:gen_Ham}),
order 2, and range $r=1$ is
\begin{equation}
\begin{split}
\widehat{H}^{\,}_{\mathrm{pbc}}\df&\,
\sum_{j=1}^{N}
\Bigg[
\sum_{a=1}^{12}
\lambda^{\,}_{a}\,
\mathrm{i}
\hat{\chi}^{\,}_{j,a}\,
\hat{\chi}^{\,}_{j+1,a}
+
\lambda^{\,}_{1,7}\,
\mathrm{i}
\hat{\chi}^{\,}_{j,1}\,
\hat{\chi}^{\,}_{j,7}
\\
&\,
+
\lambda^{\,}_{3,5}\,
\mathrm{i}
\hat{\chi}^{\,}_{j,3}\,
\hat{\chi}^{\,}_{j,5}
+
\lambda^{\,}_{2,11}\,
\mathrm{i}
\hat{\chi}^{\,}_{j,2}\,
\hat{\chi}^{\,}_{j,11}
\\
&\,
+
\lambda^{\,}_{4,6}\,
\mathrm{i}
\hat{\chi}^{\,}_{j,4}\,
\hat{\chi}^{\,}_{j,6}
+
\lambda^{\,}_{8,10}\,
\mathrm{i}
\hat{\chi}^{\,}_{j,8}\,
\hat{\chi}^{\,}_{j,10}
\Bigg]
\end{split}
\label{eq:Z2xZ2 0110 rep example gapless}
\end{equation}
with $\lambda^{\,}_{a},\lambda^{\,}_{a,b}\in\mathbb{R}$ for $a,b=1,\cdots,12$.
By construction, this Hamiltonian is gapless since
the quantum critical Kitaev chains 9 and 12 are
decoupled from all other Kitaev chains. This gapless phase
is stable to any quadratic on-site perturbation
since the only on-site quadratic terms
$
\mathrm{i}
\hat{\chi}^{\,}_{j,9}\,
\hat{\chi}^{\,}_{j,a}
$
and
$
\mathrm{i}
\hat{\chi}^{\,}_{j,12}\,
\hat{\chi}^{\,}_{j,a}
$
that could gap the quantum critical Kitaev chains 9 and 12
are odd under $G$ for any $a\neq9,12$.
The stability of this gapless phase to on-site quadratic perturbations
can thus be thought of as a consequence of Theorem
\ref{thm:LSM Theorem 1}. Theorem \ref{thm:LSM Theorem 1}
also predicts that any
$G^{\,}_{f}$-symmetric interaction of the form (\ref{eq:gen_Ham})
that opens a spectral gap in the noninteracting spectrum
must break spontaneously at least one of the
symmetries responsible
for the noninteracting spectrum being gapless.

When two copies of this $\big(0,(1,1),0\big)$ representation 
are stacked, the local fermionic Fock space $\mathcal{F}^{\,}_{j}$
of dimension $\mathcal{D}=2^{12}$ is generated by $24$ Majorana operators.
The $\mathbb{Z}^{\,}_{2}$-graded tensor product of the projective representation
(\ref{eq:Z2xZ2 0110 rep}) with itself realizes the group cohomology class
$([\nu],[\rho],\mu)=\big(0,(0,0),0\big)$,
i.e., the trivial group cohomology class.
There exists a bijective map $a\mapsto a'\df(a+12)\hbox{ mod }24$ such that
all on-site terms
$\mathrm{i}\hat{\chi}_{j,a}\,\hat{\chi}_{j,a'}$
with $a=1,\cdots,12$ can be shown to be $G^{\,}_{f}$ symmetric.
The ground-state degeneracy of any translation- and $G^{\,}_{f}$-invariant
Hamiltonian of the form (\ref{eq:gen_Ham})
can be lifted by including the 12 on-site terms
$\mathrm{i}\hat{\chi}_{j,a}\,\hat{\chi}_{j,a+12}$ with $a=1,\cdots,12$,
i.e., Theorem \ref{thm:LSM Theorem 1} is not predictive.

\subsection{One-dimensional space with the symmetry group $\mathbb{Z}^{FT}_{4}$
for an even number of local Majorana flavors}

The lattice is $\Lambda=\{1,\cdots,N\}$ with $N=2M$ an even integer
and the global fermionic Fock space $\mathcal{F}^{\,}_{\Lambda}$
is of dimension $2^{n\,M}$ with $n$ the number of local Majorana flavors.
By choosing the cardinality $|\Lambda|=2M$ to be even,
we make sure that the lattice is bipartite. This
allows to treat the two values of the index
$\mu\df\hbox{$n$ mod 2}$ in parallel.
We shall only consider the case when $n=2m$ with $m$ a positive integer
and $\mu=0$.
The symmetry group 
$G^{\,}_{f}\df\mathbb{Z}^{FT}_{4}\df\{t,t^{2},t^{3},t^{4}\}$
is the nontrivial central extension of $G\equiv\mathbb{Z}^{T}_{2}=\{t,t^{2}\}$
by $\mathbb{Z}^{F}_{2}\equiv\{p,p^{2}\}$, where the identification
$t^{2}=p$ is made. The upper index $T$
for the cyclic group $G\equiv\mathbb{Z}^{T}_{2}\equiv\{e,t\}$
refers to the interpretation of $t$ as reversal of time
(see Appendix \ref{appsubsec: ZFT4}).
As usual, $p$ denotes fermion parity.
The symmetry group $G^{\,}_{f}$ is thus
generated by reversal of time $t$,
whereby reversal of time squares to the fermion parity $p$.

The local antiunitary representation $\hat{u}^{\,}_{j}(t)$
of reversal of time generates a projective representation of the group
$\mathbb{Z}^{FT}_{4}$.
According to Appendix \ref{appsubsec: ZFT4},
all cohomologically distinct projective representations
of $\mathbb{Z}^{FT}_{4}$
are determined by the indices $[(\nu,\rho)]$, with $(\nu,\rho)\in C^{2}(G,\mathrm{U(1)})\times C^{1}(G,\mathbb{Z}^{\,}_{2})$,
through the relations\,
\footnote{
We have chosen the convention of always representing
the generator $p$ of $\mathbb{Z}^{F}_{2}$ by a Hermitian operator
according to Eq.\ (\ref{eq: def uj(g) e})} 
\begin{subequations}\label{eq:indices for ZTF4}
\begin{align}
\begin{split}
&
[(\nu,\rho)] = (\rho(t),\rho(t)),
\\
&
\hat{u}^{\,}_{j}(t)\,
\hat{u}^{\,}_{j}(p)
=
(-1)^{\rho(t)}\,
\hat{u}^{\,}_{j}(p)\,
\hat{u}^{\,}_{j}(t),
\end{split}
\label{eq:indices for ZTT4 a}
\end{align}
where
\begin{equation}
\hat{u}^{2}_{j}(t)=e^{\mathrm{i}\phi(t,t)}\,\hat{u}^{\,}_{j}(p)
\label{eq:indices for ZTF4 c}
\end{equation}
and $\phi$ is the 2-cocycle defined in Eq.\
(\ref{eq:def projective rep I}).
This gives two distinct group cohomology classes
\begin{equation}
([(\nu,\rho)],0)\in
\left\{
(0,0,0),
(1,1,0)
\right\}.
\label{eq:indices for ZTF4 d}
\end{equation}
\end{subequations}
We will start with the nontrivial projective representation
in the cohomology class $([\nu],[\rho],0)=(1,1,0)$
that we shall represent using two local Majorana flavors.
We will then construct a
projective representation in the group cohomology classes
$([\nu],[\rho],0)=(0,0,0)$
by using the graded tensor product, i.e.,
by considering 4 local Majorana flavors.
This will allow us to verify explicitly
the stacking rules of Sec.\ \ref{subsec:Stacking rules}
according to which Eq.\ (\ref{eq:stacking rel for indices})
simplifies to the rule
(see Appendix \ref{appsubsec: ZFT4})
\begin{equation}
\begin{split}
&
\left(
[(\nu,\rho)],
0
\right)= 
\left(
\rho(t),\
\rho(t),\     
0
\right),
\\
&\,\,
\rho(t)=
\rho^{\,}_{1}(t)
+
\rho^{\,}_{2}(t)
\hbox{ mod 2},
\end{split}
\label{eq:ZFT4 stacking rule}
\end{equation}
when $G^{\,}_{f}=\mathbb{Z}^{FT}_{4}$.
The indices defined in Eqs.\ (\ref{eq:indices for ZTF4})
thus form the cyclic group
$\mathbb{Z}^{\,}_{2}$
with respect to the stacking rule (\ref{eq:ZFT4 stacking rule}).

\subsubsection{Group cohomology classes
$([(\nu,\rho)],\mu)=(1,1,0)$ and $([(\nu,\rho)],\mu)=(0,0,0)$}

The local fermionic Fock space $\mathcal{F}^{\,}_{j}$
of dimension $\mathcal{D}=2$
is generated by the doublet of Majorana operators
\begin{equation}
\hat{\chi}^{\,}_{j}\equiv
\begin{pmatrix}
\hat{\chi}^{\,}_{j,1}
\\
\hat{\chi}^{\,}_{j,2}
\end{pmatrix},
\qquad
j=1,\cdots,2M.
\label{eq:fundamental Majorana doublet for ZFT4 (1,1,0)}
\end{equation}
One verifies that
\begin{subequations}\label{eq:ZFT4 110 rep}
\begin{align}
&
\hat{u}^{\,}_{j}(t)\df
\frac{1}{\sqrt{2}}
\left(
\hat{\chi}^{\,}_{j,1}
-
\hat{\chi}^{\,}_{j,2}
\right)
\mathsf{K},
\label{eq:ZFT4 110 rep a}
\\
&
\hat{u}^{\,}_{j}(p)\df
\mathrm{i}
\hat{\chi}^{\,}_{j,1}\,
\hat{\chi}^{\,}_{j,2},
\label{eq:ZFT4 110 rep b}
\end{align}
\end{subequations}
realizes the projective representation
(\ref{eq:indices for ZTF4}) with
$([(\nu,\rho)],\mu)=(1,0,0)$.
With the help of
\begin{equation}
\left[\hat{u}^{\,}_{j}(t)\right]^{-1}=
\frac{1}{\sqrt{2}}
\left(
\hat{\chi}^{\,}_{j,1}
+
\hat{\chi}^{\,}_{j,2}
\right)
\mathsf{K},
\end{equation}
one also verifies that
\begin{equation}
\hat{u}^{\,}_{j}(t)
\begin{pmatrix}
\hat{\chi}^{\,}_{j,1}
\\
\hat{\chi}^{\,}_{j,2}
\end{pmatrix}
\left[\hat{u}^{\,}_{j}(t)\right]^{-1}=
\begin{pmatrix}
-\hat{\chi}^{\,}_{j,2}
\\
+\hat{\chi}^{\,}_{j,1}
\end{pmatrix}.
\label{eq:tsf under t of Majorana doublet for 100 ZTF4}
\end{equation}

It follows from
Eq.\ (\ref{eq:tsf under t of Majorana doublet for 100 ZTF4})
that the only on-site Hermitian quadratic form
$\mathrm{i}\hat{\chi}^{\,}_{j,1}\,\hat{\chi}^{\,}_{j,2}$
is odd under reversal of time. Consequently,
\begin{align}
\widehat{H}^{\,}_{\mathrm{pbc}}\df
\sum_{j=1}^{2M}
\lambda
\left(
\mathrm{i}\,
\hat{\chi}^{\,}_{j,1}\,
\hat{\chi}^{\,}_{j+1,1}
-
\mathrm{i}\,
\hat{\chi}^{\,}_{j,2}\,
\hat{\chi}^{\,}_{j+1,2}
\right)
\end{align}
with $\lambda\in\mathbb{R}$
is the most general translation- and $G^{\,}_{f}$-invariant
Hamiltonian of the form (\ref{eq:gen_Ham})
of order 2, range $r=1$, and
whose hopping is diagonal with respect to the flavor index.
This Hamiltonian describes two Kitaev chains that have been
fine-tuned to their quantum critical point
(\ref{eq:def Kitaev chain at criticality})
between their symmetry-protected and topologically inequivalent gapped phases.
The stability of this gapless phase to on-site quadratic perturbations
can thus be thought of as a consequence of Theorem
\ref{thm:LSM Theorem 1}. Theorem \ref{thm:LSM Theorem 1}
also predicts that any
$G^{\,}_{f}$-symmetric interaction of the form (\ref{eq:gen_Ham})
that opens a spectral gap in the noninteracting spectrum
must break spontaneously at least one of the
symmetries responsible
for the noninteracting spectrum being gapless.

When two copies of the projective representation
(\ref{eq:ZFT4 110 rep})
are stacked, the local fermionic Fock space $\mathcal{F}^{\,}_{j}$
of dimension $\mathcal{D}=4$ is generated by $4$ Majorana operators.
The $\mathbb{Z}^{\,}_{2}$-graded tensor product of the projective representation
(\ref{eq:ZFT4 110 rep})
with itself realizes the group cohomology class
$([(\nu,\rho)],\mu)=\big(0,0,0\big)$,
i.e., the trivial group cohomology class.
There exists a bijective map $a\mapsto a'\df(a+2)\hbox{ mod }4$ such that
all on-site terms
$
\mathrm{i}\hat{\chi}_{j,a}\,\hat{\chi}_{j,a'}
-
\mathrm{i}\hat{\chi}_{j,a+1}\,\hat{\chi}_{j,(a+1)'}
$
with $a=1,2$ can be shown to be $G^{\,}_{f}$ symmetric.
The ground-state degeneracy of any translation- and $G^{\,}_{f}$-invariant
Hamiltonian of the form (\ref{eq:gen_Ham})
can be lifted by increasing the strength of
$
\mathrm{i}\hat{\chi}_{j,a}\,\hat{\chi}_{j,a'}
-
\mathrm{i}\hat{\chi}_{j,a+1}\,\hat{\chi}_{j,(a+1)'}
$
with $a=1,2$,
i.e., Theorem \ref{thm:LSM Theorem 1} is not predictive.

\subsection{One-dimensional space with the symmetry group $\mathbb{Z}^{F}_{2}$
for an odd number of local Majorana flavors}

The lattice is $\Lambda=\{1,\cdots,N\}$ with $N=2M$ an even integer
and the global fermionic Fock space $\mathcal{F}^{\,}_{\Lambda}$
is of dimension $2^{(2m+1)\,M}$ with $(2m+1)$
the number of local Majorana flavors, i.e., $\mu=1$.
By choosing the cardinality $|\Lambda|=2M$ to be even,
we make sure that the lattice is bipartite. 
The symmetry group is
$G^{\,}_{f}\df\{p,p^{2}\}\equiv\mathbb{Z}^{F}_{2}$
where $p$ denotes fermion parity.
If we reinterpret
$G^{\,}_{f}=G\times\mathbb{Z}^{F}_{2}$ with $G=\{e\}$ the group with
one element,
we deduce that the indices $[\nu]$ and $[\rho]$ are trivial,
i.e., $[\nu]=[\rho]=0$. The index associated with this group is then
$(0,0,1)$, for which we illustrate how 
translation symmetry prevents a nondegenerate gapped
ground state in agreement with Theorem \ref{thm:LSM Theorem 2}.

We define the $2^{(2m+1)M}$-dimensional
global Fock space $\mathcal{F}^{\,}_{\Lambda}$
using the $2(2m+1)M$ Majorana operators
obeying the algebra
\begin{equation}
\hat{\chi}^{\dag}_{j,a}=\chi^{\,}_{j,a},
\quad
\hat{\chi}^{2}_{j,a}=1,
\quad
\{\chi^{\,}_{j,a},\chi^{\,}_{j,a'}\}=2\delta^{\,}_{j,j'}\,\delta^{\,}_{a,a'},
\end{equation}
for $j,j'=1,\cdots,2M$, $a,a'=1,\cdots,2m+1$.

For simplicity, we 
first choose $2m+1=1$ local Majorana flavors. Hamiltonian
\begin{align}
\widehat{H}^{\,}_{\mathrm{pbc}}\df
\lambda
\sum_{j=1}^{2M}
\mathrm{i}\,
\hat{\chi}^{\,}_{j}\,
\hat{\chi}^{\,}_{j+1},
\qquad
\lambda\in\mathbb{R},
\label{eq:Kitaev at criticality revisited for 001 mu=1}
\end{align}
is the most general translation- and $G^{\,}_{f}$-invariant
Hamiltonian of the form (\ref{eq:gen_Ham})
of range $r=1$. It is gapless as it describes one
Kitaev chain on $M$ sites that has been
fine-tuned to its quantum critical point
(\ref{eq:def Kitaev chain at criticality})
between its symmetry-protected and topologically inequivalent gapped phases.

If we now consider $2m+1>1$ local Majorana flavors arranged into
the $(2m+1)$-multiplet $\hat{\chi}^{\,}_{j}$,
the most general translation- and $G^{\,}_{f}$-invariant
Hamiltonian of the form (\ref{eq:gen_Ham})
of order 2 and range $r=1$ is
\begin{align}
\widehat{H}^{\,}_{\mathrm{pbc}}\df
\sum_{j=1}^{2M}
\mathrm{i}\,
\hat{\chi}^{\mathsf{T}}_{j}\,
\mathsf{M}\,
\hat{\chi}^{\,}_{j+1}
\label{eq:Kitaev at criticality revisited for 001 mu=1 2m+1>1}
\end{align}
where the $(2m+1)\times(2m+1)$-dimensional matrix
$\mathsf{M}$ is real-valued and antisymmetric.
As $\mathsf{M}$ has necessarily a zero eigenvalue, the
spectrum of $\widehat{H}^{\,}_{\mathrm{pbc}}$ is gapless.

Theorem \ref{thm:LSM Theorem 2}
predicts that any
$G^{\,}_{f}$-symmetric interaction of the form (\ref{eq:gen_Ham})
that opens a spectral gap in the noninteracting spectrum
of Hamiltonian
(\ref{eq:Kitaev at criticality revisited for 001 mu=1 2m+1>1})
must break spontaneously the symmetries responsible
for the noninteracting spectrum being gapless.

After stacking two copies of the projective representation $(0,0,1)$,
it is possible to define a local fermionic Fock space with the gapless
Hamiltonian
\begin{equation}
\begin{split}
&
\widehat{H}^{\,}_{\mathrm{pbc}}\df
\sum_{j=1}^{2M}
\sum_{a=1}^{2}
\mathrm{i}\,
\hat{\chi}^{\mathsf{T}}_{j,a}\,
\mathsf{M}^{\,}_{a}\,
\hat{\chi}^{\,}_{j+1,a}
\\
&
\mathsf{M}^{\,}_{a}=-\mathsf{M}^{\mathsf{T}}_{a}\in\mathrm{Mat}(2m+1,\mathbb{R}).
\end{split}
\end{equation}
The on-site mass term
$\mathrm{i}\,
\hat{\chi}^{\mathsf{T}}_{j,1}\,
\hat{\chi}^{\,}_{j,2}$
is compatible with parity conservation.
When added to the right-hand side
so as to preserve translation symmetry,
it selects a nondegenerate gapped ground state.
There is no contradiction
with Theorem \ref{thm:LSM Theorem 2} since the
the projective representation under stacking rules is
the trivial one $(0,0,0)$. 

As was suggested just after Eq.\
(\ref{eq:Kitaev at criticality revisited for 001 mu=1}),
we can always do the reinterpretation
\begin{subequations}\label{eq:Kitaev at criticality revisited for 001 mu=1 bis}
\begin{align}
&
\widehat{H}^{\,}_{\mathrm{pbc}}\df
\lambda
\sum_{l=1}^{M}
\mathrm{i}
\left(
\hat{\chi}^{\,}_{\mathrm{o},l}\,
\hat{\chi}^{\,}_{\mathrm{e},l}
+
\hat{\chi}^{\,}_{\mathrm{o},l}\,
\hat{\chi}^{\,}_{\mathrm{e},l+1}
\right),
\\
&
\hat{\chi}^{\,}_{\mathrm{o},l}\df
\hat{\chi}^{\,}_{2l-1},
\quad
\hat{\chi}^{\,}_{\mathrm{e},l}\df
\hat{\chi}^{\,}_{2l},
\end{align}
\end{subequations}
according to which the enlarged repeat unit cell
is labeled by the odd sites $\Lambda^{\,}_{A}$ of $\Lambda$ and
there are two flavors (even and odd) of Majorana
per enlarged repeat unit cell. It is then tempting
to ask if one could use Theorem \ref{thm:LSM Theorem 1}
for some group $G$ to understand the spectrum of Eq.\
(\ref{eq:Kitaev at criticality revisited for 001 mu=1 bis}),
in which case the need for
Theorem \ref{thm:LSM Theorem 2}
would be superfluous at best or contradictory at worst.
However, there is no conflict between
Theorem \ref{thm:LSM Theorem 2}
and
Theorem \ref{thm:LSM Theorem 1},
as Theorem \ref{thm:LSM Theorem 1}
cannot be applied to understand the spectrum of Eq.\
(\ref{eq:Kitaev at criticality revisited for 001 mu=1 bis}).
To see this, we observe that the translation
by one original repeat unit cell that is presumed by
Theorem \ref{thm:LSM Theorem 2}
is represented
in Eq.\ (\ref{eq:Kitaev at criticality revisited for 001 mu=1 bis})
by
\begin{equation}
\hat{\chi}^{\,}_{\mathrm{o},l}\mapsto
\hat{\chi}^{\,}_{\mathrm{e},l},
\qquad
\hat{\chi}^{\,}_{\mathrm{e},l}\mapsto
\hat{\chi}^{\,}_{\mathrm{o},l+1},
\qquad
l=1,\cdots,M.
\end{equation}
As this transformation is not internal to the enlarged repeat unit
cell, it cannot be interpreted as the group $G$ of
on-site symmetries needed to establish Theorem \ref{thm:LSM Theorem 1}.

\section{Summary}
\label{sec:summary}

In this work, we have obtained two
Lieb-Schultz-Mattis type no-go constraints
that forbid the existence of a nondegenerate gapped ground state for
translationally invariant local lattice Hamiltonians
of the form (\ref{eq:gen_Ham}) in any dimension.
Theorem \ref{thm:LSM Theorem 1} is proved within the FMPS framework and
presumes that the repeat unit cell hosts
a finite even number of Majorana degrees of freedom that, in turn,
realize a nontrivial projective representation of
a global symmetry group $G^{\,}_{f}$.
We have extended Theorem \ref{thm:LSM Theorem 1}
to any dimension $d>1$ by using tilted and twisted boundary conditions, albeit 
with some restrictions on the global symmetry group $G^{\,}_{f}$.
Theorem \ref{thm:LSM Theorem 2}
presumes that the repeat unit cell of any $d$-dimensional lattice
hosts a finite odd number of Majorana degrees of freedom
(of course the lattice must then host an even number  of sites).
Such Lieb-Schultz Mattis-type theorems provide
non-perturbative constraints on the nature of the ground state which
are expected to have applications in fermionic models with broken $\mathrm{U(1)}$
number-conservation symmetry, but with additional global symmetries
$G^{\,}_{f}$ present. Notably, such LSM-type constraints
dictate the conditions under which a Fermi liquid can be unstable
to a low-temperature phase in which superconducting long-range order
coexists with the long-range order associated to the spontaneous breaking
of some symmetry group $G^{\,}_{f}$.

\section*{Acknowledgments}
We thank Yoshiko Ogata for the helpful comments on
the manuscript.
\"O.M.A. was supported by the 
Swiss National Science Foundation (SNSF) 
under Grant No.\ 200021 184637.
A.T. acknowledges funding by the European Union’s
Horizon 2020 research and innovation program under 
the Marie Sklodowska Curie grant agreement No 701647.

\onecolumngrid
\appendix
\addappheadtotoc

\section{Group cohomology}
\label{appsec:Group Cohomology}

We review some basic concepts in group cohomology
in Sec.\ \ref{appsubsec: Some definitions}.
We then calculate the relevant group cohomologies for
the projective representations of the group $G^{\,}_{f}$
considered in
Secs.\
\ref{appsubsec: ZT2 times ZF2}, 
\ref{appsubsec: Z2 Z2 ZF2},
and
\ref{appsubsec: ZFT4}.

\subsection{Some definitions}
\label{appsubsec: Some definitions}
\noindent Given two groups $G$ and $M$, an \textit{$n$-cochain} is the map
\begin{equation}
\begin{split}
\phi \colon G^{n} \to&\, M,
\\
(g^{\,}_{1},g^{\,}_{2},\cdots, g^{\,}_{n})\mapsto&\,
\phi(g^{\,}_{1},g^{\,}_{2},\cdots, g^{\,}_{n}),
\end{split}
\end{equation}
that maps an $n$-tuple $(g^{\,}_{1},g^{\,}_{2},\cdots, g^{\,}_{n})$
to an element $\phi(g^{\,}_{1},g^{\,}_{2},\dots,g^{\,}_{n})\in M$.
The set of all $n$-cochains from $G^{n}$ to $M$ 
is denoted by $C^{n}(G,M)$.
We define an $M$-valued 0-cochain to be an element of the group $M$
itself, i.e., $C^{0}(G,M)= M$.
Henceforth, we will denote 
the group composition rule in $G$ by $\cdot$
and the group composition rule in $M$ additively by $+$
($-$ denoting the inverse element).

Given the group homomorphism $\mathfrak{c}\colon G \to \left\{-1,1\right\}$,
for any $g\in G$, we define the group action
\begin{equation}
\begin{split} 
\mathfrak{C}^{\,}_{g}
\colon 
M \to&M,
\\
m\ \mapsto&\
\mathfrak{c}(g)\,m.
\end{split}
\end{equation}
The homomorphism $\mathfrak{c}$ indicates whether and element $g\in G$
is represented unitarily [$\mathfrak{c}(g)=+1$] or antiunitarily
[$\mathfrak{c}(g)=-1$]. 
We define the map $\delta^{n}_{\mathfrak{c}}$ 
\begin{subequations}\label{appeq:definition delta n}
\begin{equation}
\begin{split}
\delta^{n}_{\mathfrak{c}}\colon 
C^{n}(G,M)\ \to&\ C^{n+1}(G,M),
\\
\phi\ \mapsto&\ \left(\delta^{n}_{\mathfrak{c}}\phi\right),
\end{split}
\end{equation}
from $n$-cochains to $(n+1)$-cochains such that
\begin{align}
\label{eq:def coboundary operator}
\left(\delta^{n}_{\mathfrak{c}}\phi\right)
(g^{\,}_{1},\cdots,g^{\,}_{n+1})\df
\mathfrak{C}^{\,}_{g^{\,}_{1}}\!
\left(\phi(g^{\,}_{2},\cdots, g^{\,}_{n},g^{\,}_{n+1})\right)
+
\sum_{i=1}^{n}
(-1)^{i}
\phi(g^{\,}_{1},\cdots,g^{\,}_{i}\cdot g^{\,}_{i+1},
\cdots,g^{\,}_{n+1})
-
(-1)^{n}\,
\phi(g^{\,}_{1},\cdots,g^{\,}_{n}).
\end{align}
\end{subequations}
The map $\delta^{n}_{\mathfrak{c}}$
is called a \textit{coboundary operator}.  

\textbf{Example $n=2$:}
The coboundary operator $\delta^{2}_{\mathfrak{c}}$ is defined by
\begin{align}
\left(\delta^{2}_{\mathfrak{c}}\phi\right)
(g^{\,}_{1},g^{\,}_{2},g^{\,}_{3})=&\,
\mathfrak{C}^{\,}_{g^{\,}_{1}}\!
\left(\phi(g^{\,}_{2},g^{\,}_{3})\right)
+
(-1)^{1}
\phi(g^{\,}_{1}\cdot g^{\,}_{2},g^{\,}_{3})
+
(-1)^{2}
\phi(g^{\,}_{1}, g^{\,}_{2}\cdot g^{\,}_{3})
-
(-1)^{2}\,
\phi(g^{\,}_{1},g^{\,}_{2})
\nonumber\\
=&\,
\mathfrak{c}(g^{\,}_{1})\,
\phi(g^{\,}_{2},g^{\,}_{3})
-
\phi(g^{\,}_{1}\cdot g^{\,}_{2},g^{\,}_{3})
+
\phi(g^{\,}_{1}, g^{\,}_{2}\cdot g^{\,}_{3})
-
\phi(g^{\,}_{1},g^{\,}_{2}).
\label{appeq:example delta2}
\end{align}
We observe that
\begin{align}
\left(\delta^{2}_{\mathfrak{c}}\phi\right)
(g^{\,}_{1},g^{\,}_{2},g^{\,}_{3})=0
\ \Longleftrightarrow\
\phi(g^{\,}_{1},g^{\,}_{2})
+
\phi(g^{\,}_{1}\cdot g^{\,}_{2},g^{\,}_{3})=
\phi(g^{\,}_{1}, g^{\,}_{2}\cdot g^{\,}_{3})
+
\mathfrak{c}(g^{\,}_{1})\,
\phi(g^{\,}_{2},g^{\,}_{3})
\end{align}
is nothing but the 2-cocycle condition \eqref{eq:def projective rep II b}
obeyed by
$\phi$.

\textbf{Example $n=1$:}
The coboundary operator $\delta^{1}_{\mathfrak{c}}$ is defined by
\begin{align}
\left(\delta^{1}_{\mathfrak{c}}\phi\right)
(g^{\,}_{1},g^{\,}_{2})=&\,
\mathfrak{C}^{\,}_{g^{\,}_{1}}\!
\left(\phi(g^{\,}_{2})\right)
+
(-1)^{1}
\phi(g^{\,}_{1}\cdot g^{\,}_{2})
-
(-1)^{1}
\phi(g^{\,}_{1})
\nonumber\\
=&\,
\mathfrak{c}(g^{\,}_{1})\,
\phi(g^{\,}_{2})
-
\phi(g^{\,}_{1}\cdot g^{\,}_{2})
+
\phi(g^{\,}_{1}).
\label{appeq:example delta1}
\end{align}
One verifies the important identity
\begin{equation}
\Phi(g^{\,}_{1},g^{\,}_{2})\df
\left(\delta^{1}_{\mathfrak{c}}\phi\right)
(g^{\,}_{1},g^{\,}_{2})
\ \implies \
(\delta^{2}_{\mathfrak{c}}\Phi)(g^{\,}_{1},g^{\,}_{2},g^{\,}_{3})=0.
\label{appeq:one-cochain always obeys the cocycle conditions}
\end{equation} 
We observe that Eq.\ \eqref{eq:coboundary condition} 
implies that $\phi$ is the image of $\xi$ under
the coboundary operator $\delta^{1}_{\mathfrak{c}}$. 
Using the coboundary operator, we define two sets 
\begin{subequations}
\begin{align}
Z^{n}(G,M^{\,}_{\mathfrak{c}})\df
\mathrm{ker}(\delta^{n}_{\mathfrak{c}})=
\left\{
\phi \in C^{n}(G,M) \ |\
\delta^{n}_{\mathfrak{c}}\phi=0
\right\},
\end{align}
and
\begin{align}
B^{n}(G,M^{\,}_{\mathfrak{c}})\df
\mathrm{im}(\delta^{n-1}_{\mathfrak{c}})=
\left\{
\phi \in C^{n}(G,M) \ |\ 
\phi=\delta^{n-1}_{\mathfrak{c}}\phi',\,\,
\phi'\in C^{n-1}(G,M)
\right\}.
\end{align}
\end{subequations}
The cochains in $Z^{n}(G,M^{\,}_{\mathfrak{c}})$
are called \textit{$n$-cocycles}.
The cochains in $B^{n}(G,M^{\,}_{\mathfrak{c}})$
are called \textit{$n$-coboundaries}.
The action of the boundary operator on the elements of the group $M$
is sensitive to the homomorphism $\mathfrak{c}$.
For this reason, we label $M$ by
$\mathfrak{c}$ in $Z^{n}(G,M^{\,}_{\mathfrak{c}})$
and $B^{n}(G,M^{\,}_{\mathfrak{c}})$.
The importance of the coboundaries is that the identity
(\ref{appeq:one-cochain always obeys the cocycle conditions})
generalizes to
\begin{equation}
\phi=\delta^{n-1}_{\mathfrak{c}}\phi'
\ \implies\
\delta^{n}_{\mathfrak{c}}\phi=0.
\label{appeq:n-cochain always obeys the cocycle conditions}
\end{equation} 
The $n$th cohomology group is defined as the quotient of the
$n$-cocycles by the $n$-coboundaries, i.e.,
\begin{align}
H^{n}(G,M^{\,}_{\mathfrak{c}})\df
Z^{n}(G,M^{\,}_{\mathfrak{c}})/B^{n}(G,M^{\,}_{\mathfrak{c}}).
\end{align}
From now on, we omit the labels $n$ and $\mathfrak{c}$ in
$\delta^{n}_{\mathfrak{c}}$
for convenience. It should be understood that the map $\delta$
acting on a cochain $\phi$ maps it to a cochain of one higher degree. 
The $n$th cohomology group $H^{n}(G,M^{\,}_{\mathfrak{c}})$
is an additive Abelian group.
We denote its elements by $[\phi]\in H^{n}(G,M^{\,}_{\mathfrak{c}})$,
i.e., the equivalence class
of the $n$-cocycle $\phi$.

Finally, we define the following operation on the cochains.
Given two cochains $\phi\in C^{n}(G,N)$ and $\theta\in C^{m}(G,M)$,
we produce the cochain
$(\phi \cup \theta)\in C^{n+m}(G,N\times M)$
through
\begin{subequations}\label{eq:def cup product}
\begin{align}
(\phi \cup \theta)
(g^{\,}_{1},\cdots,g^{\,}_{n},g^{\,}_{n+1},\cdots,g^{\,}_{m})\df
\Big(
\phi(g^{\,}_{1},\cdots,g^{\,}_{n}),
\mathfrak{C}^{\,}_{g^{\,}_{1}\cdot g^{\,}_{2}\cdots g^{\,}_{n}}\!
\left(
\theta(g^{\,}_{n+1},\cdots,g^{\,}_{n+m})
\right)
\Big).
\label{eq:def cup product a}
\end{align}
If we compose operation (\ref{eq:def cup product})
with the pairing map $f:N\times M \to M'$ where $M'$ is an Abelian group,
we obtain the cup product
\begin{align}
(\phi \smile \theta) (g^{\,}_{1},\cdots,g^{\,}_{n},g^{\,}_{n+1},\cdots,g^{\,}_{m})
\df
f
\Bigg(
\Big(
\phi(g^{\,}_{1},\cdots,g^{\,}_{n}),
\mathfrak{C}^{\,}_{g^{\,}_{1}\cdot g^{\,}_{2}\cdots g^{\,}_{n}}\!
\left(
\theta(g^{\,}_{n+1},\cdots,g^{\,}_{n+m})
\right)
\Big)
\Bigg).
\label{eq:def cup product b}
\end{align}
\end{subequations}
Hence,
$(\phi \smile \theta)\in C^{n+m}(G,M')$.
For our purposes, both $N$ and $M$ are subsets of the integer numbers,
$M'=\mathbb{Z}^{\,}_{2}$, while the pairing map $f$ is
\begin{align}
f
\Bigg(
\Big(
\phi(g^{\,}_{1},\cdots,g^{\,}_{n}),
\mathfrak{C}^{\,}_{g^{\,}_{1}\cdot g^{\,}_{2}\cdots g^{\,}_{n}}\!
\left(
\theta(g^{\,}_{n+1},\cdots,g^{\,}_{n+m})
\right)
\Big)
\Bigg)\df
\phi(g^{\,}_{1},\cdots,g^{\,}_{n})\,
\mathfrak{C}^{\,}_{g^{\,}_{1}\cdot g^{\,}_{2}\cdots g^{\,}_{n}}\!
\left(
\theta(g^{\,}_{n+1},\cdots,g^{\,}_{n+m})
\right)
\hbox{ mod 2}
\end{align}
where multiplication of cochains $\phi$ and $\theta$ is treated 
as multiplication of integers numbers modulo $2$.
For instance, for the cup product of a 1-cochain
$\alpha\in C^{1}(G,\mathbb{Z}^{\,}_{2})$ 
and a 2-cochain $\beta \in C^{2}(G,\mathbb{Z}^{\,}_{2})$, we write
\begin{align}
(\alpha \smile \beta ) (g^{\,}_{1},g^{\,}_{2},g^{\,}_{3})=
\alpha(g^{\,}_{1})\,
\mathfrak{C}^{\,}_{g^{\,}_{1}}\!
\left(
\beta(g^{\,}_{2},g^{\,}_{3})
\right)
=
\alpha(g^{\,}_{1})\,
\beta(g^{\,}_{2},g^{\,}_{3}),
\end{align}
where the cup product takes values in $\mathbb{Z}^{\,}_{2}=\left\{0,1\right\}$
and multiplication of $\alpha$ and $\beta$ is the multiplication of integers.
In reaching the last equality,
we have used the fact that the 2-cochain $\beta(g^{\,}_{2},g^{\,}_{3})$ 
takes values in $\mathbb{Z}^{\,}_{2}$ for which 
$\mathfrak{C}^{\,}_{g^{\,}_{1}}(\beta(g^{\,}_{2},g^{\,}_{3}))=
\beta(g^{\,}_{2},g^{\,}_{3})$ for any $g^{\,}_{1}$.
The cup product defined in Eq.\ \eqref{eq:def cup product b} satisfies
\begin{align}
\delta (\phi \smile \theta)
=
\left( \delta \phi \smile \theta \right)
+
(-1)^{n}\,
\left( \phi \smile \delta\theta \right),
\label{eq:boundary of a cup product}
\end{align}
given two cochains $\phi\in C^{n}(G,N)$ and $\theta\in C^{m}(G,M)$.
Hence, the cup product of two cocycles is again a cocycle as the
right-hand side of Eq.\ \eqref{eq:boundary of a cup product} vanishes.
Having introduced the basics of group cohomology, next we shall compute the 
cohomology groups
$[\phi]\in H^{2}\big(G^{\,}_{f},\mathrm{U(1)}^{\,}_{\mathfrak{c}}\big)$
for some specific finite Abelian groups $G$ encountered in
Sec.\ \ref{sec:examples}
and whose projective representations are defined by Eqs.\
\eqref{eq:def projective rep I}
and
\eqref{eq:def projective rep II}.

\subsection{Classification of projective representations of $G^{\,}_{f}$}
\label{appsubsec: Classification of projective representations}
It was described in Sec.\
\ref{subsec:Marrying the fermion parity with the symmetry group G},
how a global symmetry group $G^{\,}_{f}$ for
a fermionic quantum system naturally contains the fermion{-}number parity
{symmetry group} $\mathbb{Z}^{F}_{2}$
in its center, i.e., it is a central
extension of a group $G$ by $\mathbb{Z}^{F}_{2}$. Such group extension
are classified by prescribing an element
$[\gamma]\in H^{2}\big(G,\mathbb{Z}^{F}_{2}\big)$,
such that we may think of $G^{\,}_{f}$
as the set of tuples $(g,h)\in G\times\mathbb{Z}^{F}_{2}$ with
composition rule as in Eq.\ \eqref{eq:def ZF2 e}.  From this
perspective there is an implicit choice of trivialization
$\tau:G^{\,}_{f}\to\mathbb{Z}^{F}_{2}$ and projection
$b: G^{\,}_{f} \to G$ such that
\begin{align}
\tau\big((g,h)\big)=h,
\qquad
b\big((g,h)\big)=g.
\end{align}
Importantly, $\tau$ is related to the extension class $\gamma$ that
defines the group extension via the relation
\begin{equation}
b^{\star}\gamma=\delta\tau
\end{equation}
where
$b^{\star}\gamma \in H^{2}\big(G^{\,}_{f},\mathbb{Z}^{F}_{2}\big)$
is the pullback of $\gamma$ via $b$.  

As explained in Sec.\ \ref{subsec:Indices}, we shall trade the
2-cocycle $\phi(g,h)\in \mathbb{Z}^{2}(G,\mathrm{U(1)}^{\,}_{\mathfrak{c}})$
with the tuple $(\nu,\rho) \in C^{2}\big(G,\mathrm{U(1)}\big)\times
C^{1}\big(G,\mathbb{Z}^{\,}_{2}\big)$ that satisfy certain
cocycle and coboundary conditions.  To this end, it is convenient to
define the modified 2-coboundary operator
\begin{align}
\mathcal{D}^{2}_{\gamma}
\left(
\nu,\,
\rho
\right)
\df
\left(
\delta \nu -\pi\,\rho \smile \gamma,
\delta \rho
\right),
\label{eq:modified 2-coboundary}
\end{align}
acting on a tuple of cochains $(\nu,\rho) \in
C^{2}\big(G,\mathrm{U(1)}\big)\times
C^{1}\big(G,\mathbb{Z}^{\,}_{2}\big)$
together with the modified
1-coboundary operator
\begin{align}
\mathcal{D}^{1}_{\gamma}
\left(
\alpha,
\beta
\right)
\df
\left(
\delta \alpha
+
\pi
\beta\smile\gamma,
\delta \beta
\right)
\label{eq:modified 1- coboundary}
\end{align}	
acting on a tuple of cochains $(\alpha,\,\beta) \in
C^{1}\big(G,\mathrm{U(1)}\big)\times C^{0}\big(G,\mathbb{Z}^{\,}_{2}\big)$.
Being a 0-cochain $\beta$ does not take any arguments and takes values
in $\mathbb{Z}^{\,}_{2}$, i.e., $\beta \in \mathbb{Z}^{\,}_{2}$. 
Note that for the 0-cochain $\beta$,
the coboundary operator \eqref{eq:def coboundary operator} acts as 
\begin{align}
(\delta^{0}_{\mathfrak{c}}\beta)(g)=
\mathfrak{C}^{\,}_{g}(\beta)
-
\beta,
\end{align}
which in fact vanishes for any $g\in G$ 
since $\beta$ takes values in $\mathbb{Z}^{\,}_{2}$ and
$\mathfrak{C}^{\,}_{g}(\beta)=\beta$.
Using
Eq.\ \eqref{eq:boundary of a cup product} and the fact that $\gamma$
is a cocycle, i.e., $\delta \gamma = 0$, one verifies that
\begin{align}
\mathcal{D}^{2}_{\gamma}\,
\mathcal{D}^{1}_{\gamma}
(\alpha,\beta)
=
\left(
0,
0
\right)
\end{align}
for any tuple
$(\alpha,\,\beta)\in
C^{1}\big(G,\mathrm{U(1)}\big)\times C^{0}\big(G,\mathbb{Z}^{\,}_{2}\big)$.

It was proved in
Ref.\ \onlinecite{Turzillo2019} that
one may assign to any 2-cocycle 
$[\phi]\in H^{2}\big(G^{\,}_{f},\mathrm{U(1)}^{\,}_{\mathfrak{c}}\big)$
an equivalence class $[(\nu,\rho)]$ of those
tuples $(\nu,\rho)\in
C^{2}\big(G,\mathrm{U(1)}\big)\times C^{1}\big(G,\mathbb{Z}^{\,}_{2}\big)$
that satisfy the cocycle condition
under the modified 2-coboundary operator \eqref{eq:modified 2-coboundary}
given by
\begin{align}
\mathcal{D}^{2}_{\gamma}
\left(
\nu,\rho
\right)
=
\left(
\delta\nu
-
\pi\,\rho\smile\gamma,\,
\delta\rho
\right)
=
(0,0).
\label{eq:Gf_cocycle_cond}
\end{align}
Indeed, two tuples $(\nu,\rho)$ and $(\nu', \rho')$
that satisfy Eq.\ \eqref{eq:Gf_cocycle_cond} are said to be
equivalent if there exists a tuple $(\alpha,\,\beta) \in C^{1}\big(G,\mathrm{U(1)}\big)\times C^{0}\big(G,\mathbb{Z}^{\,}_{2}\big)$
such that
\begin{align}
(\nu,\,\rho)
=
(\nu',\,\rho')
+
\mathcal{D}^{1}_{\gamma}
(\alpha,\,\beta)
=
(
\nu'
+
\delta \alpha
+
\pi\,
\beta\smile\gamma,\,
\delta \beta).
\label{eq:Gf_equivalence_rel}
\end{align}
In other words, using this equivalence relation we define an
equivalence class $[(\nu,\rho)]$ of the tuple $(\nu,\rho)$ as an
element of the set
\begin{align}
[(\nu,\rho)]\in
\frac{\mathrm{ker}(\mathcal{D}^{2}_{\gamma})}
     {\mathrm{im}(\mathcal{D}^{1}_{\gamma})}.
\end{align}
The proof of the one-to-one correspondence between $[\phi]$ and $[(\nu,\rho)]$
then follows in three steps which we sketch out below.
We refer the reader to Ref.\ \onlinecite{Turzillo2019} for more details.
\begin{enumerate}
\item
First, given a cocycle $\phi\in Z^{2}(G^{\,}_{f},\mathrm{U(1)}^{\,}_{\mathfrak{c}})$, one can
define $\rho\in Z^{1}\big(G,\mathbb Z_{2}\big)$ via eq.~\eqref{eq: def rho(g)}.
The fact that $\rho$ is a cocycle follows from that fact
that $\phi$ is a cocycle.
\item
Next, one can always find a representative $\phi$ in every cohomology
class $[\phi]\in H^{2}\big(G^{\,}_{f},\mathrm{U(1)}^{\,}_{\mathfrak{c}}\big)$ that satisfies the relation
$\phi=\nu+\pi\rho\smile\tau$.
\item
Finally, the fact that $\delta \phi=0$ implies that
$\delta\nu=\pi\rho\smile\gamma$.
\end{enumerate}
We note that when the $[\gamma] = 0$, i.e., the group $G^{\,}_{f}$
splits as $G^{\,}_{f} = G\times \mathbb{Z}^{F}_{2}$, the modified
coboundary operators \eqref{eq:modified 2-coboundary} and
\eqref{eq:modified 1- coboundary} reduce to the coboundary operator
\eqref{eq:def coboundary operator} with $n=2$ and $n=1$,
respectively. If so the cochains $\nu$ and $\rho$ are both cocycles,
i.e., $(\nu,\,\rho)\in Z^{2}(G,\mathrm{U(1)}^{\,}_{\mathfrak{c}})\times
Z^{1}(G,\mathbb{Z}^{\,}_{2})$. The equivalence classes $[(\nu,\rho)]$
of the tuple $(\nu,\rho)$ is then equal to the equivalence cohomology
classes of each of its components, i.e.,
\begin{align}
[(\nu,\rho)]
=
([\nu],[\rho])
\in 
H^{2}(G,\mathrm{U(1)}^{\,}_{\mathfrak{c}})
\times
H^{1}(G,\mathbb{Z}^{\,}_{2}).
\end{align}
We use the notation $([\nu],[\rho])$ for the two indices
whenever the group $G^{\,}_{f}$ splits ($[\gamma]=0$). The
notation $[(\nu,\rho)]$ applies whenever the group $G^{\,}_{f}$ does not
split ($[\gamma]\neq0$).

\subsection{The split group $\mathbb{Z}^{T}_{2}\times\mathbb{Z}^{F}_{2}$}
\label{appsubsec: ZT2 times ZF2} 
The group $\mathbb{Z}^{T}_{2}\times \mathbb{Z}^{F}_{2}$,
where the upper index $T$ for the cyclic group
$\mathbb{Z}^{T}_{2}\equiv\{e,t\}$
refers to the interpretation of $t$ as time,
is a split group. Since the group splits ($[\gamma]=0$) upon using 
Eq.\ \eqref{eq:Gf_cocycle_cond} one finds that 
$[\phi]\in H^{2}(G^{\,}_{f},\mathrm{U(1)}^{\,}_{\mathfrak{c}})$
separates into the pair of independent indices
$[\nu]\in H^{2}(\mathbb{Z}^{T}_{2},\mathrm{U(1)}^{\,}_{\mathfrak{c}})$ and 
$[\rho]\in H^{1}(\mathbb{Z}^{T}_{2},\mathbb{Z}^{\,}_{2})$.
{Since $[\nu]=0,1$ and $[\rho]=0,1$,}
\begin{equation}
H^{2}
\Big(\mathbb{Z}^{\mathrm{T}}_{2}\times\mathbb{Z}^{F}_{2},\mathrm{U(1)}^{\,}_{\mathfrak{c}}
\Big)=
\Big\{
\big([\nu],[\rho]\big)\ \Big|\
[\nu]=0,1,
\qquad
[\rho]=0,1
\Big\}.
\label{appeq:H2(ZT2timesZF2,U(1))}
\end{equation}
Below we describe how to ``measure'' these indices
as well as the product or monoidal structure that these indices satisfy.

\medskip\noindent\textbf{Claim 1:} $[\nu]=0,1$.
\begin{proof}
Any cochain $\nu$ belonging to the equivalence class $[\nu]$
is defined by the substitution $G=\mathbb{Z}^{T}_{2}$
in Eq.\ (\ref{eq:def projective rep I})
and must satisfy the cocycle and coboundary conditions
in \eqref{eq:def projective rep II b} and 
\eqref{eq:U(1) gauge equivalence}, respectively.
If one chooses $g=h=f=t$ in Eq.\ \eqref{eq:def projective rep II b}, where 
$t\in \mathbb{Z}^{T}_{2}$ is the generator of time-reversal which
is represented antiunitarily [$\mathfrak{c}(t)=-1$]. One finds 
\begin{align}
\nu(t,t)
+
\nu(e, t)=
\nu(t,e)
-
\nu(t,t)
\text{ mod $2\pi$}
\implies
\nu(t,t)=
0, \pi.
\label{appeq: TRS constraint cohomology}
\end{align}
Equation (\ref{appeq: TRS constraint cohomology})
is nothing but the statement that the representation 
of time reversal should square to either the identity or minus
the identity. These two possibilities
are not connected by a coboundary. Hence, they
correspond to different second cohomology classes. 
To see this, assume they were connected by a coboundary, i.e.,
they satisfy the equivalence condition \eqref{eq:U(1) gauge equivalence}.
On the one hand, choosing $g=t$ and $h=t$ in Eq.\
\eqref{eq:U(1) gauge equivalence}
implies that
\begin{align}
\nu(t,t)
-
\nu'(t,t)=
\varphi(t)
-
\varphi(t)
-
\varphi(e)=
-
\varphi(e)
\implies
\varphi(e)
=
\pi
\label{appeq: contradiction a}
\end{align}
if $\nu'(t,t)=\pi$ and $\nu(t,t)=0$.
However, on the other hand, choosing $g=t$ and $h=e$ in Eq.\
\eqref{eq:U(1) gauge equivalence}
implies that
\begin{align}
\nu(t,e)  
-
\nu'(t,e)=
\varphi(t)
-
\varphi(e)
-
\varphi(t)=
-
\varphi(e)
\implies
\varphi(e)
=
0,
\label{appeq: contradiction b}
\end{align}
since $\nu(g,e)=\nu(e,g)=0$ for all $g$.
Equations (\ref{appeq: contradiction a})
and (\ref{appeq: contradiction b}) contradict each other.
This contradiction implies that one cannot consistently define 
a gauge transformation $\varphi$
that interpolates between $\nu$ such that
$\nu(t,t)=\pi$ 
to $\nu'$ such that $\nu(t,t)=0$.
We denote the cases $\nu(t,t)=\pi,0$ with the equivalence 
classes $[\nu]=1,0$, respectively.
\end{proof}

\medskip\noindent\textbf{Claim 2:} $[\rho]=0,1$.
\begin{proof}
For the second index
$[\rho]\in H^{1}(\mathbb{Z}^{T}_{2},\mathbb{Z}^{\,}_{2})$,
two 1-cochains $\rho$ and $\rho'$ are equivalent
if and only if they are 1-cocycles that
differ by a coboundary of a 0-cochain.
But, by definition, a $\mathbb{Z}^{\,}_{2}$-valued 0-cochain has a vanishing coboundary. Hence, the coset
$H^{1}(\mathbb{Z}^{T}_{2},\mathbb{Z}^{\,}_{2})$
is just the set of all distinct 1-cocycles.
By definition, a 1-cocycle $\rho$ must obey
[recall Eq.\ (\ref{appeq:example delta1})]
\begin{subequations}\label{appee: proof claim 2}
\begin{align}
\rho(g)   
+
\mathfrak{c}(g)
\rho(h)
-
\rho(g\,h)=
0.
\label{appee: proof claim 2 a}
\end{align}
Choosing $g=t$ and $h=p$ delivers
\begin{align}
\rho(t)=
\rho(p)
+
\rho(t \cdot p)
=
\rho(t \cdot p),
\label{appee: proof claim 2 b}
\end{align}
\end{subequations}
where we used the fact that $\rho(p)=0$ by the definition
(\ref{eq: def rho(g)}). The value of $\rho(t)$ in $\mathbb{Z}^{\,}_{2}$
indicates whether the projective representation of
reversal of time commutes or anticommutes with
the projective representation of the
fermion parity operator $p$.
Equation (\ref{appee: proof claim 2 b}) states 
that fermion parity of the quantum representation 
of $t$ is equal to the fermion parity of
the quantum representation of 
$t\cdot p$. 
We assign the indices $[\rho]=0,1$
to the values $\rho(t)=0,1$, respectively.
\end{proof}

 Given two local projective representations $\hat{u}^{\,}_{1}$ and 
$\hat{u}^{\,}_{2}$ of the group
$G^{\,}_{f}=\mathbb{Z}^{T}_{2}\times\mathbb{Z}^{F}_{2}$ 
acting on the Fock spaces $\mathcal{F}^{\,}_{1}$
and $\mathcal{F}^{\,}_{2}$, respectively,
we now derive the indices associated with the local projective  
representation $\hat{u}$ acting on the graded tensor product
$\mathcal{F}=\mathcal{F}^{\,}_{1}\otimes^{\,}_{\mathfrak{g}}\mathcal{F}^{\,}_{2}$
of the Fock spaces $\mathcal{F}^{\,}_{1}$ and $\mathcal{F}^{\,}_{2}$.
The definition \eqref{eq:def stacked operators} implies
\begin{align}
\hat{u}(t)\df 
\hat{v}^{\,}_{1}(t)\,
\hat{v}^{\,}_{2}(t)\,
\mathsf{K},
\qquad
\hat{u}(p)\df 
\hat{v}^{\,}_{1}(p)\,
\hat{v}^{\,}_{2}(p),
\end{align}
for the representations of elements
$t\in\mathbb{Z}^{T}_{2}$ and $p\in\mathbb{Z}^{F}_{2}$.
In turn, using the relation \eqref{eq:stacked rep phase rel} and 
the definition \eqref{eq: def rho(g)}, we find
\begin{subequations}
\label{eq:Z2TZ2F stacked indices}
\begin{align}
&
\phi(g,h)=
\phi^{\,}_{1}(g,h)
+
\phi^{\,}_{2}(g,h)
+
\pi\,
\rho^{\,}_{1}(h)\,
\rho^{\,}_{2}(g)
\hbox{ mod $2\pi$},
\qquad
g,h\in G^{\,}_{f},
\\
&
\nu(t,t)=
\nu^{\,}_{1}(t,t)
+
\nu^{\,}_{2}(t,t)
+
\pi\,
\rho^{\,}_{1}(t)\,
\rho^{\,}_{2}(t)
\hbox{ mod $2\pi$},
\\
&
\rho(t)=
\frac{\phi(t,p)-\phi(p,t)}{\pi}
\hbox{ mod $2$}
\nonumber\\
&
\phantom{\rho(t)}=
\frac{1}{\pi}
\big[
\phi^{\,}_{1}(t,p)
+
\phi^{\,}_{2}(t,p)
+
\pi\,
\rho^{\,}_{1}(p)\,
\rho^{\,}_{2}(t)
-
\phi^{\,}_{1}(p,t)
-
\phi^{\,}_{2}(p,t)
-
\pi\,
\rho^{\,}_{1}(t)\,
\rho^{\,}_{2}(p)
\big]
\hbox{ mod $2$}
\nonumber\\
\hbox{\tiny $\rho^{\,}_{1}(p)=\rho^{\,}_{2}(p)=0$}
&
\phantom{\rho(t)}=
\frac{\phi^{\,}_{1}(t,p)-\phi^{\,}_{1}(p,t)}{\pi}
+
\frac{\phi^{\,}_{2}(t,p)-\phi^{\,}_{2}(p,t)}{\pi}
\hbox{ mod $2$}
\nonumber\\
&
\phantom{\rho(t)}=
\rho^{\,}_{1}(t)
+
\rho^{\,}_{2}(t)
\hbox{ mod $2$},
\end{align}
\end{subequations}
for the 2-cocycle $\nu$ and 1-cocycle $\rho$
associated with the representation $\hat{u}$.
Here, $\nu^{\,}_{1}$ and $\nu^{\,}_{2}$ are the 2-cocycles, and
$\rho^{\,}_{1}$ and $\rho^{\,}_{2}$ are the 1-cocycles associated with 
the representations $\hat{u}^{\,}_{1}$ and $\hat{u}^{\,}_{2}$, respectively.
Assignments of indices $[\nu]$ and $[\rho]$
to the local projective representations
of the group $\mathbb{Z}^{T}_{2}\times\mathbb{Z}^{F}_{2}$ (as shown above) and 
the identity \eqref{eq:Z2TZ2F stacked indices} imply 
that the indices of the tensor product representation 
are related to the indices of the constituent representations via
\begin{subequations}
\begin{align}
[\nu]=&\; [\nu^{\,}_{1}]+[\nu^{\,}_{2}]+[\rho^{\,}_{1}][\rho^{\,}_{2}] \, \quad
\text{ mod $2$}, \nonumber \\
[\rho]=&\; [\rho^{\,}_{1}]+[\rho^{\,}_{2}] \, \quad
\hspace{42pt} \text{ mod $2$},
\label{eq:Z2TZF2 stacking rule}
\end{align}
We note that for the group 
$\mathbb{Z}^{T}_{2}\times\mathbb{Z}^{F}_{2}$ the cup product 
$[\pi \rho^{\,}_{1}\smile \rho^{\,}_{2}]$ in Eq.\
\eqref{eq:stacking rel for indices} simplifies to
\begin{align}
[\pi \rho^{\,}_{1}\smile \rho^{\,}_{2}]\equiv 
[\rho^{\,}_{1}]\,
[\rho^{\,}_{2}],
\end{align}
\end{subequations}
One thus finds that different local projective representations of the group 
$\mathbb{Z}^{T}_{2}\times\mathbb{Z}^{F}_{2}$ form 
the cyclic group $\mathbb{Z}^{\,}_{4}$ 
under the stacking rule \eqref{eq:Z2TZF2 stacking rule}.

\subsection{The split group
$\mathbb{Z}^{\,}_{2}\times\mathbb{Z}^{\,}_{2}\times\mathbb{Z}^{F}_{2}$}
\label{appsubsec: Z2 Z2 ZF2}

As in Sec.\ \ref{appsubsec: ZT2 times ZF2},
the group $\mathbb{Z}^{\,}_{2}\times\mathbb{Z}^{\,}_{2}\times\mathbb{Z}^{F}_{2}$
is a split group. We denote the two generators of 
$\mathbb{Z}^{\,}_{2}\times\mathbb{Z}^{\,}_{2}$ by $g^{\,}_{1}$ and $g^{\,}_{2}$,
both of which are represented by unitary operators.
Because of the Cartesian products,
$[\phi]\in H^{2}(G^{\,}_{f},\mathrm{U(1)}^{\,}_{\mathfrak{c}})$
separates into the pair of independent indices 
$[\nu]\in H^{2}(\mathbb{Z}^{\,}_{2}\times\mathbb{Z}^{\,}_{2},
\mathrm{U(1)}^{\,}_{\mathfrak{c}})$
and 
$[\rho]\in H^{1}(\mathbb{Z}^{\,}_{2}\times\mathbb{Z}^{\,}_{2},\mathbb{Z}^{\,}_{2})$
according to Eq.\ (\ref{eq:indices if  Kuenneth formula holds}).
Since the group representation is unitary (and hence linear as opposed
to antilinear), there is no negative sign
that appears on the right-hand side of the equality in
Eq.\ (\ref{appeq: TRS constraint cohomology}).
It is not possible to constrain the possible values of 
$\nu(g^{\,}_{1},g^{\,}_{1})$ or $\nu(g^{\,}_{2},g^{\,}_{2})$
as was done in Eq.\ (\ref{appeq: TRS constraint cohomology}). 
Cocycle conditions that are akin to Eq.\
\eqref{appeq: TRS constraint cohomology}
are trivially satisfied. On the other hand, examining the algebra 
between projective representations of $g^{\,}_{1}$ and $g^{\,}_{2}$
provides useful information.
Since each symmetry transformation must have a fixed fermion parity,
the projective representations of $g^{\,}_{1}$ and $g^{\,}_{2}$ must either 
commute or anticommute with each other,
i.e.,
\begin{align}
\nu(g^{\,}_{1},g^{\,}_{2})
-
\nu(g^{\,}_{2},g^{\,}_{1})=
0,\pi.
\end{align}
These two possible values constitute the two inequivalent 
cohomology classes for the index $[\nu]$.
To show this, we consider two cochains
$\nu$ and $\nu'$ that are connected by the coboundary condition
\eqref{eq:U(1) gauge equivalence}. 
One finds
\begin{subequations}
\begin{align}
&
\nu(g^{\,}_{1},g^{\,}_{2})
-
\nu'(g^{\,}_{1},g^{\,}_{2})=
\varphi(g^{\,}_{1})
+
\varphi(g^{\,}_{2})
-
\varphi(g^{\,}_{1}\cdot g^{\,}_{2}),
\\
&
\nu(g^{\,}_{2},g^{\,}_{1})
-
\nu'(g^{\,}_{2},g^{\,}_{1})
=
\varphi(g^{\,}_{2})
+
\varphi(g^{\,}_{1})
-
\varphi(g^{\,}_{2}\cdot g^{\,}_{1}).
\end{align}
Because $G\equiv\mathbb{Z}^{\,}_{2}\times\mathbb{Z}^{\,}_{2}$ is Abelian,
this pair of equations implies that
\begin{align}
\nu(g^{\,}_{1},g^{\,}_{2})
-
\nu(g^{\,}_{2},g^{\,}_{1})  
=
\nu'(g^{\,}_{1},g^{\,}_{2})
-
\nu'(g^{\,}_{2},g^{\,}_{1}).
\end{align}
\end{subequations}
Therefore, the projective
representations of $g^{\,}_{1}\in G$ and $g^{\,}_{2}\in G$
that either commute pairwise or anticommute pairwise
must belong to distinct second cohomology classes. 
We assign the values $[\nu]=0,1$ to 
$\nu(g^{\,}_{1},g^{\,}_{2})
-
\nu(g^{\,}_{2},g^{\,}_{1})=0,\pi$, respectively.

The index $[\rho]$ characterizes whether the
representations of $g^{\,}_{1}\in G$ and $g^{\,}_{2}\in G$
commute or anticommute with the fermion parity. Note that the parity of the 
element $g^{\,}_{1}\cdot g^{\,}_{2}$ is determined 
by the parities of $g^{\,}_{1}$ and $g^{\,}_{2}$.
Therefore, $[\rho]$ retains the 
$\mathbb{Z}^{\,}_{2}\times \mathbb{Z}^{\,}_{2}$ structure.
We assign a pair of indices 
\begin{align}
[\rho]=
\big([\rho]^{\,}_{1},[\rho]^{\,}_{2}\big),
\qquad
[\rho]^{\,}_{1}=0,1,
\qquad
[\rho]^{\,}_{2}=0,1,   
\end{align}
to indicate the parities of the projective representations of 
of $g^{\,}_{1}\in G$ and $g^{\,}_{2}\in G$, respectively.
We may then write
\begin{equation}
H^{2}
\Big(\mathbb{Z}^{\,}_{2}\times\mathbb{Z}^{\,}_{2}\times\mathbb{Z}^{F}_{2},\mathrm{U(1)}^{\,}_{\mathfrak{c}}\Big)
=
\Big\{
\big([\nu],[\rho]\big)\ \Big|\
[\nu]=0,1,
\qquad
[\rho]=\big([\rho]^{\,}_{1},[\rho]^{\,}_{2}\big),
\quad
[\rho]^{\,}_{1},[\rho]^{\,}_{2}=0,1
\Big\}.
\label{appeq:H2(Z2timesZ2,U(1))}
\end{equation}

Given two local projective representations $\hat{u}^{\,}_{1}$ and 
$\hat{u}^{\,}_{2}$ of the group
$G^{\,}_{f}=\mathbb{Z}^{\,}_{2}\times\mathbb{Z}^{\,}_{2}\times\mathbb{Z}^{F}_{2}$ 
acting on two Fock spaces $\mathcal{F}^{\,}_{1}$ and $\mathcal{F}^{\,}_{2}$,
respectively,
we shall derive the indices associated with  the local projective  
representation $\hat{u}$ acting on the graded tensor 
product
$\mathcal{F}=\mathcal{F}^{\,}_{1}\otimes^{\,}_{\mathfrak{g}}\mathcal{F}^{\,}_{2}$
of Fock spaces $\mathcal{F}^{\,}_{1}$ and $\mathcal{F}^{\,}_{2}$.
The definition \eqref{eq:def stacked operators} implies
\begin{align}
\hat{u}(g^{\,}_{1})= 
\hat{v}^{\,}_{1}(g^{\,}_{1})\,
\hat{v}^{\,}_{2}(g^{\,}_{1}),
\qquad
\hat{u}(g^{\,}_{2})= 
\hat{v}^{\,}_{1}(g^{\,}_{2})\,
\hat{v}^{\,}_{2}(g^{\,}_{2}),
\qquad
\hat{u}(p) = 
\hat{v}^{\,}_{1}(p)\,
\hat{v}^{\,}_{2}(p),
\end{align}
for the representations of elements $g^{\,}_{1},g^{\,}_{2}\in 
\mathbb{Z}^{\,}_{2}\times \mathbb{Z}^{\,}_{2}$ and $p\in\mathbb{Z}^{F}_{2}$.
In turn, using the relation \eqref{eq:stacked rep phase rel} and 
the definition \eqref{eq: def rho(g)}, we find
\begin{subequations}
\label{eq:Z2Z2Z2F stacked indices}
\begin{align}
&
\nu(g^{\,}_{1},g^{\,}_{2})
-
\nu(g^{\,}_{2},g^{\,}_{1})=
\nu^{\,}_{1}(g^{\,}_{1},g^{\,}_{2})
+
\nu^{\,}_{2}(g^{\,}_{1},g^{\,}_{2})
+
\pi\,
\rho^{\,}_{1}(g^{\,}_{2})\,
\rho^{\,}_{2}(g^{\,}_{1})
-
\nu^{\,}_{1}(g^{\,}_{2},g^{\,}_{1})
-
\nu^{\,}_{2}(g^{\,}_{2},g^{\,}_{1})
-
\pi\,
\rho^{\,}_{1}(g^{\,}_{1})\,
\rho^{\,}_{2}(g^{\,}_{2})
\hbox{ mod $2\pi$}
\nonumber\\
&
\phantom{\nu(g^{\,}_{1},g^{\,}_{2})
-
\nu(g^{\,}_{2},g^{\,}_{1})}
=
\nu^{\,}_{1}(g^{\,}_{1},g^{\,}_{2})
-
\nu^{\,}_{1}(g^{\,}_{2},g^{\,}_{1})
+
\nu^{\,}_{2}(g^{\,}_{1},g^{\,}_{2})
-
\nu^{\,}_{2}(g^{\,}_{2},g^{\,}_{1})
+
\pi
\big[
\rho^{\,}_{1}(g^{\,}_{2})\,
\rho^{\,}_{2}(g^{\,}_{1})
-
\rho^{\,}_{1}(g^{\,}_{1})\,
\rho^{\,}_{2}(g^{\,}_{2})
\big]
\hbox{ mod $2\pi$},
\\
&
\rho(g^{\,}_{i})=
\frac{\phi(g^{\,}_{i},p)-\phi(p,g^{\,}_{i})}{\pi}
\hbox{ mod $2$}
\nonumber\\
&\phantom{\rho(g^{\,}_{i})}=
\frac{1}{\pi}
\left[
\phi^{\,}_{1}(g^{\,}_{i},p)
+
\phi^{\,}_{2}(g^{\,}_{i},p)
+
\pi\,
\rho^{\,}_{1}(p)\,
\rho^{\,}_{2}(g^{\,}_{i})
-
\phi^{\,}_{1}(p,g^{\,}_{i})
-
\phi^{\,}_{2}(p,g^{\,}_{i})
-
\pi\,
\rho^{\,}_{1}(g^{\,}_{i})\,
\rho^{\,}_{2}(p)
\right]
\hbox{ mod $2$}
\nonumber\\
&\phantom{\rho(g^{\,}_{i})}=
\frac{
\phi^{\,}_{1}(g^{\,}_{i},p)-\phi^{\,}_{1}(p,g^{\,}_{i})
}
{
\pi
}
+
\frac{
\phi^{\,}_{2}(g^{\,}_{i},p)-\phi^{\,}_{2}(p,g^{\,}_{i})
}
{
\pi
}
\hbox{ mod $2$}\qquad\qquad
\hbox{\tiny [where we made use of $\rho^{\,}_{1}(p)=\rho^{\,}_{2}(p)=0$]}
\nonumber\\
&\phantom{\rho(g^{\,}_{i})}=
\rho^{\,}_{1}(g^{\,}_{i})
+
\rho^{\,}_{2}(g^{\,}_{i})
\hbox{ mod $2$},
\qquad
i=1,2,
\end{align}
\end{subequations}
for the 2-cocycle $\nu$ and 1-cocycle $\rho$
associated with the representation $\hat{u}$.
Here, $\nu^{\,}_{1}$ and $\nu^{\,}_{2}$ are the 2-cocycles, and
$\rho^{\,}_{1}$ and $\rho^{\,}_{2}$ are the 1-cocycles associated with 
the representations $\hat{u}^{\,}_{1}$ and $\hat{u}^{\,}_{2}$, respectively.
Assignments of indices $[\nu]$ and $[\rho]$
to the local projective representations of the group 
$\mathbb{Z}^{\,}_{2}\times \mathbb{Z}^{\,}_{2}\times\mathbb{Z}^{F}_{2}$
(as shown above) and the identity \eqref{eq:Z2Z2Z2F stacked indices}
imply that the indices of the tensor product representation 
are related to the indices of the constituent representations via
\begin{subequations}
\begin{align}
&
[\nu]= [\nu^{\,}_{1}]+[\nu^{\,}_{2}]
+
[\rho^{\,}_{1}]^{\,}_{2}\,[\rho^{\,}_{2}]^{\,}_{1}
-
[\rho^{\,}_{1}]^{\,}_{1}\,[\rho^{\,}_{2}]^{\,}_{2} 
\quad
\hbox{ mod $2$},
\\
&
[\rho]^{\,}_{i}= [\rho^{\,}_{1}]^{\,}_{i}+[\rho^{\,}_{2}]^{\,}_{i} \, 
\qquad\qquad\qquad\qquad\qquad\,\,\,
\hbox{ mod $2$},
\quad 
i=1,2,
\end{align}
where we have denoted by
$[\rho^{\,}_{1}],[\rho^{\,}_{2}]\in 
H^{1}\big(\mathbb{Z}^{\,}_{2}\times\mathbb{Z}^{\,}_{2},\mathbb{Z}^{\,}_{2}\big)$ 
the first cohomology classes associated to the local projective 
representations $\hat{u}^{\,}_{1}$ and $\hat{u}^{\,}_{2}$,
and we used the notation
\begin{align}
[\rho^{\,}_{1}]
\equiv
\big([\rho^{\,}_{1}]^{\,}_{1},[\rho^{\,}_{1}]^{\,}_{2}\big),
\qquad
[\rho^{\,}_{2}]
\equiv
\big([\rho^{\,}_{2}]^{\,}_{1},[\rho^{\,}_{2}]^{\,}_{2}\big).
\end{align}
We note that for the group 
$\mathbb{Z}^{\,}_{2}\times\mathbb{Z}^{\,}_{2}\times\mathbb{Z}^{F}_{2}$
the cup product 
$[\pi \rho^{\,}_{1}\smile \rho^{\,}_{2}]$
in Eq.\ \eqref{eq:stacking rel for indices} simplifies to
\begin{align}
[\pi \rho^{\,}_{1}\smile \rho^{\,}_{2}] 
\equiv 
[\rho^{\,}_{1}]^{\,}_{2}\,
[\rho^{\,}_{2}]^{\,}_{1}
-
[\rho^{\,}_{1}]^{\,}_{1}\,
[\rho^{\,}_{2}]^{\,}_{2},
\end{align}
and the stacking rule \eqref{eq:stacking rel for indices} becomes
\begin{align}
\label{eq:Z2Z2ZF2 stacking rule}
\big(
[\nu],
[\rho],
0
\big)\equiv
\bigg(
[\nu],
\big([\rho]^{\,}_{1},\,[\rho]^{\,}_{2}\big),
0
\bigg)=
\bigg(
[\nu^{\,}_{1}]
+
[\nu^{\,}_{2}]
+
[\rho^{\,}_{1}]^{\,}_{2}\,
[\rho^{\,}_{2}]^{\,}_{1}
-
[\rho^{\,}_{1}]^{\,}_{1}\,
[\rho^{\,}_{2}]^{\,}_{2}\,,\,
\big([\rho^{\,}_{1}]^{\,}_{1}+[\rho^{\,}_{2}]^{\,}_{1},\,
[\rho^{\,}_{1}]^{\,}_{2}+[\rho^{\,}_{2}]^{\,}_{2}
\big),\,
0
\bigg).
\end{align}
\end{subequations}
One thus finds that different local projective representations of the group 
$\mathbb{Z}^{\,}_{2}\times \mathbb{Z}^{\,}_{2}\times\mathbb{Z}^{F}_{2}$
form the group
$\mathbb{Z}^{\,}_{2}\times \mathbb{Z}^{\,}_{2}\times \mathbb{Z}^{\,}_{2}$ 
under the stacking rule \eqref{eq:Z2Z2ZF2 stacking rule}.

\subsection{The nonsplit group $\mathbb{Z}^{FT}_{4}$}
\label{appsubsec: ZFT4}

The group $\mathbb{Z}^{FT}_{4}$ is
the nontrivial central extension of $G\equiv\mathbb{Z}^{T}_{2}$
by $\mathbb{Z}^{F}_{2}\equiv\{p,p^{2}\}$, where the upper index $T$
for the cyclic group $\mathbb{Z}^{T}_{2}\equiv\{t,t^{2}\}$
refers to the interpretation of $t$ as reversal of time.
This central extension of time reversal by fermion parity
is specified by the map $\gamma$ obeying the nonsplit
condition (\ref{eq: nonsplit condition on gamma}) because of
$\gamma(t,t)=p$ (which implies the group composition rule $t\cdot t=p$).
If so, $\nu$ is not a cocycle but a cochain with 
nonvanishing coboundary according to Eq.\
(\ref{eq:no split condition on gamma nu and rho})
and \eqref{eq:Gf_cocycle_cond}.
On the other hand, $\rho$ is a 1-cocycle. 
We thus observe that two tuples $(\nu,\rho)$ and $(\nu',\rho')$ are not in the
same equivalence class if $\rho(t)\neq \rho'(t)$.  Since $\rho(t)$ can
take two values, 0 or 1, there exist at least two distinct equivalence
classes of the tuple $(\nu,\rho)$, labeled by $\rho(t)$. Given this
value of $\rho(t)$, we shall construct the distinct equivalence
classes of $(\nu,\rho)$ corresponding to different values of $\nu\in
C^{2}(G,\mathrm{U(1)})$. Choosing $g=h=f=t$ in
Eq.\ (\ref{eq:no split condition on gamma nu and rho}) delivers
\begin{align}
\nu(t,e)
-
\nu(t,t)
-
\nu(t,t)
-
\nu(e,t)=
\pi\,
\rho(t)\,
\gamma(t,t)
\hbox{ mod $2\pi$}.
\label{eq:equation satisfied by nu(t,t) for ZFT4}
\end{align}
With the help of Eq.\ (\ref{eq:def projective rep II c})
and with the choice of the convention  $\gamma(t,t)=p\equiv 1$
for the nonsplit group $\mathbb{Z}^{FT}_{4}$, we 
we find the pair of solutions to Eq.\
(\ref{eq:equation satisfied by nu(t,t) for ZFT4})
given by
\begin{equation}
\nu(t,t)=
-\frac{\pi}{2}\,
\rho(t),
\qquad
\nu(t,t)=
-\frac{\pi}{2}
\rho(t)
+
\pi.
\label{eq:ZFT4 coboundary of nu}
\end{equation}
The multiplicative factor $\pi$ appears on the right-hand sides since
$\nu$ takes values in $\mathrm{U(1)}$ and is thus defined modulo $2\pi$. Now,
the two solutions (\ref{eq:ZFT4 coboundary of nu})
are equivalent under the equivalence relation
\eqref{eq:Gf_equivalence_rel} as can be seen by choosing
$\alpha=0$ and $\beta=p\equiv 1$ in
Eq.\ \eqref{eq:Gf_equivalence_rel}. Indeed, the term
$\pi\beta\smile\gamma=\pi$ then cancels the factor $\pi$ between the two
solutions \eqref{eq:ZFT4 coboundary of nu}.  Thus, for each value of
$\rho(t)=0,1$, there exists a single distinct equivalence class
$[(\nu,\rho)]$.  From now on, we choose the solution
$(-\pi\,\rho(t)/2,\rho(t))$ as the representative of the equivalence
class $[(\nu,\rho)]$. We assign
$[(\nu,\rho)]=(1,1)$ to the case $(\nu(t,t),\rho(t))=(-\pi/2,1)$ and
$[(\nu,\rho)]=(0,0)$ to the case $(\nu(t,t),\rho(t))=(0,0)$ and write
\begin{equation}
H^{2}\Big(\mathbb{Z}^{FT}_{4},\mathrm{U(1)}^{\,}_{\mathfrak{c}}\Big)=
\Big\{
\big[\big(\nu,\rho\big)\big]\ \Big|\
\big[\big(\nu,\rho\big)\big]=(0,0),(1,1)
\Big\}.
\label{appeq:H2(ZFT4,U(1))}
\end{equation}
Given two local projective representations $\hat{u}^{\,}_{1}$ and
$\hat{u}^{\,}_{2}$ of the group $G^{\,}_{f} = \mathbb{Z}^{FT}_{4}$
acting on two Fock spaces $\mathcal{F}^{\,}_{1}$ and
$\mathcal{F}^{\,}_{2}$, respectively, we shall derive the indices
associated with the local projective representation $\hat{u}$ acting
on the graded tensor product $\mathcal{F} =
\mathcal{F}^{\,}_{1}\otimes^{\,}_{\mathfrak{g}} \mathcal{F}^{\,}_{2}$
of Fock spaces $\mathcal{F}^{\,}_{1}$ and $\mathcal{F}^{\,}_{2}$.
The definition \eqref{eq:def stacked operators} implies
\begin{align}
\hat{u}(t)= 
\hat{v}^{\,}_{1}(t)\,
\hat{v}^{\,}_{2}(t)\,
\mathsf{K},
\qquad
\hat{u}(p)= 
\hat{v}^{\,}_{1}(p)\,
\hat{v}^{\,}_{2}(p),
\end{align}
for the representations of elements $t,p\in \mathbb{Z}^{FT}_{4}$.
In turn, using the relation \eqref{eq:stacked rep phase rel} and 
the definition \eqref{eq: def rho(g)}, we find that
if $\nu^{\,}_{1}$ and $\nu^{\,}_{2}$ are 2-cochains, and
$\rho^{\,}_{1}$ and $\rho^{\,}_{2}$ are two 1-cocycles associated with 
the representations $\hat{u}^{\,}_{1}$ and $\hat{u}^{\,}_{2}$, respectively,
then the 2-cochain $\nu$ and 1-cocycle $\rho$ associated
with the representation $\hat{u}$ are given by
\begin{subequations}\label{eq:ZFT4 stacked indices}
\begin{align}
&
\nu(t,t)=
\nu^{\,}_{1}(t,t)
+
\nu^{\,}_{2}(t,t)
+
\pi\,
\rho^{\,}_{1}(t)\,
\rho^{\,}_{2}(t)
\hbox{ mod $2\pi$},
\label{eq:ZFT4 stacked indices a}
\\
&
\rho(t)=
\rho^{\,}_{1}(t)
+
\rho^{\,}_{2}(t)
\hbox{ mod $2$},
\label{eq:ZFT4 stacked indices b}
\end{align}
\end{subequations}
respectively. Although this pair of relations
are identical to their counterparts in
Eq.\ \eqref{eq:Z2TZ2F stacked indices},
the nonsplit nature of the group $\mathbb{Z}^{FT}_{4}$
carries over to the stacked projective representation.
Indeed, inserting twice
\eqref{eq:ZFT4 coboundary of nu}
on the right-hand side of
\eqref{eq:ZFT4 stacked indices a}
gives
\begin{subequations}\label{eq:ZFT4 stacked indices bis}
\begin{align}
&
\nu(t,t)=
-\frac{\pi}{2}\,
\rho^{\,}_{1}(t)\,
-
\frac{\pi}{2}\,
\rho^{\,}_{2}(t)\,
+
\pi\,
\rho^{\,}_{1}(t)\,
\rho^{\,}_{2}(t)
\hbox{ mod $2\pi$},
\label{eq:ZFT4 stacked indices bis a}
\\
&
\rho(t)=
\rho^{\,}_{1}(t)
+
\rho^{\,}_{2}(t)
\hbox{ mod $2$}.
\label{eq:ZFT4 stacked indices bis b}
\end{align}
\end{subequations}
There are four cases to consider.
When $\big(\rho^{\,}_{1}(t),\rho^{\,}_{2}(t)\big)=(0,0)$,
$\rho(t)=0$ and $\nu(t,t)=0$.
When $\big(\rho^{\,}_{1}(t),\rho^{\,}_{2}(t)\big)=(0,1)$,
$\rho(t)=1$ and $\nu(t,t)=-\pi/2$.
When $\big(\rho^{\,}_{1}(t),\rho^{\,}_{2}(t)\big)=(1,0)$,
$\rho(t)=1$ and $\nu(t,t)=-\pi/2$.
When $\big(\rho^{\,}_{1}(t),\rho^{\,}_{2}(t)\big)=(1,1)$,
$\rho(t)=0$ and $\nu(t,t)=0$.
Hence, $\nu(t,t)$ and $\rho(t)$
on the left-hand sides of Eqs.\
(\ref{eq:ZFT4 stacked indices a})
and
(\ref{eq:ZFT4 stacked indices b}),
respectively, obey Eq.\ \eqref{eq:ZFT4 coboundary of nu}.
The stacked representation can then be labeled by
the indices
$[(\rho,\rho)]$
where $\rho=\rho^{\,}_{1}+\rho^{\,}_{2}\hbox{ mod 2}$.
At last, the stacking rule
\eqref{eq:stacking rel for indices}
takes the form
\begin{align}
\left(
[
\left(
\nu,
\rho
\right)
],
0
\right)=
\left(
[
(
\rho^{\,}_{1}+\rho^{\,}_{2}\hbox{ mod 2},\, 
\rho^{\,}_{1}+\rho^{\,}_{2}\hbox{ mod 2}
)	
],\,
0
\right).
\label{appeq:ZFT4 stacking rule}
\end{align}
One thus finds that
different local projective representations of the group
$\mathbb{Z}^{FT}_{4}$ form 
the cyclic group $\mathbb{Z}^{\,}_{2}$ under the stacking rule
\eqref{appeq:ZFT4 stacking rule}.

\section{Construction of fermionic matrix product states (FMPS)}
\label{appsec:Construction of fermionic matrix product states (FMPS)}

We review the construction
of fermionic matrix product states (FMPS). We refer the reader to Refs.\
\onlinecite{Bultinck2017,Williamson2016}
and references therein on the topic of matrix product states (MPS).
As in bosonic matrix product states (BMPS),
FMPS can be expressed as a contraction of
objects belonging to a graded tensor product of vector spaces.
The need for \textit{graded tensor product of vector spaces} stems
from the underlying fermionic algebra.

\subsection{$\mathbb{Z}^{\,}_{2}$-graded vector spaces and their
$\mathbb{Z}^{\,}_{2}$-graded tensor products}
Any fermionic Fock space $\mathcal{F}$ can be seen,
in the basis that diagonalizes the 
total fermionic number operator,
to be the direct sum over a subspace $\mathcal{F}^{\,}_{0}$
with even total fermionic number
and a subspace $\mathcal{F}^{\,}_{1}$
with odd total fermionic number. This property endows
fermionic Fock space with a natural $\mathbb{Z}^{\,}_{2}$-grading.

A $\mathbb{Z}^{\,}_{2}$-graded vector space $V$
admits the direct sum decomposition
\begin{align}
V=V^{\,}_{0}\oplus V^{\,}_{1}.
\label{eq:decomposition super vector space}
\end{align}
We shall identify
the subscripts $0$ and $1$
as the elements of the additive group $\mathbb{Z}^{\,}_{2}$.
We say that $V^{\,}_{0}$ ($V^{\,}_{1}$) has parity 0 (1).
Any vector space is
$\mathbb{Z}^{\,}_{2}$-graded since the choice
$V^{\,}_{0}=V$ and $V^{\,}_{1}=\emptyset$
is always possible.
Any subspace of $V^{\,}_{0}$ shares its parity $0$.
Any subspace of $V^{\,}_{1}$ shares its parity $1$.
A vector $\ket{v}\in V$ is called \textit{homogeneous} if
it entirely resides in either one of the subspaces $V^{\,}_{0}$
and $V^{\,}_{1}$. The parity $|v|$ of the homogeneous state $\ket{v}$
is either $0$ if $\ket{v}\in V^{\,}_{0}$ or $1$ if $\ket{v}\in V^{\,}_{1}$.
These observation on the $\mathbb{Z}^{\,}_{2}$-grading
of a vector space $V$ only become useful
when one demands that any operation acting on $V$ 
preserves the $\mathbb{Z}^{\,}_{2}$-grading.

For example, certain operations need to be defined carefully between
two $\mathbb{Z}^{\,}_{2}$-graded vector space $V$ and $W$
that preserve their $\mathbb{Z}^{\,}_{2}$ structure.
One such operation is the $\mathbb{Z}^{\,}_{2}$-graded tensor product.
Let
$V=V^{\,}_{0}\oplus V^{\,}_{1}$
and
$W=W^{\,}_{0}\oplus W^{\,}_{1}$
be two $\mathbb{Z}^{\,}_{2}$-graded vector spaces.
We define their graded tensor product as the map
\begin{subequations}\label{eq:graded tensor prod def}
\begin{align}
\otimes^{\,}_{\mathfrak{g}}
\colon
V\times W
\to
V\otimes W,
\label{eq:graded tensor prod def a}
\end{align}
such that 
\begin{align}
V^{\,}_{i}
\otimes^{\,}_{\mathfrak{g}}
W^{\,}_{j}
\subseteq
\left(V\otimes W\right)^{\,}_{(i+j)\, \text{mod}\, 2},
\qquad
i,j=0,1.
\label{eq:graded tensor prod def b}
\end{align}
By design,
the operation
$\otimes^{\,}_{\mathfrak{g}}$ carries the $\mathbb{Z}^{\,}_{2}$-grading
of $V$ and $W$ to their $\mathbb{Z}^{\,}_{2}$-graded tensor product.
In particular,
for any homogeneous vectors $\ket{v}\in V$
with parity $|v|=0,1$ and $\ket{w}\in W$ with parity $|w|=0,1$,
the graded tensor product $\ket{v}\otimes^{\,}_{\mathfrak{g}}\ket{w}$
of two homogeneous vectors has the parity
\begin{equation}
\left|
\ket{v}\otimes^{\,}_{\mathfrak{g}}\ket{w}
\right|\df
(|v|+|w|)\hbox{ mod 2}.
\label{eq:graded tensor prod def c}
\end{equation}
\end{subequations}
The connection between 
the $\mathbb{Z}^{\,}_{2}$-graded vector space
$V=V^{\,}_{0}\oplus V^{\,}_{1}$
and fermionic Fock spaces
$\mathcal{F}=\mathcal{F}^{\,}_{0}\oplus\mathcal{F}^{\,}_{1}$,
is established through the identifications
$\mathcal{F}^{\,}_{0}\to V^{\,}_{0}$
and
$\mathcal{F}^{\,}_{1}\to V^{\,}_{1}$.
However, a fermionic Fock space has more structure than a
mere $\mathbb{Z}^{\,}_{2}$-graded vector space.
Wave functions in a fermionic Fock space are fully antisymmetric
under the permutation of two fermions. This requirement
can be implemented as follows on a $\mathbb{Z}^{\,}_{2}$-graded vector space.
The exchange of two fermions can be represented by
the isomorphism
\begin{subequations}\label{eq:Exchange of particles}
\begin{equation}
R\colon 
V\otimes^{\,}_{\mathfrak{g}}W
\to
W\otimes^{\,}_{\mathfrak{g}}V,
\label{eq:Exchange of particles a}
\end{equation}
by which the graded tensor product of the
homogeneous vectors
$\ket{v}\in V$ and $\ket{w}\in W$
obeys
\begin{equation}
\ket{v}\otimes^{\,}_{\mathfrak{g}}\ket{w}
\mapsto
(-1)^{|v|\,|w|}\,
\ket{w}\otimes^{\,}_{\mathfrak{g}}\ket{v}.
\label{eq:Exchange of particles b}
\end{equation}
\end{subequations}
The map $R$ is called the \textit{reordering} operation.
It is invertible with itself as inverse since $R^{2}$ is the identity map.

For every $\mathbb{Z}^{\,}_{2}$-graded vector space $V$,
we define the dual $\mathbb{Z}^{\,}_{2}$-graded vector space $V^{*}$.
We denote an element of the dual $\mathbb{Z}^{\,}_{2}$-graded
vector space $V^{*}$
by $\bra{v}$, the dual to the vector $\ket{v}\in V$.
The dual $\mathbb{Z}^{\,}_{2}$-graded vector space $V^{*}$
inherits a $\mathbb{Z}^{\,}_{2}$
grading from assigning
the parity $|v|$ to the vector $\bra{v}\in V^{*}$
if $\ket{v}\in V$ is homogeneous with parity $|v|$.
The \textit{contraction} $\mathcal{C}$ is the map
\begin{subequations}
\begin{equation}
\begin{split}
\mathcal{C}:
V^{*}\otimes^{\,}_{\mathfrak{g}}V
\to&\,
\mathbb{C},
\\
\bra{\psi} 
\otimes^{\,}_{\mathfrak{g}}
\ket{\phi}
\mapsto&\,
\braket{\psi}{\phi},
\end{split}
\label{eq:contraction def}
\end{equation}
where $\braket{\psi}{\phi}$ denotes the scalar product
between the pair $\ket{\psi},\ket{\phi}\in V$. Hence,
\begin{align}
\mathcal{C}\left(\bra{i} 
\otimes^{\,}_{\mathfrak{g}}
\ket{j}\right)=
\delta^{\,}_{ij}
\end{align}
holds for any pair of orthonormal and homogeneous basis vectors $\ket{i},\ket{j}\in V$.
The \textit{contraction} $\mathcal{C}^{*}$ 
is the map
$\mathcal{C}^{*}:
V\otimes^{\,}_{\mathfrak{g}}V^{*}
\to
\mathbb{C}$ defined by its action
\begin{align}
\mathcal{C}^{*}
\left(
\ket{i} 
\otimes^{\,}_{\mathfrak{g}}
\bra{j}
\right)\df
\mathcal{C}
\Big(
R
\left(
\ket{i} 
\otimes^{\,}_{\mathfrak{g}}
\bra{j}
\right)
\Big)=
\mathcal{C}
\Big(
(-1)^{|i||j|}
\bra{j} 
\otimes^{\,}_{\mathfrak{g}}
\ket{i}
\Big)=
(-1)^{|i||j|}
\braket{j}{i}=
(-1)^{|i||j|}
\delta^{\,}_{ij}
\end{align}
\end{subequations}
for any pair of orthonormal basis vectors
$\ket{i},\ket{j}\in V$.
It is common practice to use the same symbol $\mathcal{C}$ for both
$\mathcal{C}$ and $\mathcal{C}^{*}$. 
Any linear operator 
\begin{subequations}
\begin{equation}
M\colon V\to V
\end{equation}
can be represented in the orthonormal and homogeneous
basis $\{\ket{i}\}$ of $V$
by the matrix
\begin{equation}
M^{\,}_{ij}=(-1)^{|i||j|}\,M^{\,}_{ji}
\end{equation}
through the expansion
\begin{align}
M\df
\sum_{i,j}
M^{\,}_{ij}\,
\ket{i} 
\otimes^{\,}_{\mathfrak{g}}
\bra{j}
\in
V\otimes^{\,}_{\mathfrak{g}}V^{*}.
\end{align}
The linear operator $M$ has a well defined parity
if and only if each term $\ket{i}\otimes^{\,}_{\mathfrak{g}}\bra{j}$
in the summation has the same parity, in which case
\begin{align}
\label{eq:parity off an operator app B}
|M|\df(|i|+|j|)\hbox{ mod 2}.
\end{align} 
\end{subequations}
More generally, if we define
\begin{subequations}\label{appeq; def parity T}
\begin{equation}
T\df
\sum_{i^{\,}_{i},\cdots,i^{\,}_{n}}
T^{\,}_{i^{\,}_{i},\cdots,i^{\,}_{n}}\,
\ket{i^{\,}_{1}} 
\otimes^{\,}_{\mathfrak{g}}
\cdots
\otimes^{\,}_{\mathfrak{g}}
\ket{i^{\,}_{n}}
\in
V^{\,}_{i^{\,}_{1}}
\otimes^{\,}_{\mathfrak{g}}
\cdots
\otimes^{\,}_{\mathfrak{g}}
V^{\,}_{i^{\,}_{n}}
\label{appeq; def parity T a}
\end{equation}
we can assign the parity
\begin{equation}
|T|\df
\left(|i^{\,}_{1}|+\cdots+|i^{\,}_{n}|\right)
\hbox{ mod 2}
\label{appeq; def parity T b}
\end{equation}
\end{subequations}
when all 
$
\ket{i^{\,}_{1}} 
\otimes^{\,}_{\mathfrak{g}}
\cdots
\otimes^{\,}_{\mathfrak{g}}
\ket{i^{\,}_{n}}
$
share the same parity.

\subsection{Fermionic matrix product state (FMPS)}
\noindent We attach to each integer $j=1,\cdots, N$
three $\mathbb{Z}^{\,}_{2}$-graded vector spaces
\begin{subequations}\label{appeq: def FMPS}
\begin{equation}
V^{\,}_{j}\df
\mathrm{span}\{\pket{\alpha}\ |\ \alpha=1,\cdots,\mathcal{D}^{\,}_{\mathrm{v},j}\},
\quad
\mathcal{F}^{\,}_{j}\df
\mathrm{span}\{\ket{\psi^{\,}_{\sigma}}\ |\ \sigma=1,\cdots,\mathcal{D}^{\,}_{j}\},
\quad
V^{*}_{j}\df
\mathrm{span}\{\pbra{\beta}\ |\ \beta=1,\cdots,\mathcal{D}^{\,}_{\mathrm{v},j}\}.
\label{appeq: def FMPS a}
\end{equation}
The basis states $\pket{\alpha}$ and $\pbra{\beta}$
of the dual pair $V^{\,}_{j}$ and $V^{*}_{j}$
of $\mathbb{Z}^{\,}_{2}$-graded vector spaces are virtual (auxiliary) states.
They are denoted by rounded kets and bras and are introduced for convenience.
Each auxiliary basis state has a well defined parity by assumption.
The basis states $\{\ket{\psi^{\,}_{\sigma}}\}$
span the physical fermionic Fock space $\mathcal{F}^{\,}_{j}$.
Each physical basis state
$\ket{\psi^{\,}_{\sigma}}$
has a well defined parity by assumption,
as follows from working in the fermion-number basis of
$\mathcal{F}^{\,}_{j}$ say.
The auxiliary $\mathbb{Z}^{\,}_{2}$-graded vector space $V^{\,}_{j}$ has
the dimension $\mathcal{D}^{\,}_{\mathrm{v},j}$.
The physical fermionic Fock space $\mathcal{F}^{\,}_{j}$ has dimension
$\mathcal{D}^{\,}_{j}$. 
A fermionic matrix product state (FMPS) takes the form
\begin{align}
\ket{\Psi}\df
\mathcal{C}^{\,}_{\mathrm{v}}
\Big(
Q(b)\,
Y\,
\mathsf{A}[1]
\otimes^{\,}_{\mathfrak{g}}
\mathsf{A}[2]
\otimes^{\,}_{\mathfrak{g}}
\cdots
\otimes^{\,}_{\mathfrak{g}}
\mathsf{A}[N]
\Big)
\label{appeq: def FMPS b}
\end{align}
and has a well defined parity provided
the objects
$Q(b)$,
$Y$,
$\mathsf{A}[1]$,
$\mathsf{A}[2]$,
$\cdots$,
$\mathsf{A}[N]$
are defined as follows.
For any $j=1,\cdots,N$,
element
$\mathsf{A}[j]\in
V^{\,}_{j}
\otimes^{\,}_{\mathfrak{g}}
\mathcal{F}^{\,}_{j}
\otimes^{\,}_{\mathfrak{g}}
V^{*}_{j+1}$
is defined by
\begin{align}
&
\mathsf{A}[j]
\df
\sum_{\alpha^{\,}_{j}=1}^{\mathcal{D}^{\,}_{\mathrm{v},j}}
\sum_{\sigma^{\,}_{j}=1}^{\mathcal{D}^{\,}_{j}}
\sum_{\beta^{\,}_{j}=1}^{\mathcal{D}^{\,}_{\mathrm{v},j+1}}
(A^{\,}_{\sigma^{\,}_{j}})^{\,}_{\alpha^{\,}_{j}\,\beta^{\,}_{j}}\,
\pket{\alpha^{\,}_{j}}
\otimes^{\,}_{\mathfrak{g}}
\ket{\psi^{\,}_{\sigma^{\,}_{j}}}
\otimes^{\,}_{\mathfrak{g}}
\pbra{\beta^{\,}_{j}}
\label{appeq: def FMPS c}
\end{align}
once the matrices $A^{\,}_{\sigma^{\,}_{j}}$,
labeled as they are by the basis elements of
the local Fock space $\mathcal{F}^{\,}_{j}$
and with the matrix elements
$(A^{\,}_{\sigma^{\,}_{j}})^{\,}_{\alpha^{\,}_{j}\,\beta^{\,}_{j}}$,
have been chosen.
The contraction $\mathcal{C}^{\,}_{\mathrm{v}}$
labeled by the lower index $\mathrm{v}$ 
is understood to be over all virtual indices 
belonging to the dual pair $(V^{*}_{j},V^{\,}_{j})$
of auxiliary $\mathbb{Z}^{\,}_{2}$-graded vector spaces,
thereby producing the tensor proportional to
\begin{equation}
T^{\,}_{\alpha^{\,}_{1}\cdots\alpha^{\,}_{N}|\beta^{\,}_{1}\cdots\beta^{\,}_{N}}\df
\delta^{\,}_{\beta^{\,}_{1}\,\alpha^{\,}_{2}}\,
\delta^{\,}_{\beta^{\,}_{2}\,\alpha^{\,}_{3}}\,
\cdots\,
\delta^{\,}_{\beta^{\,}_{N-1}\,\alpha^{\,}_{N}}\,
\delta^{\,}_{\beta^{\,}_{N}\,\alpha^{\,}_{1}}
\label{appeq: def FMPS e}
\end{equation}
if $Q(b)\in V^{\,}_{1}\otimes^{\,}_{\mathfrak{g}} V^{*}_{1}$
and
$Y\in V^{\,}_{1}\otimes^{\,}_{\mathfrak{g}}V^{*}_{1}$
were chosen to be the identity
\begin{equation}
Q(b)\equiv Y\equiv
\sum_{\alpha}
\pket{\alpha}
\otimes^{\,}_{\mathfrak{g}}
\pbra{\alpha}.
\label{appeq: def FMPS f}
\end{equation}
The integer $b=0,1$ labels the boundary conditions
selected by $Q(b)\in V^{\,}_{1}\otimes^{\,}_{\mathfrak{g}} V^{*}_{1}$.
The element $Y\in V^{\,}_{1}\otimes^{\,}_{\mathfrak{g}}V^{*}_{1}$
is needed to fix the fermion parity of $\ket{\Psi}$.
More precisely, we demand that the parity 
\eqref{eq:parity off an operator app B}
of $Q(b)\in V^{\,}_{1}\otimes^{\,}_{\mathfrak{g}} V^{*}_{1}$
and the parity (\ref{appeq; def parity T b}) of
$\mathsf{A}[j]\in
V^{\,}_{j}
\otimes^{\,}_{\mathfrak{g}}
\mathcal{F}^{\,}_{j}
\otimes^{\,}_{\mathfrak{g}}
V^{*}_{j+1}$
are \textit{both} even,
while the parity \eqref{eq:parity off an operator app B} of
$Y\in V^{\,}_{1}\otimes^{\,}_{\mathfrak{g}} V^{*}_{1}$
is \textit{either} even \textit{or} odd. Consequently, the parity of
$\ket{\Psi}$  is determined by the parity of $Y$ since
\begin{align}
|\Psi|=
\left(
|Q(b)|+|Y|+\sum_{j=1}^{N}|\mathsf{A}[j]|
\right)\hbox{ mod 2}=
|Y|.
\label{appeq: def FMPS g}
\end{align}
A prerequisite to imposing translation symmetry on any
FMPS is that all dimensions $\mathcal{D}^{\,}_{\mathrm{v},j}$
and $\mathcal{D}^{\,}_{j}$ are independent of $j=1,\cdots,N$.
Hence, we assume from now on that
\begin{equation}
\mathcal{D}^{\,}_{\mathrm{v},j}\equiv\mathcal{D}^{\,}_{\mathrm{v}},
\qquad  
\mathcal{D}^{\,}_{j}\equiv \mathcal{D}^{\,},
\qquad
j=1,\cdots,N.
\label{appeq: def FMPS i}
\end{equation} 
\end{subequations}

\subsection{Even-parity fermionic matrix product sate (FMPS)}

The FMPS
\begin{subequations}\label{eq:even parity tensors FMPS}
\begin{align}
\ket{\Psi}^{b}_{0}\df
\mathcal{C}^{\,}_{\mathrm{v}}
\Big(
Q(b)\,
Y\,
\mathsf{A}[1]
\otimes^{\,}_{\mathfrak{g}}
\cdots
\otimes^{\,}_{\mathfrak{g}}
\mathsf{A}[N]
\Big)
\label{eq:even parity tensors FMPS a}
\end{align}
is an even-parity FMPS obeying periodic ($b=0$) or antiperiodic ($b=1$)
boundary conditions if, for any $j=1,\cdots,N$,
\begin{align}
&
\mathsf{A}[j]\df
\sum_{\alpha^{\,}_{j}=1}^{\mathcal{D}^{\,}_{\mathrm{v}}}
\sum_{\sigma^{\,}_{j}=1}^{\mathcal{D}}
\sum_{\beta^{\,}_{j}=1}^{\mathcal{D}^{\,}_{\mathrm{v}}}
\big(A^{(0)}_{\sigma^{\,}_{j}}\big)^{\,}_{\alpha^{\,}_{j}\,\beta^{\,}_{j}}\,
\pket{\alpha^{\,}_{j}}
\otimes^{\,}_{\mathfrak{g}}
\ket{\psi^{\,}_{\sigma^{\,}_{j}}}
\otimes^{\,}_{\mathfrak{g}}
\pbra{\beta^{\,}_{j}},
\label{eq:even parity tensors FMPSb}
\\
&
|\mathsf{A}[j]|=
\left(
|\alpha^{\,}_{j}|
+
|\sigma^{\,}_{j}|
+
|\beta^{\,}_{j}|
\right)\hbox{ mod 2}=
0,
\label{eq:even parity tensors FMPS c}
\\
&
Y\df
\sum_{\alpha=1}^{\mathcal{D}^{\,}_{\mathrm{v}}}
\pket{\alpha}
\otimes^{\,}_{\mathfrak{g}}
\pbra{\alpha},
\\
&
Q(b=0)\df
\sum_{\alpha=1}^{\mathcal{D}^{\,}_{\mathrm{v}}}
\pket{\alpha}
\otimes^{\,}_{\mathfrak{g}}
\pbra{\alpha},
\\
&
Q(b=1)\df
\sum_{\alpha=1}^{\mathcal{D}^{\,}_{\mathrm{v}}}
(-1)^{|\alpha|}\,
\pket{\alpha}
\otimes^{\,}_{\mathfrak{g}}
\pbra{\alpha}.
\end{align}
\end{subequations}
By construction, both $Q(b)$ and $Y$ are of even parity.
Moreover,
$
\left(
|\alpha^{\,}_{j}|
+
|\sigma^{\,}_{j}|
+
|\beta^{\,}_{j}|
\right)\hbox{ mod 2}=1
$
implies that
$
\big(A^{(0)}_{\sigma^{\,}_{j}}\big)^{\,}_{\alpha^{\,}_{j}\,\beta^{\,}_{j}}=0
$.

We are going to give an alternative representation of this even-parity
FMPS under the assumption that the virtual dimension
$\mathcal{D}^{\,}_{\mathrm{v}}$
obeys the partition
$\mathcal{D}^{\,}_{\mathrm{v}}=M^{\,}_{\mathrm{e}}+M^{\,}_{\mathrm{o}}$
where
$M^{\,}_{\mathrm{e}}\equiv M$
and
$M^{\,}_{\mathrm{o}}\equiv M$ are the
numbers of even- and odd-parity virtual basis vectors, respectively.
Parity evenness of $\mathsf{A}[j]$ implies
that the $\mathcal{D}^{\,}_{\mathrm{v}}\times \mathcal{D}^{\,}_{\mathrm{v}}$ 
dimensional matrices $A^{(0)}_{\sigma^{\,}_{j}}$ 
with the matrix elements
$\big(A^{(0)}_{\sigma^{\,}_{j}}\big)^{\,}_{\alpha^{\,}_{j}\,\beta^{\,}_{j}}$
is either block diagonal
\begin{subequations}\label{appeq: A matrix decomposition even parity}
\begin{align}
A^{(0)}_{\sigma^{\,}_{j}}=
\begin{pmatrix}
B^{\,}_{\sigma^{\,}_{j}} & 0 \\
0 & C^{\,}_{\sigma^{\,}_{j}}
\end{pmatrix},
\quad
\mathrm{if}
\,\,
|\sigma^{\,}_{j}|=0,
\label{appeq: A matrix decomposition even parity a}
\end{align}
when the physical state is of even parity
[as follows from Eq.\ (\ref{eq:even parity tensors FMPS c})]
or block off diagonal
\begin{align}
A^{(0)}_{\sigma^{\,}_{j}}=
\begin{pmatrix}
0 & D^{\,}_{\sigma^{\,}_{j}}\\
F^{\,}_{\sigma^{\,}_{j}} & 0
\end{pmatrix},
\quad
\mathrm{if}
\,\,
|\sigma^{\,}_{j}|=1,
\label{appeq: A matrix decomposition even parity b}
\end{align}
when the physical state is of odd parity
[as follows from Eq.\ (\ref{eq:even parity tensors FMPS c})].
All the blocks are here $M\times M$-dimensional.
Parity evenness of $Q(b)$ with matrix elements
$\big(Q(b)\big)^{\,}_{\alpha^{\,}_{1}\,\beta^{\,}_{1}}$
and $Y$ with matrix elements
$Y^{\,}_{\alpha^{\,}_{1}\,\beta^{\,}_{1}}$
implies that
\begin{align}
Y= 
Q(b=0)=
\begin{pmatrix}
\openone^{\,}_{M} & 0\\
0 & \openone^{\,}_{M}
\end{pmatrix},
\qquad
Q(b=1)=
\begin{pmatrix}
\openone^{\,}_{M} & 0\\
0 & -\openone^{\,}_{M}
\end{pmatrix}\dfr
P.
\label{appeq: A matrix decomposition even parity c}
\end{align}
Hereby, we introduced the parity matrix $P$ that
satisfies
\begin{align}
P\,A^{(0)}_{\sigma^{\,}_{j}}\,P=
(-1)^{|\sigma^{\,}_{j}|}\,
A^{(0)}_{\sigma^{\,}_{j}}.
\label{appeq: A matrix decomposition even parity d}
\end{align}
\end{subequations}
\begin{subequations}
Inserting these explicit representations of
$Q(b)$ and $Y$
in Eq.\ (\ref{eq:even parity tensors FMPS a}) delivers
\begin{align}
\label{eq:even parity FMPS gen form}
\ket{\Psi}^{b}_{0}
\equiv
\ket{\{A^{(0)}_{\sigma^{\,}_{j}}\};b}
\df
\sum_{\bm{\sigma}}
\mathrm{tr}
\left[
P^{b+1}\,
A^{(0)}_{\sigma^{\,}_{1}}\,
A^{(0)}_{\sigma^{\,}_{2}}\,
\cdots
A^{(0)}_{\sigma^{\,}_{N}}\,
\right]
\ket{\Psi^{\,}_{\bm{\sigma}}},
\end{align}
\normalsize
where we used the shorthand notation
$\ket{\Psi^{\,}_{\bm{\sigma}}}
\df 
\ket{\psi^{\,}_{\sigma^{\,}_{1}}}
\otimes^{\,}_{\mathfrak{g}}
\ket{\psi^{\,}_{\sigma^{\,}_{2}}}
\otimes^{\,}_{\mathfrak{g}}
\cdots\,
\otimes^{\,}_{\mathfrak{g}}
\ket{\psi^{\,}_{\sigma^{\,}_{N}}}$.
The appearance of the matrix $P$ when $b=0$ is counterintuitive.
It is needed to eliminate from the sum over all physical basis states
$\{\ket{\Psi^{\,}_{\bm{\sigma}}}\}$  
those physical basis states of odd parity.
The state $\ket{\{A^{(0)}_{\sigma^{\,}_{j}}\};b}$ 
has even parity since
\begin{align}
\left(
\sum_{j=1}^{N}
|\sigma^{\,}_{j}|
\right)\hbox{ mod 2}=
\left[
\sum_{j=1}^{N}
(
|\alpha^{\,}_{j}|
+
|\beta^{\,}_{j}|
)
\right]\hbox{ mod 2}=
\left(
\sum_{j=1}^{N}
2
|\alpha^{\,}_{j}|
\right)
\hbox{ mod 2}=
0,
\end{align}
\end{subequations}
where we used
condition (\ref{eq:even parity tensors FMPS c})
to establish the first equality
and the condition
$|\beta^{\,}_{j}|=|\alpha^{\,}_{j+1}|$
that is imposed by the contractions of virtual indices
to establish the second equality.

\subsection{Odd-parity fermionic matrix product sate (FMPS)}

The FMPS
\begin{subequations}\label{eq:odd parity tensors FMPS}
\begin{align}
\ket{\Psi}^{b}_{1}\df
\mathcal{C}^{\,}_{\mathrm{v}}
\left(
Q(b)\,
Y\,
\mathsf{A}[1]
\otimes^{\,}_{\mathfrak{g}}
\cdots
\otimes^{\,}_{\mathfrak{g}}
\mathsf{A}[N]
\right),
\label{eq:odd parity tensors FMPS a}
\end{align}
is an odd-parity FMPS obeying periodic ($b=0$) or antiperiodic ($b=1$)
boundary conditions if, for any $j=1,\cdots,N$,
\begin{align}
&
\mathsf{A}[j]\df
\sum_{\alpha^{\,}_{j}=1}^{\mathcal{D}^{\,}_{\mathrm{v}}}
\sum_{\sigma^{\,}_{j}=1}^{\mathcal{D}}
\sum_{\beta^{\,}_{j}=1}^{\mathcal{D}^{\,}_{\mathrm{v}}}
\big(A^{(1)}_{\sigma^{\,}_{j}}\big)^{\,}_{\alpha^{\,}_{j}\,\beta^{\,}_{j}}\,
\pket{\alpha^{\,}_{j}}
\otimes^{\,}_{\mathfrak{g}}
\ket{\psi^{\,}_{\sigma^{\,}_{j}}}
\otimes^{\,}_{\mathfrak{g}}
\pbra{\beta^{\,}_{j}},
\label{eq:odd parity tensors FMPS b}
\\
&
|\mathsf{A}[j]| = 
\left(
|\sigma^{\,}_{j}|
+
|\alpha^{\,}_{j}|
+
|\beta^{\,}_{j}|
\right)\hbox{ mod 2}=
0,
\label{eq:odd parity tensors FMPS c}
\\
&
Y\df
\sum_{\alpha,\beta=1}^{\mathcal{D}^{\,}_{\mathrm{v}}}
Y^{\,}_{\alpha\,\beta}\,
\pket{\alpha}
\otimes^{\,}_{\mathfrak{g}}
\pbra{\beta},
\qquad
\hbox{
$Y^{\,}_{\alpha\,\beta}=0$
if
$
(
|\alpha|
+
|\beta|
)\hbox{ mod 2}=
0
$,
}
\label{eq:odd parity tensors FMPS d}
\\
&
Q(b=0)
\df
\sum_{\alpha=1}^{\mathcal{D}^{\,}_{\mathrm{v}}}
\pket{\alpha}
\otimes^{\,}_{\mathfrak{g}}
\pbra{\alpha},
\label{eq:odd parity tensors FMPS e}
\\
&
Q(b=1)\df
\sum_{\alpha=1}^{\mathcal{D}^{\,}_{\mathrm{v}}}
(-1)^{|\alpha|}\,
\pket{\alpha}
\otimes^{\,}_{\mathfrak{g}}
\pbra{\alpha}.
\label{eq:odd parity tensors FMPS f}
\end{align}
\end{subequations}
By construction, $Q(b)$ is of even parity while $Y$ is of odd parity.
Moreover,
$
\left(
|\alpha^{\,}_{j}|
+
|\sigma^{\,}_{j}|
+
|\beta^{\,}_{j}|
\right)\hbox{ mod 2}=1
$
implies that
$
\big(A^{(0)}_{\sigma^{\,}_{j}}\big)^{\,}_{\alpha^{\,}_{j}\,\beta^{\,}_{j}}=0
$.

We note that the only difference between  
definitions \eqref{eq:even parity tensors FMPS}
and \eqref{eq:odd parity tensors FMPS} is the 
choice for $Y$. In the former case its
parity is even, in the latter case its parity is odd. 
Analogously to the even FMPS case, 
we define $2M\times 2M$ dimensional matrices $A^{(1)}_{\sigma^{\,}_{j}}$ and $Y$
with the matrix elements
$\big(A^{(1)}_{\sigma^{\,}_{j}}\big)^{\,}_{\alpha^{\,}_{j}\,\beta^{\,}_{j}}$
and $Y^{\,}_{\alpha^{\,}_{1}\,\beta^{\,}_{1}}$. The parity $|Y|=1$ implies
that
\begin{subequations}\label{appeq: A matrix decomposition odd parity}
\begin{align}
Y=
\begin{pmatrix}
0 & Y^{\,}_{1}\\
Y^{\,}_{2} & 0
\end{pmatrix},
\end{align}
where $Y^{\,}_{1}$ and $Y^{\,}_{2}$ are
$M\times M$ and dimensional matrices, respectively.
Imposing translation symmetry requires that 
\begin{align}
Y\,A^{(1)}_{\sigma^{\,}_{j}}= 
A^{(1)}_{\sigma^{\,}_{j}}\, Y.
\end{align}
We choose 
\begin{align}
Y
\df
\begin{pmatrix}
0 & \openone^{\,}_{M}\\
-\openone^{\,}_{M} & 0
\end{pmatrix},
\qquad
P\,
Y\,
P 
=
- Y,
\end{align}
which implies 
\begin{align}
\label{eq:A matrix decomposition odd parity a}
&
A^{(1)}_{\sigma^{\,}_{j}}
=
\begin{pmatrix}
G^{\,}_{\sigma^{\,}_{j}} & 0 \\
0 & G^{\,}_{\sigma^{\,}_{j}}
\end{pmatrix},
\quad
\mathrm{if}
\,\,
|\sigma^{\,}_{j}|=0,
\\
\label{eq:A matrix decomposition odd parity b}
&
A^{(1)}_{\sigma^{\,}_{j}}
=
\begin{pmatrix}
0 & G^{\,}_{\sigma^{\,}_{j}}\\
-G^{\,}_{\sigma^{\,}_{j}} & 0
\end{pmatrix},
\quad
\mathrm{if}
\,\,
|\sigma^{\,}_{j}|=1,
\end{align}
\end{subequations}
where $G^{\,}_{\sigma^{\,}_{j}}$ are $M\times M$
dimensional matrices. 
Inserting these explicit representations of $Q(b)$ and $Y$
in Eq.\ 
(\ref{eq:odd parity tensors FMPS a})
delivers
\begin{subequations}
\begin{align}
\label{eq:odd parity FMPS gen form}
\ket{\Psi}^{b}_{1}
\equiv
\ket{\{A^{(1)}_{\sigma^{\,}_{j}}\};b}
\df
\sum_{\bm{\sigma}}
\mathrm{tr}
\left[
P^{b}\,
Y\,
A^{(1)}_{\sigma^{\,}_{1}}\,
A^{(1)}_{\sigma^{\,}_{2}}\,
\cdots
A^{(1)}_{\sigma^{\,}_{N}}\,
\right]
\ket{\Psi^{\,}_{\bm{\sigma}}},
\qquad
\ket{\Psi^{\,}_{\bm{\sigma}}}
\df 
\ket{\psi^{\,}_{\sigma^{\,}_{1}}}
\otimes^{\,}_{\mathfrak{g}}
\ket{\psi^{\,}_{\sigma^{\,}_{2}}}
\otimes^{\,}_{\mathfrak{g}}
\cdots\,
\otimes^{\,}_{\mathfrak{g}}
\ket{\psi^{\,}_{\sigma^{\,}_{N}}}.
\end{align}
The state $\ket{\{A^{(1)}_{\sigma^{\,}_{j}}\};b}$ has odd parity since
\begin{align}
\left(
\sum_{j=1}^{N}
|\sigma^{\,}_{j}|
\right)\hbox{ mod 2}=
\left[
\sum_{j=1}^{N}
(
|\alpha^{\,}_{j}|
+
|\beta^{\,}_{j}|
)
\right]\hbox{ mod 2}=
\left(
|\alpha|
+
|\beta|
+
\sum_{j=2}^{N-1}
2
|\alpha^{\,}_{j}|
\right)\hbox{ mod 2}=
1,
\end{align}
\end{subequations}
where we used
condition (\ref{eq:odd parity tensors FMPS c})
to establish the first equality.
For the second equality we used the conditions 
$|\alpha| = |\beta^{\,}_{N}|$ and $|\beta| = |\alpha^{\,}_{1}|$
where $|\alpha|$, $|\beta|$ are the parities of the virtual indices corresponding to 
matrix elements $Y^{\,}_{\alpha\,\beta}$, and
$|\beta^{\,}_{j}|= |\alpha^{\,}_{j+1}|$ for $j=2,\cdots,N-1$
that is imposed by the contractions of virtual indices
to establish the second equality.

\section{Proof of Theorem \ref{thm:LSM Theorem 1}}
\label{appsec:Proof of LSM}

We will prove Theorem \ref{thm:LSM Theorem 1}
for one dimensional systems within the FMPS framework.
Our proof follows closely that for
the bosonic case\,
\footnote{
Bosonic matrix products states presume that the
local Fock space $\mathcal{F}^{\,}_{\bm{j}}$
has no more than the trivial $\mathbb{Z}^{\,}_{2}$ grading, i.e.,
$\mathcal{F}^{\,}_{\bm{j}}\equiv
\mathcal{F}^{\,}_{\bm{j}\,0}\oplus\mathcal{F}^{\,}_{\bm{j}\,1}$
with
$\mathcal{F}^{\,}_{\bm{j}\,0}\equiv\mathcal{F}^{\,}_{\bm{j}}$
and
$\mathcal{F}^{\,}_{\bm{j}\,1}\equiv\emptyset$.}
introduced in Ref.\ \onlinecite{Tasaki2020}.
We will show that a parity-even or parity-odd injective FMPS
necessarily requires
the local projective representation $\hat{u}^{\,}_{j}$ of
the symmetry group $G^{\,}_{f}$ 
to have trivial second cohomology class 
$[\phi]\in H^{2}\big(G^{\,}_{f},\mathrm{U(1)}^{\,}_{\mathfrak{c}}\big)$. 
In other words, when this cohomology class is nontrivial there is no compatible
injective FMPS with even or odd parity.
The general forms \eqref{eq:even parity FMPS gen form} 
and \eqref{eq:odd parity FMPS gen form} as well as the
injectivity conditions \eqref{def:injectivity even} and 
\eqref{def:injectivity odd} are distinct for even and odd parity FMPS. 
The proofs for the even- and the odd-parity cases are thus treated
successively.
For conciseness, we are going to suppress the symbol
$\otimes^{\,}_{\mathfrak{g}}$
when working with the orthonormal and homogeneous basis
\begin{subequations}
\begin{equation}
\left\{
\ket{\Psi^{\,}_{\bm{\sigma}}}\equiv
\ket{\psi^{\,}_{\sigma^{\,}_{1}}}
\otimes^{\,}_{\mathfrak{g}}
\ket{\psi^{\,}_{\sigma^{\,}_{2}}}
\otimes^{\,}_{\mathfrak{g}}
\cdots
\otimes^{\,}_{\mathfrak{g}}
\ket{\psi^{\,}_{\sigma^{\,}_{N}}}
\right\}
\end{equation}
of the Fock space
\begin{equation}
\mathcal{F}^{\,}_{\Lambda}\equiv
\mathcal{F}^{\,}_{1}
\otimes^{\,}_{\mathfrak{g}}
\mathcal{F}^{\,}_{2}
\otimes^{\,}_{\mathfrak{g}}
\cdots
\otimes^{\,}_{\mathfrak{g}}
\mathcal{F}^{\,}_{N}.
\end{equation}
\end{subequations} 

\subsection{Even-parity FMPS}
\label{appsubsec:Even-parity FMPS}

Let [see Eq.\ (\ref{appeq: A matrix decomposition even parity})]
\begin{equation}
\ket{\{A^{(0)}_{\sigma^{\,}_{j}}\};b}\equiv
\sum_{\bm{\sigma}}
\mathrm{tr}
\left(
P^{b+1}
A^{(0)}_{\sigma^{\,}_{1}}\,
A^{(0)}_{\sigma^{\,}_{2}}\,
\cdots
A^{(0)}_{\sigma^{\,}_{N}}\,
\right)
\ket{\psi^{\,}_{\sigma^{\,}_{1}}}\,
\ket{\psi^{\,}_{\sigma^{\,}_{2}}}\,
\cdots
\ket{\psi^{\,}_{\sigma^{\,}_{N}}}
\label{eq:LSM even FMPS beginning}
\end{equation}
be a translation-invariant,
$G^{\,}_{f}$-symmetric, even-parity, and injective
FMPS obeying periodic boundary conditions when $b=0$ or
antiperiodic boundary conditions when $b=1$.
For any $g\in G^{\,}_{f}$, the global representation
$\widehat{U}(g)$ of $g$ is
defined in Eq.\ (\ref{eq: def U(g)}).
Hence, for any $g\in G^{\,}_{f}$, there exists a phase
$\eta(g;b)\in[0,2\pi)$
such that
\begin{align}
\widehat{U}(g)\,
\ket{\{A^{(0)}_{\sigma^{\,}_{j}}\};b}=
e^{\mathrm{i}\eta(g;b))}
\ket{\{A^{(0)}_{\sigma^{\,}_{j}}\};b}.
\label{eq:even FMPS unique gs condition}
\end{align} 
The action of the transformation
$\widehat{U}(g)$ on the right-hand side of 
Eq.\ \eqref{eq:LSM even FMPS beginning} gives
\begin{align}
\widehat{U}(g)\,
\ket{\{A^{(0)}_{\sigma^{\,}_{j}}\};b}=&\,
\sum_{\bm{\sigma}}
\left\{
\mathsf{K}^{\,}_{g}\,
\mathrm{tr}
\left(
P^{b+1}\,
A^{(0)}_{\sigma^{\,}_{1}}\,
A^{(0)}_{\sigma^{\,}_{2}}\,
\cdots
A^{(0)}_{\sigma^{\,}_{N}}\,
\right)
\right\}\,
\hat{v}^{\,}_{1}(g)\,
\ket{\psi^{\,}_{\sigma^{\,}_{1}}}\,
\hat{v}^{\,}_{2}(g)\,
\ket{\psi^{\,}_{\sigma^{\,}_{2}}}\,
\cdots
\hat{v}^{\,}_{N}(g)\,
\ket{\psi^{\,}_{\sigma^{\,}_{N}}}
\nonumber\\
=&\,
\sum_{\bm{\sigma}}
\left\{
\sum_{\bm{\sigma}'}
\mathrm{tr}
\left[
P^{b+1}\,
\mathsf{K}^{\,}_{g}
\left(
A^{(0)}_{\sigma'^{\,}_{1}}\,
A^{(0)}_{\sigma'^{\,}_{2}}\,
\cdots
A^{(0)}_{\sigma'^{\,}_{N}}\,
\right)
\right]
\prod_{j=1}^{N}
\bra{\psi^{\,}_{\sigma^{\,}_{j}}}
\hat{v}^{\,}_{j}(g)
\ket{\psi^{\,}_{\sigma'^{\,}_{j}}}
\right\}
\ket{\psi^{\,}_{\sigma^{\,}_{1}}}\,
\ket{\psi^{\,}_{\sigma^{\,}_{2}}}\,
\cdots
\ket{\psi^{\,}_{\sigma^{\,}_{N}}}
\label{eq:even Symmetry action on even FMPS initial step}
\end{align}
after using $N$ times the resolution of the identity,
one for each local Fock space $\mathcal{F}^{\,}_{j}$.
The right-hand side can be written more elegantly
with the definition of the $g$-dependent
$2M\times 2M$ matrix
\begin{subequations}\label{eq:even Symmetry action on even FMPS}
\begin{align}
A^{(0)}_{\sigma^{\,}_{j}}(g)\df
\sum_{\sigma'^{\,}_{j}}
\bra{\psi^{\,}_{\sigma^{\,}_{j}}}\,
\hat{v}^{\,}_{j}(g)\,
\ket{\psi^{\,}_{\sigma'^{\,}_{j}}}\,
\mathsf{K}^{\,}_{g}
\left[
A^{(0)}_{\sigma'^{\,}_{j}}
\right]
\equiv
\sum_{\sigma'^{\,}_{j}}
\left[
\mathcal{U}(g)
\right]^{\,}_{\sigma^{\,}_{j}\,\sigma'^{\,}_{j}}\,
\mathsf{K}^{\,}_{g}
\left[
A^{(0)}_{\sigma'^{\,}_{j}}
\right],
\qquad
\sigma^{\,}_{j}=1,\cdots,\mathcal{D},
\qquad
j=1,\cdots,N,
\label{eq:even Symmetry action on even FMPS a}
\end{align}
where the $\mathcal{D}\times\mathcal{D}$ matrix
$\mathcal{U}(g)$,
whose matrix elements are the complex-valued coefficients weighting
the sum over the $2M\times 2M$ matrices
$
\mathsf{K}^{\,}_{g}
\left[
A^{(0)}_{\sigma'^{\,}_{j}}
\right]
$,
acts on the local Fock space $\mathcal{F}^{\,}_{j}$ and we have defined 
\begin{align}
\mathsf{K}^{\,}_{g}\left[A^{(0)}_{\sigma^{\,}_{j}}\right]\df
\begin{cases}
A^{(0)}_{\sigma^{\,}_{j}},&\hbox{ if $\mathfrak{c}(g)=0$, }
\\&\\
\mathsf{K}\,
A^{(0)}_{\sigma^{\,}_{j}}\,
\mathsf{K},&\hbox{ if $\mathfrak{c}(g)=1$. }
\end{cases}
\label{eq:even Symmetry action on even FMPS b}
\end{align}
As usual, $\mathsf{K}$ denotes complex conjugation.
Equation (\ref{eq:even Symmetry action on even FMPS initial step})
becomes
\begin{align}
\widehat{U}(g)\,
\ket{\{A^{(0)}_{\sigma^{\,}_{j}}\};b}=
\sum_{\bm{\sigma}}
\mathrm{tr}
\left[
P^{b+1}\,
A^{(0)}_{\sigma^{\,}_{1}}(g)\,
A^{(0)}_{\sigma^{\,}_{2}}(g)\,
\cdots
A^{(0)}_{\sigma^{\,}_{N}}(g)\,
\right]
\ket{\psi^{\,}_{\sigma^{\,}_{1}}}\,
\ket{\psi^{\,}_{\sigma^{\,}_{2}}}
\cdots
\ket{\psi^{\,}_{\sigma^{\,}_{N}}},
\label{eq:even Symmetry action on even FMPS c}
\end{align}
\end{subequations}
which is nothing but the FMPS (\ref{eq:LSM even FMPS beginning})
with
$A^{(0)}_{\sigma^{\,}_{j}}$
substituted for
$A^{(0)}_{\sigma^{\,}_{j}}(g)$.
Equating the right-hand sides of Eqs.\
\eqref{eq:even FMPS unique gs condition}
and \eqref{eq:even Symmetry action on even FMPS c} implies
\begin{subequations}\label{eq:even FMPS invariance condition}
\begin{align}
\mathrm{tr}
\left[
P^{b+1}\,
A^{(0)}_{\sigma^{\,}_{1}}(g)\,
A^{(0)}_{\sigma^{\,}_{2}}(g)\,
\cdots
A^{(0)}_{\sigma^{\,}_{N}}(g)\,
\right]=
e^{
\mathrm{i}\eta(g;b)}\,
\mathrm{tr}
\left[
P^{b+1}\,
A^{(0)}_{\sigma^{\,}_{1}}\,
A^{(0)}_{\sigma^{\,}_{2}}\,
\cdots
A^{(0)}_{\sigma^{\,}_{N}}\,
\right].
\label{eq:even FMPS invariance condition a}
\end{align}
This equation is satisfied by the Ansatz
\begin{align}
A^{(0)}_{\sigma^{\,}_{j}}(g)=
e^{\mathrm{i}\theta(g)}\,
U^{-1}(g)\,
A^{(0)}_{\sigma^{\,}_{j}}\,
U(g),
\qquad
P\,U(g)\,P=
(-1)^{\kappa(g)}\,U(g),
\qquad
\theta(g)
\df
\frac{1}{N}
\left[
\eta(g;b)
-
\pi(b+1)\,
\kappa(g)
\right],
\label{eq:even FMPS invariance condition b}
\end{align}
where $\kappa(g)=0,1$ dictates if the $2M\times2M$ unitary matrix
$U(g)$ commutes or anticommutes
with the $2M\times2M$ parity matrix $P$ defined in Eq.\
(\ref{appeq: A matrix decomposition even parity c}),
since
\begin{align}
\mathrm{tr}
\left[
P^{b+1}\,
A^{(0)}_{\sigma^{\,}_{1}}(g)\,
A^{(0)}_{\sigma^{\,}_{2}}(g)\,
\cdots
A^{(0)}_{\sigma^{\,}_{N}}(g)\,
\right]
=&\,
e^{\mathrm{i}\theta(g)N}\,
\mathrm{tr}
\left[
P^{b+1}
U^{-1}(g)\,
A^{(0)}_{\sigma^{\,}_{1}}\,
A^{(0)}_{\sigma^{\,}_{2}}\,
\cdots
A^{(0)}_{\sigma^{\,}_{N}}\,
U(g)\,
\right]
\nonumber\\
\hbox{\tiny cyclicity of the trace \qquad}
=&\,
e^{\mathrm{i}\theta(g)N}\,
\mathrm{tr}
\left[
U(g)\,
P^{b+1}
U^{-1}(g)\,
A^{(0)}_{\sigma^{\,}_{1}}\,
A^{(0)}_{\sigma^{\,}_{2}}\,
\cdots
A^{(0)}_{\sigma^{\,}_{N}}\,
\right]
\nonumber\\
\hbox{\tiny
Eq.\ (\ref{eq:even FMPS invariance condition b})\qquad
}
=&\,
e^{\mathrm{i}\theta(g)N}\,
(-1)^{(b+1)\kappa(g)}\,
\mathrm{tr}
\left[
P^{b+1}
A^{(0)}_{\sigma^{\,}_{1}}\,
A^{(0)}_{\sigma^{\,}_{2}}\,
\cdots
A^{(0)}_{\sigma^{\,}_{N}}\,
\right]
\nonumber\\
=&\,
e^{\mathrm{i}\theta(g)N
+\mathrm{i}\pi(b+1)\kappa(g)}\,
\mathrm{tr}
\left[
P^{b+1}
A^{(0)}_{\sigma^{\,}_{1}}\,
A^{(0)}_{\sigma^{\,}_{2}}\,
\cdots
A^{(0)}_{\sigma^{\,}_{N}}\,
\right]
\nonumber\\
\equiv&\,
e^{\mathrm{i}\eta(g;b)}\,
\mathrm{tr}
\left[
P^{b+1}
A^{(0)}_{\sigma^{\,}_{1}}\,
A^{(0)}_{\sigma^{\,}_{2}}\,
\cdots
A^{(0)}_{\sigma^{\,}_{N}}\,
\right].
\label{eq:even FMPS invariance condition c}
\end{align}
\end{subequations}
The existence of the $2M\times2M$ invertible
matrix $U(g)$ is
guaranteed because of the injectivity of the FMPS.
In an injective even-parity FMPS, the matrices 
$A^{(0)}_{\sigma^{\,}_{1}}$, $\cdots$,
$A^{(0)}_{\sigma^{\,}_{\ell}}$
span the simple algebra of all $2M\times 2M$ matrices
for any $\ell>\ell^{\star}$ for some nonvanishing integer $\ell^{\star}$.
Hence, provided $N$ is sufficiently large,
the family of matrices
$\{A^{(0)}_{\sigma^{\,}_{1}}(g),\cdots,A^{(0)}_{\sigma^{\,}_{N}}(g)\}$
is related to the family of matrices  
$\{e^{\mathrm{i}\eta(g;b)/N}\,A^{(0)}_{\sigma^{\,}_{1}},\cdots,
e^{\mathrm{i}\eta(g;b)/N}A^{(0)}_{\sigma^{\,}_{N}}\}$
that give the \textit{same} FMPS (\ref{eq:LSM even FMPS beginning})
by the similarity transformation
[see Eqs.\ (\ref{eq:FMPs gauge transformation}) and
(\ref{eq:pre key step lemma})]
\begin{subequations}\label{eq:even FMPS gauge transformation due to inject}
\begin{align}
A^{(0)}_{\sigma^{\,}_{j}}(g)=
e^{\mathrm{i}\varphi^{(b)}_{U(g)}}\,
U^{-1}(g)\,
\left[
e^{\mathrm{i}\eta(g;b)/N}\,
A^{(0)}_{\sigma^{\,}_{j}}
\right]
U(g),
\label{eq:even FMPS gauge transformation due to inject a}
\end{align}
for some phase
$\varphi^{(b)}_{U(g)}=[0,2\pi)$
and some invertible
$2M\times2M$ matrix $U(g)$ that must also obey
\begin{align}
U(p)=P,
\qquad
P\,U(g)\,P=
(-1)^{\kappa(g)}\,U(g).
\label{eq:even FMPS gauge transformation due to inject b}
\end{align}
\end{subequations}
Here,
the map $\kappa\colon G^{\,}_{f} \to \left\{0,1\right\}$ specifies the
algebra between the similarity transformation $U(g)$
corresponding to element  $g\in G^{\,}_{f}$ and the fermion parity $P$.
The effect of the factor $(-1)^{\kappa(g)}$ is nothing but the phase
\begin{align}
\varphi^{(b)}_{U(g)}=
-
\frac{1}{N}
\pi
\left(b+1\right)
\kappa(g),
\end{align}
as follows from Eq.\ (\ref{eq:even FMPS invariance condition b}).

Equating the right-hand sides of
Eqs.\
\eqref{eq:even FMPS gauge transformation due to inject a}
and
\eqref{eq:even Symmetry action on even FMPS a} 
implies
\begin{align}
\label{eq:even Symmetry imposed relation on FMPS matrix}
e^{\mathrm{i}\theta(g)}\,
U^{-1}(g)\,
A^{(0)}_{\sigma^{\,}_{j}}\,
U(g)=
\sum_{\sigma'^{\,}_{j}}
\left[\mathcal{U}(g)\right]^{\,}_{\sigma^{\,}_{j}\,\sigma'^{\,}_{j}}\,
\mathsf{K}^{\,}_{g}
\left[
A^{(0)}_{\sigma'^{\,}_{j}}
\right],
\qquad
\sigma^{\,}_{j}=1,\cdots,\mathcal{D},
\qquad
j=1,\cdots,N.
\end{align}
We would like to isolate
$
\mathsf{K}^{\,}_{g}
\left[
A^{(0)}_{\sigma'^{\,}_{j}}
\right]
$
on the right-hand side. To this end, we do the manipulations
\begin{align}
e^{\mathrm{i}\theta(g)}
\sum_{\sigma^{\,}_{j}}
\left[
\mathcal{U}^{\dag}(g)
\right]^{\,}_{\sigma^{\prime\prime}_{j}\sigma^{\,}_{j}}
U^{-1}(g)\,
A^{(0)}_{\sigma^{\,}_{j}}\,
U(g)=&\,
\sum_{\sigma^{\vphantom{\prime}}_{j}}
\sum_{\sigma^{\prime}_{j}}
\left[
\mathcal{U}^{\dag}(g)
\right]^{\,}_{\sigma^{\prime\prime}_{j}\sigma^{\,}_{j}}
\left[\mathcal{U}(g)\right]^{\,}_{\sigma^{\,}_{j}\,\sigma'^{\,}_{j}}\,
\mathsf{K}^{\,}_{g}
\left[
A^{(0)}_{\sigma'^{\,}_{j}}
\right]
\nonumber\\
=&\,
\sum_{\sigma^{\vphantom{\prime}}_{j}}
\sum_{\sigma^{\prime}_{j}}
\bra{\psi^{\,}_{\sigma^{\prime\prime}_{j}}}
\hat{v}^{\dag}_{j}(g)\,
\ket{\psi^{\,}_{\sigma^{\,}_{j}}}\,
\bra{\psi^{\,}_{\sigma^{\,}_{j}}}
\hat{v}^{\,}_{j}(g)
\ket{\psi^{\,}_{\sigma'^{\,}_{j}}}\,
\mathsf{K}^{\,}_{g} 
\left[
A^{(0)}_{\sigma'^{\,}_{j}}
\right]
\nonumber\\
=&\,
\sum_{\sigma^{\prime}_{j}}
\bra{\psi^{\,}_{\sigma^{\prime\prime}_{j}}}
\hat{v}^{\dag}_{j}(g)\,
\hat{v}^{\,}_{j}(g)
\ket{\psi^{\,}_{\sigma'^{\,}_{j}}}\,
\mathsf{K}^{\,}_{g} 
\left[
A^{(0)}_{\sigma'^{\,}_{j}}
\right]
\nonumber\\
=&\,
\sum_{\sigma^{\prime}_{j}}
\braket{\psi^{\,}_{\sigma^{\prime\prime}_{j}}}
{\psi^{\,}_{\sigma'^{\,}_{j}}}\,
\mathsf{K}^{\,}_{g} 
\left[
A^{(0)}_{\sigma'^{\,}_{j}}
\right]
\nonumber\\
=&\,
\sum_{\sigma^{\prime}_{j}}
\delta^{\,}_{\sigma^{\prime\prime}_{j},\sigma'^{\,}_{j}}\,
\mathsf{K}^{\,}_{g} 
\left[
A^{(0)}_{\sigma'^{\,}_{j}}
\right]
\nonumber\\
=&\,
\mathsf{K}^{\,}_{g} 
\left[
A^{(0)}_{\sigma^{\prime\prime}_{j}}
\right],
\qquad
\sigma^{\prime\prime}_{j}=1,\cdots,\mathcal{D},
\qquad
j=1,\cdots,N.
\end{align}
By applying $\mathsf{K}^{\,}_{g}$ to both sides of this
equation, we obtain the self-consistency condition
\begin{subequations}
\begin{align}
A^{(0)}_{\sigma^{\,}_{j}}=&\,
\mathsf{K}^{\,}_{g} 
\left[
e^{\mathrm{i}\theta(g)}
\sum_{\sigma'^{\,}_{j}}
\left[
\mathcal{U}^{\dag}(g)
\right]^{\,}_{\sigma^{\,}_{j}\sigma^{\prime}_{j}}\,
U^{-1}(g)\,
A^{(0)}_{\sigma^{\prime}_{j}}\,
U(g)
\right]
\nonumber\\
=&\,
e^{\mathrm{i}\mathfrak{c}(g)\,\theta(g)}
\sum_{\sigma^{\prime}_{j}}
\bra{\psi^{\,}_{\sigma^{\,}_{j}}}
\left(
\hat{u}^{\dag}_{j}(g)
\ket{\psi^{\,}_{\sigma^{\prime}_{j}}}
\right)
V^{-1}(g)\,
A^{(0)}_{\sigma'^{\,}_{j}}\,
V(g),
\qquad
\sigma^{\,}_{j}=1,\cdots,\mathcal{D},
\qquad
j=1,\cdots,N,
\label{appeq:self-consistency with g}
\end{align}
with $\hat{u}^{\,}_{j}(g)$ defined in Eq.\ \eqref{eq: def U(g)},
$V(g)=U(g)$ if $\mathfrak{c}(g)=0$ and
$V(g)=U(g)\,\mathsf{K}$ if $\mathfrak{c}(g)=1$. 
We use the notation
$\left(
\hat{u}^{\dag}_{j}(g)
\ket{\psi^{\,}_{\sigma^{\prime}_{j}}}
\right)$
to indicate that the operator $\hat{u}^{\dag}_{j}(g)$ acts on the right,
an important fact to keep track of
when $\hat{u}^{\dag}_{j}(g)$ is an antiunitary operator.
Had we chosen the elements $h\in G^{\,}_{f}$ and $g\,h\in G^{\,}_{f}$,
Eq.\ (\ref{appeq:self-consistency with g})
would give the self-consistency conditions
\begin{align}
A^{(0)}_{\sigma^{\prime\prime}_{j}}=
e^{\mathrm{i}\mathfrak{c}(h)\,\theta(h)}
\sum_{\sigma^{\,}_{j}}
\bra{\psi^{\,}_{\sigma^{\prime\prime}_{j}}}
\left(
\hat{u}^{\dag}_{j}(h)
\ket{\psi^{\,}_{\sigma^{\,}_{j}}}
\right)
V^{-1}(h)\,
A^{(0)}_{\sigma^{\,}_{j}}\,
V(h),
\qquad
\sigma^{\prime\prime}_{j}=1,\cdots,\mathcal{D},
\qquad
j=1,\cdots,N,
\label{appeq:self-consistency with h}
\end{align}
and
\begin{align}
A^{(0)}_{\sigma^{\,}_{j}}=
e^{\mathrm{i}\mathfrak{c}(g\,h)\,\theta(g\,h)}
\sum_{\sigma^{\prime}_{j}}
\bra{\psi^{\,}_{\sigma^{\,}_{j}}}
\left(
\hat{u}^{\dag}_{j}(g\,h)
\ket{\psi^{\,}_{\sigma^{\prime}_{j}}}
\right)\,
V^{-1}(g\,h)\,
A^{(0)}_{\sigma^{\prime}_{j}}\,
V(g\,h),
\qquad
\sigma^{\,}_{j}=1,\cdots,\mathcal{D},
\qquad
j=1,\cdots,N,
\label{appeq:self-consistency with gh}
\end{align}
\end{subequations}
respectively.

Inserting the self-consistency condition
(\ref{appeq:self-consistency with g})
into the self-consistency condition
(\ref{appeq:self-consistency with h}) gives
\begin{align}
\label{eq:even Symmetry imposed relation with two elements}
A^{(0)}_{\sigma^{\prime\prime}_{j}}=&\,
e^{\mathrm{i}\mathfrak{c}(h)\,\theta(h)}
\sum_{\sigma^{\,}_{j}}
\bra{\psi^{\,}_{\sigma^{\prime\prime}_{j}}}
\left(
\hat{u}^{\dag}_{j}(h)
\ket{\psi^{\,}_{\sigma^{\,}_{j}}}
\right)\,
V^{-1}(h)\,
\left(
e^{\mathrm{i}\mathfrak{c}(g)\,\theta(g)}
\sum_{\sigma^{\prime}_{j}}
\bra{\psi^{\,}_{\sigma^{\,}_{j}}}\,
\left(
\hat{u}^{\dag}_{j}(g)
\ket{\psi^{\,}_{\sigma^{\prime}_{j}}}
\right)\,
V^{-1}(g)\,
A^{(0)}_{\sigma^{\prime}_{j}}\,
V(g)
\right)
V(h)
\nonumber\\
=&\,
e^{\mathrm{i}\mathfrak{c}(h)\,\theta(h)
+
\mathrm{i}\mathfrak{c}(h)\,\mathfrak{c}(g)\,\theta(g)}
\sum_{\sigma^{\,}_{j},\sigma^{\prime}_{j}}
\bra{\psi^{\,}_{\sigma^{\prime\prime}_{j}}}\,
\left(
\hat{u}^{\dag}_{j}(h)
\ket{\psi^{\,}_{\sigma^{\,}_{j}}}
\right)\,
\mathsf{K}^{\,}_{h}
\left[
\bra{\psi^{\,}_{\sigma^{\,}_{j}}}\,
\left(
\hat{u}^{\dag}_{j}(g)
\ket{\psi^{\,}_{\sigma^{\prime}_{j}}}
\right)
\right]
V^{-1}(h)\,
V^{-1}(g)\,
A^{(0)}_{\sigma^{\prime}_{j}}\,
V(g)\,
V(h)
\nonumber\\
=&\,
e^{\mathrm{i}\mathfrak{c}(h)\,\theta(h)
+
\mathrm{i}\mathfrak{c}(h)\,\mathfrak{c}(g)\,\theta(g)}
\sum_{\sigma^{\prime}_{j}}
\bra{\psi^{\,}_{\sigma^{\prime\prime}_{j}}}
\hat{u}^{\dag}_{j}(h)\,
\left(
\hat{u}^{\dag}_{j}(g)\,
\ket{\psi^{\,}_{\sigma^{\prime}_{j}}}
\right)\,
V^{-1}(h)\,
V^{-1}(g)\,
A^{(0)}_{\sigma^{\prime}_{j}}\,
V(g)\,
V(h)
\nonumber\\
=&\,
e^{\mathrm{i}\mathfrak{c}(h)\,\theta(h)
+
\mathrm{i}\mathfrak{c}(h)\,\mathfrak{c}(g)\,\theta(g)
-
\mathrm{i}\mathfrak{c}(g\,h)\phi(g,h)}
\sum_{\sigma^{\prime}_{j}}
\bra{\psi^{\,}_{\sigma^{\prime\prime}_{j}}}
\left(
\hat{u}^{\dag}_{j}(g\,h)\,
\ket{\psi^{\,}_{\sigma^{\prime}_{j}}}
\right)\,
V^{-1}(h)\,
V^{-1}(g)\,
A^{(0)}_{\sigma^{\prime}_{j}}\,
V(g)\,
V(h).
\end{align}
In reaching the penultimate and last
equalities, we used two identities.
First,
\begin{align}
\sum^{\,}_{\sigma^{\,}_{j}}
\bra{\psi^{\,}_{\sigma^{\prime\prime}_{j}}}\,
\left(
\hat{u}^{\dag}_{j}(h)\,
\ket{\psi^{\,}_{\sigma^{\,}_{j}}}
\right)\,
\mathsf{K}^{\,}_{h}
\left[
\bra{\psi^{\,}_{\sigma^{\,}_{j}}}\,
\left(
\hat{u}^{\dag}_{j}(g)\,
\ket{\psi^{\,}_{\sigma^{\prime}_{j}}}
\right)
\right]=
\bra{\psi^{\,}_{\sigma^{\prime\prime}_{j}}}\,
\left(
\hat{u}^{\dag}_{j}(h)\,
\hat{u}^{\dag}_{j}(g)\,
\ket{\psi^{\,}_{\sigma^{\prime}_{j}}}
\right)
\end{align}
is obviously true when $\mathfrak{c}(h)=1$
since $\sum^{\,}_{\sigma^{\,}_{j}}
\ket{\psi^{\,}_{\sigma^{\,}_{j}}}\,
\bra{\psi^{\,}_{\sigma^{\,}_{j}}}$
is the resolution of the identity on $\mathcal{F}^{\,}_{j}$.
When $\mathfrak{c}(h)=-1$, 
$\hat{u}^{\dag}_{j}(h)$ is antiunitary so that
\begin{align}
\sum^{\,}_{\sigma^{\,}_{j}}
\bra{\psi^{\,}_{\sigma^{\prime\prime}_{j}}}\,
\left(
\hat{u}^{\dag}_{j}(h)\,
\ket{\psi^{\,}_{\sigma^{\,}_{j}}}
\right)\,
\mathsf{K}^{\,}_{h}
\left[
\bra{\psi^{\,}_{\sigma^{\,}_{j}}}
\left(
\hat{u}^{\dag}_{j}(g)\,
\ket{\psi^{\,}_{\sigma^{\prime}_{j}}}
\right)
\right]=&\,
\sum^{\,}_{\sigma^{\,}_{j}}
\bra{\psi^{\,}_{\sigma^{\prime\prime}_{j}}}
\left(
\hat{u}^{\dag}_{j}(h)
\ket{\psi^{\,}_{\sigma^{\,}_{j}}}
\right)
\left[
\bra{\psi^{\,}_{\sigma^{\,}_{j}}}
\left(
\hat{u}^{\dag}_{j}(g)\,
\ket{\psi^{\,}_{\sigma^{\prime}_{j}}}
\right)
\right]^{*}
\nonumber\\
=&\,
\sum^{\,}_{\sigma^{\,}_{j}}
\left[
\left(
\bra{\psi^{\,}_{\sigma^{\prime\prime}_{j}}}\hat{u}^{\,}_{j}(h)
\right)\,
\ket{\psi^{\,}_{\sigma^{\,}_{j}}}
\right]^{*}\,
\left[
\bra{\psi^{\,}_{\sigma^{\,}_{j}}}
\left(
\hat{u}^{\dag}_{j}(g)\,
\ket{\psi^{\,}_{\sigma^{\prime}_{j}}}
\right)
\right]^{*}
\nonumber\\
=&\,
\left[
\sum^{\,}_{\sigma^{\,}_{j}}
\left(
\bra{\psi^{\,}_{\sigma^{\prime\prime}_{j}}}\,
\hat{u}^{\,}_{j}(h)
\right)
\ket{\psi^{\,}_{\sigma^{\,}_{j}}}
\bra{\psi^{\,}_{\sigma^{\,}_{j}}}
\left(
\hat{u}^{\dag}_{j}(g)
\ket{\psi^{\,}_{\sigma^{\prime}_{j}}}
\right)
\right]^{*}
\nonumber\\
=&\,
\left[
\left(
\bra{\psi^{\,}_{\sigma^{\prime\prime}_{j}}}\hat{u}^{\,}_{j}(h)
\right)
\left(
\hat{u}^{\dag}_{j}(g)
\ket{\psi^{\,}_{\sigma^{\prime}_{j}}}
\right)
\right]^{*}
\nonumber\\
=&\,
\bra{\psi^{\,}_{\sigma^{\prime\prime}_{j}}}\,
\left(
\hat{u}^{\dag}_{j}(h)\,
\hat{u}^{\dag}_{j}(g)
\ket{\psi^{\,}_{\sigma^{\prime}_{j}}}
\right).
\end{align}
Second, we used the projective representation
\eqref{eq:def projective rep I}
to obtain
\begin{align}
\hat{u}^{\dag}_{j}(h)\,
\hat{u}^{\dag}_{j}(g)
=&\,
\left[
\hat{u}^{\,}_{j}(g)\,
\hat{u}^{\,}_{j}(h)
\right]^{\dag}
\nonumber\\
=&\,
\left[
e^{+\mathrm{i}\phi(g,h)}\,
\hat{u}^{\,}_{j}(g\,h)
\right]^{\dag}
\nonumber\\
=&\,
\hat{u}^{\dag}_{j}(g\,h)\,
e^{-\mathrm{i}\phi(g,h)}
\nonumber\\
=&\,
e^{-\mathrm{i}\mathfrak{c}(g\,h)\,\phi(g,h)}\,
\hat{u}^{\dag}_{j}(g\,h).
\end{align}
Equating the right-hand sides of Eqs.\
\eqref{eq:even Symmetry imposed relation with two elements}
and 
\eqref{appeq:self-consistency with gh}
gives the condition
\begin{align}
e^{\mathrm{i}\mathfrak{c}(h)\,\theta(h)
+
\mathrm{i}\mathfrak{c}(h)\,\mathfrak{c}(g)\,\theta(g)
-
\mathrm{i}\mathfrak{c}(g\,h)\phi(g,h)}
V^{-1}(h)\,
V^{-1}(g)\,
A^{(0)}_{\sigma^{\prime}_{j}}\,
V(g)\,
V(h)=
e^{\mathrm{i}\mathfrak{c}(g\,h)\,\theta(g\,h)}\,
V^{-1}(g\,h)\,
A^{(0)}_{\sigma^{\prime}_{j}}\,
V(g\,h).
\end{align}
Upon using the fact that $\mathfrak{c}$ is a homomorphism so that
$\mathfrak{c}(g\,h)=\mathfrak{c}(g)\,\mathfrak{c}(h)$ holds,
we arrive at
\begin{subequations}\label{eq:even gauge transformations phase relation}
\begin{equation}
W^{-1}(g,h)\,
A^{(0)}_{\sigma^{\,}_{j}}\,
W(g,h)=
e^{-\mathrm{i}\delta(g,h;b)}\,
A^{(0)}_{\sigma^{\,}_{j}},
\qquad
\sigma^{\,}_{j}=1,\cdots,\mathcal{D},
\qquad
j=1,\cdots,N,
\label{eq:even gauge transformations phase relation a}
\end{equation}
where
\begin{equation}
W(g,h)\df
V(g)\,
V(h)\,
V^{-1}(g\,h),
\qquad
\delta(g,h;b)\df
\mathfrak{c}(g)\,\theta(h)
+
\theta(g)
-
\phi(g,h)
-
\theta(g\,h).
\label{eq:even gauge transformations phase relation b}
\end{equation}
\end{subequations}
A forteriori
\begin{align}
W^{-1}(g,h)\,
A^{(0)}_{\sigma^{\,}_{1}}\,
A^{(0)}_{\sigma^{\,}_{2}}
\cdots
A^{(0)}_{\sigma^{\,}_{\ell}}\,
W(g,h)=
e^{-\mathrm{i}\ell\,\delta(g,h;b)}\,
A^{(0)}_{\sigma^{\,}_{1}}\,
A^{(0)}_{\sigma^{\,}_{2}}
\cdots
A^{(0)}_{\sigma^{\,}_{\ell}}
\label{appeq:repeated conjugation by W}
\end{align}
holds for any positive integer $\ell$.

Injectivity of a FMPS 
implies that for some integer $\ell^{\star}>1$
and any $\ell \geq \ell^{\star}$ all
the products of the form
$A^{(0)}_{\sigma^{\,}_{1}}\,A^{(0)}_{\sigma^{\,}_{2}}\cdots A^{(0)}_{\sigma^{\,}_{\ell}}$
span the space of all $2M\times 2M$ matrices. Therefore,
Eq.\ (\ref{appeq:repeated conjugation by W})
combined with injectivity implies that the $2M\times2M$ matrix
$W(g,h)$ is an element from the center of the algebra
defined by the vector space of all $2M\times 2M$ matrices,
i.e., $\{\openone^{\,}_{2M}\}$.
Condition \eqref{appeq:repeated conjugation by W}
thus simplifies to
\begin{align}
A^{(0)}_{\sigma^{\,}_{1}}\,
A^{(0)}_{\sigma^{\,}_{2}}
\cdots
A^{(0)}_{\sigma^{\,}_{\ell}}=
e^{-\mathrm{i}\ell\,\delta(g,h;b)}\,
A^{(0)}_{\sigma^{\,}_{1}}\,
A^{(0)}_{\sigma^{\,}_{2}}
\cdots
A^{(0)}_{\sigma^{\,}_{\ell}}
\label{appeq:repeated conjugation by W bis}
\end{align}
for any $\ell\geq\ell^{\star}$. Choosing a linear combination
of
$
A^{(0)}_{\sigma^{\,}_{1}}\,
A^{(0)}_{\sigma^{\,}_{2}}
\cdots
A^{(0)}_{\sigma^{\,}_{\ell}}
$
equating the identity matrix $\openone^{\,}_{2M}$,
delivers the constraint
\begin{subequations}\label{appeq:final step proof thm 1 even case}
\begin{align}
\ell\,\delta(g,h;b)=
0,
\quad
\forall \ell > \ell^{\star}
\implies
\delta(g,h;b)=0.
\end{align}
Inserting the value of $\delta(g,h;b)$ given in Eq.\ 
\eqref{eq:even gauge transformations phase relation}
implies the final constraint
\begin{align}
\phi(g,h)=
\mathfrak{c}(g)\theta(h)
+
\theta(g)
-
\theta(g\,h).
\end{align}
\end{subequations}
This is the coboundary condition 
\eqref{eq:U(1) gauge equivalence}
when $\phi'=0$. In other words,
the local representation
$\hat{u}^{\,}_{j}$ is equivalent to the trivial projective 
representation.

\subsection{Odd-parity FMPS}
\label{appsec:Odd-parity FMPS}

Let [see Eq.\ (\ref{appeq: A matrix decomposition odd parity})]
\begin{align}
\ket{\{A^{(1)}_{\sigma^{\,}_{j}}\};b}=
\sum_{\bm{\sigma}}
\mathrm{tr}
\left[
P^{b}\,
Y\,
A^{(1)}_{\sigma^{\,}_{1}}\,
A^{(1)}_{\sigma^{\,}_{2}}\,
\cdots
A^{(1)}_{\sigma^{\,}_{N}}\,
\right]
\ket{\psi^{\,}_{\sigma^{\,}_{1}}}\,
\ket{\psi^{\,}_{\sigma^{\,}_{2}}}\,
\cdots
\ket{\psi^{\,}_{\sigma^{\,}_{N}}}
\end{align}
be a translation-invariant, $G^{\,}_{f}$-symmetric,
odd-parity (each matrix $A^{(1)}_{\sigma^{\,}_{j}}$ commutes with the matrix $Y$),
and injective FMPS obeying periodic boundary conditions when $b=0$ or
antiperiodic boundary conditions when $b=1$.  
For any $g\in G^{\,}_{f}$, the global representation
$\widehat{U}(g)$ of $g$ is
defined in Eq.\ (\ref{eq: def U(g)}).
Hence, for any $g\in G^{\,}_{f}$, there exists a phase $\eta(g;b)\in[0,2\pi)$
such that
\begin{align}
\widehat{U}(g)\,
\ket{\{A^{(1)}_{\sigma^{\,}_{j}}\};b}=
e^{\mathrm{i}\eta(g;b))}
\ket{\{A^{(1)}_{\sigma^{\,}_{j}}\};b}.
\label{eq:odd FMPS unique gs condition}
\end{align}
The counterpart to Eq.\ \eqref{eq:even Symmetry action on even FMPS}
is
\begin{subequations}\label{eq:odd Symmetry action on odd FMPS}
\begin{align}
&
\widehat{U}(g)
\ket{\{A^{(1)}_{\sigma^{\,}_{j}}\};b}=
\sum_{\bm{\sigma}}
\mathrm{tr}
\left[
P^{b}\,
Y\,
A^{(1)}_{\sigma^{\,}_{1}}(g)\,
A^{(1)}_{\sigma^{\,}_{2}}(g)\,
\cdots
A^{(1)}_{\sigma^{\,}_{N}}(g)\,
\right]
\ket{\psi^{\,}_{\sigma^{\,}_{1}}}\,
\ket{\psi^{\,}_{\sigma^{\,}_{2}}}\,
\cdots
\ket{\psi^{\,}_{\sigma^{\,}_{N}}},
\label{eq:odd Symmetry action on odd FMPS a}
\\
&
A^{(1)}_{\sigma^{\,}_{j}}(g)\df
\sum_{\sigma^{\prime}_{j}}
\bra{\psi^{\,}_{\sigma^{\,}_{j}}}\,
\hat{v}^{\,}_{j}(g)\,
\ket{\psi^{\,}_{\sigma^{\prime}_{j}}}\,
\mathsf{K}^{\,}_{g}
\left[
A^{(1)}_{\sigma^{\prime}_{j}}
\right]=
\sum_{\sigma^{\prime}_{j}}
\mathcal{U}(g)^{\,}_{\sigma^{\,}_{j},\sigma^{\prime}_{j}}\,
A^{(1)}_{\sigma^{\prime}_{j}},
\label{eq:odd Symmetry action on odd FMPS b}
\\
&
\mathsf{K}^{\,}_{g}\left[A^{(1)}_{\sigma^{\,}_{j}}\right]\df
\begin{cases}
A^{(1)}_{\sigma^{\,}_{j}},&\hbox{ if $\mathfrak{c}(g)=0$, }
\\&\\
\mathsf{K}\,
A^{(1)}_{\sigma^{\,}_{j}}\,
\mathsf{K},&\hbox{ if $\mathfrak{c}(g)=1$. }
\end{cases}
\label{eq:odd Symmetry action on odd FMPS c}
\end{align}
\end{subequations}
Odd-parity injective FMPS differ from the even ones in one crucial way.
There exists a positive integer $\ell^{\star}\geq1$ such that
for any $\ell\geq\ell^{\star}$ the products of the form
$A^{(1)}_{\sigma^{\,}_{1}}\,A^{(1)}_{\sigma^{\,}_{2}}\cdots A^{(1)}_{\sigma^{\,}_{\ell}}$
span the $\mathbb{Z}^{\,}_{2}$-graded algebra of 
$2M\times 2M$ matrices with the center
$\left\{\openone^{\,}_{2M},Y\right\}$.
Consequently, there exists a
$2M\times2M$ invertible matrix $U(g)$ and
a phase $\theta(g)\in[0,2\pi)$ such that
[recall Eq.\ (\ref{eq:FMPs gauge transformation})]
\begin{subequations}\label{eq:odd FMPS gauge transformation due to inject}
\begin{equation}
U(g)=P\,U(g)\,P,
\qquad
U(g)=(-1)^{\zeta(g)}\,Y\,U(g)\,Y,
\qquad
\zeta(g)=0,1,
\label{eq:odd FMPS gauge transformation due to inject a}
\end{equation}
with $\zeta\colon G^{\,}_{f}\to\{-1,+1\}$ a group homomorphism and
\begin{align}
A^{(1)}_{\sigma^{\,}_{j}}(g)=
e^{\mathrm{i}\theta(g)}\,
U^{-1}(g)\,
A^{(1)}_{\sigma^{\,}_{j}}\,
U(g)
\qquad
\sigma^{\,}_{j}=1,\cdots,\mathcal{D},
\qquad
j=1,\cdots,N.
\label{eq:odd FMPS gauge transformation due to inject b}
\end{align}
\end{subequations}
The same steps that lead to Eq.\ (\ref{eq:even FMPS invariance condition})
then give
\begin{subequations}\label{eq:odd FMPS invariance condition}
\begin{align}
\mathrm{tr}
\left[
P^{b}\,Y\,
A^{(1)}_{\sigma^{\,}_{1}}(g)\,
A^{(1)}_{\sigma^{\,}_{2}}(g)\,
\cdots
A^{(1)}_{\sigma^{\,}_{N}}(g)\,
\right]=
e^{
\mathrm{i}\eta(g;b)}\,
\mathrm{tr}
\left[
P^{b}\,Y\,
A^{(1)}_{\sigma^{\,}_{1}}\,
A^{(1)}_{\sigma^{\,}_{2}}\,
\cdots
A^{(1)}_{\sigma^{\,}_{N}}\,
\right]
\label{eq:odd FMPS invariance condition a}
\end{align}
with the solution
\begin{align}
A^{(1)}_{\sigma^{\,}_{j}}(g)=
e^{\mathrm{i}\theta(g)}\,
U^{-1}(g)\,
A^{(1)}_{\sigma^{\,}_{j}}\,
U(g),
\qquad
Y\,U(g)\,Y=
(-1)^{\zeta(g)}\,U(g),
\qquad
\theta(g)\df
\frac{1}{N}
\left[
\eta(g;b)
-
\pi\,
\zeta(g)
\right].
\label{eq:odd FMPS invariance condition b}
\end{align}
\end{subequations}
All the steps leading to Eq.\
(\ref{eq:even gauge transformations phase relation})
deliver
\begin{subequations}\label{eq:odd gauge transformations phase relation}
\begin{equation}
W^{-1}(g,h)\,
A^{(1)}_{\sigma^{\,}_{j}}\,
W(g,h)=
e^{-\mathrm{i}\delta(g,h;b)}\,
A^{(1)}_{\sigma^{\,}_{j}},
\qquad
\sigma^{\,}_{j}=1,\cdots,\mathcal{D},
\qquad
j=1,\cdots,N,
\label{eq:odd gauge transformations phase relation a}
\end{equation}
where
\begin{equation}
W(g,h)\df
V(g)\,
V(h)\,
V^{-1}(g\,h),
\qquad
\delta(g,h;b)\df
\mathfrak{c}(g)\,\theta(h)
+
\theta(g)
-
\phi(g,h)
-
\theta(g\,h),
\label{eq:odd gauge transformations phase relation b}
\end{equation}
\end{subequations}
and $V(g)=U(g)$ if $\mathfrak{c}(g)=0$ and
$V(g)=U(g)\,\mathsf{K}$ if $\mathfrak{c}(g)=1$.
Because $U(g)$ commutes with $P$ so does $W(g,h)$.
Because all possible products of the form
$A^{(1)}_{\sigma^{\,}_{1}}\,A^{(1)}_{\sigma^{\,}_{2}}\cdots A^{(1)}_{\sigma^{\,}_{\ell}}$
span the $\mathbb{Z}^{\,}_{2}$-graded algebra of 
$2M\times 2M$ matrices with the center
$\left\{\openone^{\,}_{2M},Y\right\}$,
$W(g,h)$ is, up to a phase factor, proportional to $\openone^{\,}_{2M}$.
The counterpart to the even-parity coboundary condition
(\ref{appeq:final step proof thm 1 even case})
then follows, thereby completing the proof of 
Theorem \ref{thm:LSM Theorem 1} for the parity-odd FMPS.

\section{Proof of Theorem \protect{\ref{thm:LSM Theorem 1}}
with twisted boundary conditions for any Abelian group $G^{\,}_{f}$
whose projective representations are all unitary}
\label{appsec:Proof Theorem 1 with twisted boundary conditions}

The lattice is $\Lambda=\{1,\cdots,N\}\equiv\mathbb{Z}^{\,}_{N}$
with $N$ an integer.
The global fermionic Fock space $\mathcal{F}^{\,}_{\Lambda}$
is of dimension $2^{mN}$
with $n=2m$ an even number of local Majorana flavors.
The local $\mathcal{F}^{\,}_{j}$ and global $\mathcal{F}^{\,}_{\Lambda}$
Fock spaces are generated by the Hermitian Majorana operators
$\hat{\chi}^{\,}_{j,a}$ obeying the Clifford algebra
\begin{equation}
\left\{
\hat{\chi}^{\,}_{j,a},
\hat{\chi}^{\,}_{j',a'}
\right\}=
2\,
\delta^{\,}_{j,j'}\,
\delta^{\,}_{a,a'},
\qquad
j,j'=1,\cdots,N,
\qquad
a,a'=1,\cdots,n=2m.
\end{equation}
The local and global fermion parity operators are
\begin{equation}
\hat{p}^{\,}_{j}\df
\prod_{a=1}^{m}\mathrm{i}\hat{\chi}^{\,}_{j,2a-1}\,\hat{\chi}^{\,}_{j,2a},
\qquad
\widehat{P}^{\,}_{\Lambda}\df
\prod_{j=1}^{N}
\hat{p}^{\,}_{j},
\end{equation}
respectively. Any polynomial 
$\hat{h}^{\,}_{j}$
in the Majorana operators that is 
of finite order,
of finite range $r$
(the integer $r$ is the maximum separation between the space
labels of the Majorana operators entering $\hat{h}^{\,}_{j}$) 
of even parity
($\widehat{P}^{\,}_{\Lambda}\,\hat{h}^{\,}_{j}\,\widehat{P}^{\,}_{\Lambda}=
\hat{h}^{\,}_{j}$),
and Hermitian
($\hat{h}^{\dag}_{j}=\hat{h}^{\,}_{j}$)
is a local Hamiltonian.
We define the unitary operator $\widehat{T}^{\,}_{\hat{1}}$ by its action
\begin{equation}
\widehat{T}^{\,}_{\hat{1}}\,
\hat{\chi}^{\,}_{j,a}\,
\widehat{T}^{-1}_{\hat{1}}
= 
\begin{cases}
\hat{\chi}^{\,}_{j+1,a},
&
\hbox{if $j=1,\cdots,N-1$ and $a=1,\cdots,n=2m$,} 
\\&\\
\hat{\chi}^{\,}_{1,a},
&
\hbox{if $j=N$ and $a=1,\cdots,n=2m$.} 
\end{cases}
\label{appeq:def T1}
\end{equation}
It follows that
\begin{equation}
\widehat{T}^{N}_{\hat{1}}=\widehat{\openone}^{\,}_{2^{mN}},
\label{appeq:T1's cylic rep ZN}  
\end{equation}
i.e., $\widehat{T}^{\,}_{\hat{1}}$ is a unitary representation of the
generator of the cyclic group $\mathbb{Z}^{\,}_{N}$.
For any Abelian central extension $G^{\,}_{f}$ of $G$ by $\mathbb{Z}^{F}_{2}$
and for any $g\in\ G^{\,}_{f}$, 
we assume the projective representation (\ref{eq:def projective rep I})
with 
\begin{equation}
\hat{u}^{\,}_{j}(g)=\hat{v}^{\,}_{j}(g)
\qquad
\hbox{ [as $\mathfrak{c}(g)=+1$ always hold by hypothesis]} 
\end{equation}
a polynomial in
$\hat{\chi}^{\,}_{j,a}$ with $a=1,\cdots,n=2m$.
We make the identifications
\begin{equation}
\begin{split}
&
\hat{v}^{\,}_{j}(e)\equiv\widehat{\openone}^{\,}_{2^{m}},
\qquad
\hat{v}^{\,}_{j}(p)\equiv\hat{p}^{\,}_{j},
\qquad
\hat{p}^{\,}_{j}\,\hat{v}^{\,}_{j}(g)\,\hat{p}^{\,}_{j}=
(-1)^{\rho(g)}\,\hat{v}^{\,}_{j}(g),
\qquad
j=1,\cdots,N,
\\
&
\widehat{U}(e)\equiv
\widehat{\openone}^{\,}_{2^{mN}},
\qquad
\widehat{U}(p)\equiv
\widehat{P}^{\,}_{\Lambda},
\qquad
\widehat{U}(g)\df
\prod_{j=1}^{N}
\hat{v}^{\,}_{j}(g),
\qquad
\forall g\in G^{\,}_{f}.
\end{split}
\end{equation}
We assume that the Hamiltonian $\hat{h}^{\,}_{j}$ is
$G^{\,}_{f}$-invariant (symmetric), i.e.,
\begin{equation}
\hat{h}^{\,}_{j}=
\widehat{U}(g)\,
\hat{h}^{\,}_{j}\,
\widehat{U}^{-1}(g),
\qquad
\forall g\in G^{\,}_{f}.
\end{equation}
By construction, the Hamiltonian defined by [recall Eq.\ \eqref{eq:gen_Ham}]
\begin{subequations}
\begin{align}
\widehat{H}^{\,}_{\mathrm{pbc}}\df
\sum_{n=1}^{N}
\left(\widehat{T}^{\vphantom{\dag}}_{\hat{1}}\right)^{n}\, 
\hat{h}^{\,}_{j}\,
\left(\widehat{T}^{\dag}_{\hat{1}}\right)^{n},
\qquad
\widehat{U}(g)\,\hat{h}^{\,}_{j}\,\widehat{U}^{-1}(g),
\qquad
\forall g\in G^{\,}_{f},
\end{align}
is translation-invariant (symmetric), 
\begin{equation}
\widehat{T}^{\,}_{\hat{1}}\,
\widehat{H}^{\,}_{\mathrm{pbc}}\,
\widehat{T}^{-1}_{\hat{1}}=
\widehat{H}^{\,}_{\mathrm{pbc}},
\end{equation}
and $G^{\,}_{f}$-invariant (symmetric), 
\begin{equation}
\widehat{U}(g)\,
\widehat{H}^{\,}_{\mathrm{pbc}}\,
\widehat{U}^{-1}(g)=
\widehat{H}^{\,}_{\mathrm{pbc}},
\qquad
\forall g\in G^{\,}_{f}.
\end{equation}
\end{subequations}
We define the family of twisted translation operators
\begin{equation}
\widehat{T}^{\,}_{\hat{1}}(g)\df
\hat{v}^{\,}_{1}(g)\,
\widehat{T}^{\,}_{\hat{1}},
\qquad
g\in G^{\,}_{f},
\qquad
\mathfrak{c}(g)=+1.
\label{appeq:def T1(g)} 
\end{equation}
Their action on the Majorana spinor
\begin{subequations}
\begin{equation}
\hat{\chi}^{\,}_{j}\df
\begin{pmatrix}
\hat{\chi}^{\,}_{j,1}
&
\cdots
&
\hat{\chi}^{\,}_{j,n}
\end{pmatrix}^{\mathsf{T}}
\end{equation}
differ from
that in Eq.\ (\ref{appeq:def T1}),
\begin{align}
\widehat{T}^{\,}_{\hat{1}}(g)\,
\hat{\chi}^{\,}_{j}\,
\widehat{T}^{-1}_{\hat{1}}(g)
= 
\begin{cases}
(-1)^{\rho(g)}\,
\hat{\chi}^{\,}_{j+1},
&
\hbox{if $j\neq N$,} 
\\&\\
\hat{v}^{\,}_{1}(g)\,
\hat{\chi}^{\,}_{1}\,
\hat{v}^{-1}_{1}(g),
&
\hbox{if $j=N$.} 
\end{cases}
\end{align}
\end{subequations}
We have the identity
\begin{align}
\left[\widehat{T}^{\,}_{\hat{1}}(g)\right]^{N}=&\,
\left[
\hat{v}^{\,}_{1}(g)\,
\widehat{T}^{\,}_{\hat{1}}
\right]
\left[
\hat{v}^{\,}_{1}(g)\,
\widehat{T}^{\,}_{\hat{1}}
\right]
\cdots
\left[
\hat{v}^{\,}_{1}(g)\,
\widehat{T}^{\,}_{\hat{1}}
\right]
\left[
\hat{v}^{\,}_{1}(g)\,
\widehat{T}^{\,}_{\hat{1}}
\right]
\nonumber\\
=&\,
\hat{v}^{\,}_{1}(g)
\left[
\widehat{T}^{\,}_{\hat{1}}\,
\hat{v}^{\,}_{1}(g)\,
\widehat{T}^{\,}_{\hat{1}}
\right]
\cdots
\hat{v}^{\,}_{1}(g)
\left[
\widehat{T}^{\,}_{\hat{1}}\,
\hat{v}^{\,}_{1}(g)\,
\widehat{T}^{\,}_{\hat{1}}
\right]
\nonumber\\
=&\,
\hat{v}^{\,}_{1}(g)
\left[
\widehat{T}^{\,}_{\hat{1}}\,
\hat{v}^{\,}_{1}(g)\,
\widehat{T}^{\,}_{\hat{1}}
\right]
\cdots
\hat{v}^{\,}_{1}(g)
\left[
\widehat{T}^{\,}_{\hat{1}}\,
\hat{v}^{\,}_{1}(g)\,
\widehat{T}^{-1}_{\hat{1}}
\right]
\widehat{T}^{2}_{\hat{1}}
\nonumber\\
\hbox{\tiny Eq.\ (\ref{appeq:def T1})\quad}
=&\,
\hat{v}^{\,}_{1}(g)
\left[
\widehat{T}^{\,}_{\hat{1}}\,
\hat{v}^{\,}_{1}(g)\,
\widehat{T}^{\,}_{\hat{1}}
\right]
\cdots
\hat{v}^{\,}_{1}(g)\,
\hat{v}^{\,}_{2}(g)\,
\widehat{T}^{2}_{\hat{1}}
\nonumber\\
=&\,
\widehat{U}(g)\,
\widehat{T}^{N}_{\hat{1}}
\nonumber\\
\hbox{\tiny Eq.\ (\ref{appeq:T1's cylic rep ZN})\quad}
=&\,
\widehat{U}(g).
\label{appeq:T1(g) raised power N is U(g)}
\end{align}
Finally, we define the family of twisted Hamiltonians
\begin{align}
\widehat{H}^{\mathrm{tilt}}_{\mathrm{twis}}(g)\df
\sum_{j=1}^{N}
\left[\widehat{T}^{\,}_{\hat{1}}(g)\right]^{j}\, 
\hat{h}^{\mathrm{tilt}}_{1}\,
\left[\widehat{T}^{-1}_{\hat{1}}(g)\right]^{j},
\qquad
\hat{h}^{\mathrm{tilt}}_{1}=
\widehat{U}(h)\,
\hat{h}^{\mathrm{tilt}}_{1}\,
\widehat{U}^{-1}(h),
\qquad
\forall h\in G^{\,}_{f}.
\label{eq:tilt twist Ham nonAbelian}
\end{align}
By design,
\begin{align}
\widehat{T}^{\,}_{\hat{1}}(g)\,
\widehat{H}^{\mathrm{tilt}}_{\mathrm{twis}}(g)\,
\widehat{T}^{-1}_{\hat{1}}(g)=&\,
\sum_{j=1}^{N-1}
\left[\widehat{T}^{\,}_{\hat{1}}(g)\right]^{j+1}\, 
\hat{h}^{\mathrm{tilt}}_{1}\,
\left[\widehat{T}^{-1}_{\hat{1}}(g)\right]^{j+1}
+
\left[\widehat{T}^{\,}_{\hat{1}}(g)\right]^{N+1}\, 
\hat{h}^{\mathrm{tilt}}_{1}\,
\left[\widehat{T}^{-1}_{\hat{1}}(g)\right]^{N+1}
\nonumber\\
\hbox{\tiny Eq.\ (\ref{appeq:T1(g) raised power N is U(g)})\qquad}
=&\,
\sum_{j=1}^{N-1}
\left[\widehat{T}^{\,}_{\hat{1}}(g)\right]^{j+1}\, 
\hat{h}^{\mathrm{tilt}}_{1}\,
\left[\widehat{T}^{-1}_{\hat{1}}(g)\right]^{j+1}
+
\widehat{T}^{\,}_{\hat{1}}(g) 
\left[
\widehat{U}(g)\,
\hat{h}^{\mathrm{tilt}}_{1}\,
\widehat{U}^{-1}(g)
\right]
\widehat{T}^{-1}_{\hat{1}}(g)
\nonumber\\
\hbox{\tiny $G^{\,}_{f}$ symmetry\qquad}
=&\,
\sum_{j=1}^{N-1}
\left[\widehat{T}^{\,}_{\hat{1}}(g)\right]^{j+1}\, 
\hat{h}^{\mathrm{tilt}}_{1}\,
\left[\widehat{T}^{-1}_{\hat{1}}(g)\right]^{j+1}
+
\widehat{T}^{\,}_{\hat{1}}(g)\,
\hat{h}^{\mathrm{tilt}}_{1}\,
\widehat{T}^{-1}_{\hat{1}}(g)
\nonumber\\
=&\,
\widehat{H}^{\mathrm{tilt}}_{\mathrm{twis}}(g).
\label{appeq:proof translation invariance}
\end{align}
We are going to derive the
important identity
\begin{subequations}\label{eq:nontrivial_symm_alg general}
\begin{align}
\widehat{U}(h)\,
\widehat{T}^{\,}_{\hat{1}}(g)\,
\widehat{U}^{-1}(h)=
e^{\mathrm{i}\chi(g,h)}\,
\widehat{T}^{\,}_{\hat{1}}(g),
\qquad
\forall g,h\in G^{\,}_{f},
\label{eq:nontrivial_symm_alg general a}
\end{align}
with the phase
\begin{align}
\chi(g,h)\df
\phi(h,g)
-
\phi(g,h)
+
\pi\,\rho(h)\,[\rho(g)+1]\,(N-1),
\qquad
\forall g,h\in G^{\,}_{f}.
\label{eq:nontrivial_symm_alg general b}
\end{align}
\end{subequations}
We shall then specify the conditions under which
the algebra defined by Eqs.\
(\ref{appeq:proof translation invariance})
and
(\ref{eq:nontrivial_symm_alg general})
guarantees that the spectrum of the twisted Hamiltonian
is degenerate.

\begin{proof}
We begin with the proof of
Eq.\ (\ref{eq:nontrivial_symm_alg general}).
We choose two elements 
$g,h\in G^{\,}_{f}$
with the local representations $\hat{v}^{\,}_{1}(g)$
and $\hat{v}^{\,}_{1}(h)$, respectively, both of 
which are unitary. 

\textbf{Step 1.}
We observe that
\begin{align}
\widehat{U}(h)\,
\hat{v}^{\,}_{1}(g)=&\,
\hat{v}^{\,}_{1}(h)\,
\hat{v}^{\,}_{2}(h)
\cdots
\hat{v}^{\,}_{N}(h)\
\hat{v}^{\,}_{1}(g).
\end{align}
We can then interchange the local operator
$\hat{v}^{\,}_{j}(h)$
and
$\hat{v}^{\,}_{j'}(g)$
pairwise
at the cost of the fermionic phase $(-1)^{\rho(h)\,\rho(g)}$
for any $j,j'=1,\cdots,N$.
This is done $(N-1)$ times
\begin{equation}
\widehat{U}(h)\,
\hat{v}^{\,}_{1}(g)=
(-1)^{\rho(g)\,\rho(h)(N-1)}
\hat{v}^{\,}_{1}(h)\,
\hat{v}^{\,}_{1}(g)\,
\hat{v}^{\,}_{2}(h)
\cdots
\hat{v}^{\,}_{N}(h).
\end{equation}
We conclude with
\begin{align}
\widehat{U}(h)\,
\hat{v}^{\,}_{1}(g)=
(-1)^{\rho(g)\,\rho(h)(N-1)}
\hat{v}^{\,}_{1}(h)\,
\hat{v}^{\,}_{1}(g)\,
\hat{v}^{\,}_{2}(h)
\cdots
\hat{v}^{\,}_{N}(h).
\label{eq:app D step 1}
\end{align}

\textbf{Step 2.}
We begin with
\begin{align}
\widehat{T}^{\,}_{\hat{1}}\,
\widehat{U}^{-1}(h)=&\,
\widehat{T}^{\,}_{\hat{1}}\,
\hat{v}^{-1}_{N}(h)\,
\hat{v}^{-1}_{N-1}(h)\,
\cdots\,
\hat{v}^{-1}_{1}(h)
\nonumber\\
=&\,
\left[
\widehat{T}^{\,}_{\hat{1}}\,
\hat{v}^{-1}_{N}(h)\,
\widehat{T}^{-1}_{\hat{1}}
\right]
\left[
\widehat{T}^{\,}_{\hat{1}}\,
\hat{v}^{-1}_{N-1}(h)
\widehat{T}^{-1}_{\hat{1}}
\right]
\,\cdots\,
\left[
\widehat{T}^{\,}_{\hat{1}}\,
\hat{v}^{-1}_{1}(h)
\widehat{T}^{-1}_{\hat{1}}
\right]
\widehat{T}^{\,}_{\hat{1}}\,
\nonumber\\
\hbox{\tiny Eq.\ (\ref{eq: def rep Gtrsl direct product a})\qquad}
=&\,
\hat{v}^{-1}_{1}(h)\,
\hat{v}^{-1}_{N}(h)\,
\cdots\,
\hat{v}^{-1}_{2}(h)\,
\widehat{T}^{\,}_{\hat{1}}.
\end{align}
Hence,
\begin{align}
\widehat{T}^{\,}_{\hat{1}}\,
\widehat{U}^{-1}(h)
=&\,
(-1)^{\rho(h)(N-1)}
\hat{v}^{-1}_{N}(h)\,
\hat{v}^{-1}_{N-1}(h)
\cdots
\hat{v}^{-1}_{1}(h)\,
\widehat{T}^{\,}_{\hat{1}},
\label{eq:app D step 2}
\end{align}
where we have reordered the factors $\hat{v}^{-1}_{j}(h)$ 
and, in doing so, obtained the coefficient $(-1)^{\rho(h)(N-1)}$
that encodes the fermionic algebra.

\textbf{Step 3.}
We combine Eqs. \eqref{eq:app D step 1} and \eqref{eq:app D step 2} into
\begin{align}
\widehat{U}(h)\,
\widehat{T}^{\,}_{\hat{1}}(g)\,
\widehat{U}^{-1}(h)
&=
(-1)^{\rho(h)\,[\rho(g)+1](N-1)}
\hat{v}^{\,}_{1}(h)\,
\hat{v}^{\,}_{1}(g)\,
\hat{v}^{\,}_{2}(h)
\cdots
\hat{v}^{\,}_{N}(h)\,
\hat{v}^{-1}_{N}(h)\,
\cdots
\hat{v}^{-1}_{1}(h)\,
\widehat{T}^{\,}_{\hat{1}}
\nonumber\\
&=
(-1)^{\rho(h)\,[\rho(g)+1](N-1)}
\hat{v}^{\,}_{1}(h)\,
\hat{v}^{\,}_{1}(g)\,
\hat{v}^{-1}_{1}(h)\,
\widehat{T}^{\,}_{\hat{1}}.
\label{eq:app D step 3}
\end{align}

\textbf{Step 4.}
We need to massage
$\hat{v}^{\,}_{1}(h)\,\hat{v}^{\,}_{1}(g)\,\hat{v}^{-1}_{1}(h)$.
To this end, we use the fact that 
the group $G^{\,}_{f}$ is Abelian to obtain
\begin{align}
\hat{v}^{\,}_{1}(h)\,\hat{v}^{\,}_{1}(g)\,
\hat{v}^{-1}_{1}(h)
=&\,
e^{\mathrm{i}\phi(h,g)}\,
\hat{v}^{\,}_{1}(h\,g)\,
\hat{v}^{-1}_{1}(h)
\nonumber\\
=&\,
e^{\mathrm{i}\phi(h,g)}\,
\hat{v}^{\,}_{1}(g\,h)\,
\hat{v}^{-1}_{1}(h)
\nonumber\\
=&\,
e^{\mathrm{i}\phi(h,g)}\,
\left[
e^{-\mathrm{i}\phi(g,h)}\,
\hat{v}^{\,}_{1}(g)\,
\hat{v}^{\,}_{1}(h)
\right]
\hat{v}^{-1}_{1}(h)
\nonumber\\
=&\,
e^{\mathrm{i}\phi(h,g)
-
\mathrm{i}\phi(g,h)}\,
\hat{v}^{\,}_{1}(g).
\end{align}
Insertion into the right-hand side of 
Eq.\ \eqref{eq:app D step 3}
delivers the result
\begin{subequations}\label{eq:nontrivial_symm_alg general proof}
\begin{align}
\widehat{U}(h)\,
\widehat{T}^{\,}_{\hat{1}}(g)\,
\widehat{U}^{-1}(h)
=&\,
(-1)^{\rho(h)\,[\rho(g)+1]\,(N-1)}
e^{
\mathrm{i}\phi(h,g)
-\mathrm{i}\phi(g,h)
}\,
\hat{v}^{\,}_{1}(g)\,
\widehat{T}^{\,}_{\hat{1}}
\equiv
e^{\mathrm{i}\chi(g,h)}\,
\widehat{T}^{\,}_{\hat{1}}(g),
\label{eq:nontrivial_symm_alg general proof a}
\end{align}
with the definition
\begin{align}
\chi(g,h)=&\,
\phi(h,g)
-
\phi(g,h)
+
\pi\,\rho(h)\,[\rho(g)+1]\,(N-1).
\label{eq:nontrivial_symm_alg general proof b}
\end{align}
\end{subequations}
\end{proof}

\textbf{Step 5.}
It is instructive to derive the transformation law of the phase
\eqref{eq:nontrivial_symm_alg general proof b}
under the global $U(1)$ gauge transformation generated by
\begin{equation}
\hat{v}^{\,}_{j}(g)\dfr
e^{\mathrm{i}\xi(g)}\,
\hat{v}^{\prime}_{j}(g),
\qquad
j=1,\cdots,N,
\qquad
\forall g\in G^{\,}_{f}.
\label{appeq:gauge transformation a}
\end{equation}
Under this transformation,
\begin{equation}
\phi'(g,h)= 
\phi(g,h)
-
\xi(g)
-
\xi(h)
+
\xi(g\,h),
\qquad
\forall g,h\in G^{\,}_{f},
\label{appeq:gauge transformation 2-cocycles}
\end{equation}
is the phase entering the projective algebra
obeyed by the operators $\{\hat{v}^{\prime}_{j}(g)\ |\ g\in G^{\,}_{f}\}$
according to Eq.\ (\ref{eq:U(1) gauge equivalence b}).
Hence, if we define
\begin{align}
\chi^{\prime}(g,h)\df
\phi^{\prime}(h,g)
-
\phi^{\prime}(g,h)
+
\pi\,\rho'(h)\,[\rho'(g)+1]\,(N-1),
\qquad
\forall g,\,h\in G^{\,}_{f},
\end{align}
we then have the relation
\begin{align}
\chi(g,h)=&\,
\phi(h,g)
-
\phi(g,h)
+
\pi\,\rho(h)\,[\rho(g)+1]\,(N-1)
\nonumber\\
=&\,
\chi'(g,h)
+
\underline{
\xi(h)
}
+
\underline{
\underline{ 
\xi(g)
}}
-
\underline{
\underline{
\underline{
\xi(h\,g)
}}}
-
\underline{
\underline{
\xi(g) 
}
}
- 
\underline{
\xi(h)
  }
+
\underline{
\underline{
\underline{
\xi(g\,h)
}}}
\nonumber\\
=&\,
\chi'(g,h),
\qquad
\forall g,\,h\in G^{\,}_{f}.
\end{align}
Hence,
$\chi(g,h)$ is gauge invariant
under the $\mathrm{U(1)}$ gauge transformation
(\ref{appeq:gauge transformation a}). The pair of
cocycles $\phi'$
and $\phi$ are equivalent
if and only if they have the same second cohomology class
$[\phi]=[\phi']\in H^{2}\big(G^{\,}_{f},\mathrm{U(1)}^{\,}_{\mathfrak{c}}\big)$,
i.e., if and only if they are related by the $\mathrm{U(1)}$ gauge transformation
(\ref{appeq:gauge transformation 2-cocycles}). The gauge invariance of
$\chi$ implies that it is independent of the choice made of $\phi$
within the equivalence class
$[\phi]\in H^{2}\big(G^{\,}_{f},\mathrm{U(1)}^{\,}_{\mathfrak{c}}\big)$.
For example, $\chi(g,h)=0$ holds for all $g,h\in G^{\,}_{f}$
for any $\phi$ belonging to 
the trivial second cohomology class $[\phi]=0$ since the function
$\phi=0$ belongs to $[\phi]=0$. 
Hence, the trivial second cohomology class $[\phi]=0$ implies
that $\chi(g,h)=0$.
Conversely, if we assume that $\chi(g,h)=0$, we find that
\begin{subequations}
\begin{align}
\phi(g,h)
=
\phi(h,g)
+
\pi\,\rho(h)\,[\rho(g)+1]\,(N-1)
\label{eq:implication of chi 0}
\end{align}
for any $g,h\in G^{\,}_{f}$. We distinguish two cases. 
First, if the number of sites $N$ is odd,
then the last term is an integer multiple of $2\pi$.
As such it can be dropped and we find that
\begin{align}
\phi(g,h)=\phi(h,g)
\label{eq:trivial second cohomology cond for Abel group}
\end{align}
for any $g,h\in G^{\,}_{f}$. 
For an Abelian group $G^{\,}_{f}$ (that is unitarily represented), Eq.\
\eqref{eq:trivial second cohomology cond for Abel group}
implies that the representations
of any two elements $g,h\in G^{\,}_{f}$
commute pairwise and therefore $[\phi]=0$. In other words, 
both $\phi(g,h)$ and $\phi(h,g)$ can be made to vanish
by an appropriate gauge transformation.
Second, if the number of sites $N$ is even, choosing $h$ to be the 
fermion parity $h=p$ in Eq.\ \eqref{eq:implication of chi 0}
implies
\begin{align}
\pi \rho(g)
=
\phi(g,p)
-
\phi(p,g)
=
\pi\,\rho(p)\,[\rho(g)+1]\,(N-1)
=
0,
\end{align}
\end{subequations}
where we used the definition \eqref{eq: def rho(g)} and 
the fact that $\rho(p)=0$ on the right-hand side of the second equality. 
Since $\rho(g)=0$ for any $g\in G^{\,}_{f}$, we can again drop the last term
on the right-hand side of
Eq.\ \eqref{eq:implication of chi 0}.
We arrive at Eq.\ 
\eqref{eq:trivial second cohomology cond for Abel group}.
Thus we have proven that $[\phi]=0$ if and only if $\chi(g,h)=0$ for any $g,h\in G^{\,}_{f}$.
As a corollary, there exists a
pair $g,h\in G^{\,}_{f}$ for which
$\chi(g,h)$ is nonvanishing
if and only if $[\phi]\neq0$.

\textbf{Step 6.}
The twisted Hamiltonian
$\widehat{H}^{\mathrm{tilt}}_{\mathrm{twis}}(g)$
is constructed so as to
commute with
the generator $\widehat{T}^{\,}_{\hat{1}}(g)$
of twisted translations and with the
representation $\widehat{U}(h)$
of any group element $h\in G^{\,}_{f}$,
whereby passing $\widehat{U}(h)$ from the left through
$\widehat{T}^{\,}_{\hat{1}}(g)$
produces the phase $\exp\big(\mathrm{i}\chi(g,h)\big)$.
If it is possible to find a pair $(g,h)$ such that
$\chi(g,h)$ is not $0$ modulo $2\pi$, then the spectrum of
$\widehat{H}^{\mathrm{tilt}}_{\mathrm{twis}}(g)$
must be degenerate. Indeed, any simultaneous eigenstate
$\ket{E(g),\exp\big(\mathrm{i}K(g)\big)}$ of
$\widehat{H}^{\mathrm{tilt}}_{\mathrm{twis}}(g)$ and
$\widehat{T}^{\,}_{\hat{1}}(g)$
must be orthogonal to the state
$\widehat{U}(h)\,\ket{E(g),\exp\big(\mathrm{i}K(g)\big)}$,
which is also an eigenstate of
$\widehat{H}^{\mathrm{tilt}}_{\mathrm{twis}}(g)$
with the energy $E(g)$ but has the eigenvalue
$
\exp\big(\mathrm{i}[K(g)+\chi(g,h)]\big)
\neq
\exp\big(\mathrm{i}K(g)\big)
$
with respect to $\widehat{T}^{\,}_{\hat{1}}(g)$.

\twocolumngrid
\bibliography{References}

\begin{thebibliography}{83}%
\makeatletter
\providecommand \@ifxundefined [1]{%
 \@ifx{#1\undefined}
}%
\providecommand \@ifnum [1]{%
 \ifnum #1\expandafter \@firstoftwo
 \else \expandafter \@secondoftwo
 \fi
}%
\providecommand \@ifx [1]{%
 \ifx #1\expandafter \@firstoftwo
 \else \expandafter \@secondoftwo
 \fi
}%
\providecommand \natexlab [1]{#1}%
\providecommand \enquote  [1]{``#1''}%
\providecommand \bibnamefont  [1]{#1}%
\providecommand \bibfnamefont [1]{#1}%
\providecommand \citenamefont [1]{#1}%
\providecommand \href@noop [0]{\@secondoftwo}%
\providecommand \href [0]{\begingroup \@sanitize@url \@href}%
\providecommand \@href[1]{\@@startlink{#1}\@@href}%
\providecommand \@@href[1]{\endgroup#1\@@endlink}%
\providecommand \@sanitize@url [0]{\catcode `\\12\catcode `\$12\catcode
  `\&12\catcode `\#12\catcode `\^12\catcode `\_12\catcode `\%12\relax}%
\providecommand \@@startlink[1]{}%
\providecommand \@@endlink[0]{}%
\providecommand \url  [0]{\begingroup\@sanitize@url \@url }%
\providecommand \@url [1]{\endgroup\@href {#1}{\urlprefix }}%
\providecommand \urlprefix  [0]{URL }%
\providecommand \Eprint [0]{\href }%
\providecommand \doibase [0]{http://dx.doi.org/}%
\providecommand \selectlanguage [0]{\@gobble}%
\providecommand \bibinfo  [0]{\@secondoftwo}%
\providecommand \bibfield  [0]{\@secondoftwo}%
\providecommand \translation [1]{[#1]}%
\providecommand \BibitemOpen [0]{}%
\providecommand \bibitemStop [0]{}%
\providecommand \bibitemNoStop [0]{.\EOS\space}%
\providecommand \EOS [0]{\spacefactor3000\relax}%
\providecommand \BibitemShut  [1]{\csname bibitem#1\endcsname}%
\let\auto@bib@innerbib\@empty
%</preamble>
\bibitem [{\citenamefont {Lieb}\ \emph {et~al.}(1961)\citenamefont {Lieb},
  \citenamefont {Schultz},\ and\ \citenamefont {Mattis}}]{Lieb1961}%
  \BibitemOpen
  \bibfield  {author} {\bibinfo {author} {\bibfnamefont {E.}~\bibnamefont
  {Lieb}}, \bibinfo {author} {\bibfnamefont {T.}~\bibnamefont {Schultz}}, \
  and\ \bibinfo {author} {\bibfnamefont {D.}~\bibnamefont {Mattis}},\
  }\href@noop {} {\bibfield  {journal} {\bibinfo  {journal} {Annals of
  Physics}\ }\textbf {\bibinfo {volume} {16}},\ \bibinfo {pages} {407}
  (\bibinfo {year} {1961})}\BibitemShut {NoStop}%
\bibitem [{\citenamefont {Affleck}\ and\ \citenamefont
  {Lieb}(1986)}]{Affleck1986}%
  \BibitemOpen
  \bibfield  {author} {\bibinfo {author} {\bibfnamefont {I.}~\bibnamefont
  {Affleck}}\ and\ \bibinfo {author} {\bibfnamefont {E.~H.}\ \bibnamefont
  {Lieb}},\ }\href {\doibase 10.1007/BF00400304} {\bibfield  {journal}
  {\bibinfo  {journal} {Letters in Mathematical Physics}\ }\textbf {\bibinfo
  {volume} {12}},\ \bibinfo {pages} {57} (\bibinfo {year} {1986})}\BibitemShut
  {NoStop}%
\bibitem [{\citenamefont {Aizenman}\ and\ \citenamefont
  {Nachtergaele}(1994)}]{Aizenman1994}%
  \BibitemOpen
  \bibfield  {author} {\bibinfo {author} {\bibfnamefont {M.}~\bibnamefont
  {Aizenman}}\ and\ \bibinfo {author} {\bibfnamefont {B.}~\bibnamefont
  {Nachtergaele}},\ }\href {\doibase 10.1007/BF02108805} {\bibfield  {journal}
  {\bibinfo  {journal} {Communications in Mathematical Physics}\ }\textbf
  {\bibinfo {volume} {164}},\ \bibinfo {pages} {17} (\bibinfo {year}
  {1994})}\BibitemShut {NoStop}%
\bibitem [{\citenamefont {Oshikawa}\ \emph {et~al.}(1997)\citenamefont
  {Oshikawa}, \citenamefont {Yamanaka},\ and\ \citenamefont
  {Affleck}}]{Oshikawa1997}%
  \BibitemOpen
  \bibfield  {author} {\bibinfo {author} {\bibfnamefont {M.}~\bibnamefont
  {Oshikawa}}, \bibinfo {author} {\bibfnamefont {M.}~\bibnamefont {Yamanaka}},
  \ and\ \bibinfo {author} {\bibfnamefont {I.}~\bibnamefont {Affleck}},\
  }\href@noop {} {\bibfield  {journal} {\bibinfo  {journal} {Physical review
  letters}\ }\textbf {\bibinfo {volume} {78}},\ \bibinfo {pages} {1984}
  (\bibinfo {year} {1997})}\BibitemShut {NoStop}%
\bibitem [{\citenamefont {Yamanaka}\ \emph {et~al.}(1997)\citenamefont
  {Yamanaka}, \citenamefont {Oshikawa},\ and\ \citenamefont
  {Affleck}}]{Yamanaka1997}%
  \BibitemOpen
  \bibfield  {author} {\bibinfo {author} {\bibfnamefont {M.}~\bibnamefont
  {Yamanaka}}, \bibinfo {author} {\bibfnamefont {M.}~\bibnamefont {Oshikawa}},
  \ and\ \bibinfo {author} {\bibfnamefont {I.}~\bibnamefont {Affleck}},\ }\href
  {\doibase 10.1103/PhysRevLett.79.1110} {\bibfield  {journal} {\bibinfo
  {journal} {Phys. Rev. Lett.}\ }\textbf {\bibinfo {volume} {79}},\ \bibinfo
  {pages} {1110} (\bibinfo {year} {1997})}\BibitemShut {NoStop}%
\bibitem [{\citenamefont {Koma}(2000)}]{Koma2000}%
  \BibitemOpen
  \bibfield  {author} {\bibinfo {author} {\bibfnamefont {T.}~\bibnamefont
  {Koma}},\ }\href {\doibase 10.1023/A:1018604925491} {\bibfield  {journal}
  {\bibinfo  {journal} {Journal of Statistical Physics}\ }\textbf {\bibinfo
  {volume} {99}},\ \bibinfo {pages} {313} (\bibinfo {year} {2000})}\BibitemShut
  {NoStop}%
\bibitem [{\citenamefont {Oshikawa}(2000)}]{Oshikawa2000}%
  \BibitemOpen
  \bibfield  {author} {\bibinfo {author} {\bibfnamefont {M.}~\bibnamefont
  {Oshikawa}},\ }\href {\doibase 10.1103/PhysRevLett.84.1535} {\bibfield
  {journal} {\bibinfo  {journal} {Phys. Rev. Lett.}\ }\textbf {\bibinfo
  {volume} {84}},\ \bibinfo {pages} {1535} (\bibinfo {year}
  {2000})}\BibitemShut {NoStop}%
\bibitem [{\citenamefont {Hastings}(2004)}]{Hastings2004}%
  \BibitemOpen
  \bibfield  {author} {\bibinfo {author} {\bibfnamefont {M.~B.}\ \bibnamefont
  {Hastings}},\ }\href {\doibase 10.1103/PhysRevB.69.104431} {\bibfield
  {journal} {\bibinfo  {journal} {Phys. Rev. B}\ }\textbf {\bibinfo {volume}
  {69}},\ \bibinfo {pages} {104431} (\bibinfo {year} {2004})}\BibitemShut
  {NoStop}%
\bibitem [{\citenamefont {Hastings}(2005)}]{Hastings2005}%
  \BibitemOpen
  \bibfield  {author} {\bibinfo {author} {\bibfnamefont {M.~B.}\ \bibnamefont
  {Hastings}},\ }\href {\doibase 10.1209/epl/i2005-10046-x} {\bibfield
  {journal} {\bibinfo  {journal} {Europhysics Letters ({EPL})}\ }\textbf
  {\bibinfo {volume} {70}},\ \bibinfo {pages} {824} (\bibinfo {year}
  {2005})}\BibitemShut {NoStop}%
\bibitem [{\citenamefont {Roy}(2012)}]{Roy2012}%
  \BibitemOpen
  \bibfield  {author} {\bibinfo {author} {\bibfnamefont {R.}~\bibnamefont
  {Roy}},\ }\href {https://arxiv.org/abs/1212.2944} {\enquote {\bibinfo {title}
  {Space group symmetries and low lying excitations of many-body systems at
  integer fillings},}\ } (\bibinfo {year} {2012}),\ \Eprint
  {http://arxiv.org/abs/1212.2944} {arXiv:1212.2944 [cond-mat.str-el]}
  \BibitemShut {NoStop}%
\bibitem [{\citenamefont {Parameswaran}\ \emph {et~al.}(2013)\citenamefont
  {Parameswaran}, \citenamefont {Turner}, \citenamefont {Arovas},\ and\
  \citenamefont {Vishwanath}}]{Parameswaran2013}%
  \BibitemOpen
  \bibfield  {author} {\bibinfo {author} {\bibfnamefont {S.~A.}\ \bibnamefont
  {Parameswaran}}, \bibinfo {author} {\bibfnamefont {A.~M.}\ \bibnamefont
  {Turner}}, \bibinfo {author} {\bibfnamefont {D.~P.}\ \bibnamefont {Arovas}},
  \ and\ \bibinfo {author} {\bibfnamefont {A.}~\bibnamefont {Vishwanath}},\
  }\href {\doibase 10.1038/nphys2600} {\bibfield  {journal} {\bibinfo
  {journal} {Nature Physics}\ }\textbf {\bibinfo {volume} {9}},\ \bibinfo
  {pages} {299–303} (\bibinfo {year} {2013})}\BibitemShut {NoStop}%
\bibitem [{\citenamefont {Watanabe}\ \emph {et~al.}(2015)\citenamefont
  {Watanabe}, \citenamefont {Po}, \citenamefont {Vishwanath},\ and\
  \citenamefont {Zaletel}}]{Watanabe2015}%
  \BibitemOpen
  \bibfield  {author} {\bibinfo {author} {\bibfnamefont {H.}~\bibnamefont
  {Watanabe}}, \bibinfo {author} {\bibfnamefont {H.~C.}\ \bibnamefont {Po}},
  \bibinfo {author} {\bibfnamefont {A.}~\bibnamefont {Vishwanath}}, \ and\
  \bibinfo {author} {\bibfnamefont {M.}~\bibnamefont {Zaletel}},\ }\href
  {\doibase 10.1073/pnas.1514665112} {\bibfield  {journal} {\bibinfo  {journal}
  {Proceedings of the National Academy of Sciences}\ }\textbf {\bibinfo
  {volume} {112}},\ \bibinfo {pages} {14551} (\bibinfo {year}
  {2015})}\BibitemShut {NoStop}%
\bibitem [{\citenamefont {Watanabe}\ \emph {et~al.}(2016)\citenamefont
  {Watanabe}, \citenamefont {Po}, \citenamefont {Zaletel},\ and\ \citenamefont
  {Vishwanath}}]{Watanabe2016}%
  \BibitemOpen
  \bibfield  {author} {\bibinfo {author} {\bibfnamefont {H.}~\bibnamefont
  {Watanabe}}, \bibinfo {author} {\bibfnamefont {H.~C.}\ \bibnamefont {Po}},
  \bibinfo {author} {\bibfnamefont {M.~P.}\ \bibnamefont {Zaletel}}, \ and\
  \bibinfo {author} {\bibfnamefont {A.}~\bibnamefont {Vishwanath}},\ }\href
  {\doibase 10.1103/PhysRevLett.117.096404} {\bibfield  {journal} {\bibinfo
  {journal} {Phys. Rev. Lett.}\ }\textbf {\bibinfo {volume} {117}},\ \bibinfo
  {pages} {096404} (\bibinfo {year} {2016})}\BibitemShut {NoStop}%
\bibitem [{\citenamefont {Qi}\ \emph {et~al.}(2017)\citenamefont {Qi},
  \citenamefont {Fang},\ and\ \citenamefont {Fu}}]{Qi2017}%
  \BibitemOpen
  \bibfield  {author} {\bibinfo {author} {\bibfnamefont {Y.}~\bibnamefont
  {Qi}}, \bibinfo {author} {\bibfnamefont {C.}~\bibnamefont {Fang}}, \ and\
  \bibinfo {author} {\bibfnamefont {L.}~\bibnamefont {Fu}},\ }\href@noop {}
  {\enquote {\bibinfo {title} {Ground state degeneracy in quantum spin systems
  protected by crystal symmetries},}\ } (\bibinfo {year} {2017}),\ \Eprint
  {http://arxiv.org/abs/1705.09190} {arXiv:1705.09190 [cond-mat.str-el]}
  \BibitemShut {NoStop}%
\bibitem [{\citenamefont {Watanabe}(2018)}]{Watanabe2018}%
  \BibitemOpen
  \bibfield  {author} {\bibinfo {author} {\bibfnamefont {H.}~\bibnamefont
  {Watanabe}},\ }\href {\doibase 10.1103/PhysRevB.97.165117} {\bibfield
  {journal} {\bibinfo  {journal} {Phys. Rev. B}\ }\textbf {\bibinfo {volume}
  {97}},\ \bibinfo {pages} {165117} (\bibinfo {year} {2018})}\BibitemShut
  {NoStop}%
\bibitem [{\citenamefont {Tasaki}(2018)}]{Tasaki2018}%
  \BibitemOpen
  \bibfield  {author} {\bibinfo {author} {\bibfnamefont {H.}~\bibnamefont
  {Tasaki}},\ }\href@noop {} {\bibfield  {journal} {\bibinfo  {journal}
  {Journal of Statistical Physics}\ }\textbf {\bibinfo {volume} {170}},\
  \bibinfo {pages} {653} (\bibinfo {year} {2018})}\BibitemShut {NoStop}%
\bibitem [{\citenamefont {Metlitski}\ and\ \citenamefont
  {Thorngren}(2018)}]{Metlitski2018}%
  \BibitemOpen
  \bibfield  {author} {\bibinfo {author} {\bibfnamefont {M.~A.}\ \bibnamefont
  {Metlitski}}\ and\ \bibinfo {author} {\bibfnamefont {R.}~\bibnamefont
  {Thorngren}},\ }\href {\doibase 10.1103/PhysRevB.98.085140} {\bibfield
  {journal} {\bibinfo  {journal} {Phys. Rev. B}\ }\textbf {\bibinfo {volume}
  {98}},\ \bibinfo {pages} {085140} (\bibinfo {year} {2018})}\BibitemShut
  {NoStop}%
\bibitem [{\citenamefont {Kobayashi}\ \emph {et~al.}(2019)\citenamefont
  {Kobayashi}, \citenamefont {Shiozaki}, \citenamefont {Kikuchi},\ and\
  \citenamefont {Ryu}}]{Kobayashi2019}%
  \BibitemOpen
  \bibfield  {author} {\bibinfo {author} {\bibfnamefont {R.}~\bibnamefont
  {Kobayashi}}, \bibinfo {author} {\bibfnamefont {K.}~\bibnamefont {Shiozaki}},
  \bibinfo {author} {\bibfnamefont {Y.}~\bibnamefont {Kikuchi}}, \ and\
  \bibinfo {author} {\bibfnamefont {S.}~\bibnamefont {Ryu}},\ }\href {\doibase
  10.1103/PhysRevB.99.014402} {\bibfield  {journal} {\bibinfo  {journal} {Phys.
  Rev. B}\ }\textbf {\bibinfo {volume} {99}},\ \bibinfo {pages} {014402}
  (\bibinfo {year} {2019})}\BibitemShut {NoStop}%
\bibitem [{\citenamefont {He}\ \emph {et~al.}(2020)\citenamefont {He},
  \citenamefont {You},\ and\ \citenamefont {Prem}}]{He2020}%
  \BibitemOpen
  \bibfield  {author} {\bibinfo {author} {\bibfnamefont {H.}~\bibnamefont
  {He}}, \bibinfo {author} {\bibfnamefont {Y.}~\bibnamefont {You}}, \ and\
  \bibinfo {author} {\bibfnamefont {A.}~\bibnamefont {Prem}},\ }\href {\doibase
  10.1103/PhysRevB.101.165145} {\bibfield  {journal} {\bibinfo  {journal}
  {Phys. Rev. B}\ }\textbf {\bibinfo {volume} {101}},\ \bibinfo {pages}
  {165145} (\bibinfo {year} {2020})}\BibitemShut {NoStop}%
\bibitem [{\citenamefont {Bachmann}\ \emph {et~al.}(2020)\citenamefont
  {Bachmann}, \citenamefont {Bols}, \citenamefont {De~Roeck},\ and\
  \citenamefont {Fraas}}]{Bachmann2020}%
  \BibitemOpen
  \bibfield  {author} {\bibinfo {author} {\bibfnamefont {S.}~\bibnamefont
  {Bachmann}}, \bibinfo {author} {\bibfnamefont {A.}~\bibnamefont {Bols}},
  \bibinfo {author} {\bibfnamefont {W.}~\bibnamefont {De~Roeck}}, \ and\
  \bibinfo {author} {\bibfnamefont {M.}~\bibnamefont {Fraas}},\ }\href
  {\doibase 10.1007/s00220-019-03537-x} {\bibfield  {journal} {\bibinfo
  {journal} {Communications in Mathematical Physics}\ }\textbf {\bibinfo
  {volume} {375}},\ \bibinfo {pages} {1249} (\bibinfo {year}
  {2020})}\BibitemShut {NoStop}%
\bibitem [{\citenamefont {Dubinkin}\ \emph {et~al.}(2021)\citenamefont
  {Dubinkin}, \citenamefont {May-Mann},\ and\ \citenamefont
  {Hughes}}]{Dubinkin2021}%
  \BibitemOpen
  \bibfield  {author} {\bibinfo {author} {\bibfnamefont {O.}~\bibnamefont
  {Dubinkin}}, \bibinfo {author} {\bibfnamefont {J.}~\bibnamefont {May-Mann}},
  \ and\ \bibinfo {author} {\bibfnamefont {T.~L.}\ \bibnamefont {Hughes}},\
  }\href {\doibase 10.1103/PhysRevB.103.125133} {\bibfield  {journal} {\bibinfo
   {journal} {Phys. Rev. B}\ }\textbf {\bibinfo {volume} {103}},\ \bibinfo
  {pages} {125133} (\bibinfo {year} {2021})}\BibitemShut {NoStop}%
\bibitem [{\citenamefont {Chen}\ \emph
  {et~al.}(2011{\natexlab{a}})\citenamefont {Chen}, \citenamefont {Gu},\ and\
  \citenamefont {Wen}}]{Chen2011b}%
  \BibitemOpen
  \bibfield  {author} {\bibinfo {author} {\bibfnamefont {X.}~\bibnamefont
  {Chen}}, \bibinfo {author} {\bibfnamefont {Z.-C.}\ \bibnamefont {Gu}}, \ and\
  \bibinfo {author} {\bibfnamefont {X.-G.}\ \bibnamefont {Wen}},\ }\href
  {\doibase 10.1103/PhysRevB.84.235128} {\bibfield  {journal} {\bibinfo
  {journal} {Phys. Rev. B}\ }\textbf {\bibinfo {volume} {84}},\ \bibinfo
  {pages} {235128} (\bibinfo {year} {2011}{\natexlab{a}})}\BibitemShut
  {NoStop}%
\bibitem [{\citenamefont {Cheng}\ \emph {et~al.}(2016)\citenamefont {Cheng},
  \citenamefont {Zaletel}, \citenamefont {Barkeshli}, \citenamefont
  {Vishwanath},\ and\ \citenamefont {Bonderson}}]{Cheng2016}%
  \BibitemOpen
  \bibfield  {author} {\bibinfo {author} {\bibfnamefont {M.}~\bibnamefont
  {Cheng}}, \bibinfo {author} {\bibfnamefont {M.}~\bibnamefont {Zaletel}},
  \bibinfo {author} {\bibfnamefont {M.}~\bibnamefont {Barkeshli}}, \bibinfo
  {author} {\bibfnamefont {A.}~\bibnamefont {Vishwanath}}, \ and\ \bibinfo
  {author} {\bibfnamefont {P.}~\bibnamefont {Bonderson}},\ }\href {\doibase
  10.1103/PhysRevX.6.041068} {\bibfield  {journal} {\bibinfo  {journal} {Phys.
  Rev. X}\ }\textbf {\bibinfo {volume} {6}},\ \bibinfo {pages} {041068}
  (\bibinfo {year} {2016})}\BibitemShut {NoStop}%
\bibitem [{\citenamefont {Po}\ \emph {et~al.}(2017)\citenamefont {Po},
  \citenamefont {Watanabe}, \citenamefont {Jian},\ and\ \citenamefont
  {Zaletel}}]{Po2017}%
  \BibitemOpen
  \bibfield  {author} {\bibinfo {author} {\bibfnamefont {H.~C.}\ \bibnamefont
  {Po}}, \bibinfo {author} {\bibfnamefont {H.}~\bibnamefont {Watanabe}},
  \bibinfo {author} {\bibfnamefont {C.-M.}\ \bibnamefont {Jian}}, \ and\
  \bibinfo {author} {\bibfnamefont {M.~P.}\ \bibnamefont {Zaletel}},\ }\href
  {\doibase 10.1103/PhysRevLett.119.127202} {\bibfield  {journal} {\bibinfo
  {journal} {Phys. Rev. Lett.}\ }\textbf {\bibinfo {volume} {119}},\ \bibinfo
  {pages} {127202} (\bibinfo {year} {2017})}\BibitemShut {NoStop}%
\bibitem [{\citenamefont {Yang}\ \emph {et~al.}(2018)\citenamefont {Yang},
  \citenamefont {Jiang}, \citenamefont {Vishwanath},\ and\ \citenamefont
  {Ran}}]{Yang2018}%
  \BibitemOpen
  \bibfield  {author} {\bibinfo {author} {\bibfnamefont {X.}~\bibnamefont
  {Yang}}, \bibinfo {author} {\bibfnamefont {S.}~\bibnamefont {Jiang}},
  \bibinfo {author} {\bibfnamefont {A.}~\bibnamefont {Vishwanath}}, \ and\
  \bibinfo {author} {\bibfnamefont {Y.}~\bibnamefont {Ran}},\ }\href {\doibase
  10.1103/PhysRevB.98.125120} {\bibfield  {journal} {\bibinfo  {journal} {Phys.
  Rev. B}\ }\textbf {\bibinfo {volume} {98}},\ \bibinfo {pages} {125120}
  (\bibinfo {year} {2018})}\BibitemShut {NoStop}%
\bibitem [{\citenamefont {Ogata}\ and\ \citenamefont
  {Tasaki}(2019)}]{Ogata2019}%
  \BibitemOpen
  \bibfield  {author} {\bibinfo {author} {\bibfnamefont {Y.}~\bibnamefont
  {Ogata}}\ and\ \bibinfo {author} {\bibfnamefont {H.}~\bibnamefont {Tasaki}},\
  }\href {\doibase 10.1007/s00220-019-03343-5} {\bibfield  {journal} {\bibinfo
  {journal} {Communications in Mathematical Physics}\ }\textbf {\bibinfo
  {volume} {372}},\ \bibinfo {pages} {951} (\bibinfo {year}
  {2019})}\BibitemShut {NoStop}%
\bibitem [{\citenamefont {Ogata}\ \emph {et~al.}(2021)\citenamefont {Ogata},
  \citenamefont {Tachikawa},\ and\ \citenamefont {Tasaki}}]{Ogata2021}%
  \BibitemOpen
  \bibfield  {author} {\bibinfo {author} {\bibfnamefont {Y.}~\bibnamefont
  {Ogata}}, \bibinfo {author} {\bibfnamefont {Y.}~\bibnamefont {Tachikawa}}, \
  and\ \bibinfo {author} {\bibfnamefont {H.}~\bibnamefont {Tasaki}},\ }\href
  {\doibase 10.1007/s00220-021-04116-9} {\bibfield  {journal} {\bibinfo
  {journal} {Communications in Mathematical Physics}\ }\textbf {\bibinfo
  {volume} {385}},\ \bibinfo {pages} {79} (\bibinfo {year} {2021})}\BibitemShut
  {NoStop}%
\bibitem [{\citenamefont {Ogata}(2020)}]{Ogata2020a}%
  \BibitemOpen
  \bibfield  {author} {\bibinfo {author} {\bibfnamefont {Y.}~\bibnamefont
  {Ogata}},\ }\href {\doibase 10.1007/s00220-019-03521-5} {\bibfield  {journal}
  {\bibinfo  {journal} {Communications in Mathematical Physics}\ }\textbf
  {\bibinfo {volume} {374}},\ \bibinfo {pages} {705} (\bibinfo {year}
  {2020})}\BibitemShut {NoStop}%
\bibitem [{\citenamefont {Matsui}(2001)}]{Matsui2001}%
  \BibitemOpen
  \bibfield  {author} {\bibinfo {author} {\bibfnamefont {T.}~\bibnamefont
  {Matsui}},\ }\href {\doibase 10.1007/s002200100413} {\bibfield  {journal}
  {\bibinfo  {journal} {Communications in Mathematical Physics}\ }\textbf
  {\bibinfo {volume} {218}},\ \bibinfo {pages} {393} (\bibinfo {year}
  {2001})}\BibitemShut {NoStop}%
\bibitem [{\citenamefont {Matsui}(2013)}]{Matsui2013}%
  \BibitemOpen
  \bibfield  {author} {\bibinfo {author} {\bibfnamefont {T.}~\bibnamefont
  {Matsui}},\ }\href {\doibase 10.1142/S0129055X13500177} {\bibfield  {journal}
  {\bibinfo  {journal} {Reviews in Mathematical Physics}\ }\textbf {\bibinfo
  {volume} {25}},\ \bibinfo {pages} {1350017} (\bibinfo {year}
  {2013})}\BibitemShut {NoStop}%
\bibitem [{\citenamefont {Prakash}(2020)}]{Prakash2020}%
  \BibitemOpen
  \bibfield  {author} {\bibinfo {author} {\bibfnamefont {A.}~\bibnamefont
  {Prakash}},\ }\href {https://arxiv.org/abs/2002.11176} {\enquote {\bibinfo
  {title} {An elementary proof of 1d lsm theorems},}\ } (\bibinfo {year}
  {2020}),\ \Eprint {http://arxiv.org/abs/2002.11176} {arXiv:2002.11176}
  \BibitemShut {NoStop}%
\bibitem [{\citenamefont {Tasaki}(2020)}]{Tasaki2020}%
  \BibitemOpen
  \bibfield  {author} {\bibinfo {author} {\bibfnamefont {H.}~\bibnamefont
  {Tasaki}},\ }\href@noop {} {\emph {\bibinfo {title} {Physics and mathematics
  of quantum many-body systems}}}\ (\bibinfo  {publisher} {Springer},\ \bibinfo
  {year} {2020})\BibitemShut {NoStop}%
\bibitem [{\citenamefont {Fannes}\ \emph {et~al.}(1992)\citenamefont {Fannes},
  \citenamefont {Nachtergaele},\ and\ \citenamefont {Werner}}]{Fannes1992}%
  \BibitemOpen
  \bibfield  {author} {\bibinfo {author} {\bibfnamefont {M.}~\bibnamefont
  {Fannes}}, \bibinfo {author} {\bibfnamefont {B.}~\bibnamefont
  {Nachtergaele}}, \ and\ \bibinfo {author} {\bibfnamefont {R.~F.}\
  \bibnamefont {Werner}},\ }\href
  {https://projecteuclid.org:443/euclid.cmp/1104249404} {\bibfield  {journal}
  {\bibinfo  {journal} {Comm. Math. Phys.}\ }\textbf {\bibinfo {volume}
  {144}},\ \bibinfo {pages} {443} (\bibinfo {year} {1992})}\BibitemShut
  {NoStop}%
\bibitem [{\citenamefont {{Kl{\"u}mper}}\ \emph {et~al.}(1992)\citenamefont
  {{Kl{\"u}mper}}, \citenamefont {{Schadschneider}},\ and\ \citenamefont
  {{Zittartz}}}]{Klumper1992}%
  \BibitemOpen
  \bibfield  {author} {\bibinfo {author} {\bibfnamefont {A.}~\bibnamefont
  {{Kl{\"u}mper}}}, \bibinfo {author} {\bibfnamefont {A.}~\bibnamefont
  {{Schadschneider}}}, \ and\ \bibinfo {author} {\bibfnamefont
  {J.}~\bibnamefont {{Zittartz}}},\ }\href {\doibase 10.1007/BF01309281}
  {\bibfield  {journal} {\bibinfo  {journal} {Zeitschrift fur Physik B
  Condensed Matter}\ }\textbf {\bibinfo {volume} {87}},\ \bibinfo {pages} {281}
  (\bibinfo {year} {1992})}\BibitemShut {NoStop}%
\bibitem [{\citenamefont {Perez-Garcia}\ \emph {et~al.}(2007)\citenamefont
  {Perez-Garcia}, \citenamefont {Verstraete}, \citenamefont {Wolf},\ and\
  \citenamefont {Cirac}}]{Garcia2007}%
  \BibitemOpen
  \bibfield  {author} {\bibinfo {author} {\bibfnamefont {D.}~\bibnamefont
  {Perez-Garcia}}, \bibinfo {author} {\bibfnamefont {F.}~\bibnamefont
  {Verstraete}}, \bibinfo {author} {\bibfnamefont {M.~M.}\ \bibnamefont
  {Wolf}}, \ and\ \bibinfo {author} {\bibfnamefont {J.~I.}\ \bibnamefont
  {Cirac}},\ }\href@noop {} {\bibfield  {journal} {\bibinfo  {journal} {Quantum
  Info. Comput.}\ }\textbf {\bibinfo {volume} {7}},\ \bibinfo {pages}
  {401–430} (\bibinfo {year} {2007})}\BibitemShut {NoStop}%
\bibitem [{\citenamefont {Schuch}\ \emph {et~al.}(2008)\citenamefont {Schuch},
  \citenamefont {Wolf}, \citenamefont {Verstraete},\ and\ \citenamefont
  {Cirac}}]{Schuch2008}%
  \BibitemOpen
  \bibfield  {author} {\bibinfo {author} {\bibfnamefont {N.}~\bibnamefont
  {Schuch}}, \bibinfo {author} {\bibfnamefont {M.~M.}\ \bibnamefont {Wolf}},
  \bibinfo {author} {\bibfnamefont {F.}~\bibnamefont {Verstraete}}, \ and\
  \bibinfo {author} {\bibfnamefont {J.~I.}\ \bibnamefont {Cirac}},\ }\href
  {\doibase 10.1103/PhysRevLett.100.030504} {\bibfield  {journal} {\bibinfo
  {journal} {Phys. Rev. Lett.}\ }\textbf {\bibinfo {volume} {100}},\ \bibinfo
  {pages} {030504} (\bibinfo {year} {2008})}\BibitemShut {NoStop}%
\bibitem [{\citenamefont {Verstraete}\ and\ \citenamefont
  {Cirac}(2006)}]{Verstraete2006}%
  \BibitemOpen
  \bibfield  {author} {\bibinfo {author} {\bibfnamefont {F.}~\bibnamefont
  {Verstraete}}\ and\ \bibinfo {author} {\bibfnamefont {J.~I.}\ \bibnamefont
  {Cirac}},\ }\href {\doibase 10.1103/PhysRevB.73.094423} {\bibfield  {journal}
  {\bibinfo  {journal} {Phys. Rev. B}\ }\textbf {\bibinfo {volume} {73}},\
  \bibinfo {pages} {094423} (\bibinfo {year} {2006})}\BibitemShut {NoStop}%
\bibitem [{\citenamefont {Hastings}(2007)}]{Hastings2007}%
  \BibitemOpen
  \bibfield  {author} {\bibinfo {author} {\bibfnamefont {M.~B.}\ \bibnamefont
  {Hastings}},\ }\href {\doibase 10.1088/1742-5468/2007/08/p08024} {\bibfield
  {journal} {\bibinfo  {journal} {Journal of Statistical Mechanics: Theory and
  Experiment}\ }\textbf {\bibinfo {volume} {2007}},\ \bibinfo {pages} {P08024}
  (\bibinfo {year} {2007})}\BibitemShut {NoStop}%
\bibitem [{\citenamefont {Chen}\ \emph
  {et~al.}(2011{\natexlab{b}})\citenamefont {Chen}, \citenamefont {Gu},\ and\
  \citenamefont {Wen}}]{Chen2011a}%
  \BibitemOpen
  \bibfield  {author} {\bibinfo {author} {\bibfnamefont {X.}~\bibnamefont
  {Chen}}, \bibinfo {author} {\bibfnamefont {Z.-C.}\ \bibnamefont {Gu}}, \ and\
  \bibinfo {author} {\bibfnamefont {X.-G.}\ \bibnamefont {Wen}},\ }\href
  {\doibase 10.1103/PhysRevB.83.035107} {\bibfield  {journal} {\bibinfo
  {journal} {Phys. Rev. B}\ }\textbf {\bibinfo {volume} {83}},\ \bibinfo
  {pages} {035107} (\bibinfo {year} {2011}{\natexlab{b}})}\BibitemShut
  {NoStop}%
\bibitem [{\citenamefont {Pollmann}\ \emph {et~al.}(2012)\citenamefont
  {Pollmann}, \citenamefont {Berg}, \citenamefont {Turner},\ and\ \citenamefont
  {Oshikawa}}]{pollmann2012}%
  \BibitemOpen
  \bibfield  {author} {\bibinfo {author} {\bibfnamefont {F.}~\bibnamefont
  {Pollmann}}, \bibinfo {author} {\bibfnamefont {E.}~\bibnamefont {Berg}},
  \bibinfo {author} {\bibfnamefont {A.~M.}\ \bibnamefont {Turner}}, \ and\
  \bibinfo {author} {\bibfnamefont {M.}~\bibnamefont {Oshikawa}},\ }\href
  {\doibase 10.1103/PhysRevB.85.075125} {\bibfield  {journal} {\bibinfo
  {journal} {Phys. Rev. B}\ }\textbf {\bibinfo {volume} {85}},\ \bibinfo
  {pages} {075125} (\bibinfo {year} {2012})}\BibitemShut {NoStop}%
\bibitem [{\citenamefont {Chen}\ \emph {et~al.}(2013)\citenamefont {Chen},
  \citenamefont {Gu}, \citenamefont {Liu},\ and\ \citenamefont
  {Wen}}]{Chen2013}%
  \BibitemOpen
  \bibfield  {author} {\bibinfo {author} {\bibfnamefont {X.}~\bibnamefont
  {Chen}}, \bibinfo {author} {\bibfnamefont {Z.-C.}\ \bibnamefont {Gu}},
  \bibinfo {author} {\bibfnamefont {Z.-X.}\ \bibnamefont {Liu}}, \ and\
  \bibinfo {author} {\bibfnamefont {X.-G.}\ \bibnamefont {Wen}},\ }\href
  {\doibase 10.1103/PhysRevB.87.155114} {\bibfield  {journal} {\bibinfo
  {journal} {Phys. Rev. B}\ }\textbf {\bibinfo {volume} {87}},\ \bibinfo
  {pages} {155114} (\bibinfo {year} {2013})}\BibitemShut {NoStop}%
\bibitem [{\citenamefont {Vishwanath}\ and\ \citenamefont
  {Senthil}(2013)}]{Vishwanath2013}%
  \BibitemOpen
  \bibfield  {author} {\bibinfo {author} {\bibfnamefont {A.}~\bibnamefont
  {Vishwanath}}\ and\ \bibinfo {author} {\bibfnamefont {T.}~\bibnamefont
  {Senthil}},\ }\href {\doibase 10.1103/PhysRevX.3.011016} {\bibfield
  {journal} {\bibinfo  {journal} {Phys. Rev. X}\ }\textbf {\bibinfo {volume}
  {3}},\ \bibinfo {pages} {011016} (\bibinfo {year} {2013})}\BibitemShut
  {NoStop}%
\bibitem [{\citenamefont {Gu}\ and\ \citenamefont {Wen}(2014)}]{Gu2014}%
  \BibitemOpen
  \bibfield  {author} {\bibinfo {author} {\bibfnamefont {Z.-C.}\ \bibnamefont
  {Gu}}\ and\ \bibinfo {author} {\bibfnamefont {X.-G.}\ \bibnamefont {Wen}},\
  }\href {\doibase 10.1103/PhysRevB.90.115141} {\bibfield  {journal} {\bibinfo
  {journal} {Phys. Rev. B}\ }\textbf {\bibinfo {volume} {90}},\ \bibinfo
  {pages} {115141} (\bibinfo {year} {2014})}\BibitemShut {NoStop}%
\bibitem [{\citenamefont {Cho}\ \emph {et~al.}(2017)\citenamefont {Cho},
  \citenamefont {Hsieh},\ and\ \citenamefont {Ryu}}]{Cho2017}%
  \BibitemOpen
  \bibfield  {author} {\bibinfo {author} {\bibfnamefont {G.~Y.}\ \bibnamefont
  {Cho}}, \bibinfo {author} {\bibfnamefont {C.-T.}\ \bibnamefont {Hsieh}}, \
  and\ \bibinfo {author} {\bibfnamefont {S.}~\bibnamefont {Ryu}},\ }\href
  {\doibase 10.1103/PhysRevB.96.195105} {\bibfield  {journal} {\bibinfo
  {journal} {Phys. Rev. B}\ }\textbf {\bibinfo {volume} {96}},\ \bibinfo
  {pages} {195105} (\bibinfo {year} {2017})}\BibitemShut {NoStop}%
\bibitem [{\citenamefont {Else}\ and\ \citenamefont
  {Thorngren}(2020)}]{Else2020}%
  \BibitemOpen
  \bibfield  {author} {\bibinfo {author} {\bibfnamefont {D.~V.}\ \bibnamefont
  {Else}}\ and\ \bibinfo {author} {\bibfnamefont {R.}~\bibnamefont
  {Thorngren}},\ }\href {\doibase 10.1103/PhysRevB.101.224437} {\bibfield
  {journal} {\bibinfo  {journal} {Phys. Rev. B}\ }\textbf {\bibinfo {volume}
  {101}},\ \bibinfo {pages} {224437} (\bibinfo {year} {2020})}\BibitemShut
  {NoStop}%
\bibitem [{\citenamefont {Ryu}\ \emph {et~al.}(2012)\citenamefont {Ryu},
  \citenamefont {Moore},\ and\ \citenamefont {Ludwig}}]{Ryu2012}%
  \BibitemOpen
  \bibfield  {author} {\bibinfo {author} {\bibfnamefont {S.}~\bibnamefont
  {Ryu}}, \bibinfo {author} {\bibfnamefont {J.~E.}\ \bibnamefont {Moore}}, \
  and\ \bibinfo {author} {\bibfnamefont {A.~W.~W.}\ \bibnamefont {Ludwig}},\
  }\href {\doibase 10.1103/PhysRevB.85.045104} {\bibfield  {journal} {\bibinfo
  {journal} {Phys. Rev. B}\ }\textbf {\bibinfo {volume} {85}},\ \bibinfo
  {pages} {045104} (\bibinfo {year} {2012})}\BibitemShut {NoStop}%
\bibitem [{\citenamefont {Kapustin}(2014)}]{kapustin2014symmetry}%
  \BibitemOpen
  \bibfield  {author} {\bibinfo {author} {\bibfnamefont {A.}~\bibnamefont
  {Kapustin}},\ }\href@noop {} {\enquote {\bibinfo {title} {Symmetry protected
  topological phases, anomalies, and cobordisms: Beyond group cohomology},}\ }
  (\bibinfo {year} {2014}),\ \Eprint {http://arxiv.org/abs/1403.1467}
  {arXiv:1403.1467 [cond-mat.str-el]} \BibitemShut {NoStop}%
\bibitem [{\citenamefont {Witten}(2016)}]{Witten2016}%
  \BibitemOpen
  \bibfield  {author} {\bibinfo {author} {\bibfnamefont {E.}~\bibnamefont
  {Witten}},\ }\href {\doibase 10.1103/RevModPhys.88.035001} {\bibfield
  {journal} {\bibinfo  {journal} {Rev. Mod. Phys.}\ }\textbf {\bibinfo {volume}
  {88}},\ \bibinfo {pages} {035001} (\bibinfo {year} {2016})}\BibitemShut
  {NoStop}%
\bibitem [{\citenamefont {Hsieh}\ \emph {et~al.}(2016)\citenamefont {Hsieh},
  \citenamefont {Hal\'asz},\ and\ \citenamefont {Grover}}]{Hsieh2016}%
  \BibitemOpen
  \bibfield  {author} {\bibinfo {author} {\bibfnamefont {T.~H.}\ \bibnamefont
  {Hsieh}}, \bibinfo {author} {\bibfnamefont {G.~B.}\ \bibnamefont {Hal\'asz}},
  \ and\ \bibinfo {author} {\bibfnamefont {T.}~\bibnamefont {Grover}},\ }\href
  {\doibase 10.1103/PhysRevLett.117.166802} {\bibfield  {journal} {\bibinfo
  {journal} {Phys. Rev. Lett.}\ }\textbf {\bibinfo {volume} {117}},\ \bibinfo
  {pages} {166802} (\bibinfo {year} {2016})}\BibitemShut {NoStop}%
\bibitem [{\citenamefont {Han}\ \emph {et~al.}(2017)\citenamefont {Han},
  \citenamefont {Tiwari}, \citenamefont {Hsieh},\ and\ \citenamefont
  {Ryu}}]{Han2017}%
  \BibitemOpen
  \bibfield  {author} {\bibinfo {author} {\bibfnamefont {B.}~\bibnamefont
  {Han}}, \bibinfo {author} {\bibfnamefont {A.}~\bibnamefont {Tiwari}},
  \bibinfo {author} {\bibfnamefont {C.-T.}\ \bibnamefont {Hsieh}}, \ and\
  \bibinfo {author} {\bibfnamefont {S.}~\bibnamefont {Ryu}},\ }\href {\doibase
  10.1103/PhysRevB.96.125105} {\bibfield  {journal} {\bibinfo  {journal} {Phys.
  Rev. B}\ }\textbf {\bibinfo {volume} {96}},\ \bibinfo {pages} {125105}
  (\bibinfo {year} {2017})}\BibitemShut {NoStop}%
\bibitem [{\citenamefont {Cheng}(2019)}]{Cheng2019}%
  \BibitemOpen
  \bibfield  {author} {\bibinfo {author} {\bibfnamefont {M.}~\bibnamefont
  {Cheng}},\ }\href {\doibase 10.1103/PhysRevB.99.075143} {\bibfield  {journal}
  {\bibinfo  {journal} {Phys. Rev. B}\ }\textbf {\bibinfo {volume} {99}},\
  \bibinfo {pages} {075143} (\bibinfo {year} {2019})}\BibitemShut {NoStop}%
\bibitem [{\citenamefont {Jian}\ \emph {et~al.}(2018)\citenamefont {Jian},
  \citenamefont {Bi},\ and\ \citenamefont {Xu}}]{Jian2018}%
  \BibitemOpen
  \bibfield  {author} {\bibinfo {author} {\bibfnamefont {C.-M.}\ \bibnamefont
  {Jian}}, \bibinfo {author} {\bibfnamefont {Z.}~\bibnamefont {Bi}}, \ and\
  \bibinfo {author} {\bibfnamefont {C.}~\bibnamefont {Xu}},\ }\href {\doibase
  10.1103/PhysRevB.97.054412} {\bibfield  {journal} {\bibinfo  {journal} {Phys.
  Rev. B}\ }\textbf {\bibinfo {volume} {97}},\ \bibinfo {pages} {054412}
  (\bibinfo {year} {2018})}\BibitemShut {NoStop}%
\bibitem [{\citenamefont {Lu}(2017)}]{Lu2017}%
  \BibitemOpen
  \bibfield  {author} {\bibinfo {author} {\bibfnamefont {Y.-M.}\ \bibnamefont
  {Lu}},\ }\href@noop {} {\enquote {\bibinfo {title} {Lieb-schultz-mattis
  theorems for symmetry protected topological phases},}\ } (\bibinfo {year}
  {2017}),\ \Eprint {http://arxiv.org/abs/1705.04691} {arXiv:1705.04691
  [cond-mat.str-el]} \BibitemShut {NoStop}%
\bibitem [{\citenamefont {Jiang}\ \emph {et~al.}(2021)\citenamefont {Jiang},
  \citenamefont {Cheng}, \citenamefont {Qi},\ and\ \citenamefont
  {Lu}}]{Jiang2021}%
  \BibitemOpen
  \bibfield  {author} {\bibinfo {author} {\bibfnamefont {S.}~\bibnamefont
  {Jiang}}, \bibinfo {author} {\bibfnamefont {M.}~\bibnamefont {Cheng}},
  \bibinfo {author} {\bibfnamefont {Y.}~\bibnamefont {Qi}}, \ and\ \bibinfo
  {author} {\bibfnamefont {Y.-M.}\ \bibnamefont {Lu}},\ }\href {\doibase
  10.21468/SciPostPhys.11.2.024} {\bibfield  {journal} {\bibinfo  {journal}
  {SciPost Phys.}\ }\textbf {\bibinfo {volume} {11}},\ \bibinfo {pages} {24}
  (\bibinfo {year} {2021})}\BibitemShut {NoStop}%
\bibitem [{Note1()}]{Note1}%
  \BibitemOpen
  \bibinfo {note} {One recognizes that Eq.\ (\ref {eq:def projective rep II b})
  is a generalization of Eq.\ (\ref {eq:def ZF2 c}) if one identifies the
  exponential of $\phi $ in Eq.\ (\ref {eq:def projective rep II b}) with
  $\gamma $ in Eq.\ (\ref {eq:def ZF2 c}) [up to the homomorphism (\ref {eq:
  def uj(g) b})].}\BibitemShut {Stop}%
\bibitem [{Note2()}]{Note2}%
  \BibitemOpen
  \bibinfo {note} {One recognizes that Eq.\ (\ref {eq:U(1) gauge equivalence
  b}) is a generalization of Eq.\ (\ref {eq:condition on kappa for group
  isomorphism}) if one identifies the exponential of $\xi $ in Eq.\ (\ref
  {eq:U(1) gauge equivalence b}) with $\kappa $ in Eq.\ (\ref {eq:condition on
  kappa for group isomorphism}) [up to the homomorphism (\ref {eq: def uj(g)
  b})].}\BibitemShut {Stop}%
\bibitem [{\citenamefont {Fidkowski}\ and\ \citenamefont
  {Kitaev}(2011)}]{Fidkowski2011}%
  \BibitemOpen
  \bibfield  {author} {\bibinfo {author} {\bibfnamefont {L.}~\bibnamefont
  {Fidkowski}}\ and\ \bibinfo {author} {\bibfnamefont {A.}~\bibnamefont
  {Kitaev}},\ }\href {\doibase 10.1103/PhysRevB.83.075103} {\bibfield
  {journal} {\bibinfo  {journal} {Phys. Rev. B}\ }\textbf {\bibinfo {volume}
  {83}},\ \bibinfo {pages} {075103} (\bibinfo {year} {2011})}\BibitemShut
  {NoStop}%
\bibitem [{\citenamefont {Bultinck}\ \emph {et~al.}(2017)\citenamefont
  {Bultinck}, \citenamefont {Williamson}, \citenamefont {Haegeman},\ and\
  \citenamefont {Verstraete}}]{Bultinck2017}%
  \BibitemOpen
  \bibfield  {author} {\bibinfo {author} {\bibfnamefont {N.}~\bibnamefont
  {Bultinck}}, \bibinfo {author} {\bibfnamefont {D.~J.}\ \bibnamefont
  {Williamson}}, \bibinfo {author} {\bibfnamefont {J.}~\bibnamefont
  {Haegeman}}, \ and\ \bibinfo {author} {\bibfnamefont {F.}~\bibnamefont
  {Verstraete}},\ }\href {\doibase 10.1103/PhysRevB.95.075108} {\bibfield
  {journal} {\bibinfo  {journal} {Phys. Rev. B}\ }\textbf {\bibinfo {volume}
  {95}},\ \bibinfo {pages} {075108} (\bibinfo {year} {2017})}\BibitemShut
  {NoStop}%
\bibitem [{\citenamefont {Williamson}\ \emph {et~al.}(2016)\citenamefont
  {Williamson}, \citenamefont {Bultinck}, \citenamefont {Haegeman},\ and\
  \citenamefont {Verstraete}}]{Williamson2016}%
  \BibitemOpen
  \bibfield  {author} {\bibinfo {author} {\bibfnamefont {D.~J.}\ \bibnamefont
  {Williamson}}, \bibinfo {author} {\bibfnamefont {N.}~\bibnamefont
  {Bultinck}}, \bibinfo {author} {\bibfnamefont {J.}~\bibnamefont {Haegeman}},
  \ and\ \bibinfo {author} {\bibfnamefont {F.}~\bibnamefont {Verstraete}},\
  }\href {https://arxiv.org/abs/1609.02897} {\enquote {\bibinfo {title}
  {Fermionic matrix product operators and topological phases of matter},}\ }
  (\bibinfo {year} {2016}),\ \Eprint {http://arxiv.org/abs/1609.02897}
  {arXiv:1609.02897 [cond-mat.str-el]} \BibitemShut {NoStop}%
\bibitem [{\citenamefont {Turzillo}\ and\ \citenamefont
  {You}(2019)}]{Turzillo2019}%
  \BibitemOpen
  \bibfield  {author} {\bibinfo {author} {\bibfnamefont {A.}~\bibnamefont
  {Turzillo}}\ and\ \bibinfo {author} {\bibfnamefont {M.}~\bibnamefont {You}},\
  }\href {\doibase 10.1103/PhysRevB.99.035103} {\bibfield  {journal} {\bibinfo
  {journal} {Phys. Rev. B}\ }\textbf {\bibinfo {volume} {99}},\ \bibinfo
  {pages} {035103} (\bibinfo {year} {2019})}\BibitemShut {NoStop}%
\bibitem [{\citenamefont {Wang}\ and\ \citenamefont {Gu}(2020)}]{Wang2020}%
  \BibitemOpen
  \bibfield  {author} {\bibinfo {author} {\bibfnamefont {Q.-R.}\ \bibnamefont
  {Wang}}\ and\ \bibinfo {author} {\bibfnamefont {Z.-C.}\ \bibnamefont {Gu}},\
  }\href {\doibase 10.1103/PhysRevX.10.031055} {\bibfield  {journal} {\bibinfo
  {journal} {Phys. Rev. X}\ }\textbf {\bibinfo {volume} {10}},\ \bibinfo
  {pages} {031055} (\bibinfo {year} {2020})}\BibitemShut {NoStop}%
\bibitem [{Note3()}]{Note3}%
  \BibitemOpen
  \bibinfo {note} {To account for the fermionic statistics, instead of the
  standard tensor product one must use a $\protect \mathbb {Z}^{\protect
  \tmspace +\thinmuskip {.1667em}}_{2}$ graded one. Fermionic Fock spaces then
  carry the structure of a $\protect \mathbb {Z}^{\protect \tmspace
  +\thinmuskip {.1667em}}_{2}$ graded vector space, also called a supervector
  space. See Appendix \ref {appsec:Construction of fermionic matrix product
  states (FMPS)} for more details on this construction.}\BibitemShut {Stop}%
\bibitem [{\citenamefont {Kapustin}\ \emph {et~al.}(2018)\citenamefont
  {Kapustin}, \citenamefont {Turzillo},\ and\ \citenamefont
  {You}}]{Kapustin2018}%
  \BibitemOpen
  \bibfield  {author} {\bibinfo {author} {\bibfnamefont {A.}~\bibnamefont
  {Kapustin}}, \bibinfo {author} {\bibfnamefont {A.}~\bibnamefont {Turzillo}},
  \ and\ \bibinfo {author} {\bibfnamefont {M.}~\bibnamefont {You}},\ }\href
  {\doibase 10.1103/PhysRevB.98.125101} {\bibfield  {journal} {\bibinfo
  {journal} {Phys. Rev. B}\ }\textbf {\bibinfo {volume} {98}},\ \bibinfo
  {pages} {125101} (\bibinfo {year} {2018})}\BibitemShut {NoStop}%
\bibitem [{Note4()}]{Note4}%
  \BibitemOpen
  \bibinfo {note} {The set $\protect \mathrm {Mat}(2M,\protect \mathbb {C})$ of
  all $2M\times 2M$ matrices is a $8M^{2}$-dimensional vector space over the
  real numbers.}\BibitemShut {Stop}%
\bibitem [{Note5()}]{Note5}%
  \BibitemOpen
  \bibinfo {note} {If $\protect \mathcal {D}=2$, $M=1$, $A^{(0)}_{1}$ is
  $\protect \mathrm {i}$ times the second Pauli matrix, and $A^{(0)}_{2}$ is
  the third Pauli matrix, then $\ell ^{\star }=4$.}\BibitemShut {Stop}%
\bibitem [{Note6()}]{Note6}%
  \BibitemOpen
  \bibinfo {note} {The basis (\ref {eq:overcomplete basis of Mat(2M,C) if
  injective a}) is in general overcomplete owing to the condition $\protect
  \mathcal {D}^{\ell ^{\star }}\geq 4M^{2}$.}\BibitemShut {Stop}%
\bibitem [{Note7()}]{Note7}%
  \BibitemOpen
  \bibinfo {note} {This anticommutation relation implies nontrivial index
  $[\rho ]$ which is defined in Sec.\ \ref {subsec:Indices}.}\BibitemShut
  {Stop}%
\bibitem [{\citenamefont {Schuch}\ \emph {et~al.}(2011)\citenamefont {Schuch},
  \citenamefont {P\'erez-Garc\'{\i}a},\ and\ \citenamefont
  {Cirac}}]{Schuch2011}%
  \BibitemOpen
  \bibfield  {author} {\bibinfo {author} {\bibfnamefont {N.}~\bibnamefont
  {Schuch}}, \bibinfo {author} {\bibfnamefont {D.}~\bibnamefont
  {P\'erez-Garc\'{\i}a}}, \ and\ \bibinfo {author} {\bibfnamefont
  {I.}~\bibnamefont {Cirac}},\ }\href {\doibase 10.1103/PhysRevB.84.165139}
  {\bibfield  {journal} {\bibinfo  {journal} {Phys. Rev. B}\ }\textbf {\bibinfo
  {volume} {84}},\ \bibinfo {pages} {165139} (\bibinfo {year}
  {2011})}\BibitemShut {NoStop}%
\bibitem [{\citenamefont {Bourne}\ and\ \citenamefont
  {Ogata}(2021)}]{Bourne2021}%
  \BibitemOpen
  \bibfield  {author} {\bibinfo {author} {\bibfnamefont {C.}~\bibnamefont
  {Bourne}}\ and\ \bibinfo {author} {\bibfnamefont {Y.}~\bibnamefont {Ogata}},\
  }\href {\doibase 10.1017/fms.2021.19} {\bibfield  {journal} {\bibinfo
  {journal} {Forum of Mathematics, Sigma}\ }\textbf {\bibinfo {volume} {9}},\
  \bibinfo {pages} {e25} (\bibinfo {year} {2021})}\BibitemShut {NoStop}%
\bibitem [{\citenamefont {Yao}\ and\ \citenamefont {Oshikawa}(2021)}]{Yao2021}%
  \BibitemOpen
  \bibfield  {author} {\bibinfo {author} {\bibfnamefont {Y.}~\bibnamefont
  {Yao}}\ and\ \bibinfo {author} {\bibfnamefont {M.}~\bibnamefont {Oshikawa}},\
  }\href {\doibase 10.1103/PhysRevLett.126.217201} {\bibfield  {journal}
  {\bibinfo  {journal} {Phys. Rev. Lett.}\ }\textbf {\bibinfo {volume} {126}},\
  \bibinfo {pages} {217201} (\bibinfo {year} {2021})}\BibitemShut {NoStop}%
\bibitem [{\citenamefont {Yao}\ and\ \citenamefont
  {Oshikawa}(2020)}]{Yao2020a}%
  \BibitemOpen
  \bibfield  {author} {\bibinfo {author} {\bibfnamefont {Y.}~\bibnamefont
  {Yao}}\ and\ \bibinfo {author} {\bibfnamefont {M.}~\bibnamefont {Oshikawa}},\
  }\href {\doibase 10.1103/PhysRevX.10.031008} {\bibfield  {journal} {\bibinfo
  {journal} {Phys. Rev. X}\ }\textbf {\bibinfo {volume} {10}},\ \bibinfo
  {pages} {031008} (\bibinfo {year} {2020})}\BibitemShut {NoStop}%
\bibitem [{\citenamefont {Hastings}\ and\ \citenamefont
  {Koma}(2006)}]{Hastings2006}%
  \BibitemOpen
  \bibfield  {author} {\bibinfo {author} {\bibfnamefont {M.~B.}\ \bibnamefont
  {Hastings}}\ and\ \bibinfo {author} {\bibfnamefont {T.}~\bibnamefont
  {Koma}},\ }\href {\doibase 10.1007/s00220-006-0030-4} {\bibfield  {journal}
  {\bibinfo  {journal} {Communications in Mathematical Physics}\ }\textbf
  {\bibinfo {volume} {265}},\ \bibinfo {pages} {781} (\bibinfo {year}
  {2006})}\BibitemShut {NoStop}%
\bibitem [{\citenamefont {Witten}(1982)}]{Witten1982}%
  \BibitemOpen
  \bibfield  {author} {\bibinfo {author} {\bibfnamefont {E.}~\bibnamefont
  {Witten}},\ }\href {\doibase https://doi.org/10.1016/0550-3213(82)90071-2}
  {\bibfield  {journal} {\bibinfo  {journal} {Nuclear Physics B}\ }\textbf
  {\bibinfo {volume} {202}},\ \bibinfo {pages} {253 } (\bibinfo {year}
  {1982})}\BibitemShut {NoStop}%
\bibitem [{Note8()}]{Note8}%
  \BibitemOpen
  \bibinfo {note} {We have chosen the convention of always representing the
  generator $p$ of $\protect \mathbb {Z}^{F}_{2}$ by a Hermitian operator
  according to Eq.\ (\ref {eq: def uj(g) e})}\BibitemShut {NoStop}%
\bibitem [{Note9()}]{Note9}%
  \BibitemOpen
  \bibinfo {note} {This is not so when $|\Lambda |=N=2M+1$ is odd.}\BibitemShut
  {Stop}%
\bibitem [{Note10()}]{Note10}%
  \BibitemOpen
  \bibinfo {note} {It is not possible to represent the group cohomology class
  $([\nu ],[\rho ],\mu )=(1,0,0)$ with a doublet of Majorana
  operators.}\BibitemShut {Stop}%
\bibitem [{\citenamefont {Fidkowski}\ and\ \citenamefont
  {Kitaev}(2010)}]{Fidkowski2010}%
  \BibitemOpen
  \bibfield  {author} {\bibinfo {author} {\bibfnamefont {L.}~\bibnamefont
  {Fidkowski}}\ and\ \bibinfo {author} {\bibfnamefont {A.}~\bibnamefont
  {Kitaev}},\ }\href {\doibase 10.1103/PhysRevB.81.134509} {\bibfield
  {journal} {\bibinfo  {journal} {Phys. Rev. B}\ }\textbf {\bibinfo {volume}
  {81}},\ \bibinfo {pages} {134509} (\bibinfo {year} {2010})}\BibitemShut
  {NoStop}%
\bibitem [{Note11()}]{Note11}%
  \BibitemOpen
  \bibinfo {note} {We have chosen the convention of always representing the
  generator $p$ of $\protect \mathbb {Z}^{F}_{2}$ by a Hermitian operator
  according to Eq.\ (\ref {eq: def uj(g) e})}\BibitemShut {NoStop}%
\bibitem [{Note12()}]{Note12}%
  \BibitemOpen
  \bibinfo {note} {It is not possible to represent the cohomology class $([\nu
  ],[\rho ],\mu )=\leavevmode@ifvmode {\setbox \z@ \hbox {\mathsurround \z@
  $\nulldelimiterspace \z@ \left (\vcenter to\@ne \big@size {}\right .$}\box
  \z@ }1,(1,0),0\leavevmode@ifvmode {\setbox \z@ \hbox {\mathsurround \z@
  $\nulldelimiterspace \z@ \left )\vcenter to\@ne \big@size {}\right .$}\box
  \z@ }$ with a doublet of Majorana operators.}\BibitemShut {Stop}%
\bibitem [{Note13()}]{Note13}%
  \BibitemOpen
  \bibinfo {note} {It is not possible to represent the cohomology class $([\nu
  ],[\rho ],\mu )=\leavevmode@ifvmode {\setbox \z@ \hbox {\mathsurround \z@
  $\nulldelimiterspace \z@ \left (\vcenter to\@ne \big@size {}\right .$}\box
  \z@ }1,(0,1),0\leavevmode@ifvmode {\setbox \z@ \hbox {\mathsurround \z@
  $\nulldelimiterspace \z@ \left )\vcenter to\@ne \big@size {}\right .$}\box
  \z@ }$ with a doublet of Majorana operators.}\BibitemShut {Stop}%
\bibitem [{Note14()}]{Note14}%
  \BibitemOpen
  \bibinfo {note} {It is not possible to represent the cohomology class $([\nu
  ],[\rho ],\mu )=\leavevmode@ifvmode {\setbox \z@ \hbox {\mathsurround \z@
  $\nulldelimiterspace \z@ \left (\vcenter to\@ne \big@size {}\right .$}\box
  \z@ }1,(0,0),0\leavevmode@ifvmode {\setbox \z@ \hbox {\mathsurround \z@
  $\nulldelimiterspace \z@ \left )\vcenter to\@ne \big@size {}\right .$}\box
  \z@ }$ with a doublet of Majorana operators.}\BibitemShut {Stop}%
\bibitem [{Note15()}]{Note15}%
  \BibitemOpen
  \bibinfo {note} {We have chosen the convention of always representing the
  generator $p$ of $\protect \mathbb {Z}^{F}_{2}$ by a Hermitian operator
  according to Eq.\ (\ref {eq: def uj(g) e})}\BibitemShut {NoStop}%
\bibitem [{Note16()}]{Note16}%
  \BibitemOpen
  \bibinfo {note} {Bosonic matrix products states presume that the local Fock
  space $\protect \mathcal {F}^{\protect \tmspace +\thinmuskip
  {.1667em}}_{\protect \bm {j}}$ has no more than the trivial $\protect \mathbb
  {Z}^{\protect \tmspace +\thinmuskip {.1667em}}_{2}$ grading, i.e., $\protect
  \mathcal {F}^{\protect \tmspace +\thinmuskip {.1667em}}_{\protect \bm
  {j}}\equiv \protect \mathcal {F}^{\protect \tmspace +\thinmuskip
  {.1667em}}_{\protect \bm {j}\protect \tmspace +\thinmuskip {.1667em}0}\oplus
  \protect \mathcal {F}^{\protect \tmspace +\thinmuskip {.1667em}}_{\protect
  \bm {j}\protect \tmspace +\thinmuskip {.1667em}1}$ with $\protect \mathcal
  {F}^{\protect \tmspace +\thinmuskip {.1667em}}_{\protect \bm {j}\protect
  \tmspace +\thinmuskip {.1667em}0}\equiv \protect \mathcal {F}^{\protect
  \tmspace +\thinmuskip {.1667em}}_{\protect \bm {j}}$ and $\protect \mathcal
  {F}^{\protect \tmspace +\thinmuskip {.1667em}}_{\protect \bm {j}\protect
  \tmspace +\thinmuskip {.1667em}1}\equiv \emptyset $.}\BibitemShut {Stop}%
\end{thebibliography}%
\end{document}